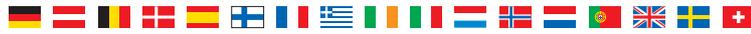

# PLANCK

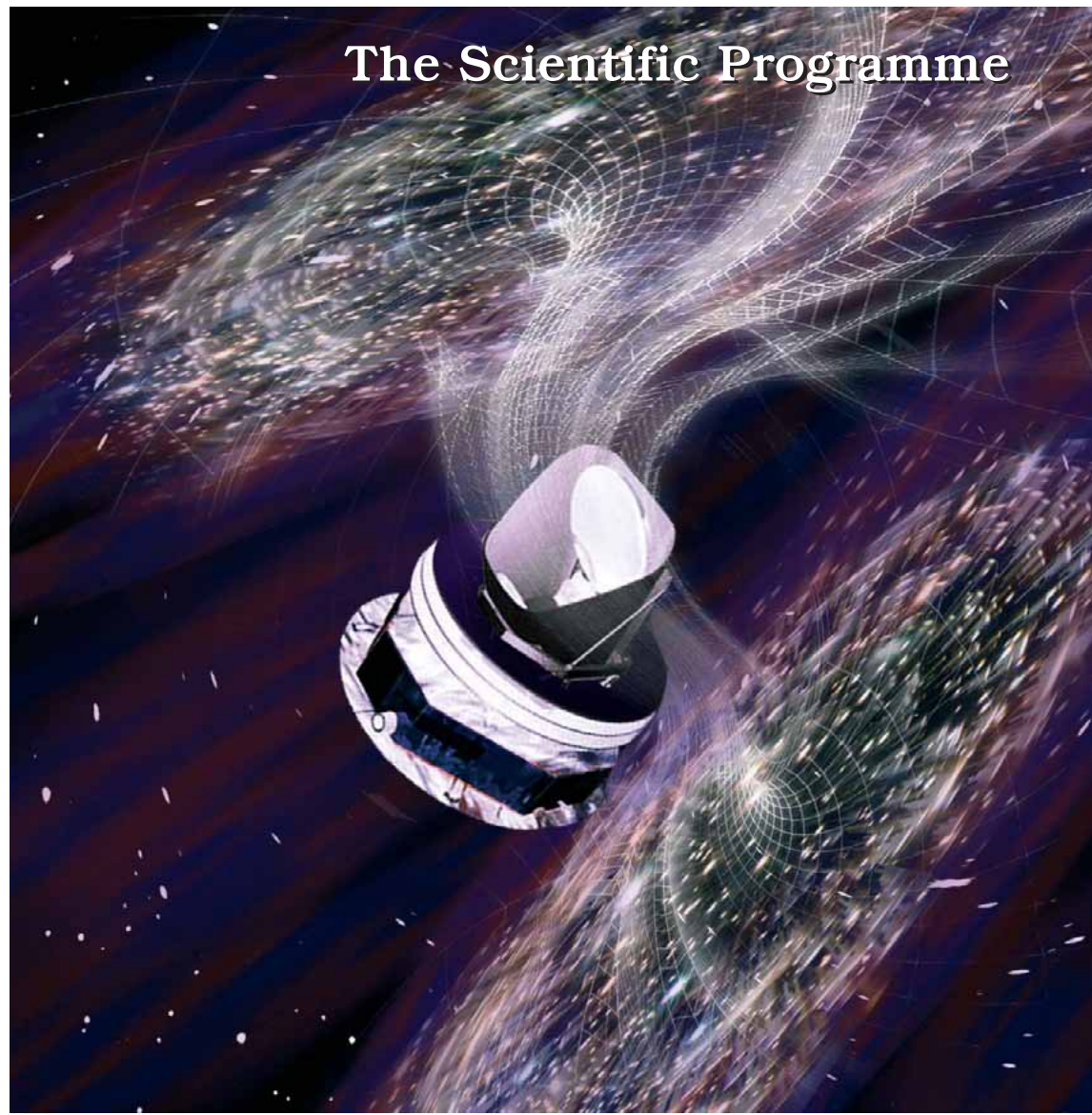

The Scientific Programme

**European Space Agency**
**Agence spatiale européenne**

# FOREWORD

Cosmology, the science of the origin, structure, and space-time relationships of the Universe, has been transformed in the last 15 years by measurements that sometimes confirmed but often challenged cherished ideas. To cite a few examples:

- Broad and deep surveys show the evolution of the large-scale distribution of matter from the distant past to the present.

- Observations of the Cosmic Microwave Background confirm that the physical processes at early times are relatively simple and well understood, and that inflation, or something very like it, occurred in the past.

- The abundances of light elements are close to theoretical models of early Universe nucleosynthesis.

- Distant supernovae show that the expansion of the Universe is accelerating.

When all available data are combined, a self-consistent picture of the Universe emerges that is radically different from the view 15 years ago: the Universe is flat, $\sim$13.7 billion years old, and contains $\sim$4% ordinary matter (only part of which is luminous), $\sim$23% 'dark' matter (of unknown type), and $\sim$73% 'dark energy' (responsible for the accelerating expansion, but also of unknown origin).

For 40 years, the cosmic microwave background (CMB) has been the most important source of information about the geometry and contents of the Universe. Even so, only a small fraction of the information available in the CMB has been extracted to date. Planck, the third space CMB mission after *COBE* and *WMAP*, is designed to extract essentially all of the information in the CMB temperature anisotropies. Planck will also measure to high accuracy the polarization of CMB anisotropies, which encodes not only a wealth of cosmological information but also provides a unique probe of the thermal history of the Universe during the time when the first stars and galaxies formed. Polarization measurements may also detect the signature of a stochastic background of gravitational waves generated during inflation, $10^{-35}$ s after the Big Bang.

This book describes the expected scientific output of the Planck mission, both cosmological and non-cosmological. Chapter 1 summarizes the experimental concept and the operation of the satellite. Chapter 2 covers the core cosmological science of the mission, describing the measurements that Planck will make, what we expect to learn from them about the geometry and contents of the Universe and about fundamental physics, and the combination of CMB data with other data to provide additional insights.

Although the primary goal of Planck is cosmology, it will survey the whole sky with an unprecedented combination of frequency coverage, angular resolution, and sensitivity, providing data valuable for a broad range of astrophysics. Chapters 3, 4, and 5 describe non-cosmological astrophysical uses of the Planck data. Chapter 3 addresses distortions of the CMB (often called "secondary anisotropies") caused by large structures of matter formed long after the CMB decoupled from the primeval plasma . They include the effects of hot gas in galaxy clusters (the Sunyaev-Zel'dovich effect), and the effects of gravitational lensing by matter concentrations along the path of CMB photons. These effects trace the formation of structure in the Universe, and therefore offer an additional and independent source of information on the early Universe.

Chapter 4 describes discrete sources of extragalactic origin whose thermal and non-thermal emission will be detected by Planck, and which can be studied also by other observatories including Herschel. Chapter 5 discusses Galactic and Solar System science. The gas and dust of the Milky Way, out of which stars are made, and which make up about 10% of the mass of the Galaxy, radiate copiously at the frequencies measured by Planck. Cloud structure and stability, star formation, and issues such as the detailed shape of the Milky Way's magnetic field, can all be addressed. Planck will also provide a new view of cool objects in our Solar System.

The guaranteed scientific return from Planck is indeed spectacular. With such a great increase in capability over previous CMB missions, however, we can also anticipate completely new science as well. As will be seen from this book, the scientific case for Planck is even stronger now than it was when the mission was first proposed a decade ago.

# Acknowledgements

This book is the result of a large effort by the Planck community (which includes a large fraction of Europe's far-infrared/submillimeter and CMB communities as well as a large number of CMB researchers from the US) to update the scientific case for the Planck mission, taking into account the spectacular achievements of the decade since the case was first made.

Planck (http://www.rssd.esa.int/Planck) is a project of the European Space Agency with instruments funded by ESA member states (in particular the Principal Investigator countries: France and Italy), and with special contributions from Denmark and the United States.

The general coordination and scientific-editorial effort of this book is due to G. Efstathiou (Survey Scientist for the HFI), C. Lawrence (Survey Scientist for the LFI), and J. Tauber (Project Scientist). The chapters were assembled from contributions by the whole Planck community, and edited by:

Chapter 1: J. Tauber, M. Bersanelli, J.-M. Lamarre

Chapter 2: G. Efstathiou, C. Lawrence, F. Bouchet, E. Martinez-González, S. Matarrese, D. Scott, M. White

Chapter 3: N. Aghanim, M. Bartelmann

Chapter 4: B. Partridge, M. Rowan-Robinson, C. Lawrence, J.-L. Puget

Chapter 5: M. Giard and R. Davis

The members of the Planck Science Team (at the time of writing M. Bersanelli, G. Efstathiou, J.-M. Lamarre, C. Lawrence, N. Mandolesi, H.-U. Norgaard-Nielsen, F. Pasian, J.-L. Puget, J.-F. Sygnet, J. Tauber) supervised the effort. The development and operation of Planck depend on the dedicated efforts of many hundreds of scientists, engineers, computer scientists, and administrators. It is not possible to name them all, but we list here some of the main entities involved:

The European Space Agency (ESA)

Alcatel Space (Cannes), ESA's prime contractor for Planck

The Istituto di Astrofisica Spaziale e Fisica Cosmica—sezione di Bologna (leading Institute for the LFI), funded mainly by the Agenzia Spaziale Italiana (ASI)

The Institut d'Astrophysique Spatiale (leading Institute for the HFI), funded mainly by the Centre National des Études Spatiales (CNES) and Centre National de la Recherche Scientifique (CNRS).

The Danish Space Research Institute (leading Institute for the Planck reflectors).

The Jet Propulsion Laboratory, funded by the United States National Aeronautics and Space Administration (NASA).

For a complete list of the more than 40 participating scientific institutes, the reader is referred to the Home Page of the Science Team of Planck (http://www.rssd.esa.int/Planck).

# TABLE OF CONTENTS

# CHAPTER 1
# THE PLANCK MISSION

## 1.1 OVERVIEW

*Planck* is a mission of the European Space Agency designed to answer key cosmological questions. Its ultimate goal is to determine the geometry and contents of the Universe, and which theories describing the birth and evolution of the Universe are correct. To achieve this ambitious objective, it will observe the cosmic microwave background radiation (CMB), emitted about 13 billion years ago, just 400,000 years after the Big Bang. Today the CMB permeates the Universe and is observed to have an average temperature of 2.726 K. Small anisotropies or deviations from this average value, observable at angular scales larger than a few arcminutes, encode a wealth of information on the properties of the Universe in its infancy. The objective of *Planck* is to measure these properties with unprecedented accuracy and level of detail.

As with all ESA scientific missions, *Planck* is being developed in a partnership with the European scientific community. Two Consortia of scientific institutes, each led by a Principal Investigator, will deliver to ESA instruments designed specifically for *Planck*. Together, these instruments will measure the CMB signal and distinguish it from other confusing sources. The instruments sit at the focus of a telescope whose mirrors are being developed in a collaborative effort between ESA and a Danish Consortium of institutes.

ESA manages the project, develops and procures the spacecraft, integrates the instruments into the spacecraft, and will launch and operate it. It is currently planned to launch *Planck* on an Ariane 5 rocket in August of 2007, together with the *Herschel Space Observatory* (see *http://www.rssd.esa.int/Herschel*). After launch, they will both be placed into orbits around a point located ∼1.5 million km from the Earth. From that far vantage point, *Planck* will sweep the sky in large swaths, and eventually cover it fully at least twice.

Each of the two instrument Consortia will operate their respective instrument and process all the data into usable scientific products. At the end of the mission (approximately in December of 2010), the Consortia will deliver the final products to ESA, who will archive them and distribute them to the wide community. Up to that time, the three scientific institutes will have exclusive access to the data for scientific exploitation.

This chapter describes the current design of the spacecraft and instruments, and outlines satellite and instrument operations, data processing, and scientific analysis. More detailed information about all aspects of *Planck* is available at *http://www.rssd.esa.int/Planck* and *http://sci.esa.int*.

## 1.2 INTRODUCTION

In 1992, the COBE team announced the detection of intrinsic temperature fluctuations in the CMB on angular scales larger than ∼7°, at a level $\Delta T/T \sim 10^{-5}$. In February 2003, the *WMAP* team announced results on scales of about 15′ with similar sensitivity. These results strongly support inflationary Big Bang models of the origin and evolution of the Universe. However, as we will show in the next chapter, many fundamental cosmological questions remain open. In particular, CMB measurements with high angular resolution and sensitivity are required to determine the initial conditions for structure evolution, the origin of primordial fluctuations, the existence of topological defects, and the nature and amount of dark matter.

In 1992, two space-based CMB experiments (COBRAS and SAMBA) were proposed to ESA. In 1996, following assessment studies, ESA selected a combined mission called CO-BRAS/SAMBA as the third Medium-Sized Mission (M3) of the Horizon 2000 Scientific Pro-gramme. COBRAS/SAMBA was subsequently renamed in honor of the German scientist Max Planck (1858-1947). Today, *Planck* forms part of ESA's "Cosmic Visions 2020" programme.

The main objective of the *Planck* mission is to measure the fluctuations of the CMB with an accuracy set by fundamental astrophysical limits. To do this, *Planck* will image the the whole sky with an unprecedented combination of sensitivity ($\Delta T/T \sim 2 \times 10^{-6}$), angular resolution (to 5′), and frequency coverage (30–857 GHz). This level of performance will enable *Planck* to measure the angular power spectrum of the CMB fluctuations to high accuracy and will allow the determination of fundamental cosmological parameters such as the cold dark matter and baryon densities with an uncertainty of order one percent or better. *Planck* will set constraints on fundamental physics at energies larger than $10^{15}$ GeV, which cannot be reached by any conceivable experiment on Earth. In addition, the *Planck* sky surveys will produce a wealth of information on the properties of extragalactic sources and on the dust and gas in our own galaxy. One specific notable result will be the measurement of the Sunyaev-Zeldovich effect in many thousands of galaxy clusters.

ESA released an Announcement of Opportunity for *Planck* instruments in October 1997. Two proposals were received: the Low Frequency Instrument (LFI, an array of receivers based on HEMT amplifiers, covering the frequency range 30–100 GHz, and operated at a temperature of 20 K, led by N. Mandolesi (INAF, Bologna); and the High Frequency Instrument (HFI, an array of receivers based on bolometers covering the frequency range 100–857 GHz, and operated at a temperature of 0.1 K), led by J. L. Puget (IAS, Orsay). After detailed review by independent scientists, both proposals were accepted by ESA in February 1999. The telescope mirrors will be provided by ESA and a Consortium of Danish institutes (referred to as DK-Planck) led by H. U. Norgaard-Nielsen (DSRI, Copenhagen).

Each of the three Consortia (LFI, HFI, and DK-Planck) has the responsibility to design, procure, and deliver to ESA a specific set of hardware. The LFI and HFI Consortia have the additional responsibilities to operate their respective instrument, and to set up Data Processing Centres to process all the scientific data into usable scientific products. ESA, as overall Project Manager, is responsible for the spacecraft, integration of instruments and spacecraft, launch, mission operations, and distribution of the scientific products to the larger scientific community. *Planck* will be launched together with ESA's Herschel Space Observatory (*http://astro.estec.esa.nl/Herschel*). *Planck* and *Herschel* will fly in different orbits around the second Lagrangian point of the Earth-Sun System.

ESA issued an Invitation to Tender in September 2000 for proposals for the design, construction, and launch of the *Herschel* and *Planck* spacecraft. Alcatel Space (Cannes, France), was selected as the Prime Contractor for the *Herschel/Planck* Project. In addition to overall management of the industrial component of *Herschel/Planck*, Alcatel is responsible for the *Planck* Payload Module. Alenia Spazio (Torino, Italy) is the main subcontractor for the procurement of the Service Module. Subsystem contractors are selected jointly by ESA and Alcatel. The launch of *Herschel* and *Planck* is currently scheduled for August of 2007.

## 1.3 PAYLOAD

Figures 1.1 and 1.2 show the main components of *Planck*: an off-axis telescope with a projected diameter of 1.5 m; a telescope baffle that simultaneously provides straylight shielding and radiative cooling; two state-of-the-art cryogenic instruments, operating at 20 K and 0.1 K; three conical "V-groove" baffles that provide thermal isolation between the warm spacecraft and the cold telescope and instruments; the service module, referred to variously as the spacecraft bus, the spacecraft, or simply the S/C; and the solar panel. In flight, the solar panel faces the Sun; everything else is always in shadow.

The cryogenic temperatures required by the instruments are achieved through a combination of passive radiative cooling and three active refrigerators. The telescope baffle and V-groove shields are key parts of the passive thermal system. The baffle (which also acts as a straylight shield) is a high-efficiency radiator of aluminium honeycomb painted black. The V-grooves are a set of three specular, conical shields with an angle of 5° between adjacent shields. This geometry provides highly efficient radiative coupling to cold space, and a high degree of thermal isolation between the warm spacecraft bus and the cold telescope, baffle, and instruments. The cooling provided by the passive system leads to a temperature of $\sim 50$ K for the telescope and baffle. The active cryocoolers further reduce the temperature to 20 K and 0.1 K for the instruments.

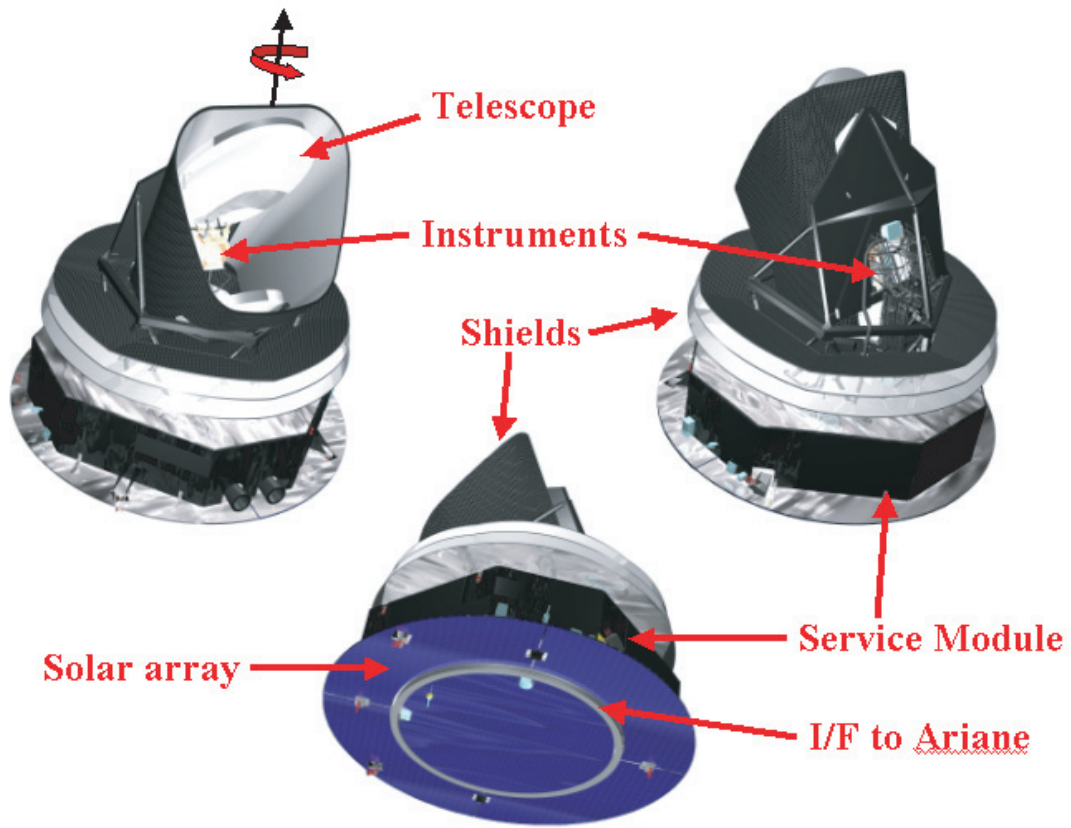

FIG 1.1.— Main elements of *Planck*. The instrument focal plane unit (barely visible) contains both LFI and HFI detectors. The function of the large baffle surrounding the telescope is to control the far sidelobe level of the radiation pattern as seen from the detectors. The specular conical shields (often called "V-grooves") thermally decouple the Service Module (which contains all warm elements of the satellite) from the Payload Module. The satellite spins around the indicated axis, such that the solar array is always exposed to the Sun, and shields the payload from solar radiation. Figures courtesy of Alcatel Space (Cannes).

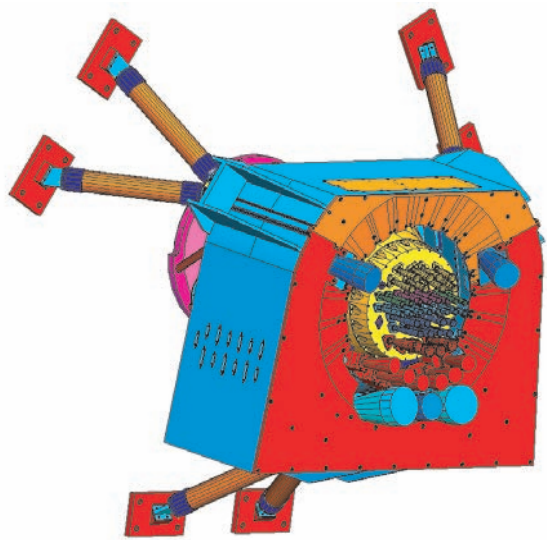

FIG 1.2.—*Planck* focal plane unit. The HFI is inserted into the ring formed by the LFI horns, and includes thermal stages at 18 K, 4 K, 2 K and 0.1 K. The cold LFI unit (20 K) is attached by bipods to the telescope structure.

The Low Frequency Instrument (LFI) covers 30–70 GHz in three bands*; the High Frequency Instrument (HFI) covers 100–857 GHz in six bands. The band centers are spaced approximately logarithmically. Performance parameters of the instruments are summarized in Table 1.1. The LFI horns are situated in a ring around the HFI. Each horn collects radiation from the telescope and feeds it to one or more detectors. As shown in Figure 1.3, there are nine frequency bands, with central frequencies varying from 30 to 857 GHz. The lowest three frequency channels are covered by the LFI, and the highest six by HFI.

TABLE 1.1

SUMMARY OF PLANCK INSTRUMENT CHARACTERISTICS

| INSTRUMENT CHARACTERISTIC | LFI | | | HFI | | | | | |
|---|---|---|---|---|---|---|---|---|---|
| Detector Technology . . . . . . . . . . . . . . . | HEMT arrays | | | Bolometer arrays | | | | | |
| Center Frequency [GHz] . . . . . . . . . . . | 30 | 44 | 70 | 100 | 143 | 217 | 353 | 545 | 857 |
| Bandwidth ($\Delta\nu/\nu$) . . . . . . . . . . . . . . . | 0.2 | 0.2 | 0.2 | 0.33 | 0.33 | 0.33 | 0.33 | 0.33 | 0.33 |
| Angular Resolution (arcmin) . . . . . . . . | 33 | 24 | 14 | 10 | 7.1 | 5.0 | 5.0 | 5.0 | 5.0 |
| $\Delta T/T$ per pixel (Stokes $I$)[a] . . . . . . . . | 2.0 | 2.7 | 4.7 | 2.5 | 2.2 | 4.8 | 14.7 | 147 | 6700 |
| $\Delta T/T$ per pixel (Stokes $Q$ &$U$)[a] . . . . . | 2.8 | 3.9 | 6.7 | 4.0 | 4.2 | 9.8 | 29.8 | . . . | . . . |

[a] Goal (in $\mu$K/K) for 14 months integration, 1$\sigma$, for square pixels whose sides are given in the row "Angular Resolution".

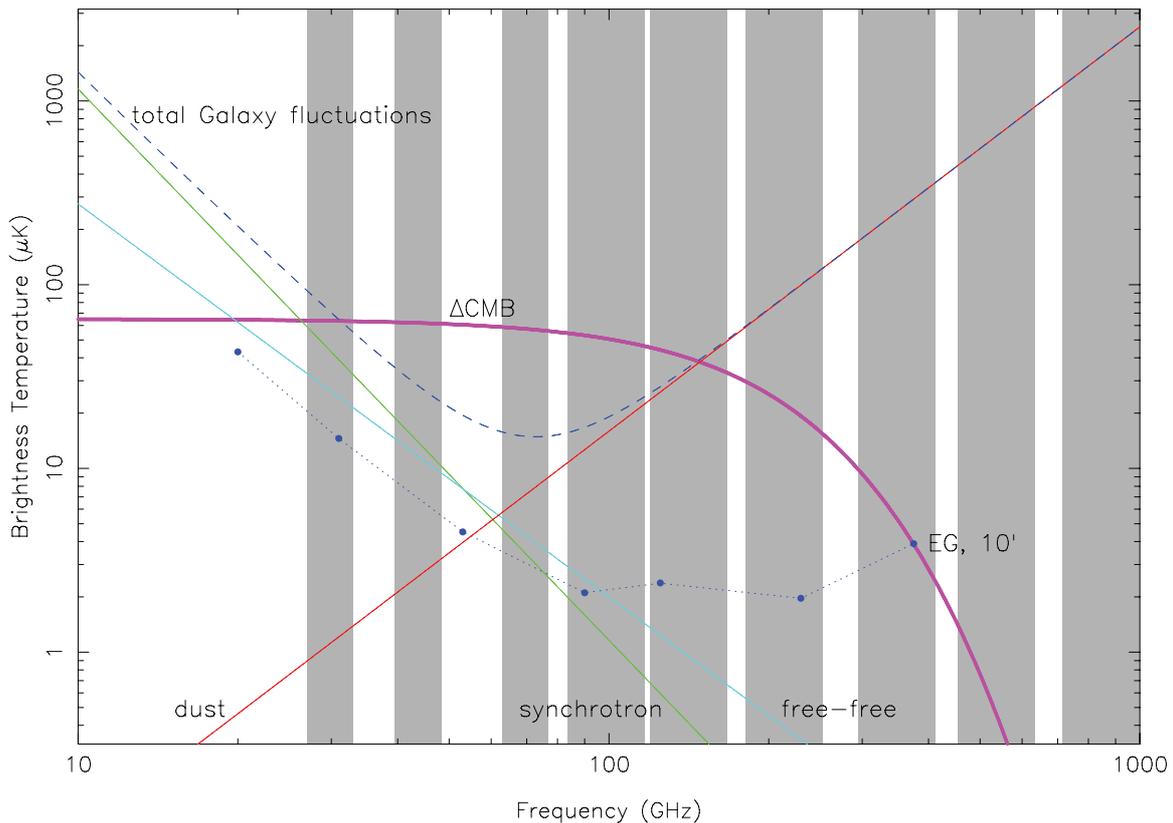

FIG 1.3.— Spectrum of the CMB, and the frequency coverage of the *Planck* channels. Also indicated are the spectra of other sources of fluctuations in the microwave sky. Dust, synchrotron, and free-free temperature fluctuation (i.e., unpolarized) levels correspond to the *WMAP* Kp2 levels (85% of the sky; Bennett et al. 2003). The CMB and Galactic fluctuation levels depend on angular scale, and are shown for ~1°. On small angular scales, extragalactic sources dominate. The minimum in diffuse foregrounds and the clearest window on CMB fluctuations occurs near 70 GHz. The highest HFI frequencies are primarily sensitive to dust.

While LFI and HFI alone have unprecedented capabilities, it is the combination of data from the two instruments that give *Planck* the imaging power, the redundancy, and the control of

*   The 100 GHz channel originally proposed was dropped in 2003 due to budget constraints.

systematic effects and foreground emissions needed to achieve the extraordinary scientific goals of the mission. In particular, the fact that systematic effects will in general produce different responses in the two instruments provides *Planck* with a powerful tool for the detection and mitigation of systematics, which will ensure that the final maps are limited only by the exquisite instrument sensitivity and unavoidable astrophysical foregrounds.

All of the LFI channels, and four of the HFI channels, can measure linear polarization as well as intensity. By combining the signals measured by several detectors, whose planes of polarisation are rotated with respect to each other in multiples of $45°$, the linear polarization of the incoming radiation can be determined fully. The horn location and orientation are chosen to optimise these measurements. Circular polarisation is not detectable by *Planck*, but is of minor interest given that the primordial CMB is not expected to be circularly polarised.

The principal observational objective of *Planck* is to produce maps of the whole sky in nine frequency channels. The telescope is an off-axis aplanatic design with a projected diameter $1.5\,m$ and an overall emissivity $\lesssim 1\%$. The FWHM of the telescope beams for the nine frequencies is given in Table 1.1.

The *Planck* maps will not only include the signal from the CMB, but also all other astrophysical foregrounds Galactic (free-free, synchrotron, and dust) and extragalactic (Figure 1.3). LFI covers well the frequency range where free-free and synchrotron emission from the Galaxy, and possibly spinning dust emission, dominate the foreground radiation. HFI covers well the frequencies where dust emission dominates the foreground radiation. Taken together, they provide the frequency coverage necessary for accurate separation of the CMB and foreground radiation. Instrumental systematic effects, as well as local uncertainties in the parameters characterizing the foregrounds, will degrade the final noise level in the maps.

The final scientific performance of the mission depends, therefore, not only on the instrumental behavior, but also on the detailed nature of astrophysical foregrounds and the behavior of many systematic effects which may produce spurious signals (such as straylight). Simulations suggest that the temperature sensitivity of *Planck* will be limited by astrophysical foregrounds rather than the instrumental performance. Our current knowledge of polarised foregrounds is not as good as that of unpolarised ones, but it is likely that foregrounds will be a significant, and potentially dominant, source of uncertainty for polarisation.

Many systematic effects can be controlled by an appropriate choice of orbit and sky scanning strategy. *Planck* will fly in a Lissajous orbit around the $L_2$ point of the Earth-Sun system (Figure 1.4). The spacecraft will spin at $\sim 1$ rpm around an axis offset by $\sim 85°$ from the telescope boresight, so that the observed sky patch will trace a large circle on the sky (Dupac and Tauber 2004). From $L_2$, the spin axis can be continuously pointed in the anti-Sun direction, and the satellite itself used to shield the payload from solar illumination. This strategy minimises potentially confusing signals due to thermal fluctuations and straylight entering the detectors through the far sidelobes. It also enables aggressive use of passive radiation to cool the payload, a key feature in the overall thermal design of *Planck*.

As the spin axis follows the Sun, the circle observed by the instruments sweeps through the sky at a rate of $1° \, day^{-1}$. The whole sky will be covered (by all feeds) in a little more than $6$ months; this operation will be repeated twice, resulting in a mission lifetime of around 15 months.

### 1.3.1 The Low Frequency Instrument (LFI)

The *Planck* Low Frequency Instrument represents the third generation microwave radiometer for space observations of CMB anisotropies, following the COBE Differential Microwave Radiometer (DMR) and the Wilkinson Microwave Anisotropy Probe (*WMAP*). The DMR, launched in 1989, detected structure in the CMB angular distribution at angular scales $\gtrsim 7°$ (Smoot et al. 1992), using two Dicke-switched radiometers at each of three frequencies, 31, 53, and 90 GHz, with noise temperatures of 250, 80, and 60 times the quantum limit, respectively, fed by pairs of feed horns pointed at the sky. *WMAP* (Bennett et al. 2003) was launched in June 2001 and is currently observing the whole sky in five frequency bands from 20 to 94 GHz with small arrays of radiatively-cooled radiometers fed by a differential two-telescope optical system. Radiometer noise temperatures are 15–25 times the quantum limit, with angular resolution ranging from $56'$ to $14'$. The LFI instrument (Bersanelli & Mandolesi 2000) with its large arrays of cryogenically cooled radiometers, represents another major advance in the state of the art. It is designed to produce images of the sky (including polarized components) at 30, 44, and 70 GHz with high sensitivity and freedom from systematic errors.

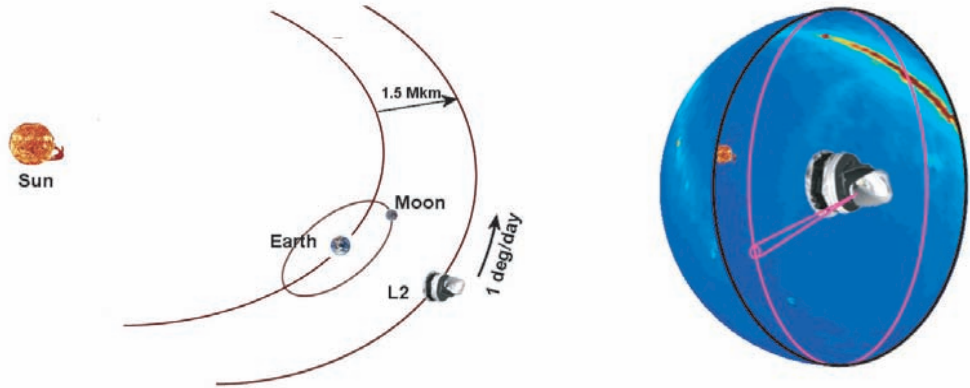

Fig 1.4.—*Planck* orbit at the 2nd Lagrangian point of the Earth-Sun system ($L_2$). The spin axis is pointed near the Sun, with the solar panel shading the payload, and the telescope sweeps the sky in large circles at 1 rpm.

The heart of the LFI instrument is a compact, 22-channel multifrequency array of differential receivers with ultra-low-noise amplifiers based on cryogenic indium phosphide (InP) high-electron-mobility transistors (HEMTs). To minimise power dissipation in the focal plane unit, which is cooled to 20 K, the radiometers are split into two subassemblies connected by a set of waveguides, as shown in Figure 1.5.

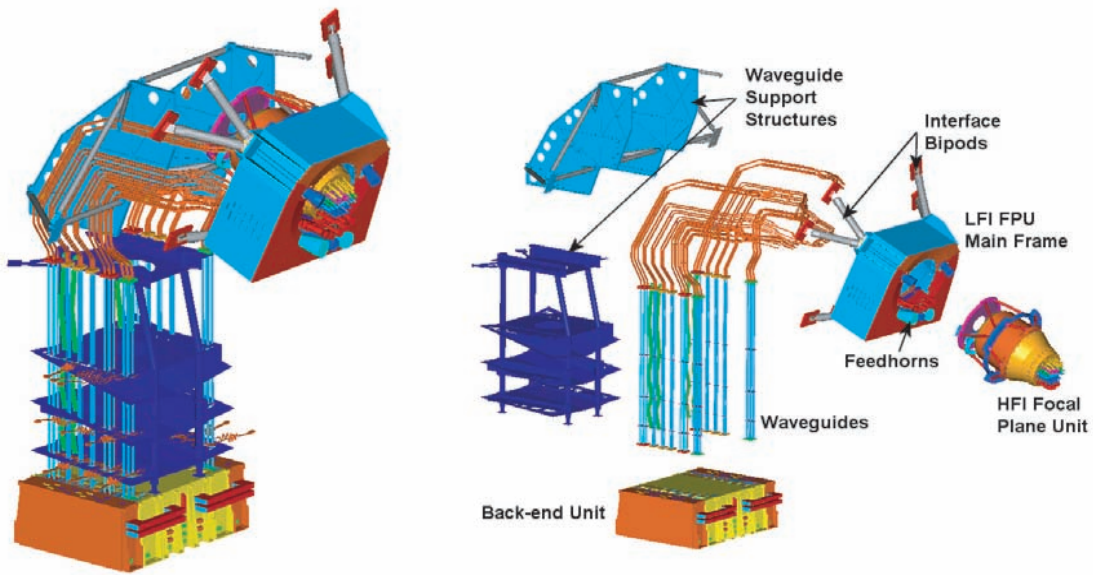

Fig 1.5.—The LFI radiometer array assembly (left), with details of the front-end and back-end units (right). The front-ends are based on wide-band low-noise amplifiers, fed by corrugated feedhorns which collect the radiation from the telescope. The waveguides transport the amplified signals from the front-end (at 20 K) to the back-end (at 300 K). They are designed to meet simultaneously radiometric, thermal, and mechanical requirements, and are thermally linked to the three V-groove thermal shields of the *Planck* payload module. The back-end unit, located on top of the *Planck* service module, contains additional amplification as well as the detectors, and is interfaced to the data acquisition electronics. The HFI is inserted into and attached to the frame of the LFI focal-plane unit.

The radiometer design is driven by the need to suppress $1/f$-type noise induced by gain and noise temperature fluctuations in the amplifiers, which would be unacceptably high for a simple total power system. A differential pseudo-correlation scheme is adopted, in which signals from the sky and from a blackbody reference load are combined by a hybrid coupler, amplified in two independent amplifier chains, and separated out by a second hybrid (Figure 1.6). The sky and the reference load power can then be measured and differenced. Since the reference signal has been subject to the same gain variations in the two amplifier chains as the sky signal, the true sky power can be recovered. Insensitivity to fluctuations in the back-end amplifiers and detectors is realized by switching phase shifters at ∼ 8 kHz synchronously in each amplifier chain. The rejection of $1/f$ noise as well as the immunity to other systematic effects is optimised if the two input signals are nearly equal. For this reason the reference loads are cooled to ∼ 4 K by mounting them on the 4-K structure of the HFI. In addition, the effect of the residual offset (< 2 K in nominal conditions) is reduced by introducing a *gain modulation factor* in the on-board processing to balance the output signal (Seiffert et al. 2002). As shown in Figure 1.6, the differencing receiver greatly improves the stability of the measured signal.

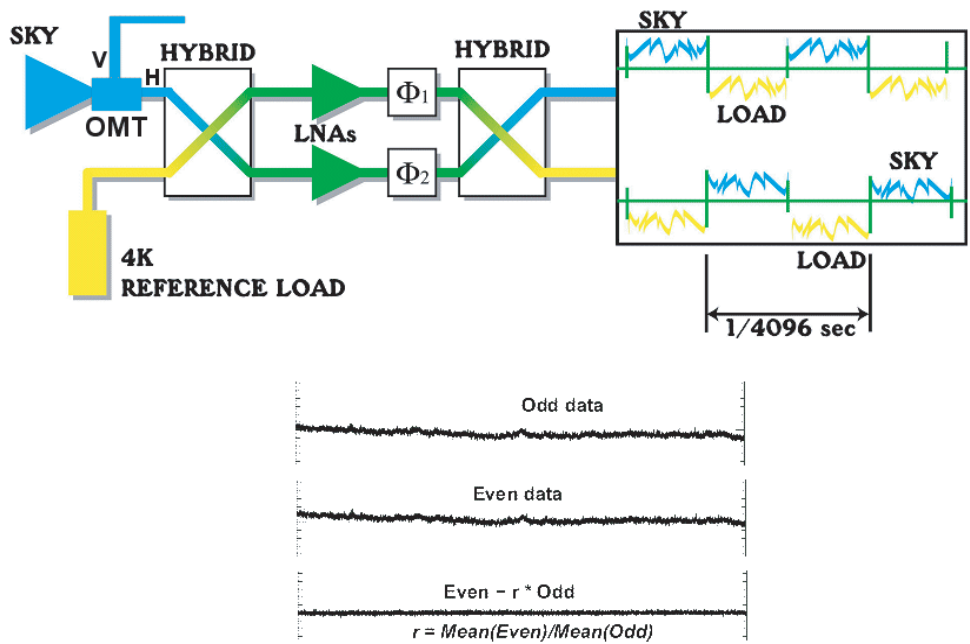

FIG 1.6.—(*Top*) Schematic of the LFI front-end radiometer. The front-end unit is located at the focus of the *Planck* telescope, and comprises: dual profiled corrugated feed horns (Villa et al. 2002); low-loss (∼ 0.2 dB), wideband (> 20%) orthomode transducers; and radiometer front-end modules with hybrids, cryogenic low noise amplifiers, and phase switches. (*Bottom*) Measured radiometer output of the Elegant Breadboard model at 30 GHz. Shown are the signals from the two detector diodes ("odd" and "even" samples), which correspond to the sky and reference load, in which the noise is dominated by a non-white, $1/f$-type component. The radiometer design, however, is such that the $1/f$ component is highly correlated in the two diodes, so that the difference signal is extremely stable and insensitive to $1/f$ fluctuations.

Major progress in the performance of cryogenic InP HEMTs has been achieved since the beginning of the LFI development. The LFI prototype radiometers establish world-record low noise performances in the 30–70 GHz range and meet or surpass the LFI requirements both for noise, bandwidth, and low power consumption.

The LFI amplifiers at 30 and 44 GHz use discrete InP HEMTs incorporated into a microwave integrated circuit (MIC). At these frequencies the parasitics and uncertainties introduced by the bond wires in a MIC amplifier are controllable and the additional tuning flexibility facilitates optimization for low noise. The LFI amplifiers have demonstrated noise temperatures < 7.5 K at 30 GHz with 20% bandwidth.

At 70 GHz there will be twelve detector chains. Amplifiers at these frequencies will use monolithic microwave integrated circuits (MMICs), which incorporate all circuit elements and the HEMT transistors on a single InP chip. At these frequencies, MMIC technology provides not only significantly better performance than MIC technology, but also allows much faster assembly and much smaller sample-to-sample variance. Given the large number of amplifiers required at 70 GHz, MMIC technology can rightfully be regarded as enabling for the LFI. Cryogenic MMIC amplifiers have been demonstrated at 75–115 GHz which exhibit < 35 K over the LFI bandwidth. The LFI will thus fully exploit both MIC and MMIC technologies at their best. The LFI perfomance is summarized in Table 1.2.

TABLE 1.2

LFI Performance Goals[a]

| Instrument Characteristic | Center Frequency [GHz] | | |
|---|---|---|---|
|  | 30 | 44 | 70 |
| InP HEMT Detector technology . . . . . . . . . . . . . . . . | MIC | | MMIC |
| Detector temperature . . . . . . . . . . . . . . . . . . . . . . . . | | 20 K | |
| Cooling system . . . . . . . . . . . . . . . . . . . . . . . . . . . . | | $H_2$ Sorption Cooler | |
| Number of feeds . . . . . . . . . . . . . . . . . . . . . . . . . . . . | 2 | 3 | 6 |
| Angular resolution [arcminutes FWHM] . . . . . . . . . . | 33 | 24 | 14 |
| Effective bandwidth [GHz] . . . . . . . . . . . . . . . . . . . . . | 6 | 8.8 | 14 |
| Sensitivity [mK Hz$^{-1/2}$] . . . . . . . . . . . . . . . . . . . . . . | 0.17 | 0.20 | 0.27 |
| System temperature [K] . . . . . . . . . . . . . . . . . . . . . . . | 7.5 | 12 | 21.5 |
| Noise per 30′ reference pixel [$\mu$K] . . . . . . . . . . . . . . | 6 | 6 | 6 |
| $\Delta T/T$ Intensity [b] [$10^{-6}$ $\mu$K/K] . . . . . . . . . . . . . . . . . | 2.0 | 2.7 | 4.7 |
| ($\Delta T/T$) Polarisation (Q and U) [b] [$\mu$K/K] . . . . . . . | 2.8 | 3.9 | 6.7 |
| Maximum systematic error per pixel [$\mu$K] . . . . . . . . . | < 3 | < 3 | < 3 |

[a] All subsystems are designed to reach or exceed the performances of this table.

[b] Average 1$\sigma$ sensitivity per pixel (a square whose side is the FWHM extent of the beam), in thermodynamic temperature units, achievable after 2 full sky surveys (14 months).

### 1.3.1.1 The 20K cooler

Cooling of the LFI front-end to 20 K is achieved with a closed-cycle hydrogen sorption cryocooler (Wade et al. 2000; Bhandari et al. 2000, 2001), which also provides 18 K precooling to the HFI. The cooler operates by thermally cycling a set of compressors filled with metal hydride to absorb and desorb hydrogen gas, which is used as the working fluid in a Joule-Thomson (JT) refrigerator. The compressor assembly is attached to a radiator at 270 ± 10 K in the warm spacecraft, and the hydrogen flow-lines are passively pre-cooled to ∼50 K before reaching the 20 K JT expansion valve. The required pressure variations can be obtained by cycling the temperature of the compressors from 270 K to ∼265 K. In the complete system six identical compressors are used, each provided with a gas-gap heat switch to optimise its thermal performance. An additional sorbent bed is used to damp pressure fluctuations of the low pressure gas. At any time, one compressor is hot and desorbing to provide the high pressure hydrogen gas, one compressor is cooling down, one is heating up, while the other three are cold and absorbing gas. This principle of operation ensures that no vibration is exported to the detectors, a unique property of this kind of cooler which is very beneficial to *Planck*.

### 1.3.1.2 Systematic effects

To meet the challenging performance goals of the LFI requires not only great sensitivity and angular resolution, but also stringent control of systematic errors. Detailed analytical and numerical studies of the main possible sources of systematic effects and their impact on the LFI observations have been carried out, including sidelobe pickup of the Galaxy and solar system bodies (Burigana et al. 2001), distorted beam shapes (Burigana et al. 1998, 2002), effects induced by temperature instability (Mennella et al. 2002), residual non-white noise components (Maino et al. 1999, 2002a), non-idealities in the radiometer (Seiffert et al. 2002), spacecraft pointing errors and nutation, and calibration accuracy (Bersanelli et al. 1996a). A major effort has been carried out during the design and development phases, and will be continued through the development of data analysis software, to ensure that systematic effects are small and well understood, and that their impact will not compromise the ultimate quality of the LFI data. The LFI goal is that the combination of all systematic effects on the final sky maps will be less than $3\,\mu$K per resolution element.

### 1.3.1.3 Polarisation

The LFI is sensitive to polarization in all channels. As shown in Table 1.2, the sensitivity to polarization fluctuations is lower by a factor $\sqrt{2}$ than the sensitivity to intensity fluctuations because the number of channels per polarization is only half as great; moreover, there will be additional complications in removing polarized foregrounds, and other potential instrumental effects will have to be considered. Nevertheless, the combination of LFI and HFI has the sensitivity required to answer some of the key questions about CMB polarization, as described in Chapter 2. In particular, in the 70 GHz range the polarised Galactic foreground component is expected to be at a minimum compared to the cosmological polarised signal, thus providing an ideal window for CMB polarisation measurements.

## 1.3.2 The High Frequency Instrument (HFI)

The *Planck* High Frequency Instrument (Figures 1.7 and 1.8; Lamarre et al. 2003) will observe the sky at six frequencies from 100–857 GHz, with sensitivity in the lower frequencies close to the fundamental limit set by the photon statistics of the background itself. The heart of the HFI is the bolometer detectors (Bock et al. 1995; Lamarre et al. 2002; Jones et al. 2003), micro-fabricated devices in which the incoming radiation is absorbed in a grid whose impedance is matched with that of vacuum, increasing the temperature of a solid-state thermometer. These detectors give extremely high performance, yet are insensitive to ionizing radiation and microphonic effects.

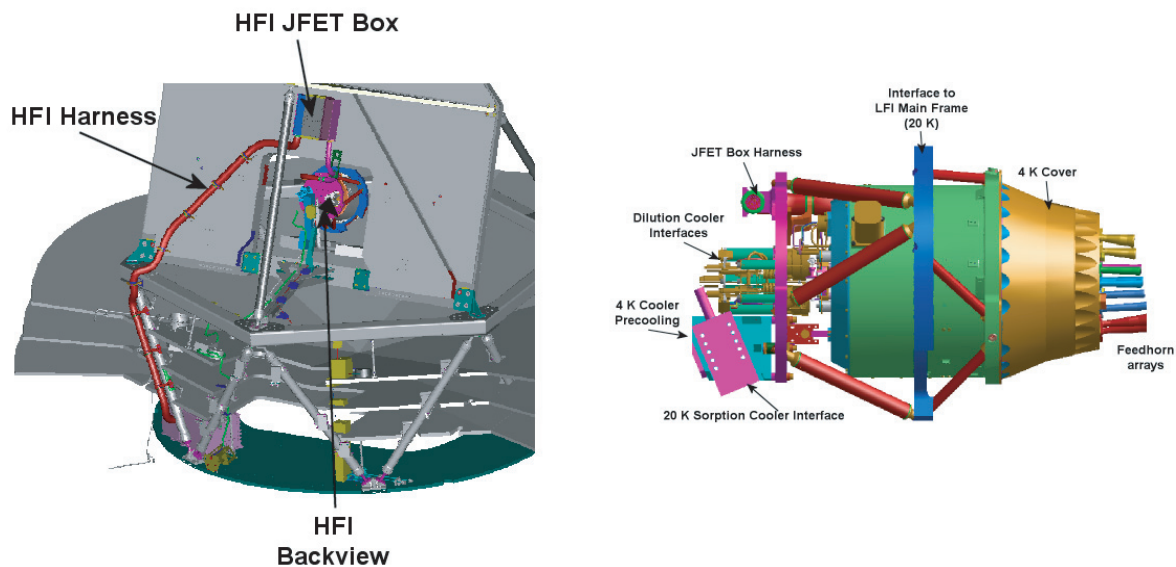

FIG 1.7.—The HFI focal plane unit. The telescope focuses radiation at the entrance of the corrugated horns. This flux is then filtered and detected by the low temperature (0.1 K) bolometers. The attachment points for the 20 K, 4 K, and 0.1 K coolers are shown, as well as the entrance point of the harness. The harness is shielded by a flexible bellows, and leads the bolometer signals to JFET-based circuits mounted in a box on the frame of the telescope. From this box, a second harness leads the signals to room-temperature electronics in the service module.

The characteristics and expected performance of the HFI are described in Table 1.3. Fifty-two bolometers are split into six channels at frequencies optimised for removal of foregrounds (mainly dust emission at these frequencies) and for the detection of the Sunyaev-Zeldovich effect (§ 3.3.2).

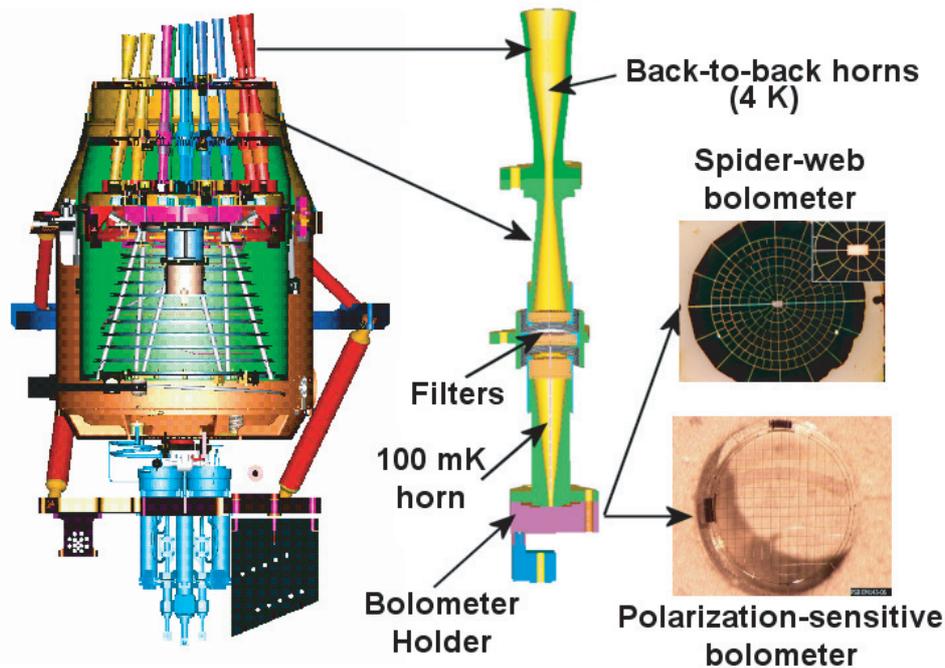

FIG 1.8.—Cutaway view of the HFI focal plane unit. Corrugated back-to-back feedhorns collect the radiation from the telescope and deliver it to the bolometer cavity through filters which determine the bandpass. The bolometers are of two kinds: (a) "spider-web" bolometers, which absorb radiation via a spider-web-like antenna; and (b) "polarisation-sensitive" bolometers, which absorb radiation in a pair of linear grids at right angles to each other. Each grid absorbs one linear polarization only. The absorbed radiant energy raises the temperature of a thermometer located either in the center of the spider-web, or at the edge of each linear grid.

Twenty bolometers are unpolarized, absorbing radiation in a grid that resembles a spider web. Thirty-two bolometers are sensitive to linear polarization (Jones et al. 2002), absorbing radiation in two orthogonal grids of parallel resistive wires, each of which absorbs only the polarized component with electrical field parallel to the wires. Combining measurements from several detectors allows a map of the sky (and CMB) polarisation to be constructed.

TABLE 1.3

HFI Performance Goals[a]

| Instrument Characteristic | Center Frequency [GHz] | | | | | |
|---|---|---|---|---|---|---|
|  | 100 | 143 | 217 | 353 | 545 | 857 |
| Spectral resolution $\nu/\Delta\nu$ ........................ | 3 | 3 | 3 | 3 | 3 | 3 |
| Detector technology ............................ | Spider-web and polarisation-sensitive bolometers | | | | | |
| Detector temperature .......................... | 0.1 K | | | | | |
| Cooling system ................................ | 20 K Sorption Cooler + 4 K J-T + 0.1 K Dilution | | | | | |
| Number of spider-web bolometers ................. | 0 | 4 | 4 | 4 | 4 | 4 |
| Number of polarisation-sensitive bolometers ........ | 8 | 8 | 8 | 8 | 0 | 0 |
| Angular resolution [FWHM arcminutes] ........... | 9.5 | 7.1 | 5.0 | 5.0 | 5.0 | 5.0 |
| Detector Noise-Equivalent Temperature [$\mu K s^{0.5}$] .... | 50 | 62 | 91 | 277 | 1998 | 91000 |
| $\Delta T/T$ Intensity [b] [$10^{-6}\mu K/K$] ................... | 2.5 | 2.2 | 4.8 | 14.7 | 147 | 6700 |
| $\Delta T/T$ Polarisation (U and Q)[b] [$10^{-6}\mu K/K$] ........ | 4.0 | 4.2 | 9.8 | 29.8 | ... | ... |
| Sensitivity to unresolved sources [mJy] ........... | 12.0 | 10.2 | 14.3 | 27 | 43 | 49 |
| ySZ per FOV [$10^{-6}$] ........................... | 1.6 | 2.1 | 615 | 6.5 | 26 | 605 |

[a] Goal sensitivities. All subsystems have been designed to reach or exceed the performances of this table, which are expected to be achieved in orbit. Sensitivity requirements are a factor of two worse, and would still achieve the core scientific objectives of the mission.

[b] Average $1\sigma$ sensitivity per pixel (a square whose side is the FWHM extent of the beam), in thermodynamic temperature units, achievable after 2 full sky surveys (14 months).

The corrugated horns at the entrance of the 4 K box ensure a well-controlled coupling of the detectors to the telescope and the sky. A set of filters, horns, and lenses (Fig. 1.8) determines the bandpass and leads the radiation to the detectors. This optical scheme has high optical efficiency compared to previous bolometer systems.

The readout electronics (Gaertner et al. 1997) are based on modulated bias and low noise lock-in amplifiers. They are able to transmit signals from DC up to 100 Hz, corresponding to the full range of angular frequencies relevant for the interpretation of CMB anisotropy. The data are compressed to an average flow of about 50 kbs for transmission to the ground.

### 1.3.2.1 Cryogenic design

To obtain the sensitivities listed in Table 1.3, the HFI bolometers must be operated at 0.1 K, obtained by a chain of three cryo-coolers (see Figure 1.9): a hydrogen sorption cooler (§ 1.3.1.1) which provides 20 K to LFI and 18 K to HFI; a Joule-Thomson refrigerator driven by mechanical compressors, precooled to 18 K by the sorption cooler, and which provides 4 K to HFI; and an open-loop $^3$He$^4$He dilution refrigerator, which provides 0.1 K. The compressor of the 4 K cooler is built in a head-to-head, momentum-compensating configuration. Vibration is reduced further by active electronic compensation of the compressor stroke.

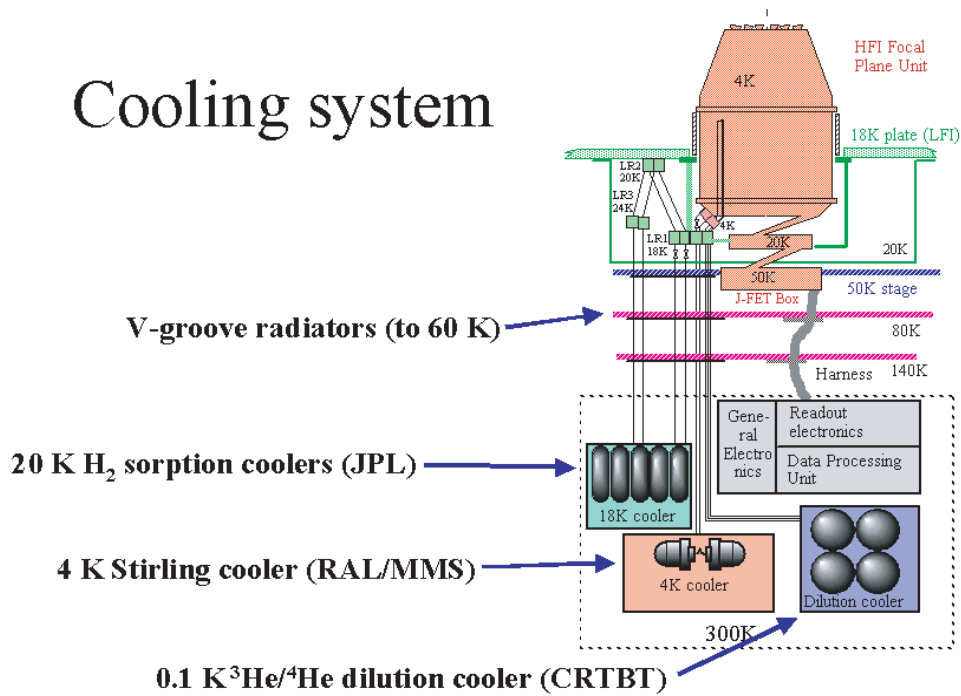

FIG 1.9.—The HFI cooling chain comprises the hydrogen sorption cooler providing 18 K, the closed-loop Joule-Thomson refrigerator providing 4 K, and a dilution refrigerator providing 0.1 K to the bolometers..

The 0.1 K cooler exploits a new dilution principle based on friction that does not need gravity to operate, and therefore makes it particularly adapted to space applications. It has been developed at the Centre de Recherches des Très Basses Températures (CRTBT) in Grenoble (France) (Benoit, Caussignac and Pujol, 1994). Gas storage capacity is sufficient for about 30 months of operations. For a total ($^3$He and $^4$He) flow rate of 12 mole/s, a cooling power of 100 nW at 0.1 K has been demonstrated. In the same process, the mixture is expanded in a Joule-Thomson valve, producing a cooling power of several hundred microwatts at 1.6 K. A prototype of this cooler has been used successfully for astronomical observations from the ground and with the balloon-borne experiment Archeops.

The cutaway view in Figure 1.8 shows the Russian-doll structure chosen to isolate the 0.1 K stage from parasitic heat input. The external box is cooled to 4.5 K by the 4 K mechanical refrigerator. Lower temperature stages thermalise different elements in the optical path. The 0.1 K stage supports the bolometer assembly and band-limiting interference filters. Further filters and a focussing lens are kept at 1.6 K. The back to back horn arrangement that couples the detectors with the telescope is at 4 K.

Passive damping and active temperature control ensure that temperature fluctuations of the various stages do not degrade significantly the intrinsic detector noise level. Special thermometers have been developed to measure and control the temperature of the 0.1 K stage.

### 1.3.3 Telescope

The telescope (Figure 1.10) is an off-axis, aplanatic design with two elliptical reflectors and a 1.5 m projected diameter. The optical system was optimized for a set of representative detectors (eight HFI and eight LFI). The performance of the aplanatic configuration is not quite as good on the optical axis as the so-called Dragone-Mizuguchi Gregorian configuration, which eliminates astigmatism on the optical axis, but is significantly better over the large focal surface required by the many HFI and LFI feeds. The main parameters of the two reflectors are given in Table 1.4.

TABLE 1.4

Telescope Mirror Parameters

| Parameter | Primary mirror | Secondary mirror |
|---|---|---|
| Radius of curvature of the ellipsoid [mm] . . . . . . . . . . . . . . . . | 1440.000 | −643.972 |
| Conic constant . . . . . . . . . . . . . . . . . . . . . . . . . . . . . . . . . . . . . | −0.86940 | −0.215424 |
| Elliptical perimeter [mm] . . . . . . . . . . . . . . . . . . . . . . . . . . . . . | 1555.98 × 1886.79 | 1050.96 × 1104.39 |
| Reflectors centre to major axis offset [mm] . . . . . . . . . . . . . . . | 1038.85 | −309.52 |

Due to stringent requirements on the level of straylight reaching the detectors, the reflectors are significantly underilluminated by the detector horns. This fact has been taken into account in the requirements on the mechanical surface errors (MSE) of the reflectors, which vary from ∼ 7.5 μm (rms) near the centre to ∼ 50 μm (rms) near the edge. These surface accuracies give an optical system with an equivalent wavefront error of ∼ 15 μm. To ensure good optical performance at operational temperatures (∼ 50 K) and low mass, the reflectors are made from a carbon fiber reinforced plastic (CFRP) honeycomb sandwich. They will be coated with aluminum to assure high (> 0.995) reflectivity at the operating frequencies. The reflectors will be supported by a carbon fiber structure which also supports the focus assembly, as well as thermal and straylight shields. A demanding cryogenic test campaign and an extensive programme of radio-frequency measurements will ensure that the *Planck* telescope lives up to its high expectations.

## 1.4 SPACECRAFT

The *Planck* spacecraft spins on a Sun-pointed axis. Its geometrical configuration is designed to prevent straylight originating from the Sun, the Earth, and the Moon from reaching the *Planck* focal plane, and to allow the greatest possible thermal isolation of the warm spacecraft from the cold payload.

*Herschel* and *Planck* are launched together on an Ariane 5 launcher. Figure 1.11 shows the launch configuration, with *Planck* inside a standard adaptor, and *Herschel* above it. Shortly after launch, *Herschel* and *Planck* will separate from the rocket and proceed independently to their respective orbits, using autonomous hydrazine-based propulsion systems. After a transit phase lasting 3-4 months, *Planck* will be injected into a Lissajous orbit around the 2nd Lagrangian point of the Sun-Earth-Moon system (Figure 1.4); this orbit subtends a maximum angle of 15° as seen from the Earth. At this location, *Planck* is able to always maintain its payload pointed towards deep space, shielded from Solar, Earth, and Lunar illumination by its solar array.

The Service Module houses all the instrument warm electronics as well as the ancillary equipment of the spacecraft (telecommunication units, solar panels and batteries, attitude control, computers, propellant tanks, etc.). It consists of an octagonal box built around a conical tube which carries the satellite loads. The propellant tanks, and the tanks containing cryogens for the HFI coolers, are located inside the tube. Most other units are located outside the tube, mounted on side panels of the octagonal box, allowing them to radiate to space.

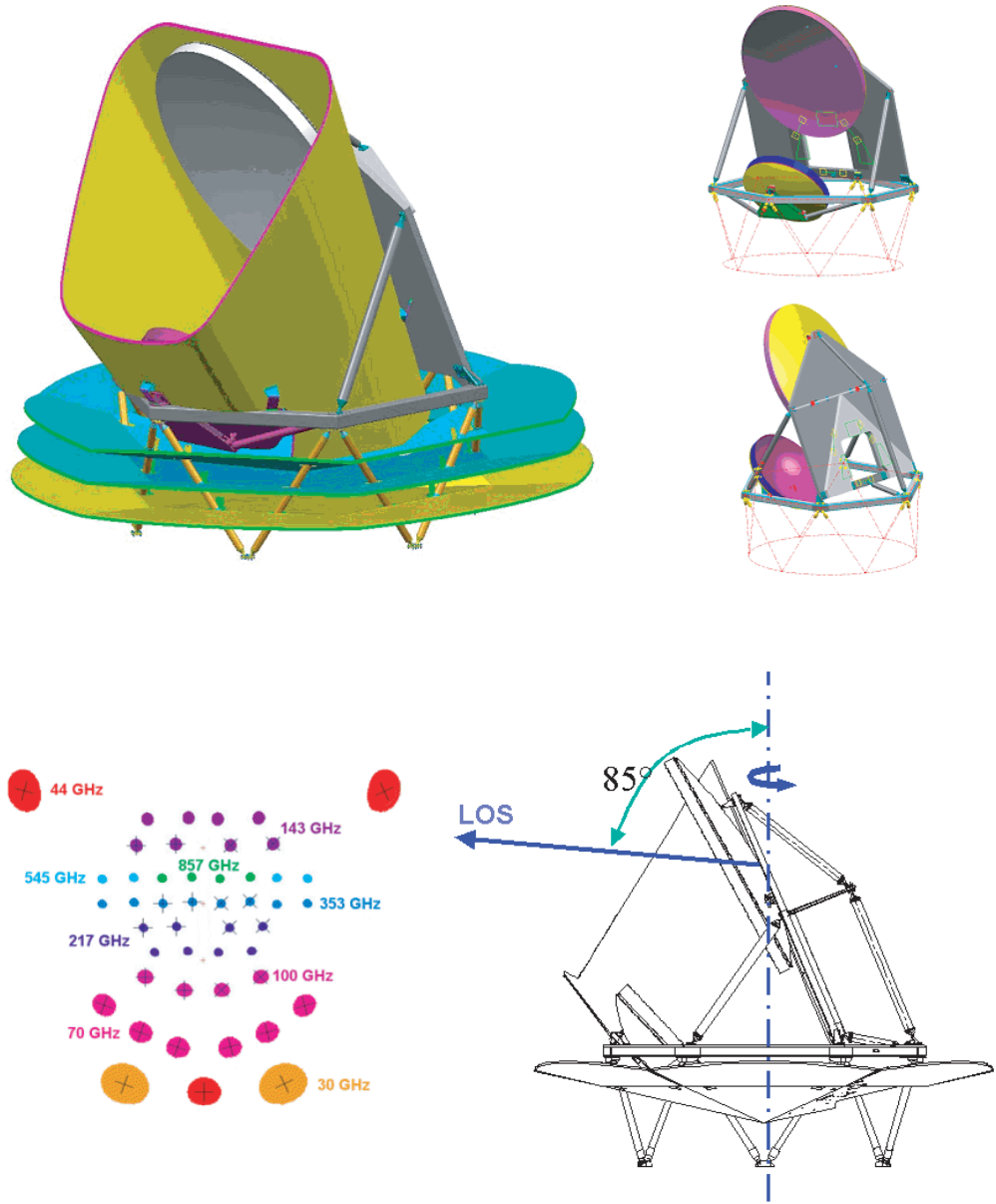

Fig 1.10.—The *Planck* optical system, including the telescope baffle and V-grooves, which provide straylight control in addition to thermal isolation. The attachment points for the LFI focal plane unit can be seen in the upper right. The reflectors are under manufacture by Astrium Gmbh (Friedrichshafen) together with the telescope support structure (under manufacture by Contraves, Zürich). Views courtesy of Alcatel Space (Cannes). The telescope field-of-view is offset from the spin axis of the satellite by an angle of 85°. The footprint of the field-of-view on the sky is seen in the lower left; it covers about 8° on the sky at its widest. Each spot corresponds to the 20 dB contour of the radiation pattern; the ellipticities seen are also representative of the shape of the beam at full-width-half-maximum. The black crosses indicate the orientation of the pairs of linearly polarised detectors within each horn. The field-of-view sweeps the sky in the horizontal direction in this diagram.

An attitude control system maintains the satellite rotation speed at 1 rpm, and keeps the Sun direction less than 10° from the spin axis, ensuring thermal stability and full electrical power for the payload. A star tracker pointed along the telescope boresight measures the spin rate, the spin axis direction, and the satellite nutation. These will be used for attitude reconstruction on the ground. The reconstruction process will achieve a detector pointing accuracy of about 0.5, and will require periodic calibration using payload detections of celestial sources.

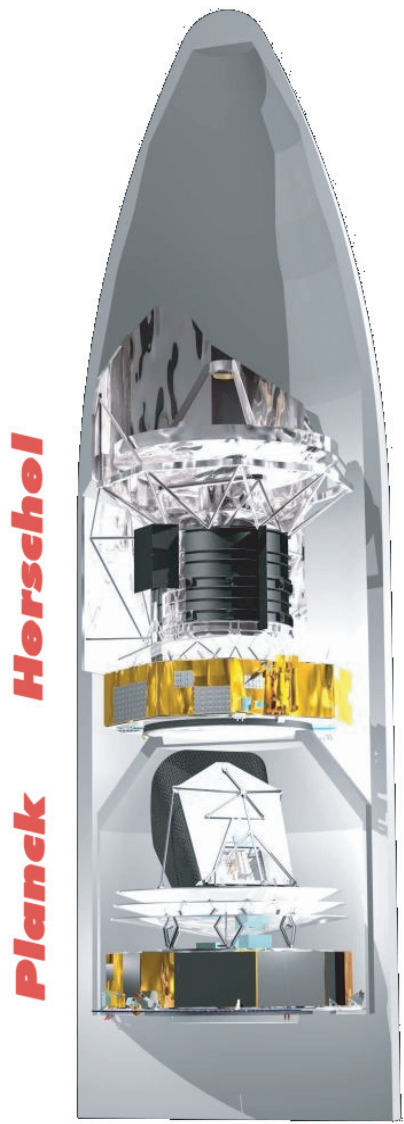

Fɪɢ 1.11.— *Planck* and *Herschel* ready for launch under the Ariane 5 fairing. The two satellites separate just after launch and proceed independently to their final orbits. View courtesy of Alcatel Space (Cannes).

The satellite will communicate with an ESA Ground Station located in New Norcia (Australia) using X-band transponders and a medium-gain antenna pointed along the spin axis and with a FWHM of 30°. Data will be transmitted to the ground during a 3-hour daily visibility window without interruption of the observations. Up to 48 hours of data can be stored in solid state memory on-board with the help of essentially lossless compression of the data by a factor of ∼3.

## 1.5 GROUND OPERATIONS AND DATA PROCESSING

Ground operations for the *Planck* satellite include three overlapping and interrelated elements: satellite and instrument operations; processing of the instrument acquired data; and scientific analysis.

### 1.5.1 Operations

The *Planck* spacecraft will be controlled from the Mission Operations Centre (MOC) in Darmstadt (Germany). The satellite will operate autonomously except for ground contact periods of ∼ 3 hours each day, during which it will downlink all the data acquired over the previous 24 hours. The instruments will be operated via the MOC by teams resident at the PI institutes, who will be responsible for the health, calibration, and optimal operation of their instruments. Survey planning and overall scientific coordination will be carried out by the *Planck* Science Office at ESTEC in The Netherlands.

## 1.5.2 Data Processing

Scientific data will be sent daily from the MOC to Data Processing Centres (DPCs) developed and operated by the Instrument Consortia. The DPCs are responsible for all levels of processing of the *Planck* data, from raw telemetry to deliverable scientific products.

The success of the mission depends on the combination of measurements from both instruments. A unifying approach is thus necessary in order to fully achieve the scientific objectives of the mission, and to obtain a single final set of data, derived optimally from all products of the two DPCs. Several coordination activities are carried out by the two DPCs, from the definition of a common system for the exchange of data and information, to the development of a common environment for the data processing pipeline, to the definition of ways for merging the products of the mission into a scientifically coherent set.

The DPCs will share a basic information management infrastructure, the *Planck* Integrated Data and Information System (IDIS, Bennett et al. 2000). IDIS is object-oriented, comprising five components: Document Management; Software Management; Process Coordination (e.g., a data pipeline manager); Data Management, covering ingestion, management, and extraction of the data; and a Federation layer, providing inter-connection among IDIS components (e.g., relating objects controlled by each component).

The DPCs are responsible for the production, delivery, and archiving of the data products of their respective instruments. Furthermore, the DPCs are responsible for the production of realistic data simulating the microwave sky observable by *Planck* and the behaviour of the two instruments in flight, and for support of instrument ground testing activities.

The DPCs will also manage a number of other data sets, containing: the characteristics of each detector and the telescope (e.g., detectivity, emissivity, time response, main beam and side lobes, etc.); calibration data sets and data at an intermediate level of processing), to be shared among members of the LFI and HFI Consortia; and ground calibration and AIV databases produced during the instrument development, including all information, data, and documents related to the payload and all systems and sub-systems. Most of this information is crucial for processing flight data and updating the instrument performance. This second set of tasks implies a high level of coordination with the Instrument Development Teams, to harmonize properly instrument knowledge and data processing.

The DPCs are organized in five levels (Figure 1.12):

- Level 1: telemetry processing and instrument control (no scientific processing of the data);
- Level 2: data reduction and calibration (requires detailed instrument knowledge);
- Level 3: component separation and optimization (requires joint analysis of HFI and LFI data);
- Level 4: generation of final products (reception, archiving, preparation of public release material);
- Level S: simulation of data acquired from the *Planck* mission on the basis of a software system agreed upon across Consortia.

To achieve cost and efficiency savings while guaranteeing redundancy, Level 4 of the DPC is shared between the two Consortia, each Level 1 stores the telemetry data of both instruments, and each Level 2 reduces the data of its own instrument. Two implementations of Level 3 will be provided by the Consortia, to guarantee the level of redundancy and cross-checking necessary for a demanding mission such as *Planck*. This data processing configuration is considered by both the LFI and HFI Consortia to be the best trade-off possible between the redundancy and cross-checking requirements, the optimisation of resources, and instrumental knowledge.

Level 1 has a direct interface with the ESA Mission Operations Centre (MOC) in Darmstadt. It downloads from the MOC telemetry data (both scientific and housekeeping), and provides it with the information necessary to operate the instruments safely and optimally. Day-to-day operations are under the responsibility of Instrument Operations Teams (one for each instrument), mainly composed of instrument developers.

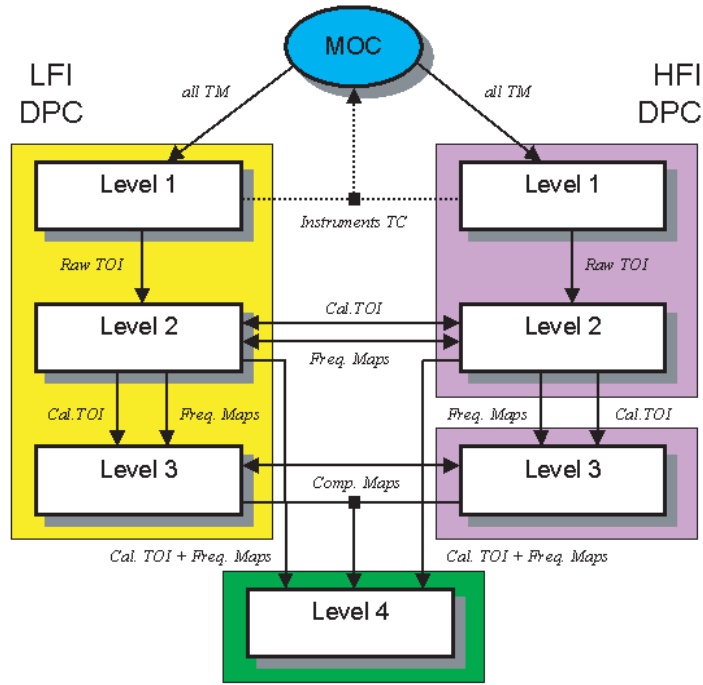

FIG 1.12.—Organization of the DPCs. Each colored box corresponds to a geographically separate location.

In its development, the DPCs receive inputs from scientists from many institutions throughout the two Consortia, who develop simulations and data processing prototypes within coordinated teams; an Integration team builds the pipeline, engineering the code to the needed robustness and reliability. Contributions of prototype software to the LFI DPC come from a variety of geographically distributed sites. LFI operations are centralized in Trieste, where they are run jointly by the OAT and SISSA. Level 4 is operated by MPA, Garching. Level S is maintained at MPA, with contributions from many parts of the teams. The HFI DPC is a more distributed system, with Levels 1 and 2 distributed among several institutes, mostly located in the Paris area, but with some contributions from the UK. HFI Level 3 processing is concentrated in the UK at Cambridge University. Levels 4 and S are shared with the LFI.

### 1.5.2.1 Computational challenges

Some steps of the processing described above are far from trivial, for example, map-making (i.e., converting the time-series of acquired data into sky maps) and component separation (i.e., separating the CMB from foreground emission). Exact algorithms for solving such problems perform matrix operations globally on the whole dataset (time series or maps), yielding $O(M^3)$ complexity (Bond et al. 1999), where $M$ is the number of elements in the matrix. This was done successfully for the COBE experiment, where the resolution was 7°; however, for *Planck* $M$ is large, up to $5 \times 10^7$ pixels per map or $\sim 10^{10}$ points in the time stream per detector per year). The exact algorithm is impossible to implement with the computational facilities that will be available on the *Planck* timescale.

Fortunately, fast hybrid approaches are possible that should give very similar results to exact solutions (Efstathiou 2004). Reasonable solutions to the map-making problem using a number of simplifying assumptions have already been developed, even covering the case of polarized radiation (e.g., de Gasperis et al. 2001). The challenge is to obtain in a reasonable time the most accurate results possible, reducing the conceptual simplifications to a minimum.

Similar considerations hold for the separation of components. A classical way of solving the problem is using maximum-entropy methods (MEM, Hobson et al. 1998), which require some a priori knowledge. Combining MEM with source detection algorithms, in particular those based on wavelets (Cayón et al. 2000) has been tested successfully (Vielva et al. 2001). However, such non-linear methods cannot produce the accurate estimates of uncertainty in the CMB at each point required by CMB science. New classes of algorithms are being developed to solve this problem. As an example, the independent component analysis (ICA) approach has provided interesting results, and is much less computationally intensive (Baccigalupi et al. 2000). An improved version of the system, FastICA (Maino et al. 2002b) has further improved the computational speed. A strength of ICA is independence of the algorithm from a priori knowledge; a weakness lies in its fragility with respect to point-like sources. Attempts are being made to combine ICA with wavelets source detection algorithms to obtain better point source extraction.

### 1.5.2.2 The Simulation Pipeline

The primary goal of *Planck* data simulation is to provide simulated data for testing *Planck* data analysis. The input sky maps, frequency dependencies, contaminating sources, beam shapes, scanning strategies, random and systematic errors can be controlled individually in simulations, and thus their impact on data reduction and results can be evaluated in detail. Conversely, this implies that the simulation software has to cover all relevant aspects of the microwave sky and of the *Planck* observing procedure at a sufficiently realistic level.

*Planck* data simulation also has secondary goals. One is that a prototype for the *Planck* data archive can be constructed using simulated data. Since the simulated data will reach or exceed the volume of the real *Planck* data, handling of and access to *Planck* data can be exercised using the simulations. Another secondary goal is that the production of simulated *Planck* data is conceptually similar to *Planck* data analysis, so that the same software environment can be used for both purposes. This implies that the *Planck* IDIS infrastructure currently under development can be realistically tested running data simulation pipelines.

The simulation pipeline is thus a testing ground for the DPC infrastructure and operations. So far, many terabytes of simulated data have been produced. The source code of all simulation modules is available through the IDIS software repository. The complete simulation package has been ported successfully to many different platforms, including Linux PCs and multi-processor shared-memory machines. New modules are regularly being integrated and embedding into the IDIS environment.

## 1.5.3 Scientific Analysis

### 1.5.3.1 Deliverable Data Products

The major deliverables of the *Planck* mission are:

- Calibrated time-ordered data
- All-sky maps at each frequency
- All-sky component maps, including the CMB itself, plus Galactic synchrotron, free-free, and dust emission, etc.
- A final compact-source catalog, including galactic and extra-galactic objects. This catalog will also include galactic clusters detected through the Sunyaev-Zeldovich effect.

These deliverables will be produced by the two DPCs jointly, and will be made publicly available two years after completion of mission operations. This period is chosen to allow one year to reduce the survey data into scientifically usable products, and one year of proprietary exploitation by the *Planck* Consortia. Once the proprietary period is finished, the final products will be delivered to ESA, which will then make them available to the public via the *Herschel* Archive. Therefore, two years after termination of survey operations, the *Planck* data will be put into the public domain.

In addition to the above formal deliverables, an early-release compact source catalog based on the first all-sky survey, will be released at an earlier time to enable follow-up observations of discrete sources with other instruments. The Early Release Compact Source Catalog will be released 9 months after completion of the first all-sky survey, i.e., approximately 16 months after the start of routine operations. While the quality of this product will not be as high as the final *Planck* source catalogues, in terms of completeness, sensitivity, positional accuracy, or calibration accuracy, it will allow the rapid follow up of interesting objects with other observatories. This will be of particular interest in the case of *Herschel*, which has a limited lifetime.

### 1.5.3.2 Data exploitation

The objectives of *Planck* will only be achieved once the data are fully analysed. As noted above, the three *Planck* Consortia have been granted a one year period to reduce the survey data into scientifically usable products, and thereafter a further one year proprietary period for their scientific exploitation.

The principal objective of *Planck* is to produce an all-sky map of the CMB. The main scientific results expected from *Planck* are cosmological, and that is where much of the attention of the *Planck* community will be focussed during the proprietary period. However, as a by-product of the extraction of the CMB, *Planck* will also yield all-sky maps of all the major sources of microwave to far-infrared emission, opening a broad expanse of other astrophysical topics to scrutiny.

In particular, the physics of dust at long wavelengths and the relative distribution of interstellar matter (neutral and ionized) and magnetic fields will be investigated using dust, free-free, and synchrotron maps. In the field of star formation, *Planck* will provide a systematic search of the sky for dense, cold condensations which are the first stage in the star formation process. One specific and local distortion of the CMB which will be mapped by *Planck* is the Sunyaev-Zeldovich (SZ) effect arising from the Compton interaction of CMB photons with the hot gas of clusters of galaxies. The well-defined spectral shape of the SZ effect allows it to be cleanly separated from the primordial anisotropy. The physics of gas condensation in cluster-size potential wells is an important element in the quest to understand the physics of structure formation and ultimately of galaxy formation.

The *Planck* all-sky surveys of foreground emission will constitute a rich astrophysical resource comparable to the IRAS and COBE-DIRBE maps at shorter wavelengths. They will cover a broad frequency range (30–857 GHz) have high angular resolution (from ∼30′ at 30 GHz to ∼5′ at 217 GHz and above), have extremely low noise, and will be well-calibrated.

Three factors contribute to the accuracy of the calibration. First, the angular responsivity of each detector will be accurately mapped to high dynamic range (as high as 100 dB), using bright unresolved sources (e.g., planets) for near-boresight angles, and the Sun, Moon, Earth and the Milky Way itself for mid- and far-boresight angles). Second, *Planck* is designed to avoid systematic effects so that the CMB signal can be recovered at microkelvin levels. Third, the absolute calibration of the *Planck* detectors at frequencies up to at least 353 GHz will be derived from the annual modulation of the CMB dipole by the Earth's orbit around the Sun, and therefore traceable to fundamental constants. The goal of *Planck* is to obtain a photometric calibration of better than 1% across all frequency channels where the CMB is strong (i.e., up to 353 GHz).

### 1.5.3.3 Scientific management

It is recognised by all participants in the mission that while *Planck* has two instruments, they constitute a single experiment dedicated to a well defined objective. Achieving the best scientific results requires a high degree of cooperation from all participants. The three *Planck* Consortia currently include about 450 scientists representing some 25 scientific institutes from all over Europe and the USA, thus covering a large fraction of the European cosmology and submm astrophysics community. Many of these scientists are actively participating in either the development of the payload hardware or that of the DPCs. After launch, they will be deeply involved in the instrument and DPC operations, and in the analysis of the data.

It is clear that the large number of people involved, and their diverse geographical distri-bution, coupled to the established delivery deadlines, require management more typical of large particle physics experiments than of astronomical observatories. This effort is being led by the *Planck* Science Team, a group of scientists chaired by the ESA Project Scientist and representing the three *Planck* Consortia. The Science Team is in charge of distributing scientific work, as well as of establishing and managing data rights and publication policy up to the end of the proprietary period.

As part of its work, the Science Team has established the Baseline Scientific Programme of *Planck*, which describes the scientific exploitation activities which will be undertaken by the *Planck* Consortia during the proprietary period. The Baseline Programme is a comprehensive plan that takes full advantage of the scientific potential of *Planck*, and that covers the research interests of the many scientists involved in *Planck*. This publication is dedicated to a detailed description of its goals and scope.

The *Planck* Scientific Programme will be revised shortly before launch to incorporate scientific developments over the next few years.

## 1.6 ACKNOWLEDGMENTS

# CHAPTER 2
# PRIMARY CMB ANISOTROPIES

## 2.1 OVERVIEW

For over forty years, observations of the CMB have had a profound influence on our knowledge of the Universe. The discovery of the CMB in 1965 established beyond reasonable doubt the paradigm of Hot Big Bang cosmology and provided firm observational evidence for an evolving Universe with a well defined beginning. The Universe cannot be in a 'steady state'—the Universe must have been hot and dense shortly after the Big Bang, with conditions more extreme than those described by any known physical theory. The discovery of the CMB also provided the first link between particle physics in the early Universe and cosmology, namely the production of light elements within the first three minutes of the Hot Big Bang. The theory of Big Bang nucleosynthesis offered a compelling explanation of the high abundance of helium in our Universe—a problem that had puzzled stellar astronomers for many decades. The discovery of the CMB also had important implications for fundamental physics. The theorems of Penrose and Hawking showed that a singularity at early times was unavoidable according to General Relativity. Evidently, classical General Relativity must break down at the Hot Big Bang, but if so, then there is a real possibility of finding observational signatures of quantum cosmology and of learning about entirely new physics.

Almost immediately after the discovery of the CMB, cosmologists realised that the fluctuations in the early Universe responsible for the structure that we see today—galaxies, clusters and superclusters—must have imprinted small differences in the temperature of the CMB (anisotropies) coming from different directions of the sky. A long series of experiments searched for these tiny variations (of order one thousandth of one percent of the CMB temperature), but they remained undiscovered until 1992, when the first results from NASA's Cosmic Background Explorer (*COBE*) were announced. The COBE discovery of primordial anisotropies received a huge amount of media coverage and was hailed by Stephen Hawking as "the greatest scientific discovery of the century." Why the interest and euphoria? The discovery of CMB anisotropies has opened up an entirely new way of studying the early Universe. The fluctuations are thought to have been generated within $10^{-35}$ seconds of the Big Bang, so by studying CMB anisotropies we probe fundamental physics at energy scales many orders of magnitude higher than those accessible to particle accelerators.

CMB anisotropies can be used to fix the values of many of the key parameters that define our Universe, in particular, its geometry, age, and composition. Following *COBE*, a remarkable series of experiments culminating in NASA's Wilkinson Microwave Anisotropy Probe (*WMAP*) established that our Universe is close to spatially flat. Furthermore, these experiments have provided compelling evidence that our Universe is dominated by dark energy and dark matter, and that present-day struture grew from nearly scale invariant primordial fluctuations.

Nevertheless, much more information about the Universe remains to be extracted from the CMB, especially its polarization. *Planck* is designed to extract essentially all information available in the temperature anisotropies and to measure CMB polarization to high accuracy. This Chapter presents the core science case for *Planck*, in particular, *Planck*'s ability to probe new physics at the time that the primordial fluctuations were generated. *Planck* is targeted towards fundamental problems in physics and cosmology: How were the fluctuations generated? Did the Universe undergo an inflationary phase and, if so, can we constrain the dynamics of inflation? Is there evidence of fundamentally new physics, e.g., the extra dimensions required by string theory? *Planck* continues the long tradition of ESA missions in the planetary sciences and astronomy aimed at understanding the origins of our Universe and the structure within it.

## 2.2 CMB Temperature Anisotropies

### 2.2.1 Introduction

The goals of *Planck* are unique amongst experiments designed to measure temperature and polarization anisotropies of the CMB. *Planck* has been designed to produce high resolution (down to 5′) maps of the temperature and polarization anisotropies, with microkelvin sensitivity per resolution element, over the entire sky. The wide frequency coverage of *Planck* (30–857 GHz) has been chosen to provide accurate discrimination of Galactic emission (particularly important for polarization, as discussed in §2.3) and to study galaxy clusters (via the Sunyaev-Zel'dovich effect, see Chapter 3) and extragalactic point sources (Chapters 4 and 5).

The vast increase in information promised by *Planck* is demonstrated in Figure 2.1, which compares how the CMB sky would be seen at the resolutions of *Planck*, *COBE*-DMR, and *WMAP*. The small scale features in the CMB sky provide a 'cosmic fingerprint' of the Universe. As this chapter will describe, deciphering this fingerprint, using precise measurements of the CMB anisotropies with *Planck*, will lead to unprecedented tests of the physics of the ultra-early Universe and the determination of cosmological parameters to extraordinarily high precision.*

### 2.2.2 The Physics of CMB Anisotropies

In the standard Hot Big Bang model, the Universe is highly ionized until a redshift $z_R \sim 1000$, the so-called recombination epoch, when protons combined with electrons to make neutral hydrogen. Prior to this epoch, photons were tightly coupled to the electrons by Thomson scattering, but once recombination was complete the Universe became transparent to radiation and photons propagated towards us along geodesics, almost unimpeded by the matter. Maps of the microwave background radiation therefore provide us with a picture of irregularities at the 'last scattering surface' when the Universe was about 400,000 years old.

In the simplest models, the observed temperature anisotropies arise from two distinct physical processes:

- *Potential Fluctuations in the Early Universe:* On angular scales $\theta \gtrsim 1°$, CMB anisotropies probe fluctuations in the gravitational potential along different lines of sight, $\Delta T/T_0 \simeq \frac{1}{3}\Delta\phi/c^2$. This is often called the Sachs-Wolfe effect (Sachs & Wolfe 1967), and is of particular importance because it links temperature anisotropies directly with potential fluctuations generated in the early Universe.

- *Sound Waves Prior to Recombination:* Small scale fluctuations in the matter-radiation fluid at the recombination epoch were causally connected and oscillated like sound waves. The small angular scale structure in the CMB (evident in Fig. 2.1) is a consequence of these oscillations. The characteristic scale of this structure at $\theta \simeq 1°$ (corresponding to a prominent acoustic peak in the CMB power spectrum at multipoles of $\ell \simeq 200$, see Fig. 2.2 below) indicates the maximum distance that a sound wave can travel up to the time of recombination. Accurate measurements of the temperature and polarization anisotropies on small angular scales thus provide information on the sound speed at the time of recombination and hence on the matter content of the Universe. Moreover, the relation between physical distance and angular separation on the sky provides information on the geometry of the Universe. Since the recombination process is not instantaneous, along any given line-of-sight we see a time average through the recombination process. This leads to suppression, or damping, of the highest frequency acoustic oscillations, corresponding to the anisotropies at the smallest angular scales.

With the high fidelity of *Planck* we can investigate these physical processes in detail. This allows us to constrain the initial conditions as well as to probe the evolutionary effects on the perturbations over the age of the Universe. Hence we can determine the characteristic *mode* of the perturbations, distinguishing between *adiabatic* perturbations (which preserve the entropy per particle and are predicted by the vast majority of theoretical models) from *isocurvature* perturbations (which arise from spatial variations in the equation of state). We can estimate the amplitude and shape of the initial spectrum of density perturbations and, through measurements of the CMB polarization, search for the imprint of primordial tensor (gravity wave) perturbations at large angular scales. Signatures of the first stars that formed in the Universe should also be detectable in the CMB polarization on the largest angular scales, and galaxy clusters at lower redshifts will be detectable in the CMB temperature anisotropies at smaller angular scales. Gravitational lensing by clustered matter at low redshifts should also be detected in the CMB temperature and polarization anisotropies. One of the attractive features of *Planck*, perhaps unique amongst space missions, is its ability to study physical effects over the entire age of the Universe, from $< 10^{-35}$ seconds after the Big Bang to the present day.

---

*   See Scientific American, Feb. 2004 for a less technical introduction to the CMB and related cosmological issues.

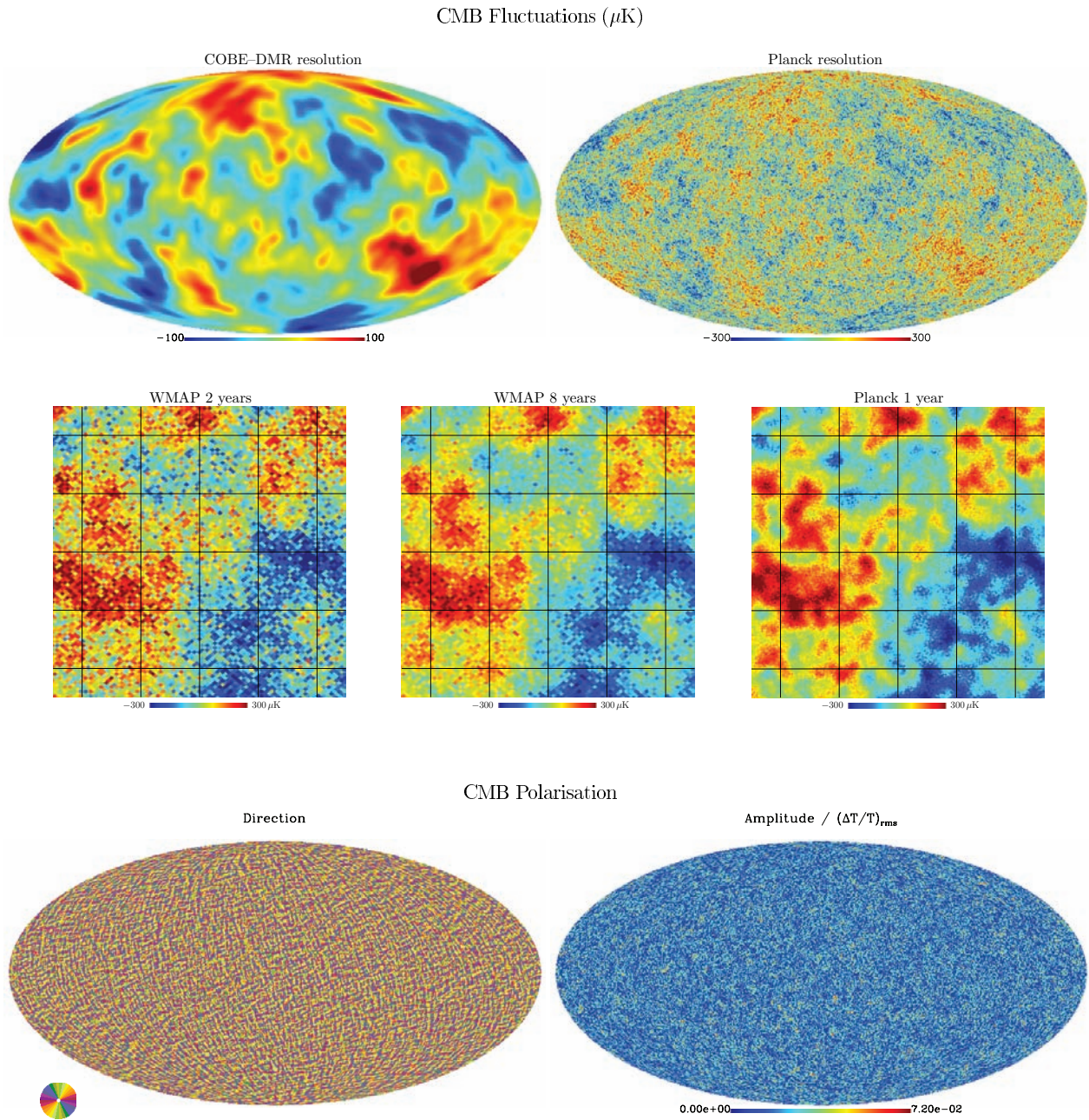

FIG 2.1.— Simulated maps of the CMB sky in inflationary CDM models. The top two pictures show temperature anisotropies over the whole sky at *COBE* and *Planck* resolutions. Such small pictures cannot show the difference in resolution between *WMAP* and *Planck*. Accordingly, the middle three pictures show an expanded view of a 5° × 5° patch of sky at *WMAP* (94 GHz, 15′ FWHM) and *Planck* (217 GHz, 5′ FWHM) resolutions, with noise calculated for 2 and 8 years for *WMAP* and 1 year for *Planck*. The significant differences between *WMAP* and *Planck* in resolution and noise (even for an 8-year *WMAP* mission) are shown in a different way in Fig. 2-8. The lower two pictures show the direction and amplitude of polarization anisotropies at *Planck* resolution for a pure scalar fluctuation mode.

The evolution of CMB anisotropies involves well-understood physics in the linear regime and hence is on a very sound theoretical basis. It is therefore possible to make precise predictions of how the CMB anisotropies depend on the basic quantities that characterise our Universe. The statistics of the temperature anisotropies can therefore be used to derive key parameters, such as the spatial curvature, dark energy content, and the baryon and dark matter densities. *One of the principal goals of Planck is to determine these fundamental parameters to unprecedented precision of better than a percent.*

With the accuracy achievable with *Planck*, these cosmological parameters will be determined with such high precision that tests of new physics will become possible. For example: Does the spectrum of (scalar) fluctuations deviate from a scale-invariant form? Is there any evidence for departures from a pure power-law spectrum? Is there a curl-component (*B*-mode) to the polarization fluctuations, indicative of a tensor mode generated during inflation? What is the mass of the heaviest neutrino? Is the global topology of the Universe non-trivial? Do the physical constants depend on epoch? Does the dark energy density vary with time as in quintessence models? These and other questions are discussed in the rest of this chapter.

### 2.2.3 Statistics of the CMB Sky

The temperature anisotropies of the sky can be expanded in spherical harmonics,

$$\Delta T = \sum_{\ell m} a_{\ell m} Y_{\ell m}(\theta, \phi). \tag{2.1}$$

According to a wide class of models of the early Universe, the temperature anisotropies should obey Gaussian statistics to high accuracy. If this is the case, then *all* of the statistical properties of the temperature anisotropies can be computed from a single function of multipole index $\ell$, the *temperature power spectrum*

$$C_\ell^T = \langle |a_{\ell m}|^2 \rangle. \tag{2.2}$$

More generally, the CMB anisotropies are expected to be linearly polarized, giving two other measurable quantities (the Stokes $Q$ and $U$ parameters) at each point on the sky. These can be decomposed in terms of $E$- and $B$-type polarization patterns (also known as grad and curl modes). The temperature and polarization anisotropies in Gaussian models can then be characterised by four power spectra, $C^T$, $C^E$, $C^B$, and $C^{TE}$, describing the temperature anisotropies, $E$- and $B$-type polarization, and the temperature-polarization cross correlations. (More detailed descriptions of these power spectra are given in § 2.3.3).

Because of the ubiquity of Gaussian theories, and the lack of any convincing experimental evidence of non-Gaussianity (implying that the vast majority of the information content is in the power spectrum), a large amount of effort has gone in to measuring and interpreting the CMB power spectra. Accordingly, measurement of the CMB temperature and polarization power spectra to high precision up to multipoles $\ell \gtrsim 2500$ is one of the primary scientific goals of the *Planck* mission.

However, it would be a mistake to regard *Planck* simply as an experiment designed to measure power spectra to high precision. The Gaussian assumption is such a critical feature of theoretical models that it needs to be tested experimentally to high precision. This should be achieved by *Planck*. Its high sensitivity, wide frequency, and complete sky coverage, mean that it is well suited to detecting small departures from Gaussianity and unusual rare events (cosmic defects, for example). All theories predict non-Gaussianity at some level, but in some specific models, including cosmic defects, curvaton, and multi-field inflation models, the amplitude of the non-Gaussianity can be well above the detection threshold for *Planck*. Clearly, the detection of even a small degree of primordial non-Gaussianity in the CMB would have potentially revolutionary implications for early Universe physics. The motivation for non-Gaussian models, and the ability of *Planck* to detect non-Gaussianity, is discussed in detail in § 2.5.

The Gaussian assumption also leads to the idea of *cosmic variance*, which limits our ability to constrain theoretical models. Each of the $a_{\ell m}$ coefficients in the expansion of Equation 2.1 is a random variate with mean zero and variance $C_\ell$. Thus each $C_\ell$ is distributed like $\chi^2$ with $(2\ell + 1)$ degrees of freedom. Since our observable patch of the Universe is only a single realisation of this stochastic process, there is a limit to how well we can constrain theoretical models, independently of how well we measure our actual sky. The fractional variance on each $C_\ell$ is approximately $(1/f_{\text{sky}})[2/(2\ell+1)]$, where $f_{\text{sky}}$ is the fraction of the sky that is uncontaminated by foregrounds. *Planck* is designed to be cosmic-variance limited across essentially the entire primary CMB temperature anisotropy power spectrum.

Since the discovery of primordial CMB anisotropies by the *COBE* satellite (Smoot et al. 1992), the tremendous potential of CMB anisotropies to revolutionize cosmology has been convincingly demonstrated by many ground-based and balloon-borne experiments and most recently by the *WMAP* satellite (Bennett et al. 2003). Figure 2.2 shows a summary of the CMB anisotropy measurements shortly before and after the first *WMAP* results. These diagrams plot estimators of the anisotropy amplitude $(\Delta T_\ell)^2$, which is related to the power spectrum $C_\ell$ according to

$$(\Delta T_\ell)^2 = \frac{1}{2\pi}\ell(\ell+1)C_\ell^T. \tag{2.3}$$

Notice that prior to the announcement of the *WMAP* results, the position of the first acoustic peak at $\ell \simeq 200$ had already been well established. This provided evidence that the Universe is close to spatially flat (though see §2.2.6). The CMB data prior to *WMAP* also provided strong evidence for the detection of a second and hints of a third acoustic peak, setting crude but useful constraints on the physical densities in baryons and cold dark matter.* More importantly, the presence of multiple, coherent, acoustic peaks provided powerful evidence that the primordial fluctuations were most likely adiabatic perturbations of the sort generated during an inflationary phase in the early Universe.

### 2.2.4 Current Knowledge of CMB Anisotropies

The righthand panel of Figure 2.2 shows the temperature anisotropy power spectrum determined from the first year's data from the *WMAP* satellite (Bennett et al. 2003). Notice the dramatic improvement in the precision of the measurements. The first and second acoustic peaks are measured with exquisite precision, though the errors become large at $\ell \gtrsim 500$ because of *WMAP*'s limited angular resolution and sensitivity. The solid lines in both panels show the predictions of the best-fit $\Lambda$-dominated Cold Dark Matter cosmology (henceforth referred to as $\Lambda$CDM). This model is spatially flat, with $\Omega_\Lambda = 0.73$, $\omega_c = 0.116$, $\omega_b = 0.024$, $h = 0.72$, and has scale-invariant adiabatic fluctuations. The parameters of this model are consistent with many other astronomical data-sets, including the 2dF and SDSS galaxy redshift surveys, the magnitude-redshift relation for Type Ia supernovae, direct measurements of the Hubble constant, and primordial Big-Bang nucleosynthesis. For these reasons, this model is often referred to as the *concordance* $\Lambda$CDM model (see e.g., Bahcall et al. 1999). At the time of writing, the concordance $\Lambda$CDM model appears to be consistent with *all* astrophysical data-sets, and it is clear from Figure 2.2 that it fits the CMB anisotropy data to very high precision.†

Despite the tremendous success of the concordance $\Lambda$CDM model, it is worth emphasising that some of the parameters of the model seem extraordinarily strange. For example, Figure 2.3 summarises our present views on the major constituents of the Universe. Ordinary baryons, the only major constituent that has been directly detected in the laboratory, account for only 4% of the mass-density budget of the Universe. Even here, only about a tenth of the ordinary baryons have condensed into stars—the rest are dark baryons, namely, the highly ionized gas in the diffuse intergalactic medium and the hotter denser gas in the central regions of clusters and groups of galaxies. Some other form(s) of weakly interacting matter accounts for the remaining 23% or so of the *clustered* matter in the Universe. We know neither the nature of this dark matter nor its evolutionary history.

---

* We will denote the physical densities of baryons and cold dark matter as $\omega_b = \Omega_b h^2$ and $\omega_c = \Omega_c h^2$, where $\Omega$ is the density in units of the critical density and $h$ is Hubble's constant in units of $100\,\text{km}\,\text{s}^{-1}\text{Mpc}^{-1}$.

† It has been argued that some astrophysical data-sets combined with the *WMAP* data require small deviations from the concordance $\Lambda$CDM model, but this evidence is currently far from conclusive, as discussed in §2.3.

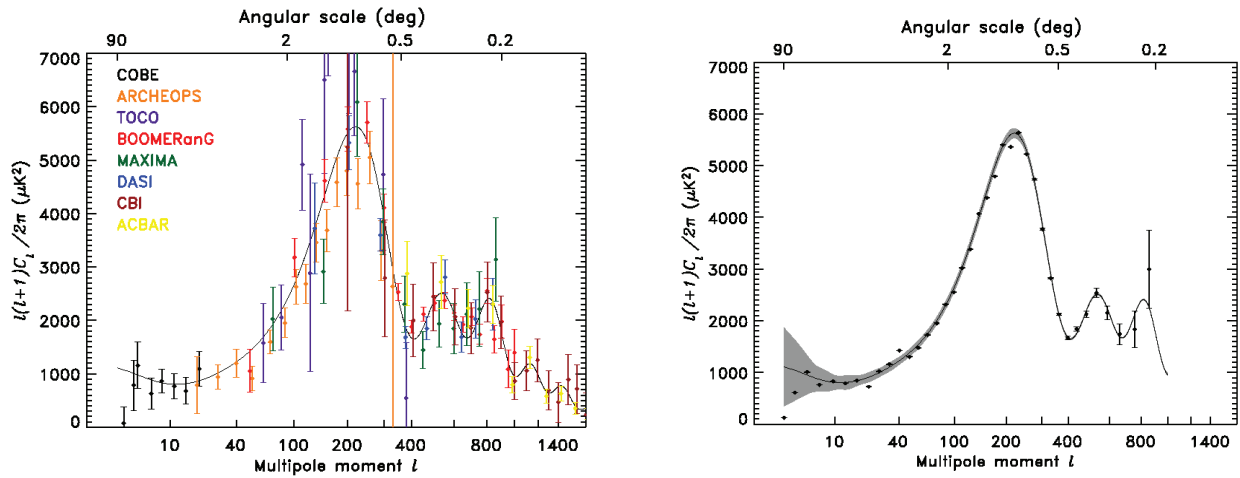

FIG 2.2.— The left panel (courtesy of the *WMAP* Science Team) shows a summary of CMB anisotropy measurements from various experiments prior to the release of the first year results from *WMAP*. The references to the experimental data are as follows: *COBE* (Tegmark 1996), Archeops (Benoit et al. 2003), TOCO (Miller et al. 2002), Boomerang (Ruhl et al. 2003), Maxima (Lee et al. 2001), DASI (Halverson et al. 2002), CBI (Pearson et al. 2003) and ACBAR (Kuo et al. 2004). The right panel shows results from the first year of *WMAP* data.

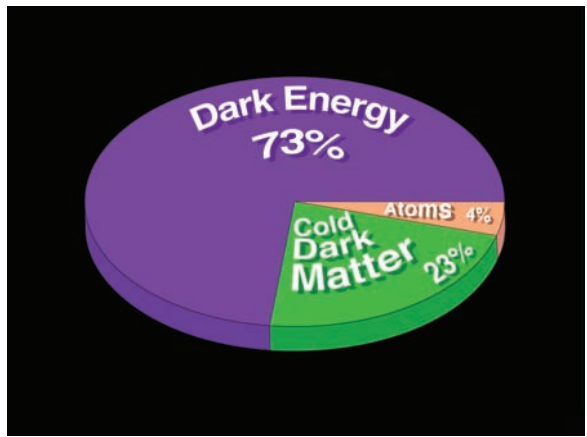

FIG 2.3.— Current picture for the composition of the Universe (from the *WMAP* web-site).

But the biggest surprise of all is the evidence that most of the mass density of the Universe is contributed *by empty space itself*. Apparently, the sea of virtual particles that, according to quantum field theory, pervades the vacuum, contributes by far the largest contribution to the energy budget of the Universe. This phenomenon is even less well understood than the weakly interacting dark matter. A naive calculation of the contributions of quantum zero-point fluctuations of fundamental fields would suggest a value for the cosmological constant of $\Lambda \simeq (c^3/\hbar G) \sim (L_{\mathrm{Pl}})^{-2} \sim 10^{69}$ m$^{-2}$, where $L_{\mathrm{Pl}}$ is the Planck length. Observations, however, suggest a value of $\Lambda = (3\Omega_\Lambda H_0^2/c^2) \sim (L_0)^{-2} \sim 10^{-52}$ m$^{-2}$, where $L_0 = c/H_0$ is our present Hubble radius. These two estimates disagree by over 120 orders of magnitude. This exceedingly small but non-zero value of the vacuum energy is one of the major unsolved problems in fundamental physics. Another way of phrasing the discrepancy is to note that the observed value of the cosmological constant introduces a characteristic energy scale of a few meV, much smaller than the Planck scale of $m_{\mathrm{Pl}} \sim 10^{19}$GeV.

In summary, over the last decade or so cosmologists have accumulated evidence using a variety of methods for a concordance ΛCDM cosmology with the parameters shown in Figure 2.3. The parameters of this model, however, are not understood even remotely. The physics behind 96% of the mass density of the Universe is not known—and in the case of the dark energy we do not even have a firm theoretical framework for tackling the problem. There may also be implications for the very foundations of fundamental physics. For example, it is not known how to formulate string theory within the asymptotically de-Sitter universe implied by Figure 2.3.

The other exciting result from *WMAP* is the measurement of the polarization of the CMB anisotropies. Since Thomson scattering is anisotropic, the CMB photons are expected to show a linear polarization of order a few percent. (The physics of CMB polarization, and in particular the ability of *Planck* to detect the polarization signature of gravitational waves, is discussed in detail in § 2.3.3.) The first detection of CMB polarization was made by the Degree Angular Scale Interferometer (DASI, Kovac et al. 2002). Figure 2.4 shows the temperature-polarization cross power spectrum $C^{TE}$ measured by *WMAP*. These results show the characteristic oscillatory behaviour expected of adiabatic initial fluctuations generated during an inflationary phase in the early Universe. At $\ell \lesssim 20$, *WMAP* detected a signal well above the adiabatic prediction. Such an excess can be produced by Thomson scattering at recent epochs. In fact, the *WMAP* results suggest a Thomson optical depth for such secondary reionization of $\tau \simeq 0.17$, indicating that the inter-galactic medium was reionized considerably earlier than previously thought. To achieve such a high optical depth requires reionization at a redshift $z > 10$ (perhaps as high as $z \sim 20$), implying that non-linear structures had already formed at these high redshifts. As we will see in § 2.3.3, this reionization bump in the polarization power spectrum significantly enhances the prospects of detecting the polarization signal of gravitational waves generated during inflation. Finding the objects whose radiation ended the cosmological dark ages is an important driver for many of the most ambitious optical, infra-red and radio astronomical facilities planned for the next decade. The CMB polarization signal provides measurements which are complementary to these experiments.

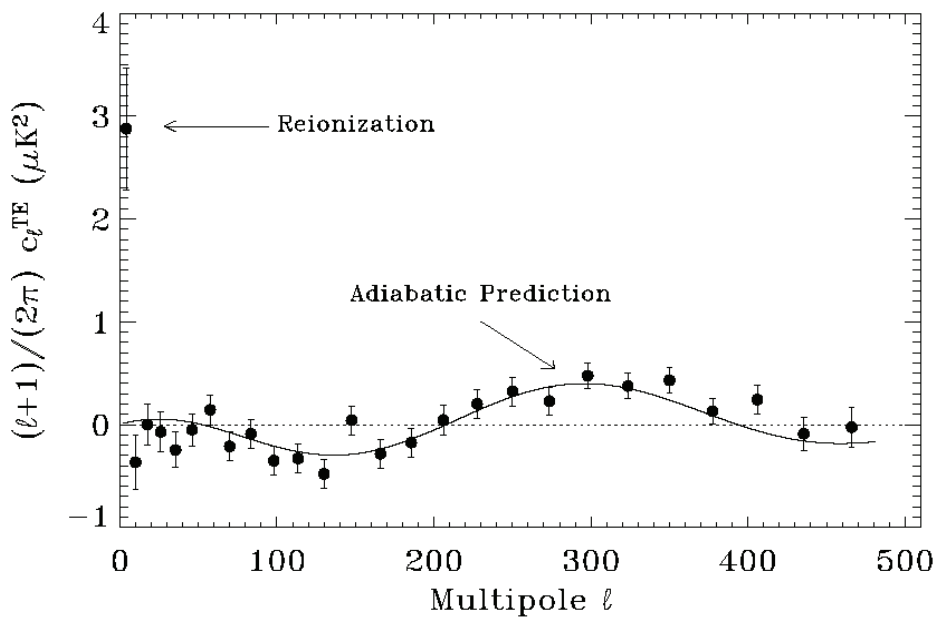

Fɪɢ 2.4.—The polarization-temperature cross power spectrum measured by *WMAP* (from Hinshaw et al. 2003). The solid line shows the predicted oscillatory signal for purely adiabatic fluctuations. The excess signal at $\ell < 20$ provides evidence that the Universe was reionized at early times.

## 2.2.5 Beyond WMAP

Given that *WMAP* has already been extremely successful, and with several more years of data yet to come, it is natural to ask whether anything remains to be done. The answer is resoundingly 'yes'. As explained below, *WMAP* can measure less than 10% of the information contained in the CMB temperature anisotropies, and only a tiny fraction of the information contained in the polarization anisotropies. *Planck* will measure essentially all of the temperature information in the CMB, and a substantial fraction of the polarization information. This translates into huge gains in our knowledge of the early Universe, the epoch of recombination and structure formation at late times. We present here an overview of some of *Planck*'s specific advantages.

*Planck* has three times the angular resolution of *WMAP*, an order of magnitude lower noise at the optimal frequency bands around 100 GHz (even for an 8-year *WMAP* mission), and frequency coverage from 30–857 GHz. The improved angular resolution and sensitivity will permit much more accurate estimates of the CMB power spectra. For the $C_\ell^T$ power spectrum alone, *Planck* is expected to be cosmic-variance limited to at least a factor of four higher in $\ell$ leading to a gain of at least a factor of 16 in the sheer number of modes measured with signal-to-noise ratio $\sim 1$. The wider frequency coverage ensures that *Planck* will be able to measure and separate the CMB from confusing foregrounds with the accuracy required by its low noise level. This will be especially important in the case of polarization. All of this results in a dramatic improvement in constraints on cosmological parameters, as discussed in the next section, and more importantly, in *Planck*'s unique new capabilities for constraining cosmology. These include:

- *Improved tests of inflationary models of the early Universe. Planck* has the sensitivity, polarization capability, and spatial dynamic range required to place strong limits on the slope of the primordial power spectrum, and any change or "running" of the slope with scale. Most inflationary models predict small deviations from an exact scale-invariant spectrum. These deviations are difficult for *WMAP* to measure, but easily detectable by *Planck* (see §2.3). The measurement of deviations from a pure scale invariant power spectrum would provide a firm probe of the *dynamics* of an inflationary phase.

- *Accurate estimates of cosmological parameters. Planck* will be the first high resolution, all-sky mission to map the entire CMB. This will enable precise measurements of the higher peaks in the spectrum, which in turn translate into accurate determinations of the values of important cosmological parameters. For example, by measuring the amplitudes of the higher peaks with high accuracy *Planck* will improve our knowledge of the baryon and dark matter densities by an order of magnitude. Similarly, *Planck* will lead to more accurate estimates of the distance to the last scattering surface and of the size of the sound-horizon at last scattering. These measurements can be combined with the next generation of observations of distant Type Ia Supernovae, large galaxy lensing surveys and galaxy redshift surveys, to learn more about the mysterious dark energy that dominates our Universe.

- *Estimates of the polarization power spectra. WMAP* has poor sensitivity to CMB polarization and will provide only crude measurements of the E-mode polarization spectrum, $C^E$, and is unlikely to detect the B-mode spectrum $C^B$. In contrast, *Planck* should be able to measure the E-mode spectrum to multipoles $\ell \sim 1500$ with high precision and good control over polarized foreground contamination. These measurements are critical for, amongst other things, determining the allowable modes of the primordial fluctuations (adiabatic vs various types of isocurvature mode), determination of the reionization history of the Universe, and setting constraints on features in the primordial power spectrum. *Planck* may also detect the B-mode polarization anisotropies, if tensor modes contribute at a level of a few percent or more of the amplitude of the scalar modes. The amplitude of the tensor component remains a free parameter of inflationary models of the early Universe, but amplitudes of this order are predicted in an important class of models, namely the 'chaotic' inflationary models advocated by Linde (see, e.g., Linde 1983, 1991). The detection of primordial gravity waves would provide unassailable proof that the Universe went through a period of inflation and would establish the energy scale of the inflationary phase.

- *Tests of non-Gaussianity.* As mentioned earlier in this Section, in most inflationary models the fluctuations should be accurately Gaussian. However, a number of authors have claimed to find evidence for non-Gaussianity in the first year's data from *WMAP*. These include anomalies in the orientations of the low order multipoles, departures from isotropy in the CMB temperature power spectrum, and peculiarities in the statistics of hot and cold spots (see §2.5). At present, it is not known whether these effects are real features of the early Universe or caused by systematic errors in the *WMAP* data. It is possible that further observations by *WMAP* will resolve these discrepancies and confirm that the fluctuations are close to Gaussian. Alternatively, mysterious anomalies may remain which defy theoretical explanation. Either way, *Planck* has something to offer, firstly as an independent all-sky experiment with different systematics from *WMAP*, and secondly providing greater sensitivity to any non-Gaussian signatures that may exist.

- *Significant new secondary science probes.* A host of new higher-order effects can be measured because of the sensitivity, angular resolution, and frequency coverage of *Planck*. Examples include the thermal and kinetic Sunyaev-Zel'dovich effects from rich clusters of galaxies, lensing of the CMB to measure the matter power spectrum and potentially the neutrino mass, the integrated Sachs-Wolfe effect as a probe of low-$z$ physics (perhaps constraining fluctuations in the dark energy), and a range of related cross-correlation studies which could further constrain the properties of the dark sector of the Universe.

## 2.3 Cosmological Parameters from Planck

This section presents a detailed analysis of cosmological parameter estimation with *Planck*, and contrasts *Planck* with other CMB experiments and *WMAP* in particular. §2.3.1 introduces some of the problems caused by parameter degeneracies, i.e., parameter combinations that produce almost identical CMB anisotropy spectra. §2.3.2 shows how *Planck* measurements will break parameter degeneracies and set constraints on the dynamics of inflation. This section also summarizes inflationary models and demonstrates the effectiveness of *Planck* in constraining the form of the primordial fluctuation spectrum *independently of any other astrophysical data*. The importance of CMB polarization measurements is discussed in §2.3.3, where we present forecasts of what *Planck* might achieve. A more complete analysis of parameter degeneracies, contrasting *WMAP* and *Planck* is presented in §2.3.4.

### 2.3.1 Parameter Degeneracies

It is well known that parameters estimated from CMB anisotropies show strong degeneracies (e.g., Bond et al. 1997; Zaldarriaga, Spergel & Seljak 1997; Efstathiou & Bond 1999). This section describes a few examples of degeneracies that affect *WMAP* and other current CMB data.

The best known example is the *geometrical degeneracy* between the matter density, vacuum energy, and curvature. Models will have nearly identical CMB power spectra if they have identical initial fluctuation spectra, reionization optical depth, and if they have identical values of $\omega_{\rm b}$, $\omega_{\rm c}$ and acoustic peak location parameter

$$\mathcal{R} = \frac{\Omega_{\rm m}^{1/2}}{|\Omega_k|^{1/2}} \sin_k \left[ |\Omega_k|^{1/2} y \right], \tag{2.4a}$$

where

$$y = \int_{a_{\rm r}}^{1} \frac{da}{[\Omega_{\rm m} a + \Omega_k a^2 + \Omega_\Lambda a^4]^{1/2}}, \tag{2.4b}$$

measures the distance to last scattering, with $a_{\rm r}$ the scale factor at recombination normalised to unity at the present day, and the function $\sin_k$ depends on the curvature $\Omega_k$ ($\equiv 1 - \Omega_{\rm m} - \Omega_\Lambda$):

$$\sin_k x = \begin{cases} \sinh x, & \Omega_k > 0; \\ x, & \Omega_k = 0; \\ \sin x, & \Omega_k < 0. \end{cases} \tag{2.4c}$$

This geometrical degeneracy is almost exact and precludes reliable estimates of either $\Omega_\Lambda$ or the Hubble parameter $h$ from measurements of the CMB anisotropies alone. As an example,

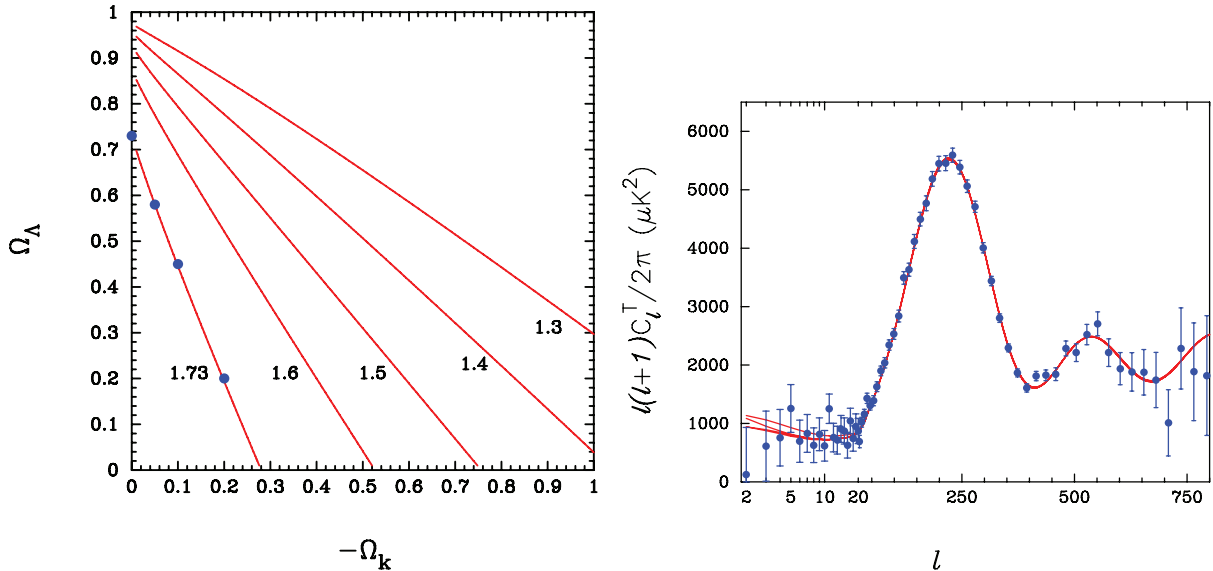

FIG 2.5.—The geometrical degeneracy. The left panel shows contours in the $\Omega_\Lambda$–$\Omega_k$ plane for which the parameter $\mathcal{R}$ defined in Equation 2.4a is constant, with values of $\mathcal{R}$ indicated next to each curve. Models with identical initial fluctuation spectra, matter content, and $\mathcal{R}$ will have almost identical CMB power spectra. This is illustrated in the right panel, which shows the *WMAP* points and CMB power spectra for models with values of $\Omega_\Lambda$ and $\Omega_k$ corresponding to the four points in the left hand panel. These models are highly degenerate, showing that the CMB anisotropy data alone cannot strongly constrain both the geometry of the Universe and the value of $\Omega_\Lambda$.

Figure 2.5 shows the temperature power spectrum measured by *WMAP* together with a scale-invariant $\Lambda$CDM model with $\omega_b$, $\omega_c$, $h$ and $\tau$ fixed to the *WMAP* best fit values. The other curves in the figure show nearly degenerate closed models with $\Omega_k = -0.05$, $-0.10$ and $-0.20$.

The left panel of Figure 2.5 shows lines of constant $\mathcal{R}$ in the $\Omega_\Lambda$–$\Omega_k$ plane and the four blue dots show the parameters of the degenerate curves plotted in the right panel. Notice that this degeneracy is essentially exact, except for minor differences at low multipoles which are dwarfed by the large cosmic variance of the measurements. It is often stated that the location of the CMB acoustic peaks constrains the Universe to be nearly spatially flat. However, as this example shows, *CMB anisotropy data alone* cannot break the geometrical degeneracy between $\Omega_\Lambda$ and $\Omega_k$. The example shown in Figure 2.5 illustrates that universes with substantial spatial curvature are compatible with the *WMAP* CMB data.

The geometrical degeneracy illustrated above can, of course, be broken by combining CMB data with other cosmological data-sets, as illustrated in Figure 2.6. Examples of these other constraints include:

- Direct estimates of Hubble's constant, such as the constraint from the HST Key Project team, $H_0 = 72 \pm 8 \, \text{km s}^{-1} \, \text{Mpc}^{-1}$ (Freedman et al. 2001).

- Determination of the magnitude-redshift relation for Type Ia supernovae (Knop et al. 2004; Riess et al. 2004).

- Measurements of the power spectrum of the galaxy distribution determined from large galaxy redshift surveys (Efstathiou et al. 2002; Percival et al. 2002; Tegmark et al. 2004).

The geometrical degeneracy discussed above is an extreme example because it is almost exact. There are many more partial degeneracies between parameters determined from CMB anisotropies alone. Two examples are shown in Figure 2.7. The left panel shows the degeneracy between the scalar mode spectral index $n_S$, baryon density $\omega_b$, and optical depth $\tau$ for the *WMAP* data. Clearly, without invoking additional data the *WMAP* constraints on $n_S$ are nowhere near the accuracy required (better than $\pm 0.05$) to place interesting constraints on inflationary models of the early Universe (see §2.3.2). Furthermore, there is an unfortunate feature in that the value of a parameter that is sensitive to early universe physics, $n_s$, is strongly degenerate on another parameter, $\tau$, which depends on complex and uncertain late time physics (the formation of the first stars). The right panel of Figure 2.7 shows the constraints on the amplitude of the scalar fluctuations (expressed by the parameter $\sigma_8$, which is the rms matter density contrast at the present epoch averaged in spheres of radius $8h^{-1}$ Mpc) and the matter density. As can be seen, these parameters are poorly determined because $\sigma_8$ is strongly affected by uncertainties in $\tau$ and $n_S$, while $\Omega_m$ is strongly affected by the geometrical degeneracy.

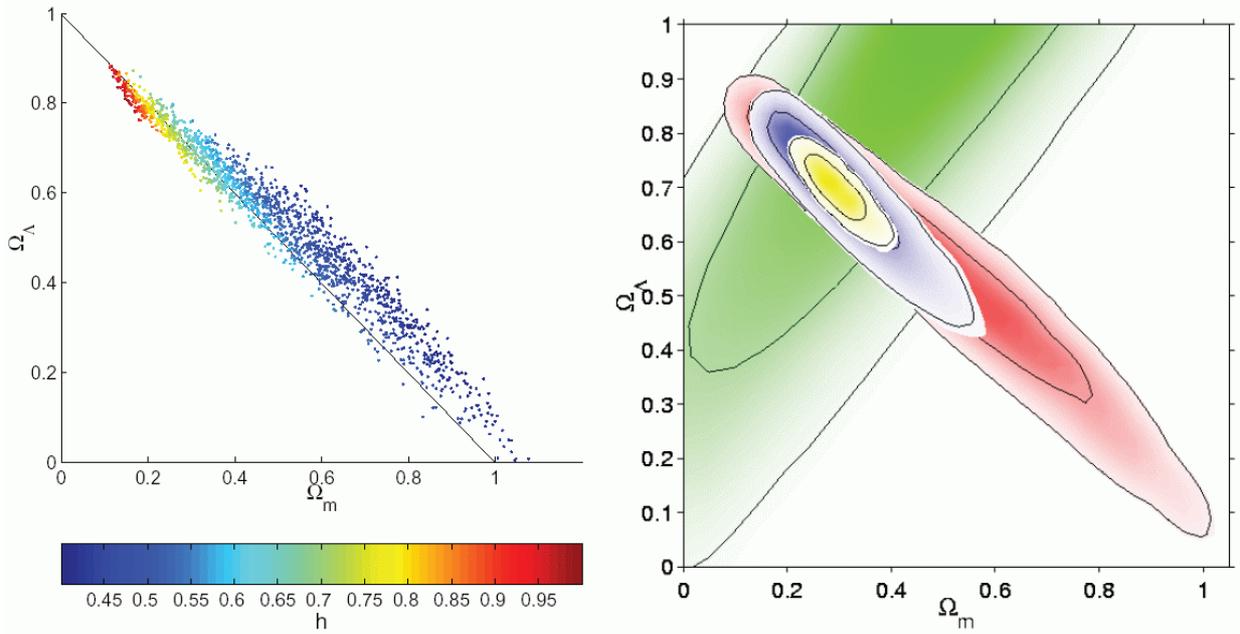

FIG 2.6.—Breaking the geometric degeneracy (from Lewis & Bridle 2002). The left panel shows samples of the posterior distribution in the $\Omega_m$–$\Omega_\Lambda$ plane from the pre-*WMAP* CMB data. The points are colour coded according to the value of the Hubble parameter $h$. The right panel shows 68% and 98% confidence contours for various data-sets. The supernovae constraints alone are shown by the green contours; the CMB data alone are shown by the red contours; CMB+HST key project constraints on $h$ are shown by the blue contours; finally, the yellow contours show the constraints derived by combining adding additional constraints from the distribution of galaxies in the 2dF Galaxy Redshift Survey and from primordial nucleosynthesis (in the form of a Gaussian prior of $\omega_b = 0.020 \pm 0.002$).

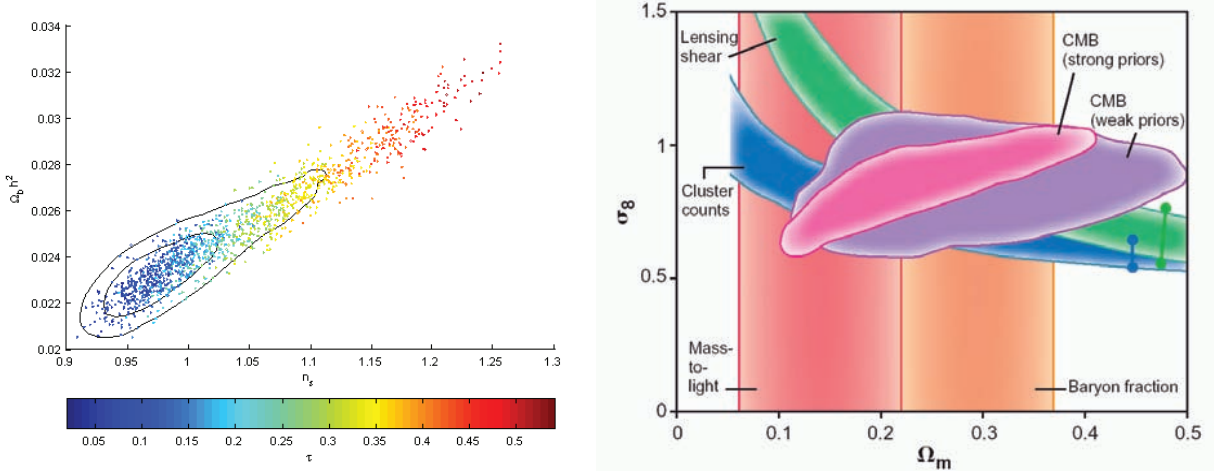

FIG 2.7.—The left panel shows constraints from *WMAP* in the $\omega_b$–$n_S$ plane (points, colour coded according to optical depth $\tau$) and from *WMAP* combined with CMB anisotropy measurements at smaller scales and with 2dF galaxy clustering data. The right panel shows constraints on the amplitude of the fluctuations, $\sigma_8$, and $\Omega_m$ for several data-sets. The contours labelled CMB show the constraints imposed by *WMAP* with weak priors (i.e., assuming a $\Lambda$CDM universe with a pure power law scalar spectral index and no tensor component) and with strong priors (i.e., where all parameters except $h$, $\omega_b$ and $n_S$ are assumed known). The green and blue contours show constraints from weak gravitational lensing surveys and from estimates of the abundance of rich clusters of galaxies; the widths indicate the ranges of results reported in the literature. The vertical bands show constraints on $\Omega_m$ derived from the mass-to-light ratios and gas content of rich clusters of galaxies (From Bridle et al. 2002).

The results from *WMAP* and other CMB anisotropy experiments have clearly opened up the era of precision cosmology and demonstrate the promise of the CMB for probing fundamental physics. Nevertheless, the CMB anisotropy measurements are strongly affected by parameter degeneracies, leading to the following two consequences:

- Accurate estimates of cosmological parameters require the addition of astrophysical constraints, for example from Type Ia supernovae, large galaxy redshift surveys, or fluctuations in the Ly $\alpha$ absorption lines in the spectra of distant quasars. However, these astrophysical data-sets are usually much more complex than observations of the CMB and often have unquantified systematic errors. Are Type Ia supernovae really standard candles from $z = 0$ to high redshift? Do galaxies accurately trace the matter fluctuations on large scales? Can the Ly $\alpha$ absorption lines really be used to trace the distribution of matter on scales of $\sim 1\,\mathrm{Mpc}$? Systematic errors in complex astrophysical data-sets can, potentially, lead to serious biases in cosmological parameters (e.g., Seljak et al. 2003).

- The degeneracies between parameters become more severe as the number of parameters is increased. In other words, estimates of cosmological parameters are sensitive to theoretical assumptions. Adding a gravitational wave component to the CMB anisotropies, admitting non-power law fluctuation spectra, or a quintessence-like component for the dark energy, can significantly affect the values of cosmological parameters. In extreme cases, for example allowing an arbitrary admixture of isocurvature modes (see § 2.4.3), the cosmological parameters from a mission with the sensitivity and resolution of *WMAP* become quite indeterminate (e.g., Bucher, Moodley & Turok 2001).

These problems can be largely solved by making higher sensitivity measurements of the CMB temperature and polarization anisotropies. The left panel of Figure 2.8 shows a simulated CMB power spectrum for the concordance $\Lambda$CDM model after four years observation with *WMAP*. The panel to the right shows the same realisation observed by *Planck* for one year. The improvement in accuracy is dramatic and comparable to the huge improvement achieved by *WMAP* over previous CMB experiments illustrated in Figure 2.2. According to the instrumental specifications of *Planck*, the CMB power spectrum can be measured to about cosmic variance accuracy (§ 2.2.3) over the full range of multipoles $2 \lesssim \ell \lesssim 2500$. In practice, various systematic effects will need to be accounted for (e.g., diffuse and unresolved foregrounds, beam calibration errors) and may dominate the error budget over certain multipole ranges. However, the instruments and data processing have been designed to minimize such residual systematic errors. Obtaining a nearly cosmic-variance-limited power spectrum over such a broad multipole range has been one of the main science drivers of the *Planck* mission. Such measurements open up a wide range of new science. In particular, *Planck* will be able to:

- set tight constraints on cosmological parameters with far less reliance on theoretical assumptions and complex astrophysical data;

- study the ionization history of the Universe (§ 2.3.3);

- probe the dynamics of the inflationary era (§ 2.3.2);

- test fundamental physics beyond inflation, e.g., brane-world or pre-Big Bang cosmologies (§ 2.4.4).

These science goals are discussed in greater detail below.

### 2.3.2  Planck and Inflation

The origin of the fluctuations that grew to make clusters, galaxies and all of the other structure in our Universe poses a fundamental problem. Until recently, there was no plausible physical mechanism that could explain where these fluctuations came from. It is easy to see why exotic physics needs to be invoked: a map of the CMB temperature anisotropies provides a snapshot of the fluctuations at the time of recombination, when the Universe was only $t_{\mathrm{r}} \sim 400{,}000$ years old. However, the distance that light could have travelled within this timescale, $L \sim ct_r$, would subtend an angle of *only about a degree* on the sky. The very existence of the CMB temperature anisotropies observed by *COBE* , on angular scales $\gtrsim 7°$, points to new physics, since the fluctuations could not, according to this naive argument, have been in causal contact at the time of recombination.

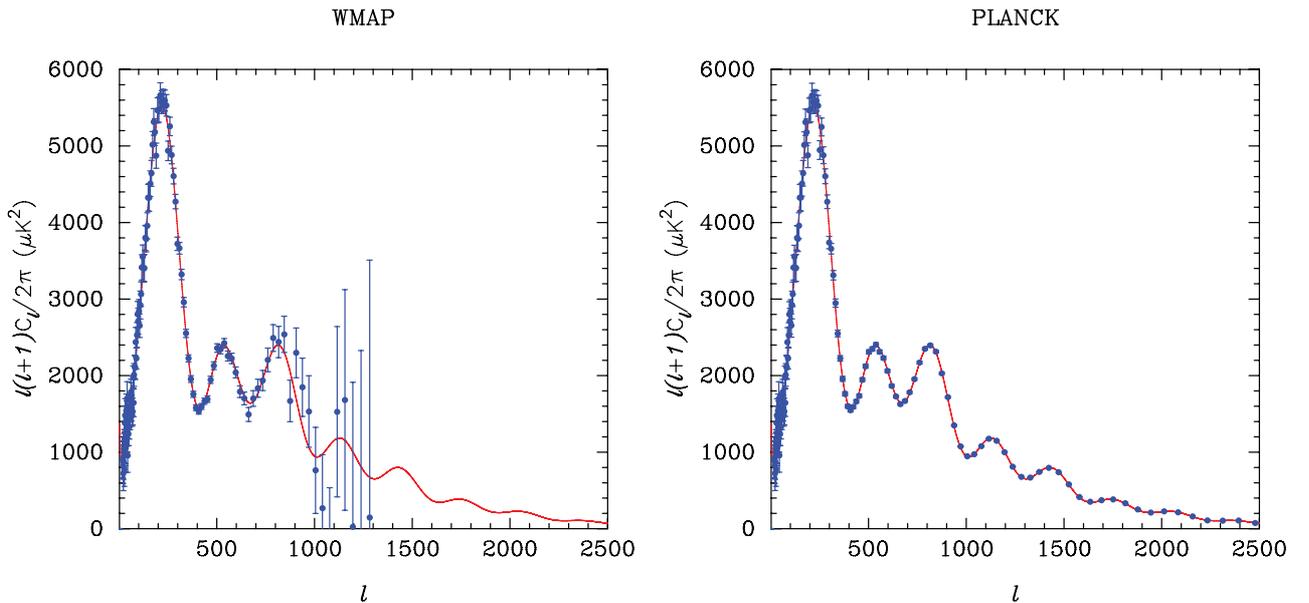

FIG 2.8.—The left panel shows a realisation of the CMB power spectrum of the concordance ΛCDM model (red line) after 4 years of *WMAP* observations. The right panel shows the same realisation observed with the sensitivity and angular resolution of *Planck*.

Inflation offers a solution to this apparent paradox. The usual Friedman equation for the evolution of the cosmological scale factor $a(t)$ is

$$H^2 = \left(\frac{\dot{a}}{a}\right)^2 = \frac{8\pi G}{3}\rho - \frac{k}{a^2}, \tag{2.5}$$

where dots denote differentiation with respect to time and the constant $k$ is positive for a closed universe, negative for an open universe and zero for a flat universe. Local energy conservation requires that the mean density $\rho$ and pressure $p$ satisfy the equation

$$\dot{\rho} = -3\left(\frac{\dot{a}}{a}\right)(\rho + p). \tag{2.6}$$

Evidently, if the early Universe went through a period in which the equation of state satisfied $p = -\rho$, then according to Equation 2.6 $\dot{\rho} = 0$, and Equation 2.5 has the (attractor) solution

$$a(t) \propto \exp(Ht), \qquad H \simeq \text{constant}. \tag{2.7}$$

In other words, the Universe will expand nearly exponentially. This phase of rapid expansion is known as *inflation*. During inflation, neighbouring points will expand at superluminal speeds and regions which were once in causal contact can be inflated in scale by many orders of magnitude. In fact, a region as small as the Planck scale, $L_{\text{Pl}} \sim 10^{-35}$ m, could be inflated to an enormous size of $10^{10^{12}}$ m—many orders of magnitude larger than our present observable Universe ($\sim 10^{26}$ m)!

As pointed out forcefully by Guth (1981), an early period of inflation offers solutions to many fundamental problems. In particular, inflation can explain why our Universe is so nearly spatially flat without recourse to fine-tuning, since after many e-foldings of inflation the curvature term $(k/a^2)$ in Equation 2.5 will be negligible. Furthermore, the fact that our entire observable Universe might have arisen from a single causal patch offers an explanation of the so-called horizon problem (e.g., why is the temperature of the CMB on opposite sides of the sky so accurately the same if these regions were never in causal contact?). But perhaps more importantly, inflation also offers an explanation for the origin of fluctuations.

In the simplest inflationary models, the accelerated expansion is driven by a single scalar field, $\phi$, sometimes known as the *inflaton*. Its energy and pressure are given by

$$\rho = \frac{1}{2}\dot{\phi}^2 + V(\phi), \quad p = \frac{1}{2}\dot{\phi}^2 - V(\phi), \qquad (2.8)$$

where we have assumed for simplicity that $\phi$ is a real scalar field with potential $V(\phi)$. The equation of motion of the scalar field is given by

$$\ddot{\phi} + 3H\phi = -\frac{\partial V}{\partial \phi}. \qquad (2.9)$$

We can see immediately from Equations 2.8 that if the field evolves slowly ($\dot{\phi} \ll V(\phi)$), then the equation of state satisfies $p \simeq -\rho$ and an inflationary phase is possible with

$$H^2 = \frac{8\pi G}{3}V(\phi) = \frac{8\pi}{3m_{\mathrm{Pl}}^2}V(\phi), \qquad (2.10)$$

where, henceforth, we will use Planck units. Evidently, in this simple model, inflation can only occur if the scalar field satisfies certain so-called slow-roll conditions. These conditions can be expressed conveniently in terms of the slow-roll parameters $\epsilon$ and $\eta$:

$$\epsilon = \frac{m_{\mathrm{Pl}}^2}{16\pi}\left(\frac{V'}{V}\right)^2, \quad \eta = \frac{m_{\mathrm{Pl}}^2}{8\pi}\left(\frac{V''}{V}\right). \qquad (2.11)$$

Generally, a successful model of inflation requires $\epsilon$ and $\eta$ to be less than unity.

Any scalar field will have quantum fluctuations $\delta\phi$. These quantum fluctuations will oscillate for wavenumbers $k < aH$, i.e., while the physical size of the fluctuation is smaller than the Hubble radius $ct$. However, during the inflationary phase, fluctuations will be inflated in scale to sizes many orders of magnitude larger than the Hubble radius. When $k > aH$, the fluctuations freeze and become classical perturbations described by General Relativity rather than by quantum field theory. According to these ideas, all of the fluctuations seen in the Universe today, from the scales of grains of sand to the super-Hubble radius anisotropies first detected by *COBE*, originated as quantum fluctuations in the very early Universe.

During the inflationary phase, the fluctuations in $\delta\phi$ generate scalar perturbations in the curvature. It is these scalar fluctuations that grow to produce the structure that we see today. In addition, gravitational waves are generated during inflation which introduce a tensor contribution to the fluctuations. The power spectra of the fluctuations in scalar and tensor components at the end of inflation can be expressed as

$$P_{\mathrm{S}}(k) \simeq \left(\frac{H^2}{16\pi^3\dot{\phi}^2}\right)_{(k=aH)} \qquad \text{(scalar perturbations)}, \qquad (2.12a)$$

$$P_{\mathrm{T}}(k) \simeq \left(\frac{H^2}{4\pi^2 m_{\mathrm{Pl}}^2}\right)_{(k=aH)} \qquad \text{(tensor perturbations)}, \qquad (2.12b)$$

and each of these modes will generate a corresponding anisotropy pattern in the CMB. An example for a simple scale-invariant inflationary model is shown in Figure 2.9.

Notice that the scalar and tensor CMB power spectra have almost identical shapes at multipoles $\ell \lesssim 20$. However, the tensor component dies away at higher multipoles and does not show the acoustic peak structure of the scalar power spectrum. This is because the gravitational wave perturbations decay when their physical scale becomes smaller than the Hubble radius.

Although inflation generically predicts a mixture of scalar and tensor perturbations, their relative amplitudes are highly model dependent. The amplitude of the tensor component depends on the fourth power of the energy scale of inflation (via Equations 2.12b and 2.10). The

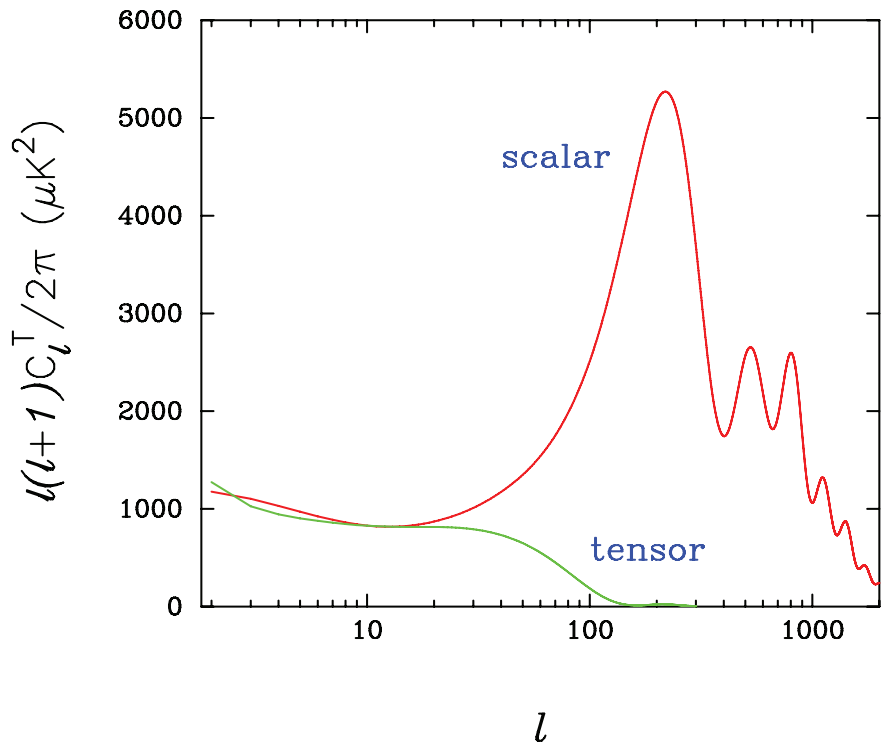

FIG 2.9.—The contributions to CMB anisotropies from scalar and tensor (gravitational wave) perturbations generated during inflation. The concordance ΛCDM model has been assumed with exactly scale-invariant initial fluctuations. The amplitudes of the tensor and scalar power spectra have been chosen arbitrarily to be equal at $\ell = 10$.

observed CMB anisotropies at low multipoles constrain this energy scale in simple inflationary models to be

$$V_{\text{inf}} \lesssim (3.5 \times 10^{16} \text{GeV})^4,$$ (2.13)

where $V_{\text{inf}}$ is the amplitude of the inflaton potential when fluctuations of wavelength comparable to the size of our present Hubble radius crossed the horizon during the inflationary phase. Thus the energy scale of inflation must be more than three orders of magnitude lower than the Planck scale. Generally, the ratio of amplitudes of the scalar and tensor CMB power spectra at low multipoles is given by

$$r \equiv \frac{C_\ell^{\text{T}}}{C_\ell^{\text{S}}} \sim 10\epsilon,$$ (2.14)

where the exact numerical factor on the right-hand side of this equation depends on the multipole at which the ratio is evaluated and on the values of certain cosmological parameters, in particular the cosmological constant $\Lambda$.

The spectral indices for the scalar and tensor power spectra, $n_{\text{S}}$ and $n_{\text{T}}$, can be expressed to first order in terms of the two slow-roll parameters $\epsilon$ and $\eta$ as

$$n_{\text{S}} - 1 \simeq -6\epsilon + 2\eta,$$ (2.15a)

and

$$n_{\text{T}} \simeq -2\epsilon.$$ (2.15b)

Furthermore, these spectra will not necessarily be pure power laws. The deviation from a pure power law can be expressed in terms of the following derivative, sometimes called the *run* in the spectral index:

$$\frac{dn_{\text{S}}}{d\ln k} \simeq -16\epsilon\eta + 24\epsilon^2 + 2\xi^2,$$ (2.16)

where the parameter $\xi$ is

$$\xi^2 = \frac{m_{\text{Pl}}^4}{(8\pi)^2} \left( \frac{V'V'''}{V^2} \right).$$ (2.17)

As can be seen from these equations, the run in the spectral index is of second order in the slow-roll parameters and so is much less than unity in most models of inflation.

Although inflation provides an attractive framework for solving some of the major puzzles in cosmology, there is no accepted particle physics *model* of inflation. At present, there is a large literature on phenomenological models of inflation, but no consensus on the underlying particle physics and few constraints from observations. To navigate through the range of possible models, it is useful to adopt a simple classification scheme (e.g., Kinney 2002):

**Large Field Models ($|\epsilon| < \eta$)**—The chaotic inflationary class of models pioneered by Linde (1983) falls into this category. In this type of model, the inflaton field begins at values much greater than the Planck scale $\phi \gg m_{\mathrm{Pl}}$. If the potential is steep enough, inflation can begin at Planck energy scales and the entire Universe can inflate from a single Planck-sized patch. Inflation ends when the slow-roll conditions are no longer satisfied, typically when the inflaton field drops to a value of a few $m_{\mathrm{Pl}}$. The perturbations on scales of galaxies and clusters are generated towards the end of the inflationary phase. The archetypal chaotic inflationary model has a power-law potential, e.g.,

$$V(\phi) = \frac{\lambda}{4}\phi^4, \tag{2.18}$$

though there is no compelling theoretical reason to favour any particular form of potential.

Chaotic inflationary models generally predict small departures from a scale-invariant fluctuation spectrum and a high amplitude for the tensor component. For example, for the quartic potential of Equation 2.18,

$$\epsilon \simeq 0.02, \quad \eta \simeq 0.03, \quad n_{\mathrm{S}} \simeq 0.06, \quad \frac{dn_{\mathrm{S}}}{d\ln k} \simeq 0.001, \quad r \simeq 0.2, \tag{2.19}$$

where we have assumed $N = 50$ e-foldings between the time that cluster-scale fluctuations crossed the horizon and the time that inflation ends.

**Small Field Models ($\eta < -\epsilon$)**—The high field models described above require $\phi > m_{\mathrm{Pl}}$. As it is not clear whether standard quantum field theory is applicable at such high field values, a large number of authors have investigated whether inflation can be realised in a low-energy effective field theory. The archetype for such 'small field' inflationary models is the new inflationary model, proposed by Linde (1982) and Albrecht & Steinhardt (1982), in which the effective inflaton potential arises from a spontaneous grand-unified symmetry breaking. Typically, in this type of model, the field $\phi$ starts at $\phi = 0$ and rolls slowly down a shallow potential, for example,

$$V(\phi) = V_0(1 - \mu\phi^p). \tag{2.20}$$

Such models are plagued by unnatural small coupling constants. Furthermore, there are problems with initial conditions. For example, if the scale $V_0$ is associated with a GUT scale of $\lesssim 10^{16}\mathrm{GeV}$, the Universe must already have achieved a size of $\gtrsim 10^6$ Planck lengths before inflation could begin. What determined this scale and how did the Universe achieve this state? In addition, to explain the amplitude of the fluctuations seen today, the inflaton field must have been weakly coupled to other forms of matter, in which case why should the field have started at $\phi \simeq 0$? As a result of these and other problems, it has proved difficult to find a compelling particle physics motivation for this type of inflationary scenario. Nevertheless, such models generically predict nearly scale invariant fluctuations, with a negligible tensor mode contribution.

**Hybrid Models ($0 < \epsilon < \eta$)**—In this type of model, the inflaton field typically rolls from a high value (possibly greater than $m_{\mathrm{Pl}}$) towards $\phi = 0$. This set-up requires a mechanism to end inflation, e.g., when the field rolls to some critical value $\phi = \phi_{\mathrm{c}}$. The archetypal hybrid inflation model uses two coupled scalar fields with a potential (e.g., Linde 1991)

$$V = V_0 + \frac{1}{2}m^2\phi^2 - \frac{1}{2}m_\psi^2\psi^2 + \frac{1}{4}\lambda\psi^4 + \frac{1}{2}\lambda'\psi^2\phi^2. \tag{2.21}$$

In this model, most of the energy density during inflation is supplied by the field $\psi$. Once $\phi$ rolls down to a critical value $\phi_c = m_\psi / \sqrt{\lambda'}$, the $\psi$ field is destabilised and rolls down to its true vacuum, ending inflation. Viable hybrid inflationary scenarios can be constructed with apparently 'natural' (i.e., not finely tuned) parameters. This is attractive, because it offers the possibility that a particle physics explanation for hybrid inflation might be found (see, e.g., the review by Lyth & Riotto 1999). A different mechanism for realising hybrid inflation, motivated by brane interactions in string theory, is discussed in § 2.4.4.

These three classes of inflationary model are plotted in the $r$–$n_S$ plane in Figure 2.10. At present, the parameters of inflationary models are poorly constrained by observations. The *WMAP* data, combined with data on large-scale structure, give a weak constraint $r \lesssim 0.8$. The $2\sigma$ constraint on $n_S$ is shown in Figure 2.7 and spans the broad range $0.93 \lesssim n_S \lesssim 1.1$. Evidently, much tighter constraints on these parameters are required before we can genuinely probe the dynamics of inflation.

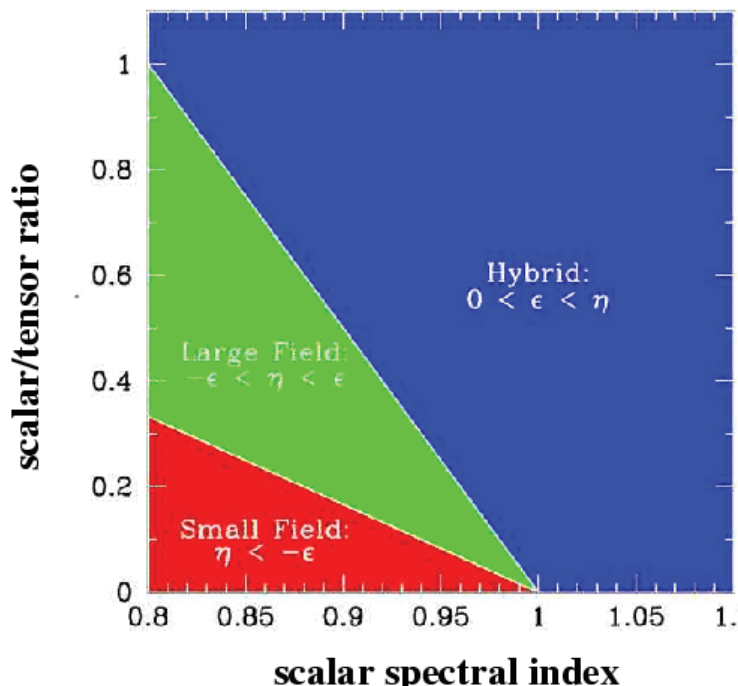

FIG 2.10.—The three classes of inflationary models discussed in the text, delineated in the plane of the tensor-to-scalar ratio $r$ (Eq. 2.14) versus scalar spectral index $n_S$ (adapted from Kinney, Melchiorri and Riotto, 2001)

As an illustration of what can be achieved by *Planck*, Figure 2.11 shows simulated CMB power spectra for a $\Lambda$CDM model with a spectral index of $n_S = 0.95$. The solid lines in the upper panels show the concordance $\Lambda$CDM model with an exactly scale-invariant spectrum, $n_S = 1$. On the scale of these plots, the differences between the $n_S = 1$ and $n_S = 0.95$ models are hardly visible. The lower panels therefore show the differences between the two power spectra. The left-hand panels show simulated *WMAP* data, and the right-hand panels show simulated *Planck* data. As can be seen, *WMAP* can barely distinguish between these two models even if all other cosmological parameters are assumed to be known. In contrast, *Planck* can easily distinguish between them (at several thousand standard deviations).

One of the most controversial results from the *WMAP* first year analysis is the possible evidence for a non-zero run in the spectral index (Spergel et al. 2003). By combining *WMAP*, other CMB data at smaller angular scales, 2dF data on galaxy clustering, and quasar Ly $\alpha$ data on the clustering of matter on small scales, there is a hint of a run in the spectral index of $dn_S / d\ln k = -0.031^{+0.016}_{-0.017}$. This result is heavily dependent on the accuracy of the Ly $\alpha$ constraints (e.g., Croft et al. 2002), which may be strongly affected by a variety of systematic errors. As a result, the claim for a detection of a run has not gained widespread acceptance. Furthermore, a run as large as claimed by the *WMAP* team seems strange from the theoretical point of view, since it should generally be of second order in the slow-roll parameters (equations 2.16 and 2.19). Although it is possible to construct inflationary models which can match these results, the parameters of such models are far from natural.*

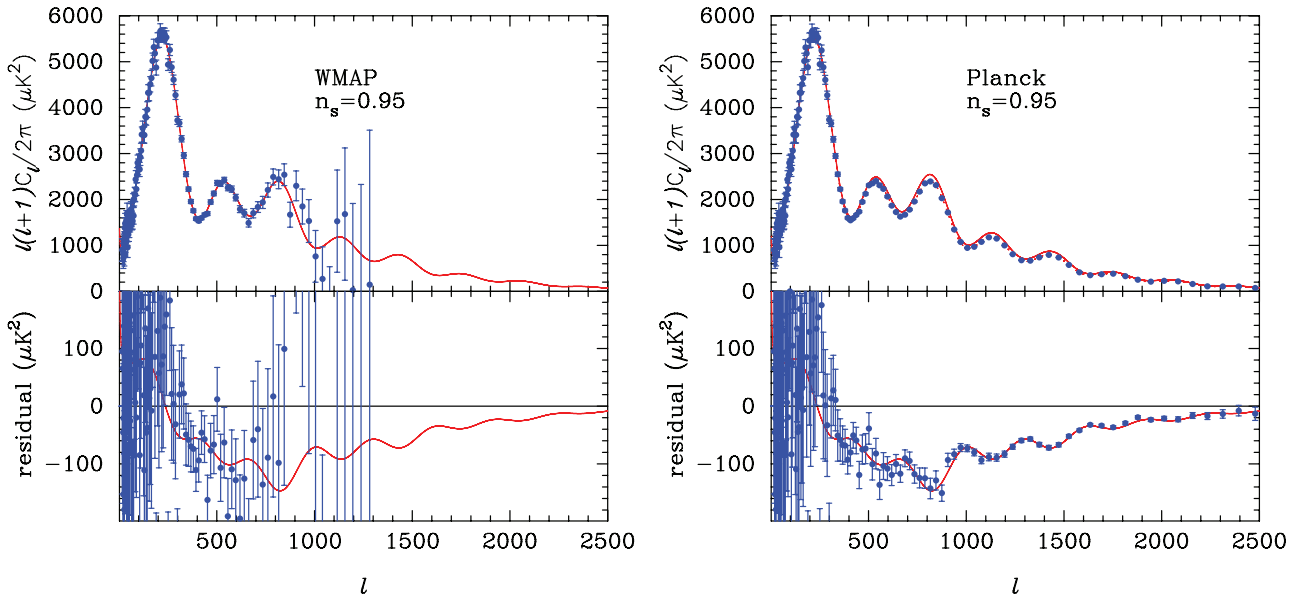

FIG 2.11.—The solid lines in the upper panels of these figures show the power spectrum of the concordance ΛCDM model with an exactly scale invariant power spectrum, $n_S = 1$. The points, on the other hand, have been generated from a model with $n_S = 0.95$ but otherwise identical parameters. The lower panels show the residuals between the points and the $n_S = 1$ model, and the solid lines show the theoretical expectation for these residuals. The left and right plots show simulations for *WMAP* and *Planck*, respectively.

Figure 2.12 shows an analogous plot to Figure 2.11, but now illustrating how well *WMAP* and *Planck* can distinguish between a model with a pure power law spectrum $n_S = 0.95$ (zero run), and a model with $n_S = 0.95$ and a run of $-0.03$, as claimed from *WMAP*. As this figure shows, *WMAP* has neither the lever-arm of a large range in $\ell$, nor the sensitivity, to distinguish between these models. Any detection of a run from *WMAP* must therefore rely on other data. In contrast, *Planck* has both the lever-arm and sensitivity to detect deviations from a pure power-law spectrum of this order, *without recourse to any other astrophysical data*.

The examples given in this sub-section have been chosen to show that *Planck* can easily distinguish between certain types of inflationary model. A more detailed analysis, properly accounting for degeneracies between cosmological parameters, is given in § 2.3.4. But first we discuss the prospects of testing inflationary models using CMB polarization measurements with *Planck*.

### 2.3.3 Planck and Polarization of the CMB

As mentioned in § 2.2, Thomson scattering of anisotropic radiation at last scattering gives rise to linear polarization in the microwave background (Rees 1968). The polarization of the CMB was first detected by the DASI experiment (Kovac et al. 2002). A further indirect detection, via the cross-correlation with temperature anisotropies, has been measured from the one-year *WMAP* data (Kogut et al. 2003) as shown in Figure 2.4. The first full-sky polarization maps from *WMAP* are expected to be released shortly.

The CMB polarization signal is predicted to have an rms of $\sim 5\,\mu$K, peaking at multipoles $\ell \sim 1000$ (the angle subtended by the photon mean free path at last scattering). The polarization signal depends more directly than the temperature signal on the fluctuations at the last scattering surface and thus encodes a wealth of cosmological information, some of which is complementary to the temperature anisotropies. In addition, large-angle polarization is gen-erated by subsequent rescattering as the Universe reionizes, providing a unique probe of the thermal history of the Universe at the end of the dark ages (redshifts $z \lesssim 50$) when the first stars and galaxies formed. Such large angle polarization is seen in the *WMAP* one-year data (see Figure 2.4), and the inferred optical depth, $\tau \sim 0.17$, provides strong evidence for an early reionization epoch. Furthermore, the presence of acoustic oscillations in the *WMAP* and DASI temperature-polarization power spectra, with precisely the phase relation to the temperature spectrum that is predicted for passive, adiabatic fluctuations, lends further support to this paradigm for structure formation.

---

*   Consistent with this general level of skepticism, two new analyses of the matter power spectrum from Ly$\alpha$ lines appeared during the final editing of this chapter and both *fail* to confirm any evidence for a run in the spectral index (Viel et al. 2004; Seljak et al. 2004).

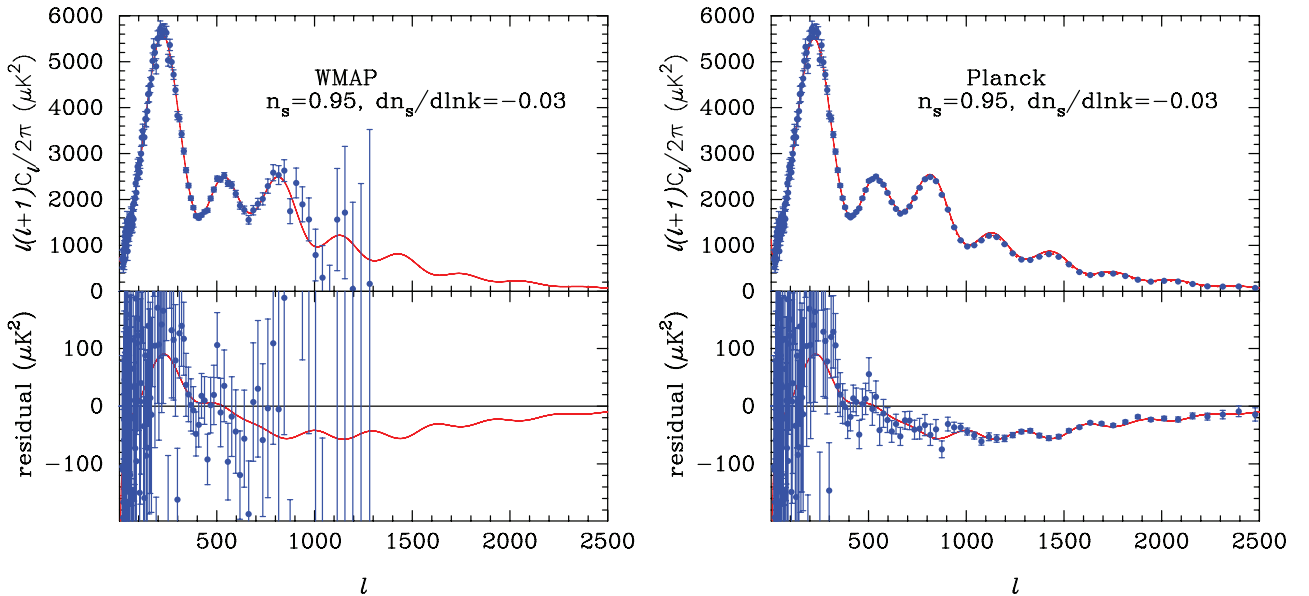

FIG 2.12.—Same as Figure 2.11, but now comparing the concordance $\Lambda$CDM model, having $n_\mathrm{S} = 0.95$ and zero run (solid line), with a realisation of a model having with $n_\mathrm{S} = 0.95$ (at a fiducial wavenumber of $k_0 = 0.05\ \mathrm{Mpc}^{-1}$) and a run of $dn_\mathrm{S}/d\ln k = -0.03$.

Looking to the future, further potential returns from polarization observations include:

- improved limits on cosmological parameters, in particular the lifting of several degeneracies that involve the optical depth through reionization;

- a probe of the detailed ionization history beyond a single optical depth estimate;

- the clean signature of a stochastic background of gravitational waves generated during inflation;

- evidence for weak gravitational lensing through the distortion of the CMB polarization on small scales.

### 2.3.3.1 Decomposition of polarization maps into E and B components

Partially-polarized radiation is conveniently described by the four Stokes parameters $I$, $Q$, $U$ and $V$ (e.g., Chandrasekhar 1950). The $V$ Stokes parameter is expected to vanish for the microwave background, since it describes circular polarization which is not generated by Thomson scattering. The $Q$ and $U$ parameters, describing linear polarization, form the components of the second-rank, symmetric and trace free, linear polarization tensor $\mathcal{P}_{ab}$. In general, the polarization tensor can be expanded in terms of two scalar fields $P_E$ and $P_B$, the electric (or gradient) and magnetic (or curl) polarization respectively. The decomposition of $\mathcal{P}_{ab}$ into $E$- and $B$-modes is unique on the full sphere and is analogous to the decomposition of a vector field into a gradient and a divergence-free vector.

The $E$- and $B$-mode polarization can be expanded in spherical harmonics as

$$P_E = \sum_{\ell \geq 2} \sum_{|m| \leq \ell} \sqrt{\frac{(\ell-2)!}{(\ell+2)!}} a_{\ell m}^E Y_{\ell m}, \quad P_B = \sum_{\ell \geq 2} \sum_{|m| \leq \ell} \sqrt{\frac{(\ell-2)!}{(\ell+2)!}} a_{\ell m}^B Y_{\ell m}, \quad (2.22)$$

which defines the $E$ and $B$ multipoles, $a_{\ell m}^E$ and $a_{\ell m}^B$ respectively, in analogy with the spherical harmonic coefficients $a_{\ell m}$ introduced in Equation 2.1 for the total temperature anisotropy. Rotational invariance demands the correlation structure

$$\langle a_{\ell m}^{Y*} a_{\ell' m'}^{Y'} \rangle = C_\ell^{YY'} \delta_{\ell\ell'} \delta_{mm'}, \qquad (2.23)$$

where $Y$ and $Y'$ can be $T$ (temperature or total intensity), $E$, or $B$. Furthermore, if the initial fluctuations are Gaussian, the temperature and polarization anisotropies can be described completely by the power spectra $C^{TT}$, $C^{EE}$, $C^{BB}$, $C^{TE}$, $C^{TB}$, $C^{EB}$. In the absence of any parity-violating processes, $C_\ell^{TB} = C_\ell^{EB} = 0$ and hence the anisotropies can be characterised fully by only four power spectra.* Of course this assumption can be tested using *Planck* data, if $B$-modes are detectable at all.

The cosmological importance of the $E$–$B$ decomposition stems from the result that linear, scalar perturbations do not produce $B$-mode polarization (Kamionkowski et al. 1997; Seljak & Zaldarriaga 1997; Hu & White 1997). Linear polarization produced by Thomson scattering is locally $E$-mode and quadrupolar in character at the scattering event. For scalar perturbations the spatial pattern of the polarization field at last scattering is curl-free; free streaming in the linear regime projects a curl-free spatial pattern onto a curl-free angular distribution. For tensor modes the polarization is also $E$-mode and quadrupolar in character at last scattering, but the spatial distribution has non-vanishing curl. Projection along the line-of-sight now produces $B$-polarization. Vector modes at last scattering would also produce $B$-polarization on the celestial sphere, but these modes are only expected to be significant in models with active generation of perturbations (e.g., through topological defects). Such models are effectively ruled out as the main source of structure formation by current CMB temperature anisotropy data, though cosmic strings may be present at a sub-dominant level according to some inflationary models (§ 2.4.5).

### 2.3.3.2 *Temperature-polarization cross-correlation*

For scalar perturbations the polarization field at last scattering traces the gradient of the photon dipole field at the previous scattering event. In models with a dominant perturbation mode (e.g., the adiabatic mode) the relative phase between the photon dipole and density fields is fixed by the continuity equation to be $\pi/2$ at each wavenumber. This gives rise to pronounced oscillations with $\ell$ in the temperature-polarization cross-correlation spectrum $C_\ell^{TE}$ (Coulson et al. 1994). An example is shown in Figure 2.13 for a pure adiabatic mode $\Lambda$CDM model. The tensor-to-scalar ratio has been set arbitrarily to $r = 0.1$, and both primordial spectra are assumed to be scale-invariant. The optical depth for secondary reionization was set to $\tau = 0.17$. Note the rise in power on large scales due to reionization, and the prominent acoustic oscillations.

Figure 2.13 also shows forecasts for the $\pm 1\sigma$ errors on $C_\ell^{TE}$ that should be obtainable with four years of *WMAP* data, the 2003 flight of BOOMERanG (B2K), and with *Planck*. For *WMAP* we have used only the 90 GHz channel for polarization, and assumed a polarization sensitivity $\sqrt{2}$ worse than the quoted temperature sensitivity at this frequency. The B2K forecasts are based on the six usable polarization-sensitive bolometers in the 145 GHz channel, using the in-flight instantaneous sensitivity of $180 \,\mu\mathrm{K\,s}^{1/2}$. We assumed a six-day deep survey covering $80\,\mathrm{deg}^2$, and a three-day shallow survey covering $1020\,\mathrm{deg}^2$. For *Planck* we have used only the 100, 143, and 217 GHz HFI channels. We have further assumed that 65% of the sky will be usable for *WMAP* and *Planck*, corresponding to a $\pm 20°$ Galactic cut. In the absence of systematic effects, the errors $\Delta C_\ell^X$ are estimated from

$$(\Delta C_\ell^{TE})^2 \simeq \frac{1}{(2\ell+1)f_{\mathrm{sky}}}[(C_\ell^T + w_T^{-1} W_\ell^{-2})(C_\ell^E + w_P^{-1} W_\ell^{-2}) + (C_\ell^{TE})^2]. \qquad (2.24)$$

---

* To simplify the notation, we will henceforth write $C^T$ for $C^{TT}$, $C^E$ for $C^{EE}$ and $C^B$ for $C^{BB}$, but we will retain the notation $C^{TE}$ for consistency with the *WMAP* papers.

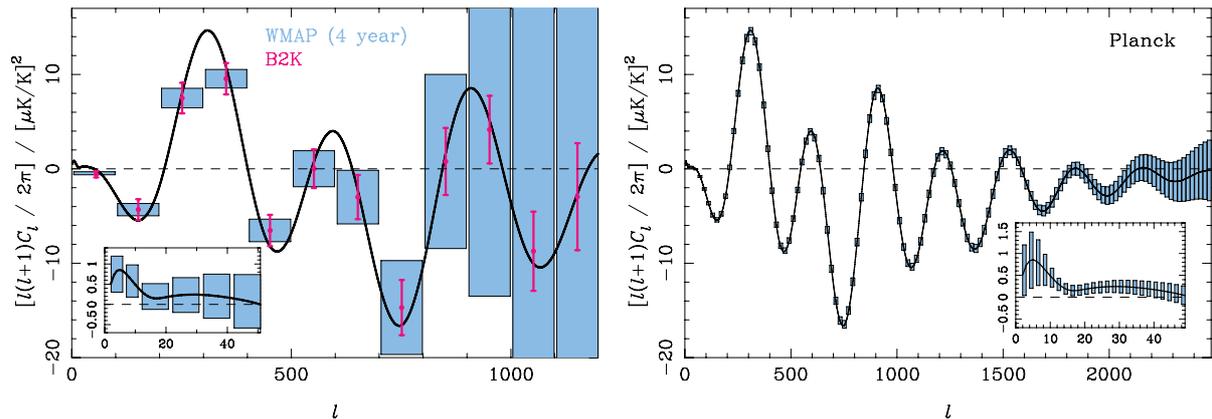

FIG 2.13.—Forecasts for the $\pm 1\sigma$ errors on the temperature-polarization cross-correlation power spectrum $C_\ell^{TE}$ in a $\Lambda$CDM model (with $r = 0.1$ and $\tau = 0.17$) from *WMAP* (4 years of observation) and BOOMERanG2K (left) and *Planck* (right). In the left-hand plot, flat band powers are estimated with $\Delta\ell = 100$ for both experiments for ease of comparison. The inset shows the *WMAP* forecasts on large angular scales with a finer $\Delta\ell$ resolution. For *Planck*, flat band powers are estimated with $\Delta\ell = 20$ in the main plot, but with $\Delta\ell = 2$ in the inset on large scales.

Here $w_T = (\sigma_{\mathrm{p},T}\,\theta_{\mathrm{FWHM}})^{-2}$ and $w_P = (\sigma_{\mathrm{p},P}\,\theta_{\mathrm{FWHM}})^{-2}$ are the weights per solid angle for temperature and polarization, respectively, for an experiment observing a fraction $f_{\mathrm{sky}}$ of the sky with noise per resolution element $\theta_{\mathrm{FWHM}} \times \theta_{\mathrm{FWHM}}$ of $\sigma_{\mathrm{p},T}$ for temperature and $\sigma_{\mathrm{p},P}$ for $Q$ and $U$. The beam window function is $W_\ell = \exp[-\ell(\ell+1)/(2\ell_{\mathrm{beam}}^2)]$ for a Gaussian beam with $\ell_{\mathrm{beam}} = \sqrt{8\ln 2}\,(\theta_{\mathrm{FWHM}})^{-1}$, and has approximately the same form in polarization for a pure co-polar beam. For B2K we combined the errors from the two surveys optimally, assuming correlations between the observed regions were negligible. The generalisation of Equation 2.24 to multiple channels is straightforward (e.g., Zaldarriaga et al. 1997; Kinney 1998).

Note also that Equation 2.24 only takes account approximately of mode-coupling due to the limited sky coverage. In particular it does not take account of effects that arise when one attempts to separate $E$ and $B$-polarization with partial sky coverage; Zaldarriaga 2001; Lewis 2002; Bunn 2002); such corrections are only significant for very small survey areas.

It is clear from Figure 2.13 that by the time *Planck* flies the oscillations in the temperature-polarization cross-power spectrum will have been mapped out with coarse $\Delta\ell$ resolution by *WMAP*, B2K and, very likely, a number of other ground and balloon experiments. However, only *Planck*'s unique combination of resolution, sensitivity, frequency coverage, and sky coverage will allow a near cosmic-variance-limited reconstruction of $C_\ell^{TE}$ over the full range of scales of primary cosmological interest.

### 2.3.3.3 *E*-mode polarization

The ability of *Planck* to measure the electric polarization power spectrum $C_\ell^E$ is compared to *WMAP* and B2K in Figure 2.14. The cosmological model and assumptions about the instrument characteristics are as described in the previous subsection. The $1\sigma$ errors on $C_\ell^E$ are given approximately by

$$(\Delta C_\ell^E)^2 \simeq \frac{2}{(2\ell+1)f_{\mathrm{sky}}}(C_\ell^E + w_P^{-1}W_\ell^{-2})^2. \qquad (2.25)$$

Direct detection of $E$-mode polarization, via its power spectrum $C_\ell^E$, is more challenging than statistical detection using the cross-correlation with the temperature anisotropies, since the expected $E$-polarization signal is much weaker than the correlated part of the temperature. After four years, *WMAP* should barely make a detection in a few broad bands around $\ell = 400$, and on the largest scales where reionization dominates (provided Galactic polarized foregrounds can be removed). However, *Planck* should be able to map out $C_\ell^E$ on all scales up to and beyond the peak of the spectrum at $\ell \sim 1000$.

Critically, *Planck* should be able to resolve accurately the large-angle polarization signal arising from the epoch of reionization, constraining both the height and position of the peak at $\ell \sim 5$. The large-angle $E$-mode polarization power spectrum is more sensitive to the reionization history than its cross-correlation with temperature. In particular, the height of the peak scales as the square of the optical depth to reionization, and, in models with abrupt reionization, the position of the peak can be used to constrain the reionization epoch. The high value of $\tau$ implied by the one-year *WMAP* data, when combined with observations of the Gunn-Peterson trough in high-redshift quasar spectra (e.g., Becker et al. 2002; Fan et al. 2002; Songaila 2004), suggests an extended period of partial ionization, rather than abrupt reionization. Figure 2.15 (modified from Holder et al. 2003) shows the ionization histories of three physically-motivated models of reionization, all constructed to have the same optical depth together with their resulting large-angle $E$-mode polarization power spectra. The three models assume different efficiencies for star formation in dark halos at high redshift and different metallicities of these early stars. Although the main reionization peak is similar in these models, the secondary peak structure near $\ell = 20$ differs by more than cosmic variance showing that CMB polarization can probe more than a single optical depth parameter. For *Planck*, the uncertainty in $C_\ell^E$ arising from instrument noise is comparable to the cosmic variance at around $\ell = 20$ in these models. Nevertheless, it should be possible to extract valuable information on the reionisation history beyond a simple sharp transition (Holder et al. 2003; Hu & Holder, 2003).

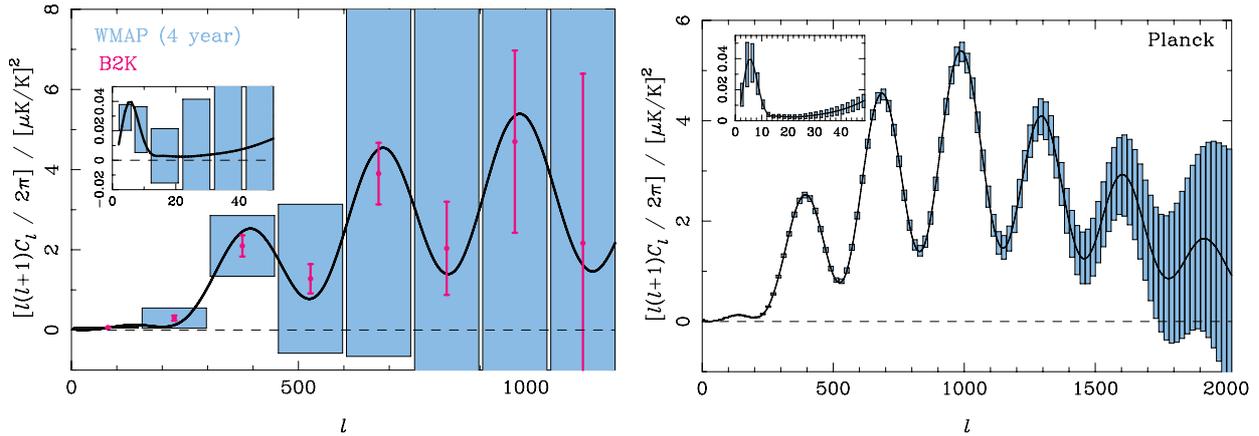

FIG 2.14.—Forecasts for the $\pm 1\sigma$ errors on the $E$-mode polarization power spectrum $C_\ell^E$ from *WMAP* and B2K (left) and *Planck* (right). The cosmological model, and the assumptions about instrument characteristics, are the same as in Figure 2.13. For *WMAP* and B2K, flat band powers are estimated with $\Delta\ell = 150$ (with finer resolution on large scales for *WMAP* in the inset). For *Planck* we have used the same $\ell$-resolution as in Figure 2.13.

The ability of large-angle polarization observations to constrain the optical depth to reionization breaks important parameter degeneracies present in measurements of the temperature anisotropies alone. For example, as shown in Figure 2.7, the scalar spectral index $n_S$ is strongly degenerate with the optical depth parameter $\tau$. More troubling is the near-exact degeneracy involving the tensor to scalar ratio $r$, the optical depth, and the scalar normalisation $A_S$. As explained in § 2.3.2, breaking these degeneracies is essential if one is to attempt to discriminate between the many proposed inflationary models. Accurate measurements of the $E$-mode polarization can improve constraints in the $r$–$n_S$ plane by partially lifting degeneracies involving $\tau$. To improve constraints on $r$ further, it is necessary to get around the problem that the tensor contribution to the temperature and polarization is only significant on large scales ($\ell < 100$), and so is generally lost in the cosmic variance of any scalar contribution. A decomposition of polarization measurements into $E$ and $B$-modes is therefore essential for detecting tensor modes generated during inflation. Since scalar perturbations do not contribute to the $B$-mode of polarization in linear theory, $B$-mode polarization can in principle provide direct constraints on $r$, limited only by our ability to deal with foreground and secondary polarization.

### 2.3.3.4 *B-mode polarization with Planck*

The most ambitious goal of CMB polarimetry experiments is to map the $B$-mode polarization. A detection of a large-angle signal with a thermal spectrum would provide a smoking-gun signature of a stochastic background of gravitational waves. In models of inflation, the amplitude of the $B$-mode of polarization is a direct measure of the inflationary energy scale, and so a detection

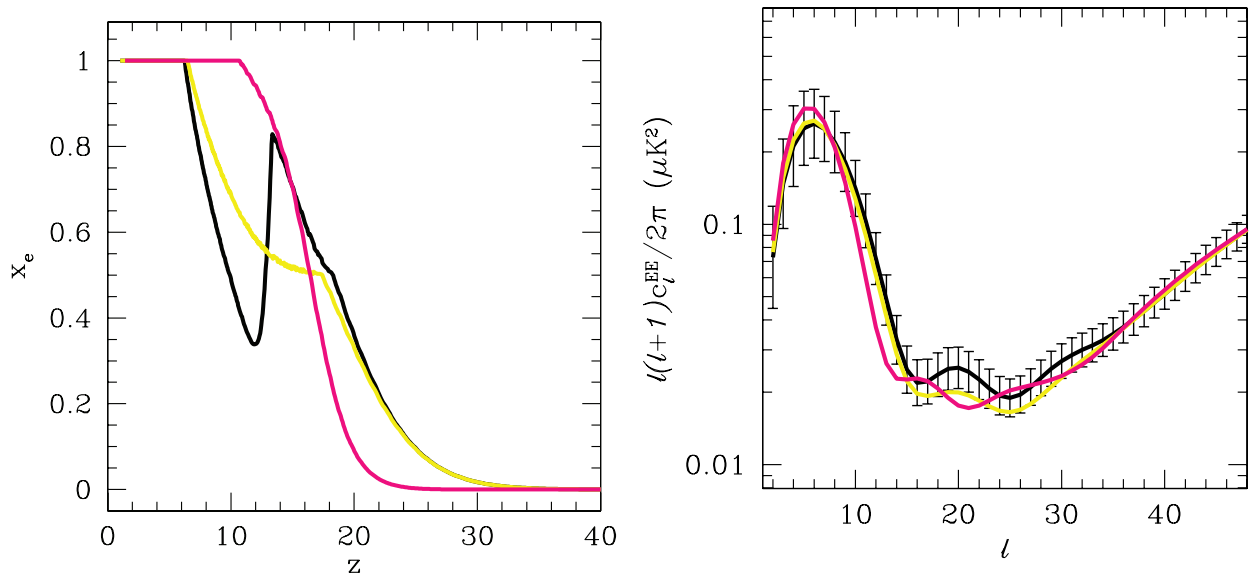

FIG 2.15.—Left: ionization histories for three physically-motivated models of reionization, each having the same optical depth ($x_e$ is the fractional abundance of ionized hydrogen). Right: large-scale $E$-mode polarization power spectra for the different ionization histories, with all other parameters held fixed. Cosmic variance errors for a full-sky experiment are plotted for the model shown in black. (Figures modified from Holder et al. 2003).

would provide a firm observational link with physics of the early Universe.

Sensitivity to $B$-polarization allows for rigorous hypothesis testing for the existence of tensor modes. In the presence of the much larger $E$-polarization signal, the most reliable way to conduct such a test is to adopt a set of variables that unambiguously probes only the $B$-polarization field. This procedure is non-trivial if parts of the sky are excluded, e.g., because of strong Galactic contamination, since the decomposition of the polarization field into $E$ and $B$ parts is ambiguous when there are boundaries. Methods have now been developed to overcome this obstacle (e.g., Chiueh & Ma 2002; Lewis, Challinor & Turok 2002; Bunn et al. 2003). Results of applying the methods of Lewis (2003) to simulated *Planck* data are presented in Figure 2.16, which shows the probability of detecting a given tensor amplitude (at 95% confidence) with *Planck* observations using 65% of the sky, for different values of the epoch of (abrupt) reionization. The same assumptions about instrument characteristics have been made as in previous sections. It is striking how the additional large-angle power arising from early reionization improves the detectability of tensor modes. In particular, for $\tau = 0.17$, as suggested by *WMAP*, the power of the null hypothesis test that there are no tensor modes exceeds 80% for *Planck* if $r > 0.05$. This value corresponds to an energy scale of inflation $V^{1/4} = 1.6 \times 10^{16}$ GeV (see §2.3.2).

Figure 2.17 shows the forecasted errors on $C_\ell^B$ for a model with $r = 0.1$ and $\tau = 0.17$. These errors are modelled by Equation 2.25 with $C_\ell^B$ replacing $C_\ell^E$. Note that with such early reionization almost half the total power in primordial $B$-polarization is generated at reionization. The figure suggests that for $r = 0.1$ *Planck* can characterise the primordial $C_\ell^B$ in around four bands. For this model, the $B$-polarization generated by weak gravitational lensing (Zaldarriaga & Seljak 1998) of the much-larger $E$-mode polarization signal dominates above $\ell \sim 150$. However, the amplitude of the lensing contribution to $C_\ell^B$ scales as $A_S^2$, and so the expected amplitude can be predicted with confidence. On large scales, the pixel noise of *Planck* is around a factor of ten higher than the rms of lensing-induced $B$-polarization, so lensing will not confuse the detection of any primordial $B$-mode signal. However, the high-$\ell$ side of Figure 2.17 shows that *Planck* should make a detection of lensing through its effect on the $B$-mode power spectrum on small scales. This would be of considerable cosmological interest in its own right.

The results presented in this section show that *Planck* is capable of making extremely accurate measurements of the polarization power spectra. The $TE$ and $EE$ power spectra should be measurable to near cosmic variance up to multipoles $\ell \sim 1000$, provided systematic errors and foreground contamination can be kept under control. Fortunately, *Planck* has polarization sensitivity over the range 30–353 GHz in seven separate frequency channels using two instruments. This range spans the frequency $\sim 80$ GHz where the contaminating signal of Galactic polarization is expected to be at its minimum. Since *Planck* samples the sky with a large number of detectors over a wide frequency range, it will be possible to perform many redundancy checks on the polarization maps.

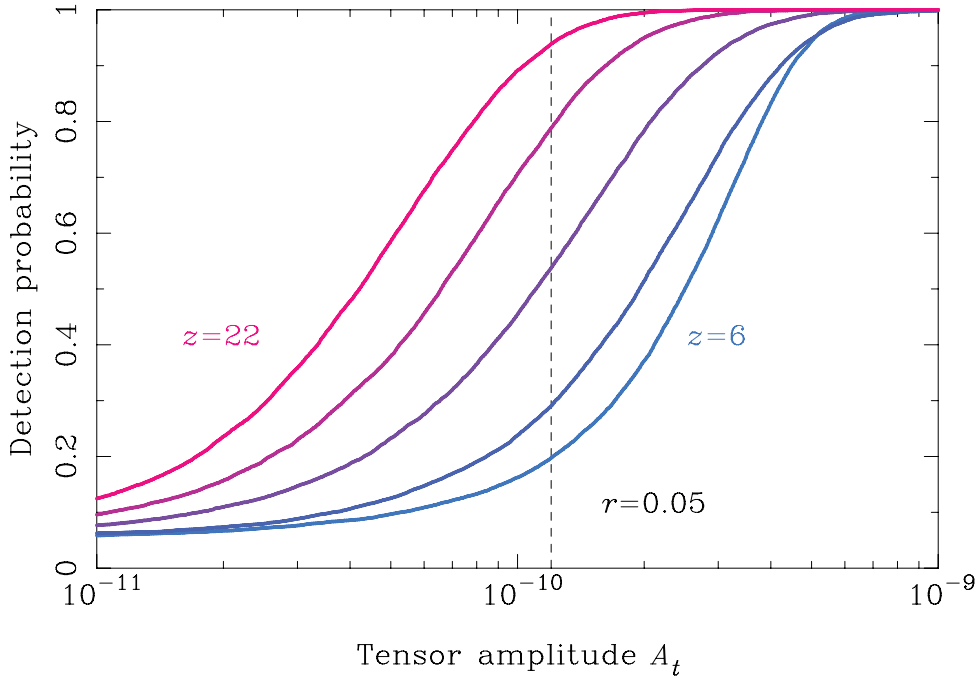

FIG 2.16.—The probability of detecting $B$-mode polarization at 95% confidence as a function of $A_{\mathrm{T}}$, the amplitude of the primordial tensor power spectrum (assumed scale-invariant), for *Planck* observations using 65% of the sky. The curves correspond to different assumed epochs of (instantaneous) reionization: $z = 6$, 10, 14, 18 and 22. The dashed line corresponds to a tensor-to-scalar ratio $r = 0.05$ for the best-fit scalar normalisation, $A_{\mathrm{S}} = 2.7 \times 10^{-9}$, from the one-year *WMAP* observations.

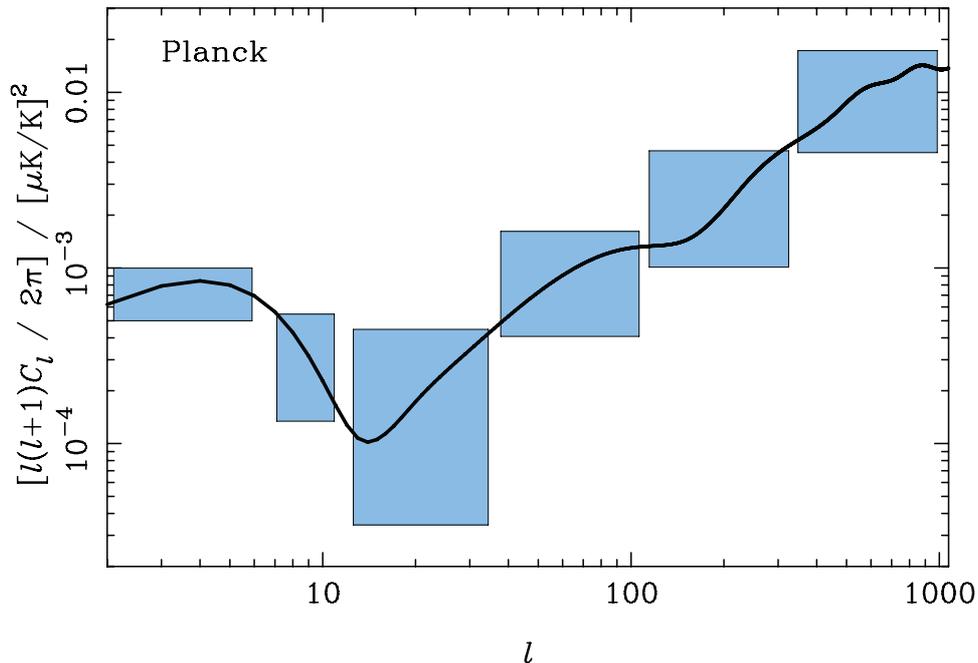

FIG 2.17.—Forecasts for the $\pm 1\sigma$ errors on the $B$-mode polarization power spectrum $C_\ell^B$ from *Planck* (for $r = 0.1$ and $\tau = 0.17$). Above $\ell \sim 150$ the primary spectrum is swamped by weak gravitational lensing of the $E$-polarization produced by the dominant scalar perturbations. The cosmological model, and the assumptions about instrument characteristics, are the same as in Figure 2.13.

The results presented in Figures 2.16 and 2.17 show that *Planck* will be sensitive to a *B*-mode polarization signal if the scalar-tensor ratio *r* is greater than a few per cent. This marks an important science goal for *Planck*, for if it can be achieved it will be possible to confirm or rule out an important class of inflationary models—the chaotic inflationary models advocated by Linde and collaborators. If *Planck* fails to detect tensor modes, then this would imply that the energy scale for inflation is low. A detection of tensor modes, however, would imply in a model independent way that the inflaton must have changed by factors of order unity or more in Planck units from the time that the fluctuations were generated until the end of inflation. If this is indeed the case, then it is unlikely that inflation can be understood in terms of a low energy effective field theory.

### 2.3.4 Cosmological Parameters: Comparison of WMAP and Planck

In this section, which is based largely on the work of Bond et al. (2004), we apply Monte Carlo Markov Chain (MCMC; e.g., Knox, Christensen & Skordis 2001; Lewis & Bridle 2002) methods to assess the accuracy with which cosmological parameters can be determined from *Planck* and other CMB experiments. These analyses account for parameter degeneracies and so give a realistic impression of how accurately individual parameters can be determined.

We assume the concordance $\Lambda$CDM model as the input model and consider the set of seven parameters: $\omega_b$, $\omega_c$, $n_S$, $\tau$, $h$, $A_S$ (the amplitude of the scalar mode), and $n_{run} \equiv dn_S/d\ln k$. In all cases, weak priors on the Hubble constant $(0.45 < h < 0.9)$ and on the age of the Universe $(t_0 > 10\,\text{Gyr})$ are applied and the equation of state of the dark energy is fixed at $w_Q = -1$ (i.e., a cosmological constant).

Figure 2.18 shows MCMC forecasts of 1 and $2\sigma$ contours for a variety of cosmological parameters, contrasting *WMAP* after 4 years observation (assuming only the 94 GHz channel) with *Planck* after 1 year of observations (assuming only the 143 GHz channel). The figure illustrates dramatically how *Planck* can break degeneracies between cosmological parameters. In particular, the scalar spectral index and the running of the spectral index will be tightly constrained by *Planck*, setting stringent constraints on the dynamics of inflation. Of course, other CMB anisotropy data will be available in addition to *WMAP*, and so the comparison in Figure 2.18 is conservative.

Table 2.1 lists uncertainties in cosmological parameters for various experiments. The column labelled input gives the input value of each parameter. The column labelled June03 lists the $1\sigma$ uncertainty on each parameter derived for the CMB data available at this time, i.e., *WMAP* after 1 year, together with BOOMERanG, CBI, VSA, ACBAR, DASI, and MAXIMA. The column labelled June03+2dF gives the constraints from these CMB data combined with data from the 2dF galaxy redshift survey. The columns labelled *WMAP4* and *Planck* list constraints from *WMAP* alone after 4 years of observation and from *Planck* alone after 1 year of observation. The final column lists forecasts for *WMAP* after 4 years combined with a hypothetical high resolution angular-scale experiment similar to ACT or the SPT,* but assuming optimistically that such an experiment would measure temperature and polarization anisotropies at high multipoles up to $\ell \sim 2000$, over 2.4% of the sky, entirely free of systematic errors.

Two sets of forecasts are given: the set labelled 'flat+weak priors' lists the constraints for a six parameter model $(\omega_b, \omega_c, n_S, \tau, h, A_S)$ assuming a spatially flat universe and with the weak priors on $h$ and $t_0$ described above. The set labelled '+running' adds a run in the spectral index, $n_{run}$, as an additional parameter. In all cases *Planck* provides the most precise constraints on cosmological parameters, in particular for the parameters $n_S$ and $n_{run}$ that are critical for testing inflationary models. This is true even if *Planck* is compared to the optimistic case of *WMAP4*+ACT/SPT.

Although we can expect a significant improvement in the CMB anisotropy data up to $\ell = 2000$ before *Planck* flies, the all-sky coverage of *Planck* gives it a large impact on parameter precision, especially for detecting small deviations of the scalar power spectrum from a pure power law. This is illustrated graphically in Figure 2.19, which shows $1\sigma$ and $2\sigma$ error ellipses for various parameter combinations for WMAP4, WMAP4+ACT/SPT, and *Planck*. *Planck* will remain extremely competitive against any foreseeable developments from ground and balloon experiments, especially given that *Planck* has the broad frequency coverage to subtract foregrounds which will certainly be important for polarization measurements.

---

* ACT: the Atacama Cosmology Telescope. see http://www.hep.upenn.edu/ angelica/act/act.html; SPT:South Pole Telescope, see http://astro.uchicago.edu/spt.

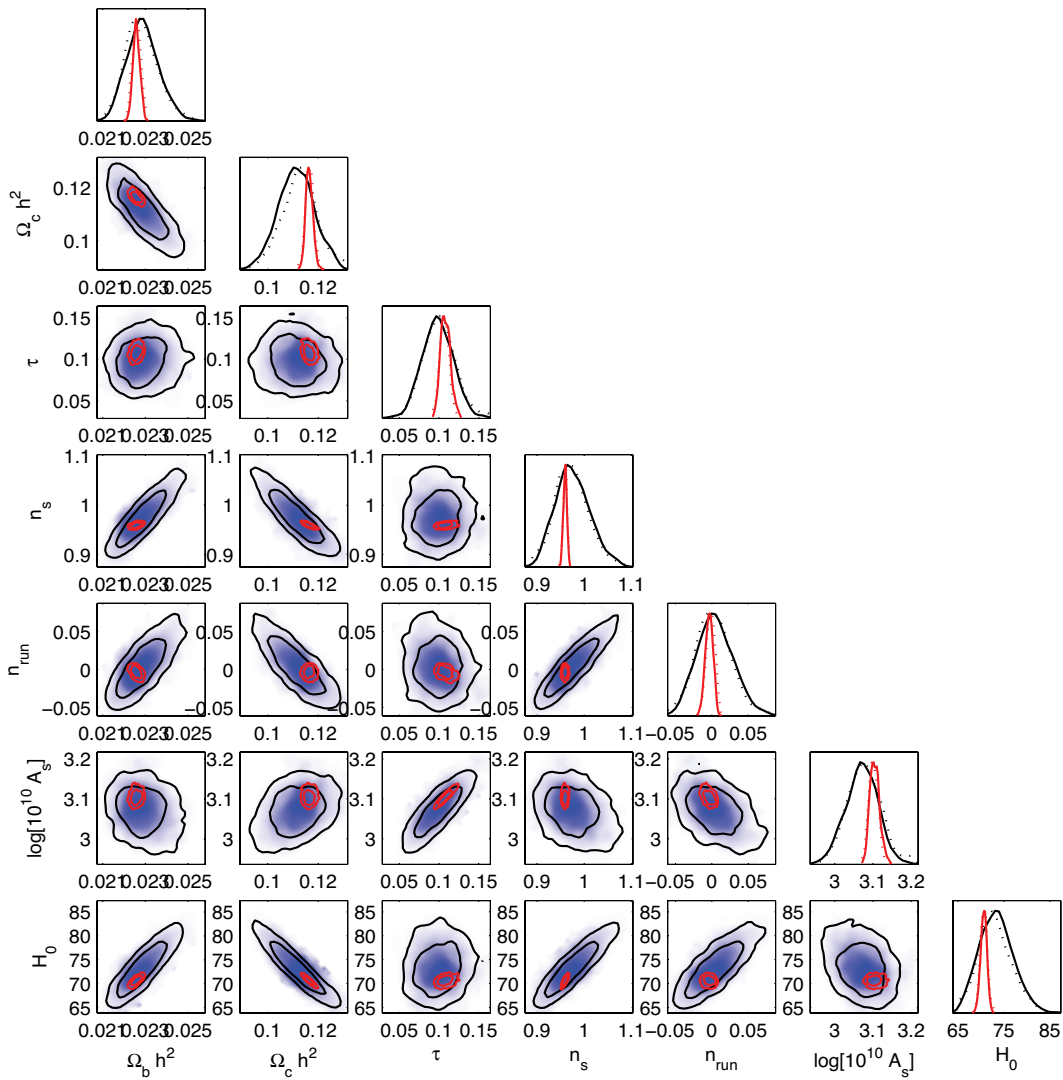

FIG 2.18.—Forecasts of 1 and $2\sigma$ contour regions for various cosmological parameters when the spectral index is allowed to run. Blue contours show forecasts for *WMAP* after 4 years of observation and red contours show results for *Planck* after 1 year of observations. The curves show marginalized posterior distributions for each parameter.

## 2.4 Probing Fundamental Physics with Planck

Figure 2.20 shows a schematic diagram of the evolution and thermal history of the Universe from the Planck time to the present. Since the *COBE* maps were first published, there have been spectacular advances in our cosmological understanding, in large part due to measurements of the CMB anisotropies. Fundamental questions relating to the geometry of the Universe, its composition, and its age can now be answered in fairly precise terms, using several complementary astrophysical techniques together with observations of the CMB.

Nevertheless, despite this spectacular recent progress in measuring the geometry and contents of the Universe, we are far from understanding why it is the way it is, and precisely how structure formed within it. Empirical progress on these questions requires much more precise measurements, which is precisely what *Planck* was designed to do. In particular, unresolved questions connected with the early Universe include:

• What is the dark energy that appears to be causing the Universe to accelerate at late times?

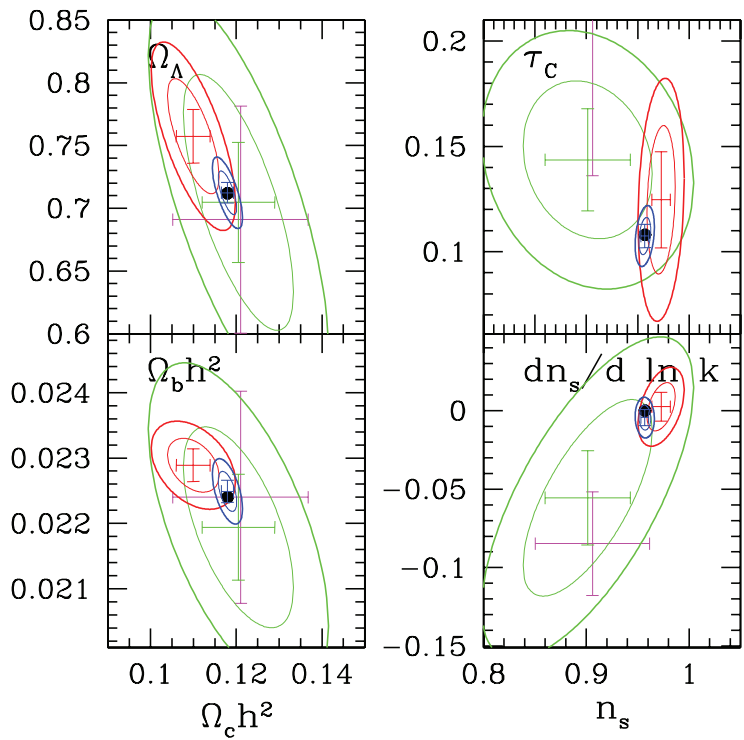

Fig 2.19.—Forecasts of 1 and $2\sigma$ contour regions for *WMAP* 4 years (green), for *Planck* 1 year (blue) and *WMAP* 4 years +ACT/SPT (red, see text) . The input values of the parameters are given by the black dots. The error bars in magenta show the precision from current CMB data when the spectral index is allowed to run.

TABLE 2.1

PARAMETER FORECASTS FOR WMAP AND PLANCK

| Parameter | Input Value | June'03 | June'03 +2dF | WMAP₄ | Planck | WMAP₄ ACT/SPT |
|---|---|---|---|---|---|---|
| **Flat+weak priors** | | | | | | |
| $\omega_b$ . . . . . . . . . | 0.2240 | 0.00095 | 0.00090 | 0.00047 | 0.00017 | 0.00025 |
| $\omega_c$ . . . . . . . . . | 0.1180 | 0.011 | 0.007 | 0.0039 | 0.0016 | 0.0035 |
| $n_S$ . . . . . . . . . | 0.9570 | 0.026 | 0.024 | 0.0125 | 0.0045 | 0.0080 |
| $\tau$ . . . . . . . . . . | 0.108 | 0.059 | 0.056 | 0.020 | 0.005 | 0.021 |
| **+running** | | | | | | |
| $\omega_b$ . . . . . . . . . | 0.2240 | 0.00162 | 0.00090 | 0.00047 | 0.00017 | 0.00025 |
| $\omega_c$ . . . . . . . . . | 0.1180 | 0.0158 | 0.007 | 0.0039 | 0.0016 | 0.0035 |
| $n_S(k_n)$ . . . . . . | 0.9570 | 0.055 | 0.024 | 0.0125 | 0.0045 | 0.0080 |
| $n_{run}$ . . . . . . . | 0.0 | 0.033 | 0.029 | 0.025 | 0.005 | 0.0092 |
| $\tau$ . . . . . . . . . . | 0.108 | 0.112 | 0.074 | 0.019 | 0.006 | 0.0266 |

What is its physical origin? Why do we exist at the epoch when the dark energy is just becoming dominant?

- What is the nature of the dark matter that dominates the present matter density? Is it made up of relic super-symmetric particles? Does it have properties which affect structure formation?

- Do we have a full inventory of the matter content of the Universe? What is the net contribution of massive neutrinos? Are we missing other components, for example, singlet neutrinos?

- When and how were the baryons created? Why is $\Omega_b \simeq \Omega_c/10$?

- Are there observational signatures of string theory, the most promising candidate for a complete theory of physics which includes quantum gravity?

- How far back can we extrapolate using classical General Relativity? Do physical constants,

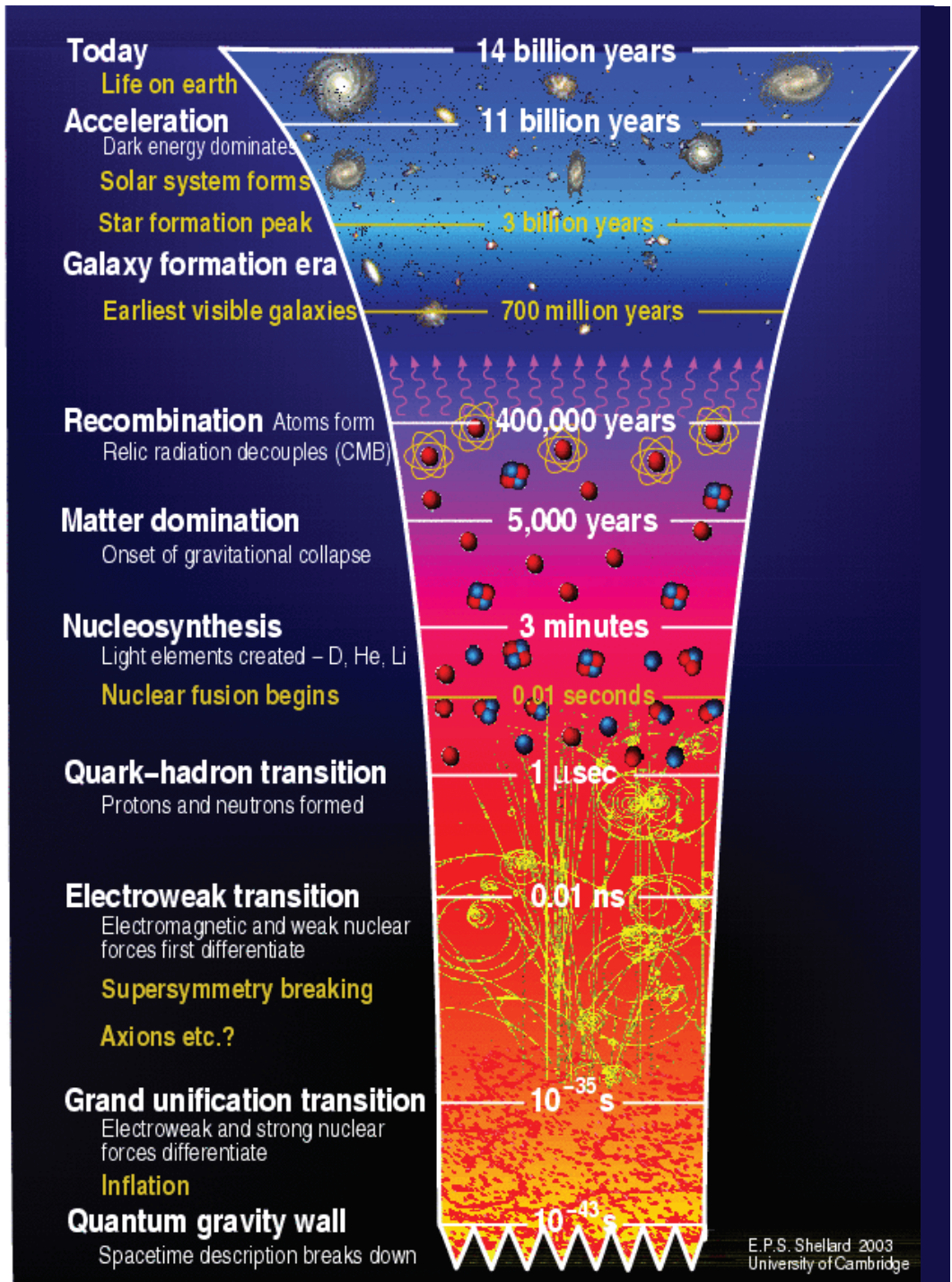

FIG 2.20.—Schematic diagram of the history of the Universe from the Planck time to the present.

such as the fine structure constant $\alpha$, vary with time? Are there deviations from the usual Friedmann equations as predicted in some brane-world scenarios?

• What is the physics behind inflation? Are the initial perturbations purely adiabatic, or are there isocurvature perturbations as well? Are cosmic defects produced at the end of inflation? Can inflation be realised in string theory? Is inflation eternal?

• Are there signatures of physics at the Planck scale or beyond imprinted on the fluctuation spectra?

• How did the Universe begin? Can string theory resolve the problem of the initial Big Bang singularity? Can we probe through the Big Bang to a previous phase of the Universe's history?

• What physics selects the vacuum solution for our Universe? String theory appears to have an

enormous number of vacua (perhaps $10^{10^{100}}$, see e.g., Susskind 2003). Can we experimentally probe any part of the landscape of string vacua?

Possible answers to these questions are still highly speculative, particularly those involving the nascent field of string cosmology. Since this is a rapidly developing field, in the rest of this section we present only brief descriptions of some key ideas and how they might relate to the *Planck* mission. The essential point is that whatever turns out to be the correct theoretical framework, high precision CMB anisotropy measurements are the most promising empirical probe of the fundamental physics of the early Universe.

## 2.4.1 Neutrino Physics

The last few years have seen important advances in our understanding of neutrinos. Clear evidence for oscillations among the three neutrino flavours has been found from experiments designed to detect solar neutrinos and atmospheric neutrinos (e.g., Ahmad et al. 2001; Fukuda et al. 2002, ). The atmospheric neutrino experiments require a mass difference $(\delta m_{\mathrm{atm}})^2 \simeq 3 \times 10^{-3}\,\mathrm{eV}$, while the solar neutrino experiments require a mass difference $(\delta m_{\mathrm{sol}})^2 \simeq 5 \times 10^{-5}\,\mathrm{eV}$. However, the absolute masses of the three neutrino mass eigenstates ($m_1$, $m_2$, $m_3$) are not constrained by these experiments. The possibilities are: (i) a hierarchical ordering, $m_1 \sim 0$, $m_2 \sim \delta m_{\mathrm{sol}}$, $m_3 \sim \delta m_{\mathrm{atm}}$; (ii) an inverted hierarchy, $m_3 \sim 0$, $m_2 \sim \delta m_{\mathrm{atm}}$, $m_1 \sim \delta m_{\mathrm{atm}}$; (iii) degenerate neutrino masses $m_1 \sim m_2 \sim m_3 > \delta m_{\mathrm{atm}}$. These possibilities can lead to different cosmological implications (see e.g., Dolgov 2002 for an overview). In the first two cases, the heaviest neutrino would have a mass of $\sim 0.06\,\mathrm{eV}$ and neutrinos would make a small contribution to the cosmic mass density

$$\omega_\nu = \Omega_\nu h^2 = \frac{\sum m_\nu}{94.2\,\mathrm{eV}} \sim 0.0007, \qquad (2.26)$$

i.e., about 20% of the baryon density contributed by ordinary stars. In the case of nearly equal masses, neutrino oscillation experiments imply even greater cosmic neutrino densities.

The strongest constraints on the absolute values of neutrino masses come from cosmological observations. Free-streaming associated with a non-zero neutrino mass suppresses the amplitude of the matter fluctuations at wavenumbers greater than $k \sim 0.03(m_\nu/\mathrm{eV})\Omega_{\mathrm{m}}^{1/2} h\,\mathrm{Mpc}^{-1}$ (Bond, Efstathiou & Silk, 1980; Doroshkevich et al. 1980). This effect can, in principle, be detected by combining CMB anisotropy measurements with estimates of the matter power spectrum from galaxy redshift surveys (or some other probe such as gravitational lensing). Current constraints from the CMB combined with the 2dF and SDSS redshift surveys, lead to a bound of $m_\nu \lesssim 0.7\,\mathrm{eV}$ (Elgaroy et al. 2002; Spergel et al. 2003; Tegmark et al. 2004)

The direct effect of neutrino masses on the CMB power spectra is too small to be detected by *WMAP*. Instead, CMB anisotropies constrain other parameters (such as the scalar spectral index), reducing the effects of parameter degeneracies in analysing the matter power spectrum from galaxy surveys. By greatly improving the constraints on these other parameters, *Planck* will improve the ability of galaxy surveys to constain the sum of neutrino masses. Eisenstein et al. (1999) forecast that the sum of neutrino masses can be determined to $0.2\,\mathrm{eV}$ by the combination of *Planck* and SDSS.

Eisenstein et al. (1999) also found that the *Planck* satellite can measure neutrino mass with an error of $0.26\,\mathrm{eV}$ without incorporating galaxy survey data. This sensitivity limit is related to the temperature at which the plasma recombines and the photons last scatter off the free electrons, $T_{\mathrm{dec}} \simeq 0.3\,\mathrm{eV}$. Neutrinos with $m_\nu \lesssim T_{\mathrm{dec}}$ do not leave any imprint on the last-scattering surface that would distinguish them from $m_\nu = 0$.

Through gravitational lensing, however, CMB observations are also sensitive to the matter power spectrum at intermediate redshifts and in particular the suppression in power at scales below the free-streaming length. Including the gravitational lensing effect, Kaplinghat et al. (2003) found that the *Planck* error forecast improves to $0.15\,\mathrm{eV}$. Such a limit is very interesting because it will strongly constrain the level of degeneracy in the spectrum of neutrino masses. Clearly one cannot have $\delta m \sim 0.06\,\mathrm{eV}$ for neutrinos each with $m \gg \delta m$ (i.e., degenerate) and yet have the sum of their masses be less than $0.15\,\mathrm{eV}$.

In the standard model, the effective number of neutrino species $N_\nu$ at the time of Big Bang nucleosynthesis (BBN) is $N_\nu = 3.04$. However, other uncoupled relativistic species (e.g., gravitons, gravitinos, Goldstone bosons) could contribute to the radiation density at the time of BBN, effectively raising the value of $N_\nu$. Neutrino asymmetry could also raise $N_\nu$, though the existence of neutrino oscillations limits the contribution to $\Delta N_\nu \lesssim 0.1$. Furthermore, it may be dangerous to assume that $N_\nu$ is constant between the redshift of BBN ($z_{BBN} \simeq 10^8$–$10^9$) and the redshifts probed by the CMB ($z \simeq 10^3$–$10^4$), since relativistic particles could be created within this redshift interval from the decays of massive particles.

At present, $N_\nu$ is poorly constrained. For example, *WMAP* and the 2dF survey give $N_\nu = 3.1^{+3.9}_{-2.8}$ (Hannestad 2002). If all of the relativistic matter is assumed to be present prior to BBN, then adding in the observational constraints on deuterium and helium abundances tightens this constraint to $N_\nu = 2.6^{+0.4}_{-0.3}$.

Relativistic particles produce distinctive features in the CMB power spectra. For example, Bashinsky & Seljak (2004) show that perturbations of relativistic neutrinos generate a unique phase shift in the positions of the acoustic peaks. This effect could potentially be detected in a high sensitivity CMB experiment. For *Planck*, $\Delta N_\nu$ can be constrained to within 0.24 independent of any observational constraints from BBN. The combination of constraints on $\sum m_\nu$ and $N_\nu$, together with measurements from direct neutrino experiments, can help limit additional sterile neutrino flavours as well as constrain the individual masses.

## 2.4.2 The Nature of Dark Energy

One of the most challenging problems in modern cosmology is to provide an explanation for the accelerated expansion of the Universe implied by observations of the magnitude-redshift relation for distant Type Ia supernovae ( Riess et al. 1998; Perlmutter et al. 1999). These observations reopened the quest for the cosmological constant, which was introduced by Einstein in 1917 but later abandoned (Einstein 1931) and infamously cited as his greatest scientific blunder (Gamow 1970). As pointed out in § 2.2, the small value of the cosmological constant is extremely puzzling and attempts to explain its value from particle physics have been spectacularly unsuccessful (e.g., Weinberg 1989).

A classical cosmological constant (characterised by a constant equation of state $w \equiv p/\rho = -1$) is by no means the only possibility for the dark energy. Examples of alternative dark energy models include quintessence, k-essence, and frustrated networks of topological defects (see Carroll 2001 for a review). Quintessence models have attracted considerable attention and are based on very light slowly rolling scalar fields. In such models, the equation of state of the dark energy component differs, in general, from $w = -1$ and varies as a function of time. General dark energy models affect the CMB anisotropies is several ways: (i) via changes to the angular diameter distance to the last scattering surface, which alters the locations of the acoustic peaks; (ii) from the late-time integrated Sachs-Wolfe (ISW) effect, which alters the amplitudes of the low CMB multipoles; and (iii) via perturbations of the scalar field itself (e.g., DeDeo, Caldwell & Steinhardt 2003; Weller & Lewis 2003).

The current constraints on a general equation of state (assumed constant) are shown in Figure 2.21. This diagram utilises a number of data-sets including the CMB anisotropy data (as of June 2003), Type 1a supernovae, and the 2dF redshift survey, as well as constraints on $H_0$ and from BBN. Fluctuations in the quintessence field were included in this analysis, though the equation of state was assumed to be constant. Despite the large number of data-sets, the constraints on $w$ are still relatively poor. Interestingly, the favoured value is close to $w = -1$, corresponding to a cosmological constant. However, values as high as $w \simeq -0.7$ are allowed by the data. Values of $w < -1$ are also allowed. This region of parameter space is particularly interesting, since dark energy with $w < -1$ violates the weak energy condition and may, therefore, be unstable.

How will *Planck* help to constrain dark energy? CMB anisotropy alone suffers from significant degeneracies with respect to $w$ and other parameters. This is illustrated in the left hand panel of Figure 2.22 (from Huterer & Turner 2001), which shows forecasts of constraints on $w$ and $\Omega_m$. The green region shows the $2\sigma$ area excluded from *WMAP* alone. The purple region shows the area excluded from observations of *Planck* alone. This region is much narrower, because *Planck* provides an accurate measurement of the distance to the last scattering surface via the positions of the acoustic peaks in the CMB power spectrum. The black region shows the constraints expected from a large survey of supernovae anticipated with the SNAP* satellite ($\sim 1000$ supernovae at a typical redshift of $z \sim 0.7$). As is well known, the constraints from Type 1a SNe are also degenerate in this diagram, but at a different angle from those derived from the CMB anisotropies. The combination of accurate CMB measurements and SNe data can, therefore, set tight constraints on the equation of state. Also shown in this diagram are projected constraints on $\Omega_m$ from the final SDSS galaxy redshift survey. Any astrophysical constraints that narrow the range of allowed values of $\Omega_m$ will break degeneracies in the $w$–$\Omega_m$ plane. Examples of astrophysical data that have been proposed as possible constraints on $\Omega_m$ include galaxy redshift surveys, lensing surveys, and measurements of the space densities and evolution of rich clusters of galaxies.

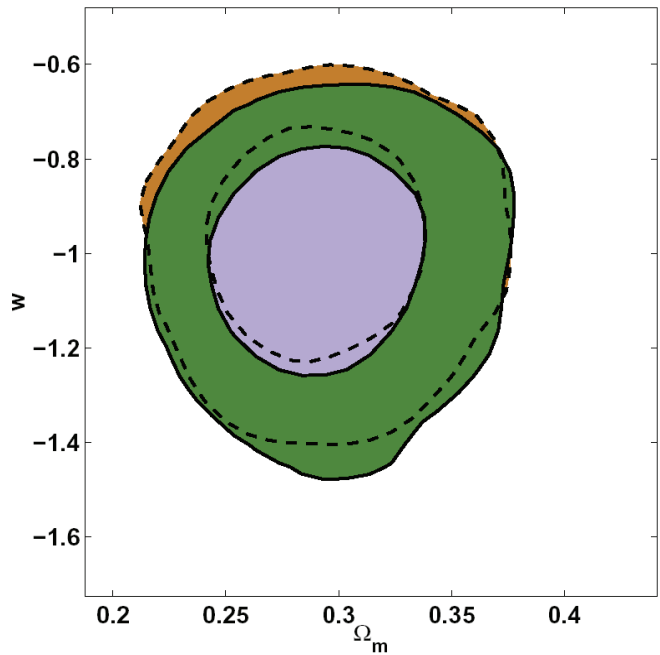

FIG 2.21.—1 and $2\sigma$ contours showing constraints on a generalised equation of state from current (June 2003) CMB data combined with data from 2dF, Supernovae, $H_0$ and BBN. The solid contours show constraints for a sound speed $c_s^2 = 1$ and the dotted contours show constraints marginalizing over an arbitrary sound speed. (From Weller & Lewis 2003).

Future surveys will also provide constraints on any possible *evolution* of the equation of state parameter with redshift. For example, let us parameterise the evolution of $w$ at low redshift as $w_0 + w_1 z$. Projected constraints on $w_0$ and $w_1$ are shown in the right hand diagram in Figure 2.22 (from Seo & Eisenstein 2003). The inner red contours show the $1\sigma$ contour in the $w_0$–$w_1$ plane expected by combining constraints from *Planck*, SDSS, SNe data from SNAP, and a set of deep galaxy redshift surveys (labelled LSS) at redshifts $z \sim 1$ and $z \sim 3$. Clearly, setting constraints on the time evolution of $w(z)$ is a challenging but extremely important task. Another quantity which can be constrained is the dark energy sound speed (i.e., perturbations in the scalar field), which is also a challenging measurement but may be possible through correlating the low-$\ell$ *Planck* data with surveys of large-scale structure (Bean & Doré 2004). Detection of a deviation of $w$ from $w = -1$, or for a time evolution of $w$, or a sound speed in the dark energy, would raise challenging questions for fundamental physics, e.g., what fixes the the energy and mass scales associated with a quintessence field?

Finally, we note that all constraints on quintessence models discussed so far have referred to a minimally coupled scalar field, i.e., the quintessence field is assumed to be coupled to the rest of the matter of the Universe via gravity alone. However, very little is known about the dark sector of the Universe and the assumption of minimal coupling may not be true. For example, a quintessence field may be coupled to the dark matter via Yukawa type couplings. Such interactions may produce fifth-force-type effects, which can alter the growth rate of fluctuations and introduce imprints in the CMB anisotropies (e.g., Farrar & Peebles 2003). As experiments improve, there may well be unexpected surprises in the physics of the dark sector.

---
* SNAP: Supernovae/Acceleration Probe, see http://snap.lbl.gov

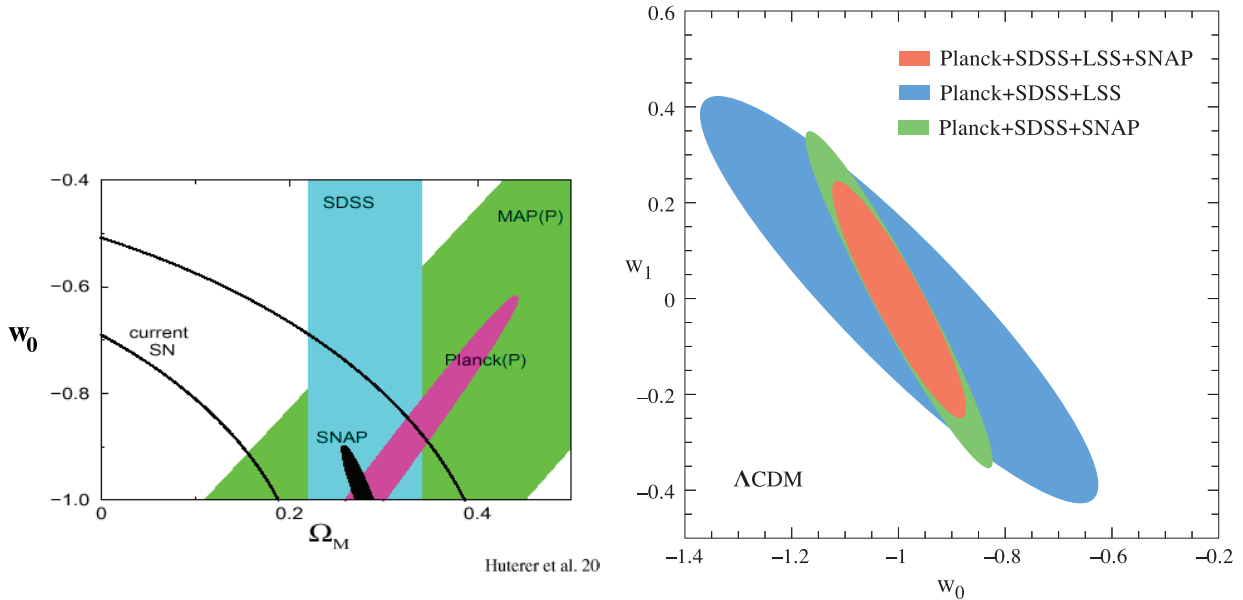

Fig 2.22.—The left panel (from Huterer & Turner 2001) shows forecasts of constraints on the dark energy equation of state parameter $w$ and $\Omega_m$ for various experiments including *Planck*. The right panel (from Seo & Eisenstein 2003) shows forecasts of constraints on the time evolution of $w$, parameterised through $w \equiv w_0 + w_1 z$, for *Planck* combined with various redshift surveys and SNe observations from SNAP (see text for details).

### 2.4.3 Isocurvature (entropy) perturbations

One of the goals of CMB anisotropy observations is to establish the underlying character of the primordial cosmological perturbations. As mentioned in § 2.3.2, the simplest perturbations are *adiabatic*, for which $\delta\rho/\rho$ for matter is 4/3 of that for radiation, preserving the entropy per particle. *Isocurvature* or entropy perturbations are also possible, for which matter perturbations compensate those of radiation, conserving total energy density. Examples of isocurvature perturbations include modes where the number density of the baryons (Peebles 1987) or of the CDM (Efstathiou & Bond 1986) is varied with respect to the photon number density. While initially there is no perturbation in the total density, as the Universe evolves, the excess or deficit of matter creates variations in the gravitational potential, which cause all the components to cluster.

The simplest single-field models of inflation discussed in § 2.3.2 predict that the primordial perturbations were strictly adiabatic. However, multiple scalar field inflationary models generically lead to a mixture of adiabatic and isocurvature perturbations (see Lyth & Riotto 1999 for an overview of inflationary models). Another possible source of isocurvature perturbations is fluctuations in the size of any extra dimensions during inflation (e.g., Kofman 2003). These lead to spatial inhomogeneity in the coupling parameters of the theory which could cause spatial fluctuations in the proportion of the inflaton that decays into baryons and/or CDM.

For an admixture of an adiabatic and a single isocurvature mode, the power spectrum of the CMB anisotropies can be written as

$$C_\ell = (1 - \alpha)C_\ell^{\mathrm{ad}} + \alpha C_\ell^{\mathrm{iso}} + 2\beta\sqrt{\alpha(1-\alpha)}C_\ell^{\mathrm{corr}}, \tag{2.27}$$

where $C_\ell^{\mathrm{ad}}$ and $C_\ell^{\mathrm{iso}}$ are the power spectra of the adiabatic and isocurvature modes and $C_\ell^{\mathrm{corr}}$ is the power spectrum of their cross-correlation (e.g., Amendola et al. 2002). The relative amplitudes of the two modes and their correlations are highly model dependent, depending for example, on the details of reheating after inflation and whether one or more scalar fields decays after inflation (as in curvaton inflationary models, e.g., Lyth & Wands 2002, 2003). These uncertainties are encapsulated by the parameters $\alpha$ and $\beta$ in Equation 2.27. More generally in a universe with baryons, photons, leptons, neutrinos, and a weakly-interacting CDM component, there are five possible modes: an adiabatic mode; a baryon isocurvature mode; a CDM isocurvature mode; and two neutrino isocurvature modes, corresponding to density and velocity fluctuations (Bucher, Moodley & Turok 2001). The most general Gaussian perturbation may then be characterized by generalizing the usual scalar power spectrum $P(k)$ to a $5 \times 5$, symmetric, positive-definite matrix-valued function $P_{IJ}(k)$ of the wavenumber, where the indices $I$, $J$ run over the five modes.

For the case of a single CDM isocurvature mode, current constraints on the parameters $\alpha$ and $\beta$ from *WMAP* combined with the 2dF survey are shown in Figure 2.23. The constraints are relatively poor and allow a large contribution from an isocurvature mode, with $\alpha < 0.47$ at the $2\sigma$ level.

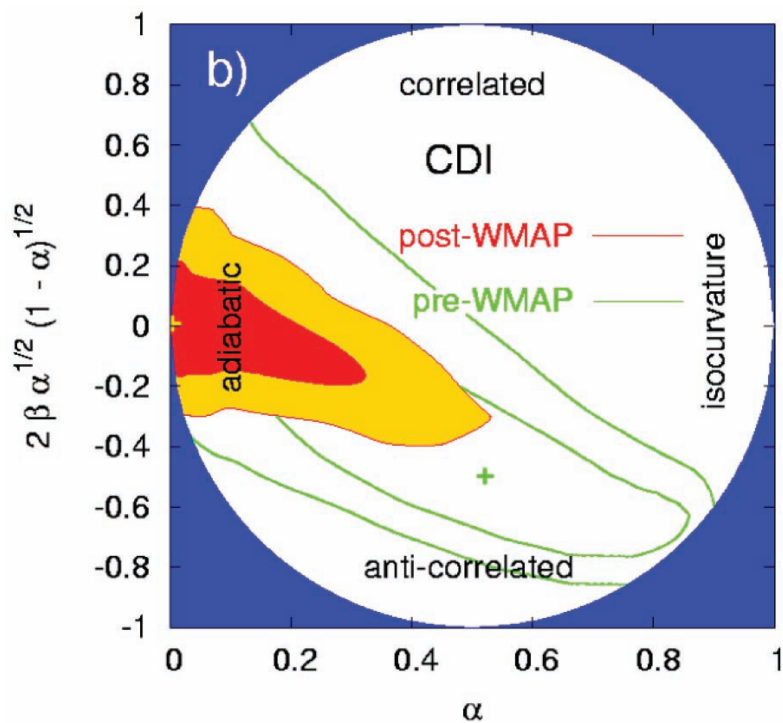

FIG 2.23.—$1\sigma$ and $2\sigma$ contours on the parameters $\alpha$ and the cross-correlation coefficient $2\beta\sqrt{\alpha(1-\alpha)}$ for an admixture of adiabatic and CDM isocurvature modes. These constraints use CMB anisotropy data combined with the 2dF galaxy survey. The green contours show the constraints using CMB data prior to *WMAP* and the filled contours show the post-*WMAP* constraints. (From Crotty et al. 2003).

Isocurvature modes introduce significant degeneracies with other cosmological parameters. Table 2.2 (adapted from Bucher et al. 2001) shows the ability of *WMAP* and *Planck* to constrain cosmological parameters together with the contributions of the various isocurvature modes and their correlations. Errors are indicated as percentages, where the strength of an isocurvature mode is indicated as its fractional contribution to the total CMB power. *WMAP* cannot rule out large, even dominant, admixtures of such modes. *Planck* with polarization, however, will be able to establish stringent limits on the presence of isocurvature modes, enabling tests of adiabaticity to high precision (at least an order of magnitude better than with *WMAP*). Models with isocurvature perturbations can also produce non-Gaussian signatures in the CMB temperature distribution of much higher amplitude than expected in single field inflationary models (see § 2.5). Tests for non-Gaussian signatures can therefore provide additional constraints on isocurvature modes.

TABLE 2.2

Testing Isocurvature Modes with WMAP and Planck

| | WMAP T adia only | WMAP TP adia only | WMAP T all modes | WMAP TP all modes | Planck T adia only | Planck TP adia only | Planck T all modes | Planck T+P all modes | Planck TP all modes |
|---|---|---|---|---|---|---|---|---|---|
| $\delta h/h$ . . . . . . . . . . | 12.37 | 7.42 | 175.74 | 20.40 | 11.50 | 3.71 | 94.67 | 7.75 | 4.53 |
| $\delta\Omega_\mathrm{b}/\Omega_\mathrm{b}$ . . . . . . . | 27.76 | 13.34 | 325.21 | 28.57 | 22.89 | 7.34 | 182.01 | 15.44 | 8.98 |
| $\delta\Omega_k$ . . . . . . . . . | 9.79 | 82.72 | 75.30 | 4.55 | 5.93 | 1.86 | 50.58 | 3.95 | 2.30 |
| $\delta\Omega_\Lambda/\Omega_\Lambda$ . . . . . . . | 12.92 | 85.02 | 123.61 | 18.53 | 2.67 | 1.20 | 9.79 | 2.36 | 1.48 |
| $\delta n_\mathrm{S}/n_\mathrm{S}$ . . . . . . . . | 7.02 | 81.62 | 89.87 | 6.53 | 0.74 | 0.37 | 4.89 | 0.87 | 0.70 |
| $\tau$ . . . . . . . . . . . . . | 37.39 | 81.81 | 104.83 | 2.23 | 10.16 | 0.41 | 72.53 | 0.77 | 0.57 |
| $\langle$NIV, NIV$\rangle$ . . . . . | ... | ... | 114.34 | 11.47 | ... | ... | 80.87 | 1.36 | 1.14 |
| $\langle$BI, BI$\rangle$ . . . . . . . | ... | ... | 573.46 | 29.71 | ... | ... | 56.72 | 6.31 | 4.27 |
| $\langle$NID, NID$\rangle$ . . . . . | ... | ... | 351.72 | 29.87 | ... | ... | 42.05 | 4.73 | 2.40 |
| $\langle$NIV, AD$\rangle$ . . . . . | ... | ... | 434.70 | 44.06 | ... | ... | 212.80 | 8.19 | 4.69 |
| $\langle$BI, AD$\rangle$ . . . . . . . | ... | ... | 1034.79 | 59.25 | ... | ... | 94.11 | 14.97 | 9.05 |
| $\langle$NID, AD$\rangle$ . . . . . | ... | ... | 1287.47 | 67.49 | ... | ... | 179.17 | 13.68 | 5.85 |
| $\langle$NIV, BI$\rangle$ . . . . . . | ... | ... | 601.83 | 32.29 | ... | ... | 79.07 | 7.63 | 3.68 |
| $\langle$NIV, NID$\rangle$ . . . . . | ... | ... | 743.93 | 46.46 | ... | ... | 133.88 | 7.42 | 2.98 |
| $\langle$BI, NID$\rangle$ . . . . . . | ... | ... | 534.33 | 39.11 | ... | ... | 115.54 | 7.68 | 4.70 |

$1\sigma$ errors (given as percentages of input model values) on cosmological parameters and isocurvature mode amplitudes anticipated for the *WMAP* and *Planck* satellite. In the column headers, 'T' denotes constraints inferred from temperature measurements alone, 'TP' those from the complete temperature and polarization measurements, and 'T+P' those inferred if temperature and polarization information is used separately without including the cross-correlation. Here the spectral indices for isocurvature modes have been fixed to their scale-invariant values, which would yield a flat CMB spectrum asymptotically at low $\ell$. 'AD' refers to adiabatic modes, 'NIV' and 'NID' to neutrino velocity and density isocurvature modes, and 'BI' to baryon isocurvature modes. (Adapted from Bucher et al. 2001).

## 2.4.4 String Cosmology

String theory is currently the most promising candidate for a fundamental quantum theory that incorporates gravity with the other forces of nature. String theory is not yet properly understood, though much progress has been made over the last twenty years. The five ten-dimensional anomaly-free superstring theories are now thought to occupy different regions of the parameter space of a single eleven-dimensional theory called M-theory. The string coupling strength in M-theory determines the size of the eleventh dimension, which is small if the string coupling strength is less than the Planck scale.

If string theory is indeed a true fundamental theory of nature, then it must necessarily address some of the oustanding questions of cosmology raised in the introduction to this section. String cosmology, however, is still in its infancy. Given our lack of knowledge of M-theory, many aspects of string cosmology are necessarily speculative. String-inspired cosmological models are of interest not so much because they might provide a compelling description of our Universe—string theory is not well enough understood to achieve this yet—but because they can address problems which have not been tractable before. These models give us an idea of the range of new physics that might apply at early times and of the prospects for testing string theory via cosmological observations.

We will focus in this section on a few specific applications of string theory to cosmology (see Quevedo 2002 for a good introductory review). The next two subsections are based on the idea that our Universe resides on a membrane (generically called a brane) embedded within a higher dimensional space in which some or all of the extra spatial dimensions are compactified. In such brane-world cosmologies, the particles and interactions of the standard model correspond to open string modes in which the end points are constrained to move on the brane, while the graviton corresponds to a closed string mode which can propagate into the extra dimensions. The final subsection discusses some speculative models that trace cosmology through the Big Bang to a pre-big bang era.

### 2.4.4.1 Randall-Sundrum and brane world cosmologies

Randall & Sundrum (1999a) considered a particular scenario (RSI model) with two 3-branes, one with positive tension and one with negative tension, embedded in a slice of a 5D warped anti-de Sitter (AdS) spacetime. The motivation for this model was to solve the large hierarchy between the electroweak scale ($\sim 1\,\text{TeV}$) and the Planck scale ($10^{19}\,\text{GeV}$), which could be achieved by adjusting the separation between the branes. Soon afterwards, these authors showed that if the negative tension brane is placed at infinity, Newtonian gravity can be recovered on the positive tension brane if the curvature scale of the AdS is chosen to be suitably small (RSII model). This raised the possibility that some of the the extra dimensions in string theory could be noncompact, perhaps even infinite.*

Brane-world cosmologies based on the Randall-Sundrum models are interesting in several respects (see Brax & van de Bruck 2003 for a review). Most importantly, the Einstein equations on the 4D brane show deviations from the standard FRW behaviour. For example, the Hubble constant is related to the energy density $\rho$ on the 4D brane according to

$$H^2 = \frac{8\pi G}{3}\rho\left(1 + \frac{\rho}{2\sigma}\right) + \frac{\Lambda_4}{3}, \tag{2.28}$$

where $\sigma$ is the brane tension and $\Lambda_4$ is the value of the cosmological constant on the 4D brane. Thus at early times, $H^2 \propto \rho^2$ if $\rho \gg \sigma$, compared to the usual FRW relation $H^2 \propto \rho$ (Eq. 2.5). As a result, early universe cosmology in brane-world cosmologies is different to that in FRW models. In particular, the dynamics of inflation differs. Furthermore, perturbations in the bulk geometry can source perturbations on the 4D brane and so the full 5D theory needs to be understood to compute the evolution of perturbations in the CMB. These types of model can easily be made more complicated, for example, by adding scalar fields to the AdS bulk. At present, a large amount of research is being done to understand inflation and perturbation theory in these models and to compute the effects of a large dimension on the CMB. Since detailed predictions are model dependent, and in some cases controversial, they will not be summarized here; however, this work suggests that under some circumstances large extra dimensions could produce observable effects in the CMB at a level detectable by *Planck* (e.g., Leong et al. 2002: Rhodes et al. 2003)

### 2.4.4.2 Brane inflation

The realisation that our entire Universe may be confined to a brane has stimulated a lot of research on whether brane interactions can produce inflation. In the first example (Dvali and Tye 1999), it was argued that dilaton exchange between two branes leads to an attractive force balanced by the repulsive force between charges on the branes. After supersymmetry breaking, a non-zero potential develops that could perhaps drive inflation. Since then, a number of other models have been developed that might realise brane inflation. These include branes intersecting at nontrivial angles, configurations of branes and antibranes, and brane configurations in warped compactified geometries (see Quevedo 2003, for a review). At present, these models are speculative, but they share some interesting features that may be generic. Firstly, in these models, the inflaton field is identified with the separation between brane (or brane-antibrane) pairs. Thus inflation in these models has a geometrical interpretation. Secondly, open strings must end on the brane pairs as in Figure 2.24. At some critical separation, the masses of these string modes become negative (tachyonic) at which point inflation ends, just as in the hybrid inflationary models discussed in §2.3.2. The ability to generate a natural end to inflation is an attractive feature of some models of brane inflation. Thirdly, cosmic strings can form at the end of inflation. The expected string tensions are model dependent, but are thought to lie most plausibly in the range

$$10^{-6} \gtrsim G\mu \gtrsim 10^{-11}, \tag{2.29}$$

---

\* Prior to this work, it had been thought that since gravity propagates in all dimensions, agreement with Newtonian $1/r^2$ gravity required the sizes of any extra dimensions to be less than a mm.

(Pogosian et al. 2003, and references therein). If the string tension lies towards the upper end of this range, then strings should be easily detectable in the CMB, both as non-Gaussian features in the temperature maps and through their distinctive B-mode polarization power spectrum (Fig. 2.24), which is very different from the B-mode spectrum expected from tensor modes (Fig. 2.9). If the characteristic string tension lies towards the lower end of this range, then the strings would be difficult to detect in the CMB but may be may be detectable via their gravitational radiation. Furthermore, these cosmic strings should show a spectrum of tensions (Jones et al. 2003), and so are distinguishable from 'ordinary' cosmic strings associated with symmetry breaking in an abelian Higgs-like model, which have a fixed string tension. The detection of cosmic strings with a spectrum of string tensions would provide strong evidence to support string theory and the existence of extra dimensions.

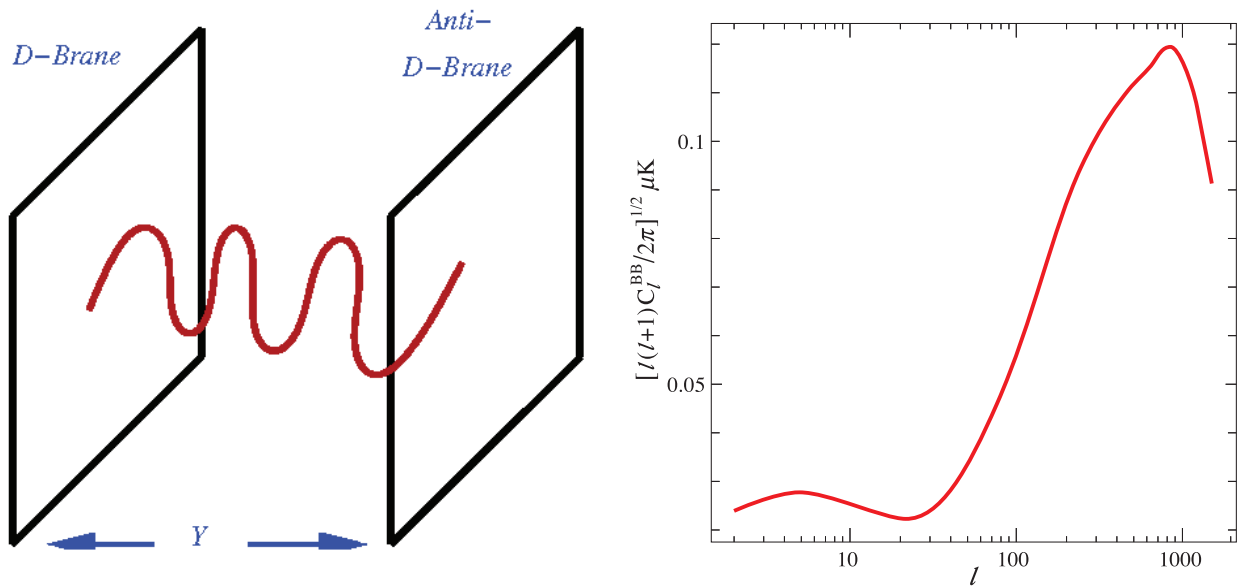

FIG 2.24.— The left panel shows an open string state with end points fixed on a brane and antibrane. At some critical separation the string state becomes massless and at smaller separations becomes tachyonic. At this critical point inflation ends, as in hybrid models of inflation (from Quevedo 2003). The right panel shows the B-mode polarization CMB power spectrum expected from cosmic strings (adapted from Pogosian et al. 2003)

The models of brane inflation discussed above assumed that the sizes of the compactified dimensions (moduli) of the theory were stabilised by some unknown mechanism. As a result these models are not self-consistent. Recently, significant progress has been made towards understanding the stabilisation of moduli in string theories. Kachru et al. (2003a) have demonstrated that long-lived de Sitter vacuua can be constructed in type IIB string theories by adding anti-D3 branes to highly warped compactifications with fluxes. This has led to the development of more realistic models of brane inflation within these meta-stable de Sitter vacua (Kachru et al. 2003b; Hsu & Kalosh 2004; Blanco-Pillado et al. 2004). Brane inflation appears to be possible in these models under certain special circumstances, though the fine-tunings required may not be a problem if anthropic arguments are invoked to select our Universe. These models make different predictions for the CMB fluctuation spectra and so the constraints on inflationary parameters from *Planck* described in § 2.3.2 should have a major impact on the search for string-based inflation.

### 2.4.4.3 Pre-big bang and ekpirotic scenarios

These models attempt to eliminate the need for inflation and to address the question of what happened before the Big Bang. The pre-big bang model of Veneziano and collaborators (see Gasperini and Veneziano 2003 for a review) is motivated by the application of T-duality to string theory in a Friedmann-Robertson-Walker-like background. The basic idea is that the Big Bang can be traced smoothly (though non-perturbatively) through $t = 0$ to a pre-big bang phase at $t < 0$. The perturbations, instead of being generated during an inflationary phase of expansion at $t > 0$, are generated during the pre-big bang phase as the universe contracts towards $t = 0$. However, calculations of the scalar and tensor fluctuation spectra suggest that they should be strongly 'blue' in conflict with observational constraints from the CMB. It is not yet clear whether these (and many other) difficulties are fatal to the model or whether they can be resolved with a more complete understanding of string theory. Nevertheless, the model provides an indication of how string theory might eliminate a singularity at $t = 0$ and lead to observational contact with a pre-big bang phase in the history of the Universe.

The ekpyrotic and cyclic models (e.g., Khoury et al. 2001; Steinhardt & Turok 2002; Khoury et al. 2003), in their simplest versions, utilise a 5D world consisting of two parallel 4D-branes separated by a gap. Particles on the branes can be produced if and when the branes collide. The Hot Big Bang is therefore identified with a collision between the two branes. As the branes approach each other prior to the collision, an observer on one of the branes would experience a contracting universe. There are some similarities with the pre-big bang scenario described above, in that the Big Bang is preceded by a pre-big bang contracting phase. The geometrical interpretation is, however, very different. Fluctuations in the inter-brane separation can be generated prior to the collision. Although controversial, it is thought that after the collision these fluctuations produce a nearly scale-invariant spectrum of scalar fluctuations from which structure forms, as in the standard Hot Big Bang cosmology. There is no high-energy inflationary phase in this theory and no tensor modes are produced. In the cyclic variant, the inter-brane potential is chosen so that the branes attract each other after the collision, leading to an endless cycle of Big Bangs. As with the Veneziano pre-big bang model, many aspects of the ekpyrotic theory are controversial and poorly understood (for a critique see Linde 2003).

## 2.5 Non-Gaussianity: Beyond the Power Spectrum

### 2.5.1 Non-Gaussian primordial models

As discussed in § 2.3.2, inflation has become the dominant paradigm for understanding the initial conditions for structure formation. A consequence of the assumed flatness of the inflaton potential is that intrinsic non-linear (hence non-Gaussian) effects during slow-roll inflation are generally small, although they are non-zero and calculable (e.g., Falk et al. 1993; Gangui et al. 1994; Wang & Kamionkowski 2000; Acquaviva et al. 2003; Maldacena 2003). Thus the adiabatic perturbations originated by the quantum fluctuations of the inflaton field during inflation are essentially Gaussian distributed, i.e., they have no (or very little) phase-coherence at this earliest stage. The statistical description of a Gaussian perturbation field is as simple as possible: the power spectrum (or its real-space counterpart, the two-point correlation function) encodes all information. Higher-order statistical moments either vanish (as is the case of odd-order moments such as the *bispectrum*, the Fourier transform of the three-point correlation function) or can be trivially expressed in terms of the second-order one (as in the case of even-order moments).

Non-linear gravitational evolution on super-horizon scales after inflation significantly enhances the original non-Gaussianity to an expected present level of order $10^{-5}$ of the rms CMB temperature anisotropy. This is undetectable in current CMB data including *WMAP*, and about an order of magnitude below the expected sensitivity of *Planck* to non-Gaussianity (see Komatsu & Spergel 2001), as measured by the bispectrum (see below).

Despite the simplicity of the inflationary paradigm, however, the mechanism by which adiabatic (curvature) perturbations are generated is not yet fully established. An alternative to the standard scenario which has recently gained increasing attention is the *curvaton mechanism* (see also § 2.4.3), according to which the final perturbations are produced from an initial isocurvature perturbation associated with quantum fluctuations of a light scalar field, the so-called curvaton, whose energy density is negligible during inflation. These curvaton isocurvature perturbations are transformed into adiabatic ones when the curvaton decays into radiation long after the end of inflation. Another recently proposed idea is the inhomogeneous reheating scenario which acts if super-horizon spatial fluctuations in the decay rate of the inflaton field are induced, causing adiabatic perturbations in the final reheating temperature in different regions of the Universe.

Both the curvaton and inhomogeneous reheating scenarios may lead naturally to much higher levels of non-Gaussianity than standard single-field inflation. Observable levels of non-Gaussianity are also predicted in a number of theoretical variants of the simplest inflationary models, including: generalised multi-field models, in which the final density perturbation is either strongly or mildly non-Gaussian, and characterized by a cross-correlated mixture of adiabatic and isocurvature perturbation modes (Wands et al. 2002 and references therein); cosmic defects, which would be observable via specific non-Gaussian signatures (e.g., Landriau & Shellard 2003); and late time phase transitions, which have been proposed for a variety of theoretical reasons. In each case, a different type of non-Gaussianity is expected. Therefore, in addition to developing generic statistical tests for non-Gaussianity, custom statistical tests are needed with which the *Planck* maps can be scrutinised for specific non-Gaussian signatures.

In summary, the inflationary fluctuation generation mechanism provides a remarkable link between quantum mechanics, general relativity, and the structure of the Universe on the largest visible scales. A complete confirmation of this mechanism would certainly qualify as one of the greatest achievements in science. It is therefore important to quantify or constrain the amount of primordial non-Gaussianity present in CMB data, as this will represent a crucial observational discriminant between competing models for primordial perturbation generation.

### 2.5.1.1 Testing specific inflationary models

An exact calculation within second-order perturbation theory of nonlinear effects during inflation has been performed only recently (Acquaviva et al. 2003; Maldacena 2003). To describe the theoretical findings quantitatively, let us introduce a useful parameterization of non-Gaussianity according to which the primordial gravitational potential $\Phi$ is given by a linear Gaussian term $\phi_G$ plus a quadratic contribution (e.g., Verde et al. 2000)

$$\Phi(\mathbf{x}) = \phi_G(\mathbf{x}) + f_{\mathrm{NL}}\phi_G^2(\mathbf{x}), \qquad (2.30)$$

where the dimensionless parameter $f_{\mathrm{NL}}$ sets the strength of the non-Gaussianity term. The second-order calculation gives $f_{\mathrm{NL}} \sim 10^{-2}$, but non-linear gravitational corrections after inflation enhance the non-Gaussianity level to $f_{\mathrm{NL}} \sim 1$, almost independent of the detailed inflationary dynamics. Any hope of observing this level of non-Gaussianity with *Planck* will require exploring different statistical estimators specifically designed to search for non-Gaussianity of this type, with the optimum method being tuned using simulated maps of *Planck*-resolution. Such simulations for this linear + quadratic potential model are shown in Figure 2.25 (from Ligouri et al. 2003).

There is no compelling theoretical motivation to restrict oneself to single-field, slow-roll inflationary models, since most unified models of gravity and elementary particles include many candidate fields. One can then straightforwardly construct models which produce non-Gaussian fluctuations, for example where the density perturbations are proportional to the square of a Gaussian random field (e.g., Allen, Grinstein, & Wise 1987). Other examples include non-Gaussian isocurvature perturbations produced by the excitation of Goldstone modes during inflation and non-Gaussian signatures in correlated mixtures of adiabatic and isocurvature modes in multiple-field inflation models (e.g., Bartolo et al. 2002). Values of $f_{\mathrm{NL}} \gg 1$ can arise easily in this framework.

Several investigators (e.g., Lyth, Ungarelli & Wands 2003; Zaldarriaga 2004) have computed the level of non-Gaussianity in the curvaton and inhomogeneous reheating scenarios, finding that values of $f_{\mathrm{NL}} \gg 1$ are possible. Finally, in the so-called *ghost inflation* picture the de Sitter expansion phase is driven by a ghost-condensate (Arkani-Hamed et al. 2004) whose energy density perturbations are characterized by a detectable level of non-Gaussianity, $f_{\mathrm{NL}} \sim 100$.

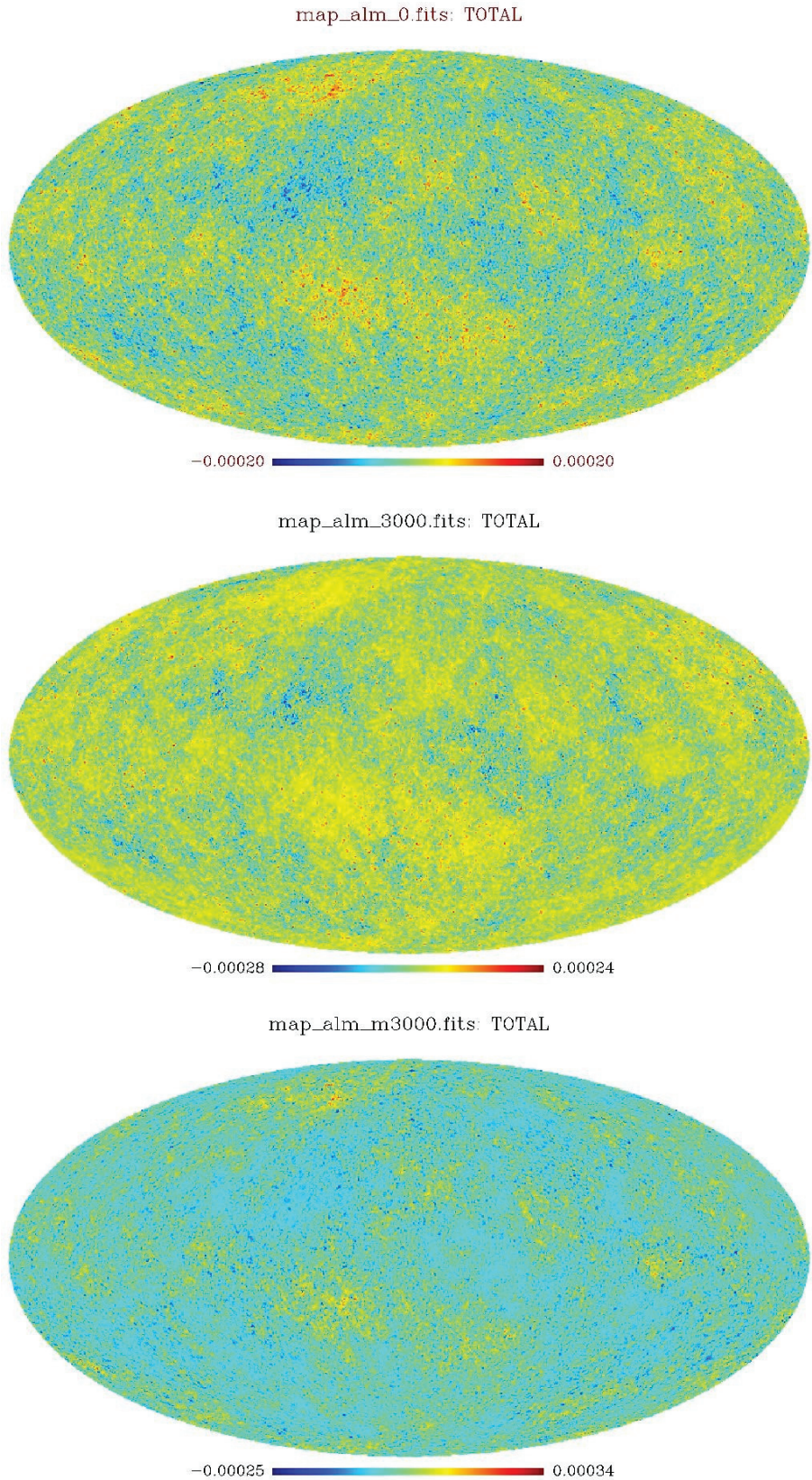

FIG 2.25.—From top to bottom, the panels show a Gaussian CMB map and two non-Gaussian simulations of an inflationary motivated linear plus quadratic potential model at approximately *Planck* resolution. The non-Gaussian maps are characterized respectively by $f_{\rm NL} = +3000$ and $f_{\rm NL} = -3000$ (these high values are needed in order to make the non-Gaussianity effects visible by eye). The non-Gaussian multipoles for the maps have been obtained by the algorithm described in Liguori et al. (2003).

### *2.5.1.2 Testing non-inflationary models*

There are, of course, more exotic pictures that can produce non-Gaussianity. For example, it is possible to closely mimic the CMB anisotropies of inflationary models by invoking a suitable source of stress energy (roughly speaking, spherically symmetric exploding shells of matter), using only causal physics operating within the standard big bang (Turok 1996a; Hu, Spergel, & White 1997). However, such models have an additional parameter which controls the Gaussianity of the resulting sky pattern.

Topological defects could produce distinctive observational signatures in the cosmic microwave sky. For example, cosmic strings, whose production has some theoretical motivation in superstring-inspired brane inflation models (§ 2.4.4.2), produce linear discontinuities (Kaiser & Stebbins 1984), for which gradient-detection and signal-whitening methods have been proposed. Cosmic textures produce hot and cold spots and global monopoles produce correlated pairs of hot spots. as shown in Figure 2.26. These signals may be detectable, in principle, even if the associated symmetry breaking scales are well below GUT scales, that is, even if cosmic defects are *not* the primary seeds for large-scale structure formation. To detect these defects, it is necessary to develop a variety of custom statistics, including various linear and nonlinear filtering schemes, resulting in optimal tests for each type of cosmic defect (see Landriau & Shellard 2003, 2004).

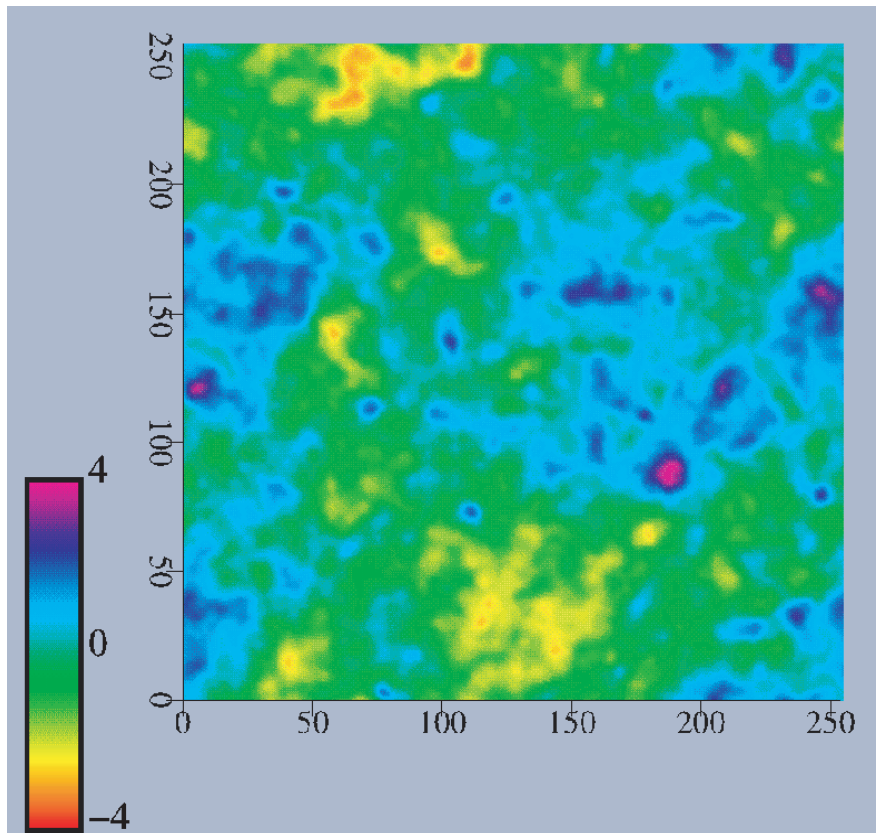

FIG 2.26.—Theoretical realisation of CMB anisotropy expected from global monopoles. The temperature anisotropy is plotted in units of one standard deviation. A $4\sigma$ peak is clearly visible, which would be extremely rare in a Gaussian model. The map is ten degrees on a side.

Late-time phase transitions may also manifest themselves on the CMB sky. One theoretical possibility is that the dark energy changed to its current value in a series of jumps, which has some motivation in M theory (e.g., Feng et al. 2001). *Planck* will be able to search our horizon volume for the resulting bubbles of lower vacuum density.

These models are far from exhaustive. In the period remaining up to the launch of *Planck*, it is quite possible that new theoretical developments will occur, leading to new suggestions for observational tests.

### 2.5.2 Large Scale Geometry and Topology

#### 2.5.2.1 Large scale homogeneity/isotropy

The CMB sky provides an area for testing the assumption that on the largest scales space-time asymptotically approaches the Friedmann-Robertson-Walker metric. Small deviations from this metric lead to observable signatures in the CMB. For example, open or flat models with global rotation or shear will exhibit a spiral pattern of anisotropy. Additional geodesic focussing can also create hot-spots in open models, while closed models exhibit a pure quadrupole pattern. The *Planck* CMB maps can be compared with templates generated for non-maximally symmetric models (e.g., Bianchi models) to provide sensitive constraints. Although the expected anisotropic patterns are on relatively large-scales, so that constraints have already established with *COBE* (e.g., Kogut, Hinshaw & Banday 1997), systematic limits are dominated by the ability to subtract chance cross-correlations with the Galactic foregrounds. The exquisite signal-to-noise ratio and frequency coverage of *Planck* will improve this situation. An additional and unique aspect of the *Planck* mission will be its ability to measure polarization, adding a new aspect to the search for global patterns related to deviations from a simple FRW metric.

#### 2.5.2.2 Large scale topology

Because General Relativity is a local theory, the global topology of the Universe remains theoretically undetermined, and it is an important challenge to constrain it (see e.g., Levin 2002). The CMB offers a unique probe of a topologically compact universe, since the compactness leads to fluctuations which are described by an anisotropic Gaussian random field. After averaging over an ensemble of observers, the net effect is that the CMB sky appears to be non-Gaussian. One can therefore hope to observe non-trivial topologies via their non-Gaussian signatures (Inoue 2001). In addition, a topology scale less than the horizon scale produces circles in the sky and the truncation in the angular power spectrum and correlation function of temperature fluctuations. The topological choices simplify in a flat universe where there are six physically allowed compact orientable topologies to consider, each of which leaves a distinct pattern on the CMB.

The first year of *WMAP* data allows constraints to be placed on topology using the matched circles method (Cornish et al. 2004). *Planck* will be able to set stronger constraints than this, through searches for non-Gaussian patterns imprinted on the CMB sky in wider classes of background models, even if the topological scale exceeds the horizon scale.

### 2.5.3 The Lowest Multipoles

Data from the first year of *WMAP* appear to confirm an effect seen at lower significance in the *COBE* data, namely that the CMB looks smooth on the largest angular scales (Spergel et al. 2003). There is no doubt that there is a lack of large-angle correlation in the *WMAP* data, although there is disagreement about the significance to attach to this result (e.g., Efstathiou et al. 2004; de Oliveira-Costa et al. 2004). This effect may be a signature of new physics and many ideas have already been proposed, including non-trivial topology and features in the primordial power spectrum.

What is certainly true is that one must be extremely vigilant about foreground and other systematic effects at the largest angular scales. Therefore if these results are to be checked independently it is important to do so using an experiment with quite a different frequency coverage and scanning strategy than *WMAP*, and to cover the entire sky in order to detect the lowest multipoles. *Planck* is the only experiment currently being built which can make such independent measurements of the quadrupole, octopole, and other low multipoles.

### 2.5.4 Non-Gaussianity from Secondary Anisotropies

It is important to be aware of other sources of non-Gaussianity in real CMB maps, which could confuse the search for primordial non-Gaussianity. The temperature anisotropies of the CMB encompass both the primary cosmological signal, directly related to the initial density fluctuations, and the foreground contributions amongst which are the secondary anisotropies. The latter are generated after matter-radiation decoupling and arise from the interaction of the CMB photons with the matter. These interactions are commonly given different names in different regimes, including the Rees-Sciama effect (Rees & Sciama 1968), the Sunyaev-Zel'dovich (SZ) effects (Sunyaev & Zel'dovich 1980), the Ostriker-Vishniac effect (Ostriker & Vishniac 1986), and inhomogeneous re-ionization of the Universe (e.g., Aghanim et al. 1996).

The secondary fluctuations associated with cosmic structures induce non-Gaussian signatures in addition to those from the primary CMB anisotropies (see Chapter 3 for a more detailed description of these phenomena and their effects). The SZ effect due to galaxy clusters is one of the most important sources of secondary anisotropies. Owing to its peculiar spectral signature, the thermal effect (induced by the CMB photon scattering off free electrons in the hot intra-cluster gas) will be separated to very good accuracy. However, the contribution from the kinetic SZ effect (due to the Doppler shift of the photons when the clusters move with respect to the CMB rest frame), which is spectrally indistinguishable from the primary anisotropies, will not be easily subtracted. The SZ effect of galaxy clusters is by nature a non-Gaussian process. The non-Gaussian signature can be characterised in wavelet space or in real space to design optimum detection schemes.

On its way from the last-scattering surface to us, CMB photons pass mass inhomogeneities which deflect their paths by gravitational lensing. This process can be described approximately as a random walk of light through a continuous field of mass inhomogeneities, and thus leads to a diffusion process. The diffusion tends to broaden structures in the CMB on angular scales smaller than $\sim 10'$ (see e.g., Bartelmann & Schneider 2001). A purely Gaussian CMB remains approximately Gaussian if deflected by density inhomogeneities which can be described as a Gaussian random field. However, non-Gaussianity is imposed on the CMB by lensing caused by non-linear structure along the line-of-sight. Non-linear evolution skews the lensing distribution by creating a tail of high magnification factors at the expense of magnification factors below unity. This leads, for example, to a non-Gaussian redistribution of hot spots on the CMB (e.g., Takada & Futamase 2001). Minkowski functionals can be useful for detecting the non-Gaussian effect of weak lensing in maps of the size and resolution expected from *Planck*.

### 2.5.5 Recent Searches for non-Gaussianity

Searches for non-Gaussianity in CMB data have been performed mostly on the 4-year *COBE*-DMR maps and the 1-year *WMAP* maps, the only whole-sky data-sets presently available. Both data-sets have been found to be compatible with the Gaussian hypothesis by the majority of the statistical tests applied. Monte Carlo simulations, taking into account the instrumental and observational constraints of the data under analysis, are usually performed to estimate distributions of the testing statistic as well as confidence levels. Tests which have been performed include: 1-point moments and cummulants, such as skewness and kurtosis; 3-point correlation functions; higher order correlations using the Edgeworth expansion; extrema distributions and correlations; genus statistics and other Minkowski functionals; partition functions; principal component analyses; phase correlations; gradient and higher-order derivatives; wavelet coefficients; and various forms of the bispectrum, trispectrum, etc.

In the case of *WMAP* (where the limits are about 30 times better than those achieved from *COBE*), Komatsu et al. (2003) analysed the temperature maps with angular bispectrum and Minkowski functional techniques and compared the results with simulations of the non-Gaussian model of Equation 2.30. They found that the *WMAP* data are consistent with Gaussian primordial fluctuations, and established the most stringent limits on $f_{\rm NL}$ obtained so far, namely $-58 < f_{\rm NL} < 134$, at the 95% confidence level. Other ground- and balloon-based experiments which are ongoing or planned could shed more light on the Gaussian hypothesis before *Planck* launches. However, these experiments will suffer from their ability to cover only a small fraction of the sky over a restricted range of frequencies, and are therefore unlikely to make a significant impact.

Both the *COBE* and *WMAP* CMB maps show some weak evidence for non-Gaussianity.

For *COBE* there were claims of correlations between modes, particularly at the multipole $\ell = 16$ (e.g., Ferreira & Magueijo 1997; Ferreira et al. 1998). Further study by Banday et al. (2000) showed that the signal came mainly from a specific period of time and from one frequency channel alone, and therefore could not have a cosmological origin. Hints of non-Gaussianity have also been reported for *WMAP* (Chaing et al. 2003; Park 2004; Eriksen et al. 2004; Vielva et al. 2004; Hansen et al. 2004; Larson & Wandelt 2004). Although the safe bet is that all of these effects are systematic in origin, it is nevertheless important to carry out more stringent tests of all such effects. Compared with *WMAP*, *Planck*'s greater sensitivity and frequency coverage will allow greatly improved foreground removal, thereby eliminating many local sources of non-Gaussianity.

# CHAPTER 3
# SECONDARY ANISOTROPIES

## 3.1 OVERVIEW

According to the standard cosmology outlined in Chapter 2, photons decoupled from baryons at the so-called epoch of recombination when the Universe was about 400,000 years old. It is customary to refer to CMB fluctuations generated prior to recombination as primary anisotropies, and those generated after recombination was complete as secondary anisotropies. These definitions are unambiguous for small-scale anisotropies, but the distinction becomes blurred for large-angle anisotropies for which the light travel time across a perturbation is of order the age of the Universe.

Secondary CMB anisotropies contain a wealth of information on physical processes that are difficult to study in any other way. For example, although the Universe became neutral at a redshift of about 1000, we know from the spectra of quasars that it must have been highly ionized prior to a redshift of about 6. But when was it reionized and what kind of objects were responsible for the reionization? As described in § 2.3, the temperature-polarization cross-correlation on large angular scales measured by *WMAP* indicates a high optical depth for reionization; *Planck* measurements will constrain the ionization history.

It is difficult to heat the intergalactic medium to a temperature higher than about $10^4$ K by photoionization. However, as massive non-linear structures developed in the Universe, the temperature of the gas within them would have begun to rise. Within clusters of galaxies—the most massive non-linear structures in the present Universe—the hot gas attains temperatures of $10^8$ K or more. A CMB photon traversing a cluster can gain energy by scattering off a much more energetic electron. This so-called thermal Sunyaev-Zel'dovich effect causes a distortion of the spectrum of the CMB in the direction of a cluster. Because of its spectral signature, the thermal Sunyaev-Zeldovich effect can be separated from primary CMB anisotropies using observations at several frequencies. This effect can be used in a number of ways to study the evolution of cosmic structure. For example, the number of clusters detected via the Sunyaev-Zel'dovich effect as a function of redshift provides a sensitive probe of the spectrum of fluctuations and of the cosmology. CMB observations of clusters can be combined with other data, (in particular, X-ray observations) to learn about the formation of clusters and the galaxies withing them.

According to the General Theory of Relativity, photons do not travel along straight lines but follow geodesic paths which are distorted in response to the curvature of space-time. Specifically, light rays are bent towards matter overdensities and away from matter underdensities in a way analogous to the effects of convex or concave optical lenses. Since the matter in the Universe is clustered, the paths of CMB photons travelling across the Universe are not straight. The resulting characteristic patterns in the CMB can therefore be used to trace out the large-scale matter distribution. In addition to gravitational this lensing effect, inhomogeneities can affect the CMB anisotropies in another way. CMB photons will gain energy as they fall into an overdensity but they will lose energy as they climb back out. For a static potential, the two effects cancel exactly, but any change in the gravitational potential during the time it takes for a photon to cross the perturbation will produce a residual temperature change in the CMB. This effect is sometimes called the integrated Sachs-Wolf effect for large-scale linear perturbations and the Rees-Sciama effect for small-scale non-linear perturbations.

This chapter describes in detail the secondary anisotropies of the microwave background caused by the three main effects, the Sunyaev-Zel'dovich effect, gravitational lensing, and the integrated Sachs-Wolfe effect.

## 3.2 GALAXY CLUSTERS

### 3.2.1 Cosmological Importance

Galaxy clusters were first discovered as regions on the sky where galaxies are overdense compared to their neighbourhood (Figure 3.1). They contain between a few hundred and a few thousand galaxies. Zwicky (1933) was the first to realize that the galaxies move so fast that the sum of their own masses would be insufficient by a wide margin to keep clusters gravitationally bound. Adding up the galaxy light and converting it to mass with the mass-to-light ratio typical for galaxies yields only a few per cent of the mass required to bind the galaxies gravitationally. This was one of the first hints that massive objects are dominated by some form of dark matter. Typical total cluster masses range between $10^{14}$ and $10^{15}$ solar masses.

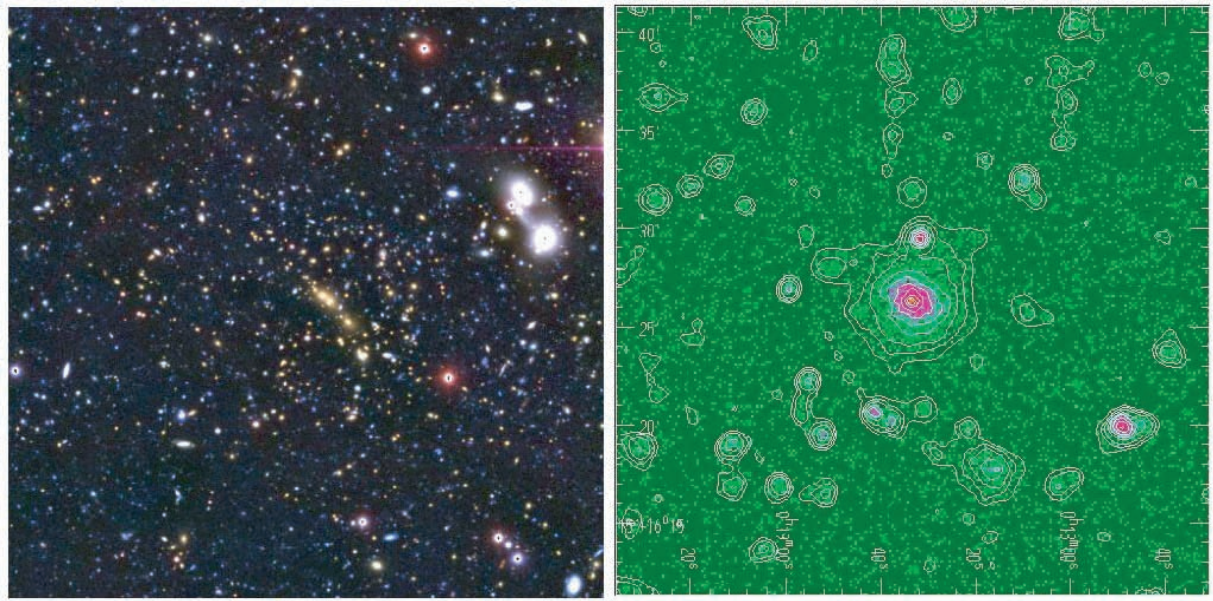

FIG 3.1.— Cluster Cl 0016+16 in visible light (Clowe et al. 2000; left panel), the Sunyaev-Zel'dovich effect(Hughes and Birkinshaw 1998; colours, right panel), and X-rays (contours, right panel).

Since galaxy clusters are the most massive virialized objects in the Universe, they are relatively rare. The theory of structure formation from a Gaussian random density field predicts that the number distribution of objects in mass, $N(M)$, should fall off exponentially above the so-called non-linear mass scale, $M_*$, which is at present of order $10^{13}$ solar masses,

$$N(M)\mathrm{d}M = \frac{\alpha\bar{\rho}}{\sqrt{2\pi}} \left(\frac{M}{M_*}\right)^{\alpha/2} \exp\left[-\frac{1}{2}\left(\frac{M}{M_*}\right)^{\alpha}\right] \frac{\mathrm{d}M}{M^2}\,,$$

where $\bar{\rho}$ is the mean cosmic matter density and $\alpha$ is related to the logarithmic slope $n$ of the dark-matter power spectrum at the mass scale $M$ by $\alpha = 1 + n/3 \approx 2/3$ for galaxy clusters (Press and Schechter 1974). The spatial number density of galaxy clusters reflects the amplitude of fluctuations in the dark-matter density, which is quantified by the normalisation of the dark-matter power spectrum. The exponential decline in the number density of clusters with increasing mass implies that small changes in the power-spectrum amplitude lead to large changes in the number density of massive clusters. In other words, the dark-matter density fluctuations in the Universe can be calibrated sensitively by determining the spatial cluster density (White, Efstathiou, and Frenk 1993; Viana and Liddle 1996; Eke, Cole, and Frenk 1996).

In hierarchical models of structure formation, for which the Cold Dark Matter (CDM) model is the prototypical example, the first structures to form have low masses (characteristic of dwarf galaxies or less). Progressively more massive objects, such as Milky-Way type galaxies, form later, and the most massive gravitationally bound objects in the Universe, the galaxy clusters, form later still. How exactly the formation and evolution of structures proceeds depends sensitively on the cosmological model. If the model has a high matter density, the expansion is strongly decelerated, and structures can grow against the global expansion. Such models permit continuous structure growth, and galaxy clusters formed very recently in them. Models with a low matter density and a cosmological constant expand rapidly, thus structures must form early or they cannot collapse against the expansion. In such models, galaxy clusters need to form substantially earlier than in high-density models. The evolution of the cluster population therefore depends sensitively on the cosmological density of matter and the cosmological constant (cf. Figures 3.2 and 3.3).

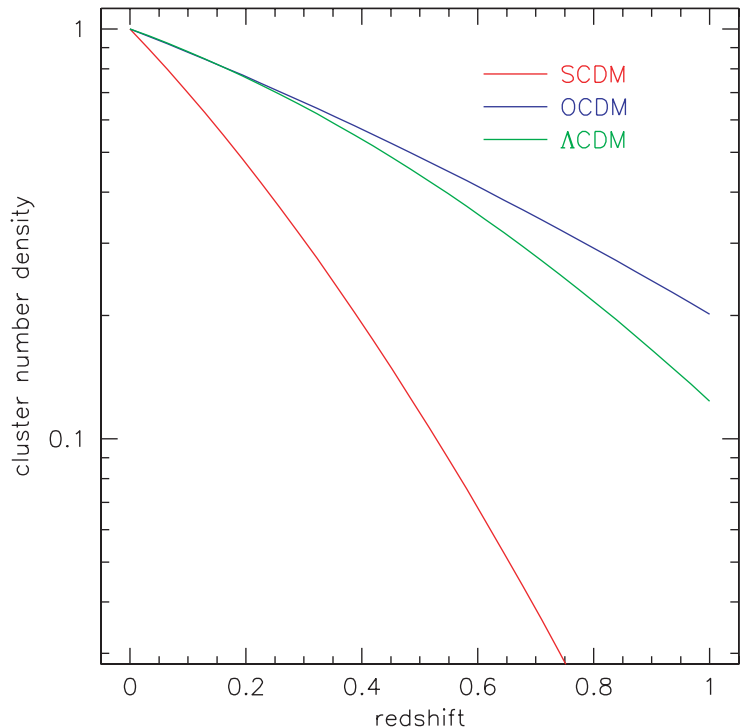

FIG 3.2.—The evolution of the cluster population with increasing redshift is shown for three different cosmological models. ΛCDM has low matter density ($\Omega_m = 0.3$) and is flat ($\Omega_\Lambda = 0.7$), OCDM has low matter density ($\Omega_m = 0.3$) and is open ($\Omega_\Lambda = 0$), and SCDM has critical matter density ($\Omega_m = 1$). The curves show how the number density of clusters with $M \geq 10^{14}\,h^{-1}M_\odot$ falls off towards higher redshift compared to its present value.

It is a generic feature of Gaussian random fields that gravitational collapse leads to highly anisotropic structures called sheets and filaments. Clusters occur at the intersections between filaments. As such, they trace the network of cosmic structure, but in a highly biased way because they mark the knots of the network (see Figure 3.3). Clusters are thus not randomly distributed, but highly correlated with each other. The exact properties of that correlation, for instance its amplitude, dependence on spatial scale, and variation with cluster mass, all depend on the cosmological model.

The density fluctuation field can be inferred from the peculiar motions relative to the Hubble flow. According to the gravitational instability scenario, mass density fluctuations and peculiar velocities both evolve under the effect of gravity in an expanding universe. On large scales, where linear theory is valid, the divergence of the velocity field $\vec{v}$ is simply proportional to the density contrast $\delta$, where the constant of proportionality is a power of the cosmic matter density parameter $\Omega_m$,

$$\nabla \cdot \vec{v} = -H_0 \Omega_m^{0.6}\, \delta \ .$$

Thus, peculiar velocities directly probe the mass distribution in the universe, and methods have been developed for reconstructing the matter density and three-dimensional velocity fields, and determining the cosmic density parameter (Bertschinger and Dekel 1989; Dekel et al. 1993; Zaroubi et al. 1999; Zaroubi 2002).

The inferred matter distribution reflects the initial conditions and the dynamical evolution of cosmic structures. Combining cosmic velocity fields and CMB fluctuations can thus strongly

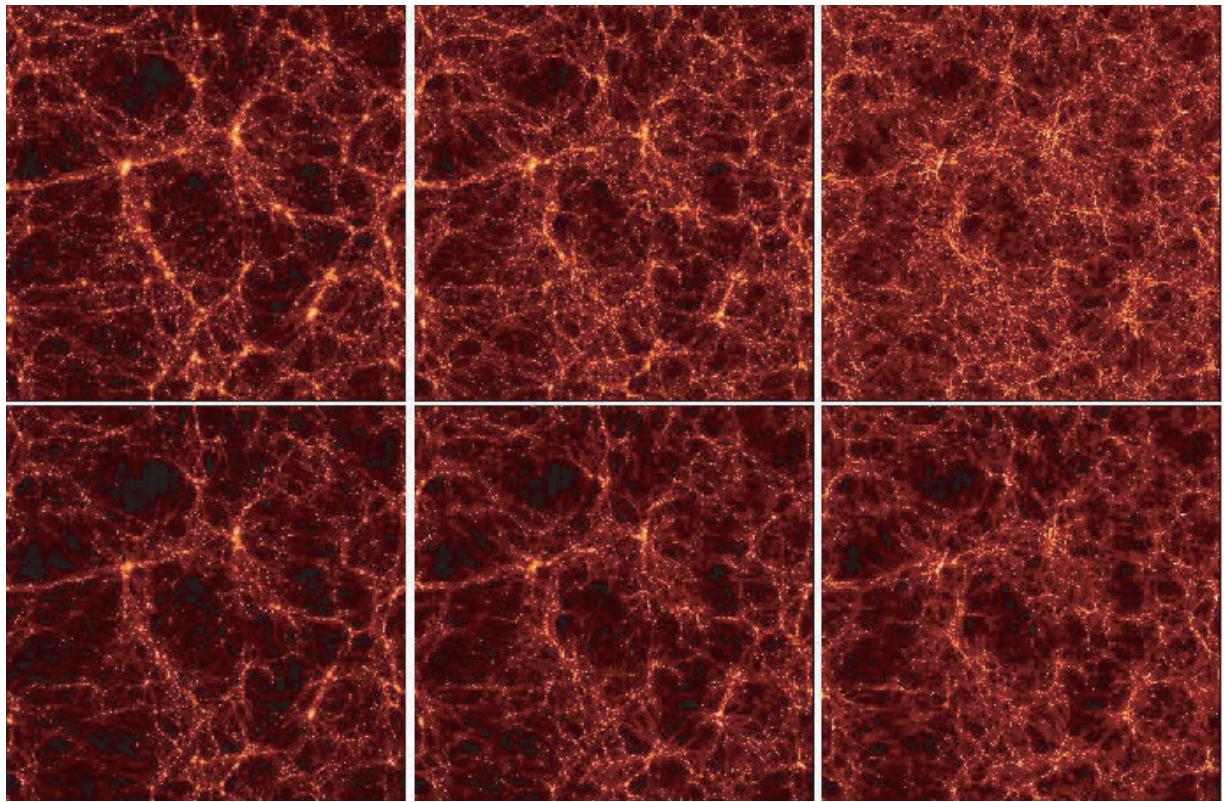

Fig 3.3.— Structure growth in two simulations of low-density cosmological models, one spatially flat (ΛCDM, upper panels) and one open (OCDM, lower panels). From left to right, the three columns are at $z = 0$, $z = 1$ and $z = 3$, respectively. Structure grows faster in the ΛCDM than in the OCDM model. (The VIRGO Collaboration; Jenkins et al. 1998)

constrain theories of structure formation and evolution, and cosmological models in general. Measuring cosmic velocity fields using isolated galaxies suffers from considerable systematic uncertainties because their distances need to be known before their velocities can be deduced. However, as we shall see below, velocity fields can be measured directly using galaxy clusters.

### 3.2.2 Hot Gas in Clusters

It was recognised soon after the first X-ray observations in space that galaxy clusters are the brightest X-ray sources in the sky, with luminosities of order $10^{44}$ erg s$^{-1}$. This radiation turned out to have a thermal spectrum, i.e., a spectral energy distribution which decreased exponentially with energy. Typical temperatures range between a few to just above ten keV, or $10^7$–$10^8$ K. At such temperatures, the emitting gas is fully ionised, and the electrons scattered off the ions emit bremsstrahlung in the X-ray regime. Apart from a few individual X-ray sources, the X-ray radiation is diffuse and spread across the entire cluster, indicating that the deep cluster potential wells which bind the cluster are filled with a hot, intracluster plasma. Typical X-ray surface brightnesses and total luminosities indicate that the gas is dilute, reaching densities of $10^{-4}$–$10^{-3}$ particles per cm$^3$ in cluster cores. The intracluster gas contributes of order ten per cent of the total cluster mass within the virial radius.

As an early test for the presence of a hot, thermal plasma filling entire clusters, Sunyaev and Zel'dovich (1972) suggested that the plasma electrons should have a unique effect on photons of the cosmic microwave background which pass through clusters on their way to us. Some of the photons are scattered off the hot electrons, which are by orders of magnitude more energetic than the photons. The photons are therefore inversely Compton scattered to much higher energies. Microwave experiments in a given, low-frequency wave band would therefore see fewer photons from the direction to a galaxy cluster; in other words, clusters would cast shadows on the microwave sky (see Figure 3.4). The inverse Compton-scattering does not change the photon number, however, so the scattered photons have to re-appear at higher frequencies.

Thus, experiments at higher frequencies would see the microwave sky brighter at the locations of galaxy clusters.

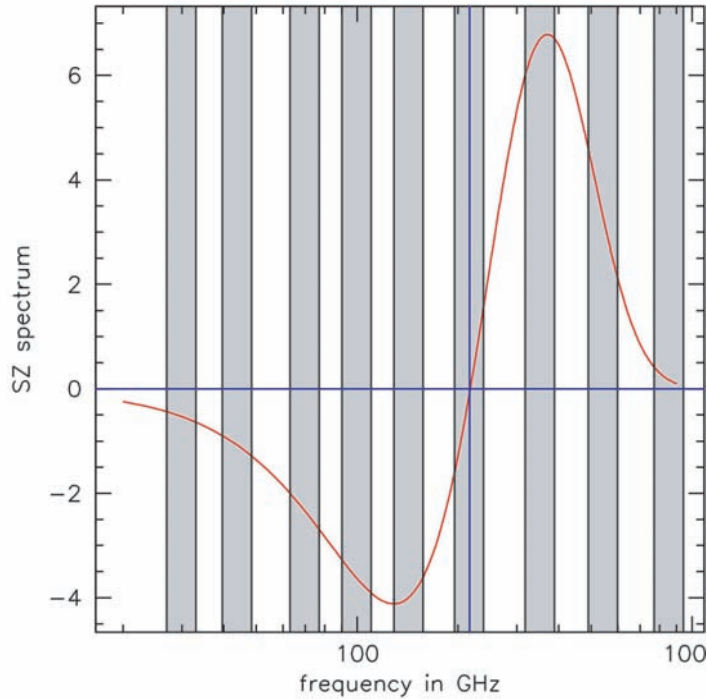

Fig 3.4.— Spectrum of the thermal Sunyaev-Zel'dovich effect (red curve, arbitrary units), overlaid with *Planck*'s frequency channels (grey bands). The CMB intensity is increased at frequencies above 217 GHz (i.e., to the right of the vertical blue line) at the expense of lower frequencies. In other words, galaxy clusters cast shadows below 217 GHz and shine above.

This is the thermal Sunyaev-Zel'dovich (tSZ) effect, which is one of the most important physical effects contributing to *Planck*'s foreground signal. It is quantified by the so-called Compton-*y* parameter,

$$y = \frac{kT}{m_e c^2}\, \sigma_T \int n_e \mathrm{d}l \equiv \frac{kT}{m_e c^2}\, \tau_e \; ,$$

where $\sigma_T$ is the Thomson cross section, $n_e$ is the electron density, and $\tau_e$ is the optical depth (the probability that a CMB photon is scattered as it passes through the cluster). The ratio of the thermal energy $kT$ of the electrons to their rest-mass energy $m_e c^2$ is the mean relative energy change of the photon in one scattering process. In effect, the tSZ effect measures the thermal electron pressure $n_e kT$, integrated along the line-of-sight through the cluster. The total tSZ signal, integrated on the sky across the face of the cluster, measures the total thermal energy of the intracluster electrons.

The shape of the tSZ spectrum caused by a purely thermal electron population does not depend on the electron temperature, but only on the temperature of the photons. This gives rise to the important feature of the tSZ effect that it is independent of cluster redshift. Although clusters at higher redshift see a hotter CMB, the scattered CMB photons are redshifted as they continue propagating to us. The two effects must cancel exactly. Hence, the observed frequency dependence of the tSZ effect is the same for all clusters, irrespective of the cluster redshift. The frequency dividing between loss and gain of photons is 217 GHz (see Figure 3.4).

This consideration rests on the non-relativistic theory of Compton scattering. It is appropriate as long as the thermal electron energy is negligible compared to the electron rest mass. For very hot clusters, a small fraction of the electrons reach relativistic energies, and those electrons give rise to relativistic modifications of the tSZ spectrum. The zero-point frequency of 217 GHz is shifted to slightly larger energies, and the amplitude of the high-energy tail of the tSZ spectrum increases. Both effects depend on the cluster temperature.

Another effect discovered by Sunyaev-Zel'dovich is caused by the relative motion of clusters with respect to the rest frame of the CMB. As discussed above, this motion is a necessary consequence of the fact that clusters trace the large-scale matter distribution, and so are moving with the peculiar velocity of the network of cosmic structures. Their electron population acts much like a reflecting screen. First, it imparts a Doppler effect on the CMB photons passing through it, and thus leads to an energy shift in the photon spectrum. The radiation intensity at frequency $\nu$ changes by

$$\Delta I = -\tau_{\mathrm{e}}\,\frac{v}{c}\,I_0\,\frac{x^4 \mathrm{e}^x}{(\mathrm{e}^x - 1)^2}\;,$$

where $x = h\nu/kT$ is the ratio between the photon energy and the thermal electron energy, $I_0$ is the incoming radiation intensity, and $v$ is the line-of-sight velocity of the cluster. This is called the kinetic Sunyaev-Zel'dovich (kSZ) effect (Sunyaev and Zel'dovich 1980). Typical peculiar cluster velocities are of order a few hundred $\mathrm{km\,s^{-1}}$, so the relative frequency shift is of order a few times $10^{-4}$. The kSZ effect provides one of the most direct ways for measuring radial peculiar velocities because it rests upon a well-understood physical effect and does not depend on cluster distance. As a result, large systematic uncertainties in peculiar velocity measurements can be avoided.

Since Compton scattering is anisotropic, unpolarised infalling radiation acquires a net linear polarisation if it has at least a quadrupole anisotropy relative to the scattering cluster. The microwave background radiation scattered by a moving cluster will therefore be polarised in a way which reflects the motion of the cluster transverse to the line-of-sight. In contrast, the kSZ effect measures the cluster velocity along the line-of-sight.

### 3.2.3 Clusters in Different Wave Bands

Wide-area surveys at optical and near infrared bands will provide spectroscopic and photometric redshifts for large samples of clusters distributed across a substantial fraction of the sky. Optical and infrared data are essential for studying how the galaxy population in clusters depends on redshift, cluster mass, total gas content etc.

The hot gas in galaxy clusters emits X-rays due to thermal bremsstrahlung. The X-ray emissivity is

$$j \propto n_{\mathrm{e}}^2 \sqrt{T}\;,$$

i.e., quadratic in the electron density $n_{\mathrm{e}}$ and therefore dominated by the core, where the gas density is highest. Although it seemed for a long time as if clusters were generally relaxed in their cores, recent high-resolution X-ray imaging and spectroscopy by the Chandra and XMM telescopes shows that regions of cool gas are immersed in hot surroundings, shocks are frequent, and the temperature distribution, in some cases, is far from isothermal. Cooling and heating processes in the intracluster plasma are still too poorly understood to model them in any detail in numerical simulations. Interpretation of the cluster X-ray emission is therefore difficult. This is one of the reasons why a considerable improvement in our understanding of galaxy clusters is expected by combining SZ and X-ray observations, since the two techniques sample the baryons in different ways. While the tSZ effect integrates over the thermal energy density in clusters, the total X-ray luminosity integrates the squared electron density and therefore depends much more strongly on the total baryonic cluster mass. Being dominated by the cluster cores, cluster detection in the X-rays is much less prone to projection effects than other methods for defining and detecting clusters.

It is not yet clear how the intracluster gas was heated to temperatures around and above $10^7$ K. It was possibly preheated before clusters formed by the energetic radiation of early objects like galaxies or quasars. It is unknown how the thermal energy content of clusters evolves at moderate and high redshifts. Likewise, the relation between cluster temperature and X-ray luminosity is known at low redshift, but it remains to be investigated if and how this relation changes towards higher redshift. Moreover, the relationship between baryonic and total cluster mass remains to be clarified, for which large cluster samples are crucial.

### 3.2.4 Lensing clusters

Mass deflects light. A light ray passing a mass with Newtonian gravitational potential $\Phi$ is deflected by an angle

$$\vec{\alpha} = \frac{2}{c^2} \int \vec{\nabla}_\perp \Phi \, dl,$$

determined by the integrated pull of the potential perpendicular to the line-of-sight. In revealing mass, this gravitational lensing has the great advantage of being entirely independent of the physical state of the cluster matter. On the other hand, the lensing effect integrates the deflection by matter inhomogeneities along the entire light path, and thus is fairly sensitive to chance projections of physically unrelated matter inhomogeneities onto each other.

Detection of lensing depends on the existence of a suitable class of objects whose average properties allow the identification of differential light deflection. For instance, the images of faint background galaxies display a preferentially tangential alignment with respect to cluster centres which maps the gravitational tidal field of the clusters (Figure 3.5). Since the tidal field and the projected surface mass density are related by an underlying scalar potential $\psi$,

$$\psi = \frac{D_{\mathrm{ds}}}{D_{\mathrm{d}} D_{\mathrm{s}}} \frac{2}{c^2} \int \Phi \, dl \, ,$$

where $D_{\mathrm{d}}$, $D_{\mathrm{ds}}$, and $D_{\mathrm{s}}$ are the distances between the observer and the lens, the lens and the source, and the observer and the source, respectively, it is possible to reconstruct the cluster mass distribution from the observable distortions measured on background galaxy images (Kaiser and Squires 1993).

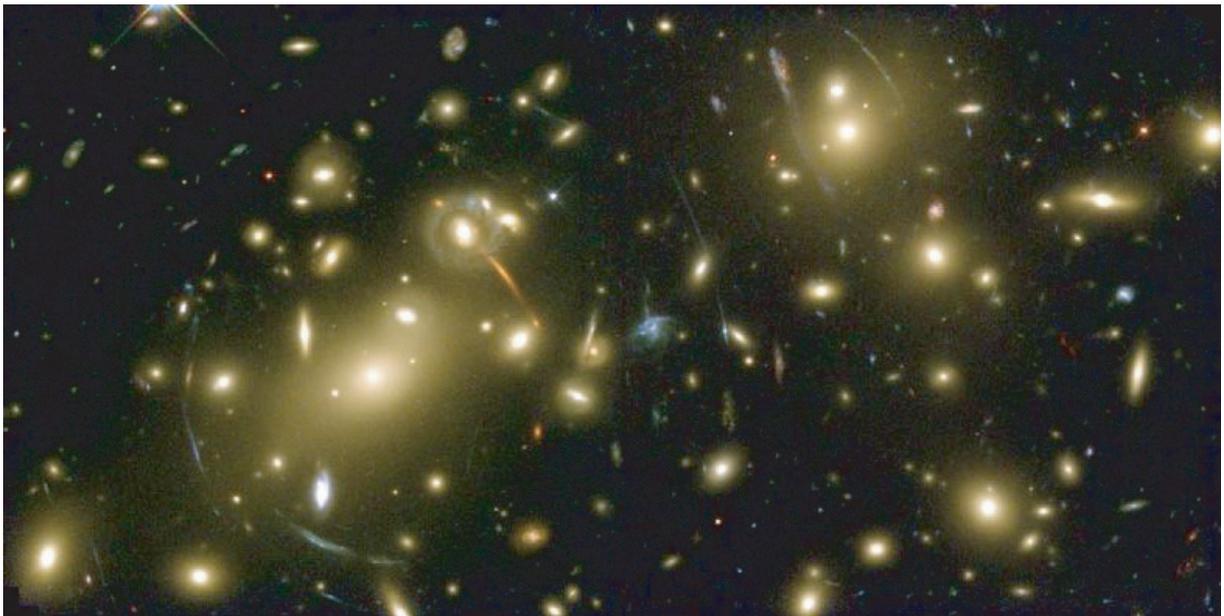

FIG 3.5.— Image of the rich galaxy cluster Abell 2218 showing many arcs and arclets that are images of background galaxies distorted by the gravitational tidal field of the cluster mass. (W. Couch, R. Ellis, NASA)

Gravitational lensing depends not only on the absolute surface mass density of the lens, but also on the distances within the lens system consisting of the sources, the lens, and the observer. A lens is most efficient if placed approximately half way between the sources and the observer, and the efficiency drops to zero as the lens is shifted either towards the observer or the sources. This is quantified by the critical surface mass density $\Sigma_{\mathrm{cr}}$ required for strong lensing,

$$\Sigma_{\mathrm{cr}} = \left[ \frac{4\pi G}{c^2} \frac{D_{\mathrm{d}} D_{\mathrm{ds}}}{D_{\mathrm{s}}} \right]^{-1} .$$

In other words, for a determination of absolute cluster masses to be possible, the distances of the cluster and the sources from the observer need to be known. Besides redshift measurements, this knowledge depends on the cosmological model.

While converting relative to absolute cluster masses can be difficult in individual cases, average properties of a large cluster sample can be more reliably determined. Detecting potential clusters as tSZ sources, confirming them through X-ray and optical information, and estimating photometric redshifts through their member galaxies not only strongly reduces projection effects, but also provides much of the information required to convert a lensing signal to a cluster mass estimate. While the exact redshift distribution of galaxies behind individual clusters can vary quite substantially, the average properties of the source galaxy redshift distribution are much better constrained.

Gravitational lensing, combined with the tSZ effect, X-ray emission, and optical information, thus provides a technique for determining the mass distribution of clusters and for characterising the relations between optical cluster light, X-ray emission, thermal cluster energy, and cluster mass. Moreover, even clusters at fairly low and fairly high redshifts can produce a significant lensing signal. Thus it will be possible to study the evolution of these relations with cluster redshift.

Optical observations of cluster lensing generally require deep images of clusters to be taken. It would be impractical to observe individually all of the clusters expected to seen by *Planck* to a depth where sources lensed by the clusters are detected. The Sloan Digital Sky Survey* does not go this deep; however, deeper surveys like VISTA† or PRIME‡ will be of great importance for cluster studies.

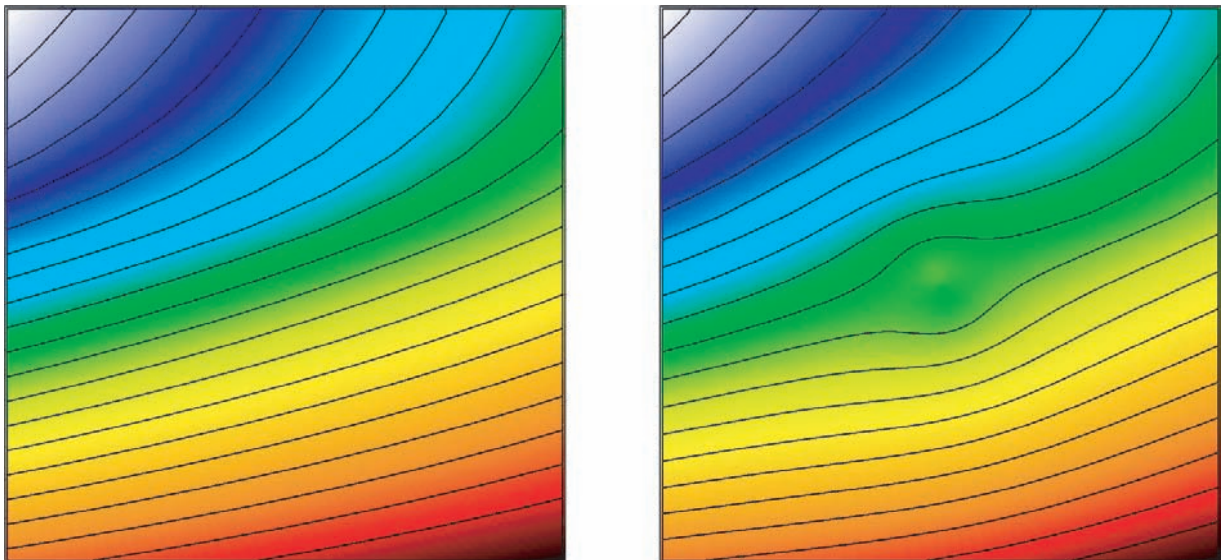

FIG 3.6.— Example of a CMB temperature field of area $10' \times 10'$ (left panel) lensed by a cluster of mass $10^{15} h^{-1}$ Mpc at redshift 0.4 in a $\Lambda$CDM universe (right panel). There is very little structure in the unlensed map, and lensing by the cluster creates a characteristic distortion most easily seen in the contours in the right panel.

Moreover, clusters also lens the microwave background itself. Since lensing conserves surface brightness, a homogeneous microwave background would be unchanged by lensing. On the angular scale of galaxy clusters, (i.e., on scales of at most a few arc minutes), the microwave background is almost featureless. Inhomogeneities in the temperature distribution simply appear as gentle temperature gradients behind clusters. Lensing by clusters imprints a highly specific structure upon this gradient. If the intrinsic temperature of the CMB behind a cluster is $\tilde{T}(\vec{\theta})$, the lensed temperature is, to first order,

$$T(\vec{\theta}) = \tilde{T}(\vec{\theta}) - \vec{\alpha}(\vec{\theta}) \cdot \vec{\nabla}\tilde{T}(\vec{\theta}) \, ,$$

where $\vec{\alpha}$ is the deflection angle (Seljak and Zaldarriaga 2000). The effect is illustrated in Figure 3.6.

---

*    http://www.sdss.org/

†    http://www.vista.ac.uk/

‡    http://prime.pha.jhu.edu/

On scales greater than *Planck*'s resolution of 5′, the lensing signal by individual clusters becomes very weak. However, *Planck* will detect a huge cluster sample, whose maps can be suitably stacked to enhance the signal-to-noise ratio of lensing features in the outskirts of clusters. It may thus be possible to trace the average cluster density profile far from cluster centres by searching for the characteristic imprint of gravitational lensing on the microwave background itself.

### 3.2.5 *Planck*'s cluster sample

*Planck*'s frequency channels were chosen with the tSZ effect in mind. The characteristic spectral signature of the tSZ effect (Figure 3.4) means that it is possible to identify clusters and to distinguish them from other unresolved sources. *Planck*'s best angular resolution of 5′ is lower by a factor of $\sim 5$ than that of the interferometers and bolometer arrays used to study the SZ effect from the ground. As detailed below, this has an impact on the SZ science that *Planck* can do. On the negative side, *Planck* will be less sensitive to lower mass, higher-redshift, clusters than many ground-based experiments, and its SZ detections of clusters will lack the angular resolution needed to study the details of the cluster gas profile, as required for the estimation of the Hubble constant from individual clusters. However, on the positive side, *Planck*'s will produce a large *all-sky* sample of clusters with easily computable selection criteria.

Although the change in surface brightness caused by the tSZ effect is independent of cluster redshift, the tSZ signal integrated across the observing beam falls like the inverse-squared cluster distance $D$,

$$Y = \int y(\vec{\theta})\,\mathrm{d}^2\theta = \frac{kT}{m_e c^2}\,\frac{\sigma_T}{D^2}\,N_e \,,$$

where $N_e$ is the total number of thermal electrons in the cluster. For clusters at moderate and high redshift, convolution with the beam leads to a reduction of the signal amplitude. Clusters below the detection threshold contribute to a tSZ background, whose fluctuations within the beam profile define the noise level. Taking distance effects, beam dilution, and background noise into account, we estimate that *Planck* should see about 30,000 galaxy clusters over the whole sky via the tSZ effect. The median of the cluster redshift distribution should be near 0.3, and a significant fraction of clusters will be near or beyond redshift unity (cf. Figures 3.7 and 3.8).

The mass distribution of *Planck*'s clusters is expected to exhibit a pronounced feature which directly reflects the baryon fraction and the thermal history of the intracluster medium. The integrated tSZ signal is determined by the total thermal energy of the cluster gas, and thus proportional to the mean gas temperature. According to the spherical collapse model for the formation of haloes, the temperature is expected to increase with redshift, i.e., more distant clusters of a given mass $M$ should be hotter,

$$T \propto M^{2/3}\,(1+z)\,.$$

The angular-diameter distance increases with redshift, reaches a flat maximum near redshift unity and then drops towards higher redshift, which is a consequence of space-time curvature. This implies that the inverse-squared distance dependence of the integrated tSZ effect has a fairly mild or even negligible effect on high-redshift clusters, while the cluster temperature keeps increasing with redshift (cf. Figure 3.9). Sufficiently massive clusters can thus be seen to arbitrarily high redshift. The mass distribution of tSZ clusters should therefore show a pronounced peak at a certain mass, beyond which it should fall off exponentially (see Figure 3.10). The location and height of this peak reflect the total thermal energy in clusters at high redshift, and thus constrain the baryon mass and the thermal history of clusters.

Besides the cluster counts and their distributions in redshift and mass, their correlation properties can also be predicted, and the comparison between the observed and predicted auto-correlation function provides a further important test for cosmological models and our understanding of cluster formation and evolution (Moscardini et al. 2002).

Galaxy clusters frequently host radio galaxies, whose emission can contaminate the measured, integrated tSZ signal at low microwave frequencies. Specifically at low frequencies,

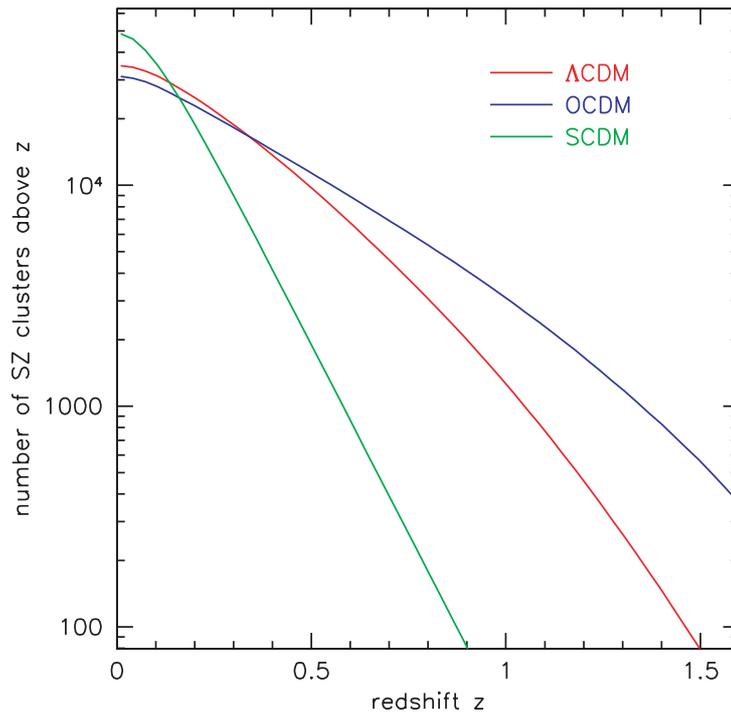

FIG 3.7.—The number of clusters *Planck* should detect at $\sim 3\sigma$ over the full sky at redshifts exceeding $z$, for three different cosmologies. For this calculation, the assumptions about the amount and structure of the intracluster medium were chosen to reproduce the SZ properties of observed low-redshift clusters. Abundance evolution is then based on large-scale numerical simulations (based on Bartelmann 2001).

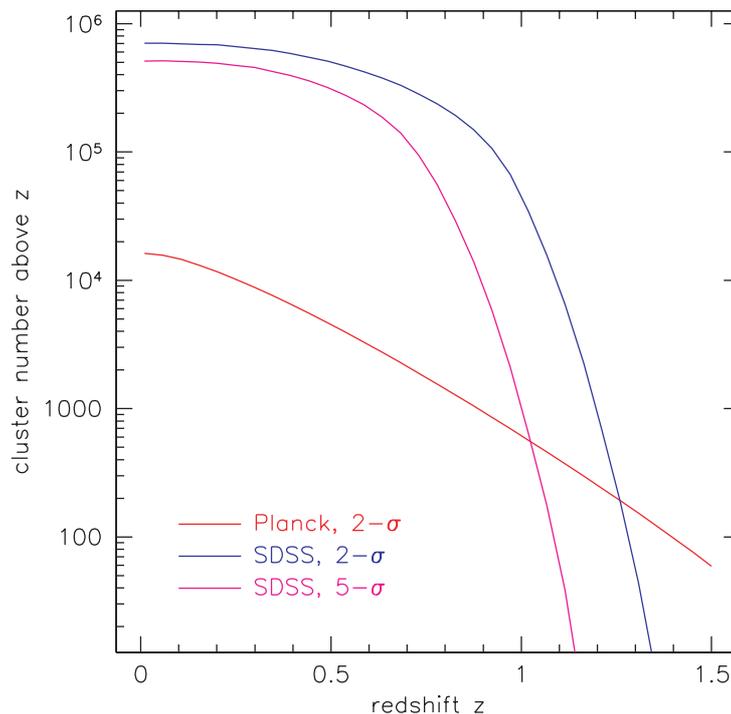

FIG 3.8.—The number of clusters expected to be visible in the optical for the Sloan Digital Sky Survey and in the microwave regime for *Planck* is shown as a function of cluster redshift. In the SDSS area, *Planck* should detect more than $10^4$ clusters, almost all of which will also be detected by SDSS. Those *Planck* clusters which will not appear in the SDSS will be at high redshift (adapted from Bartelmann and White 2002).

*Planck*'s beam does not permit any spatial resolution even of low-redshift clusters, so that the tSZ signal is completely mixed with the radio signal from cluster galaxies. However, radio emission dies off quite rapidly towards higher frequency. The low-frequency part of the tSZ spectrum can thus be contaminated, while the high-frequency part will remain unchanged; in other words, the spectrum can be distorted, so that cluster identification from their microwave signal alone can be hampered. A realistic and detailed assessment on the amount and the consequences of the radio contamination needs to be done, but it is clear that the combination of microwave data with observations in other wavebands will greatly help unambiguous cluster detection and measurement.

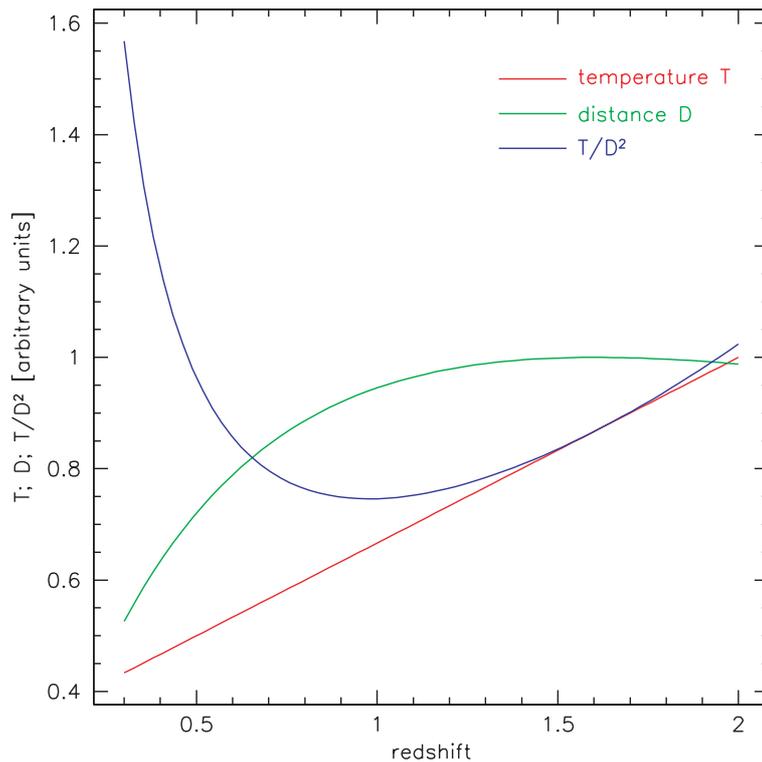

FIG 3.9.— Schematic plot illustrating the fact that the thermal Sunyaev-Zel'dovich effect of sufficiently massive clusters is expected to be visible to arbitrary redshifts: The temperature of halos of fixed mass is expected to increase linearly with redshift, while the angular-diameter distance reaches a broad maximum near redshift unity and then gently decreases. Consequently, the solid-angle-integrated Compton $y$ parameter reaches a minimum near redshift unity and increases towards higher redshift. Thus, clusters sufficiently massive to be visible to redshift unity are even more easily detectable at higher redshifts.

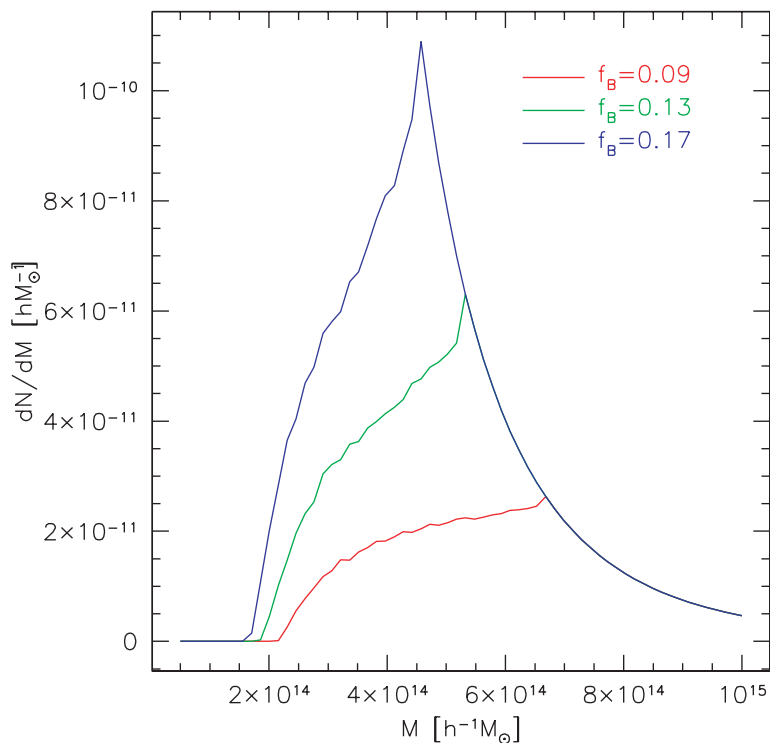

FIG 3.10.— The fact that the thermal Sunyaev-Zel'dovich effect of sufficiently massive clusters is detectable at arbitrary redshifts leads to a characteristic peak in the mass distribution of detectable clusters. Above the peak, the mass distribution has an exponential tail whose amplitude reflects the amplitude of the dark-matter power spectrum (from Bartelmann 2001).

Relativistic corrections to the tSZ spectrum increase the frequency at which the tSZ effect vanishes (Figure 3.4), but this shift will have negligible effect on the *Planck* measurements. However, the relativistic corrections also increase the amplitude of the high-frequency tail of the tSZ spectrum. While this effect is small for individual clusters, the size of *Planck*'s cluster sample should be large enough to detect this effect in combined spectra of a suitable cluster subsample. In principle, a measurement of the relativistic tSZ corrections would allow constraints to be placed on the cluster temperature (Enßlin and Kaiser 2000).

### 3.2.6 Cluster motion

The radial velocity component of galaxy clusters can be determined using the ratio of the tSZ and kSZ effects, combined with estimates of the intracluster temperature that could be obtained either through X-ray observation or directly using *Planck* SZ data (Pointecouteau et al. 1998). Both the X-ray and thermal SZ information can be used to estimate the optical depth per cluster to Thomson scattering, or at least average optical depths within suitably defined cluster samples. The velocity measurement requires a good separation of the tSZ and kSZ effects, which can be achieved through accurate multi-frequency observations in the millimetre and sub-millimetre range.

*Planck*'s sensitivity for measurements of radial peculiar cluster velocities has been estimated using an adaptive spatial filter and assuming that the cluster temperature is known (Aghanim et al. 1997). The observed temperature fluctuation towards any given cluster is contaminated by the CMB itself, the background fluctuations due to the unresolved clusters population, any residual astrophysical foreground contribution, and the instrumental noise.

Using the adaptive spatial filter, the uncertainty expected in individual velocity measurements on bright clusters falls between 500 and $1000\,\mathrm{km\,s^{-1}}$, depending on the cluster core radius (cf. Figure 3.11). The main sources of error are the CMB itself, instrumental noise, and residuals from component separation. The latter currently appear to dominate, but may be reduced by better foreground-separation techniques.

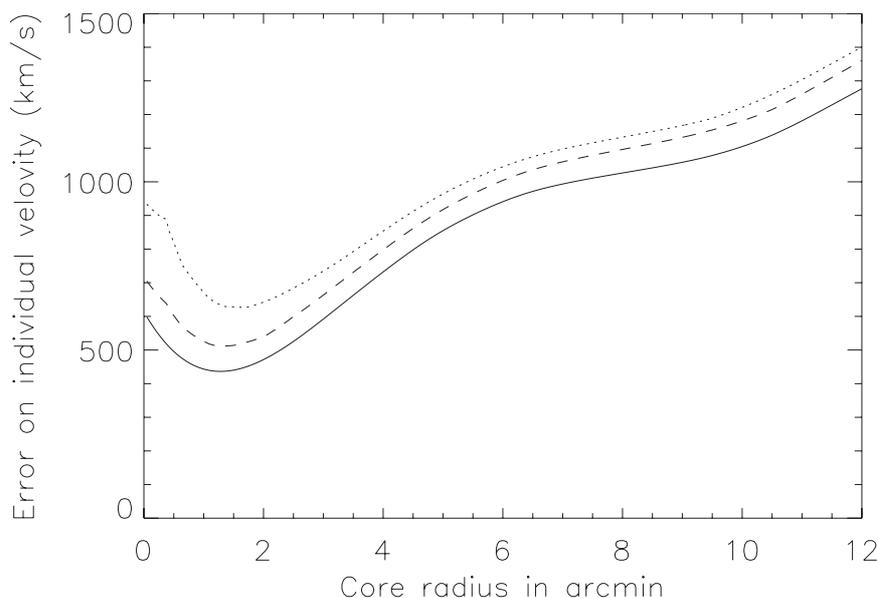

FIG 3.11.—The total rms error in individual cluster velocities, including all contributions (primary CMB, background kinetic SZ, residuals from component separation, Planck-like instrumental noise). The solid, dotted, and dashed lines represent SCDM, OCDM and ΛCDM models, respectively (Aghanim, Górski, and Puget 2001)

Although the expected uncertainties for individual clusters are large, measurement of the cosmic velocity field by averaging over many clusters appears promising. Expected bulk-flow accuracies are of order several hundred $\mathrm{km\,s^{-1}}$ for low-redshift clusters. At redshift unity, they are of order $70\,\mathrm{km\,s^{-1}}$. The kSZ effect might be the only way to measure large-scale bulk motions accurately at high redshifts (cf. Figure 3.12).

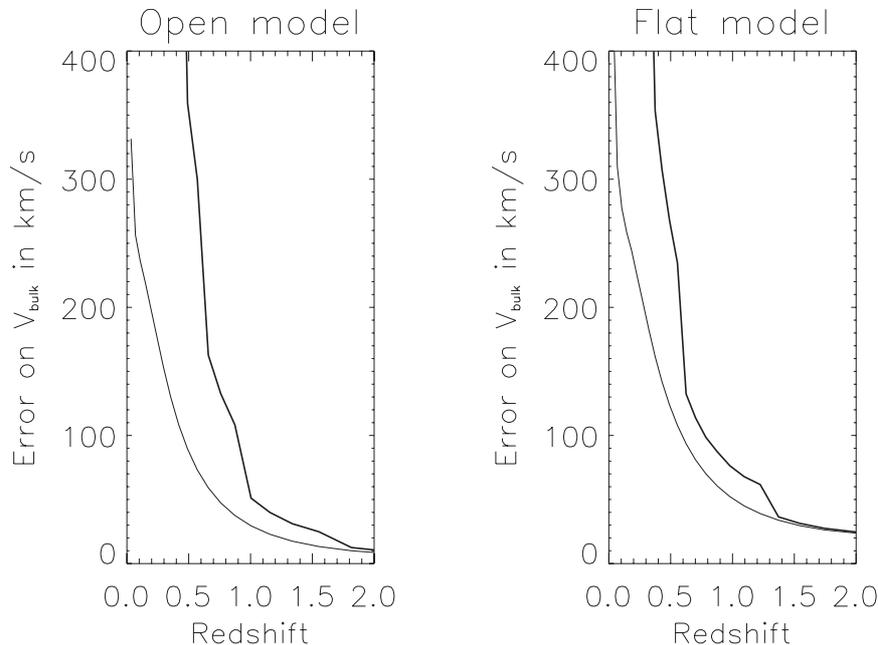

FIG 3.12.—The *rms* error in cluster bulk velocities is shown for two different cosmological models, taking into account (heavy line) or ignoring (thin line) the degradation due to source confusion (Aghanim, Górski and Puget 2001). While the error is large for low redshifts, it drops rapidly towards high redshift.

The evolution of the amplitude of the bulk motions can be used as a cosmological test. It can be studied by combining *Planck* SZ data with measurement of low-redshift bulk flows from more traditional techniques. The evolution of bulk motions depends essentially on the cosmic matter density parameter $\Omega_m$, and thus provides complementary cosmological information. The *rms* peculiar velocity within a window $W_R$ of radius $R$ is, in linear theory,

$$\sigma_v = \frac{H_0 \Omega_m^{0.6}}{\sqrt{2}\pi} \left[ \int_0^\infty |W_R(k)|^2 \, P(k) \, \mathrm{d}k \right]^{1/2} ,$$

where $P(k)$ is the dark-matter power spectrum.

The velocity field can be measured by averaging temperature anisotropies in a pixel-like approach, rather than by averaging over individual detected clusters (Kashlinsky and Atrio-Barandela 2000). At each cluster location, the measured temperature fluctuation will be a combination of the tSZ and kSZ effects, the CMB, and instrumental noise. When averaged over many clusters, the latter three will average down, while the first contribution will be dominated by a possible bulk flow component. This way, *Planck* could probe bulk flows on large scales down to amplitudes below $30\,\mathrm{km\,s^{-1}}$.

The CMB scattered off the intracluster plasma becomes linearly polarised if the CMB illuminating the cluster has a quadrupole moment, and the amplitude of the quadrupole moment depends on the cluster motion. This kinetic polarisation effect could allow measurement of the transverse velocity component, thus providing a direct test of the assumed irrotational nature of the cosmic velocity field on large scales. Detection of a rotational component would have significant implications for structure formation models, and for matter-reconstruction methods mostly based on the assumption that the cosmic velocity field is a potential flow.

The kinetic polarisation for individual clusters is below the average nominal *Planck* sensitivity, even for clusters with high optical depth and transverse velocity (Audit and Simons 1999;

Sazonov and Sunyaev 1999). However, depending on the scanning strategy, the signal-to-noise ratio can be up to ten times higher in regions of high redundancy, where the kinetic polarisation could be detectable for a few individual clusters. Moreover, stacking data from neighbouring clusters will increase the kinetic polarised signal for coherent cluster flows, while reducing it in regions where cluster velocities are uncorrelated. Thus, it could be a promising tool for mapping large-scale transverse velocities.

### 3.2.7 External information

The previous sections illustrate the huge potential for cluster science with *Planck*. Although tSZ observations are routinely undertaken from the ground already, *Planck* will offer two important advantages: full-sky coverage and wide frequency coverage. However, while ground-based observations, in particular interferometric ones, can resolve clusters, the majority of *Planck*'s clusters will be unresolved. The exploitation of the *Planck* data for cluster science therefore rests on their combination and correlation with other data sets.

Cross-correlation with optical and near-infrared survey data can directly constrain the ratio between the thermal energy of the intracluster gas and the amount of light emitted by cluster galaxies. In deep surveys, cluster lensing provides a way for measuring cluster masses. Lensing by clusters of the CMB itself may allow the average outer density profiles to be constrained from stacked cluster maps. The determination of the amount, evolution, and physical properties of the intracluster gas requires the combination of X-ray and tSZ data. Finally, the contamination of tSZ observations by radio emission from cluster galaxies can be assessed and corrected only if radio data are used in the analysis of tSZ data. While sufficiently deep optical and near-infrared surveys will certainly exist by the time *Planck* will fly, a new full-sky X-ray survey is not being planned. It will thus be necessary to reanalyse existing X-ray data, in particular the ROSAT All-Sky Survey.

## 3.3 REIONIZATION

The Universe reionized some time between the epoch of recombination ($z \approx 1000$) and $z \approx 6$, where quasar spectra show that intergalactic hydrogen is highly ionized. As a consequence of reionization, some CMB photons must have suffered Thomson scattering as they travelled from the last scattering surface to us. In addition, the anisotropy of Thomson scattering means that the CMB polarisation must have been affected as well (see Figure 3.13). Therefore, a detailed knowledge of the ionisation history of the universe is crucial for the interpretation of the power spectrum in terms of cosmological parameters. Moreover, there is a chance of finding imprints of the dark ages in the CMB, provided reionisation of IGM took place early enough.

The process of galaxy formation is not well-understood either observationally or theoretically. It is impossible to understand galaxy formation theoretically without detailed knowledge of many individual complicated processes such as star formation, feedback by supernovae, and the radiative transfer of UV light emitted by stars into the IGM. The current treatment of galaxy formation is quite phenomenological. The initial mass function of stars, the star formation rates, and the escape fraction of UV photons from galaxies are all model parameters which have to satisfy the observational constraints. However there are no observations to restrict the parameters beyond redshifts of five. Perhaps only the CMB can provide galaxy formation models with direct observational constraints.

The two most important phases in the ionisation history of the Universe are when protons and electrons first combined at $z \sim 1,000$, and later, when nonlinear gravitational collapse induced reionisation. The standard description of hydrogen recombination has improved in recent calculations which include the treatment of recombination from many hydrogen levels, leading to changes of a few per cent in the anisotropy power spectrum, well within the sensitivity of *Planck* (Seager et al. 2000). More sophisticated models of helium recombination modify the damping length at early times and can also have significant effects on the power spectrum, particularly at small angular scales affecting the amplitude of the acoustic peaks.

These calculations need to be further improved, along with an accurate assessment of the potential uncertainties. However, we note that the recombination process is well understood only in certain regimes accessible in the laboratory, with extrapolations required for the high temperatures and diffuse densities probed by primordial recombination. Modifications to the standard calculation could also be necessary as a result of variations in the coupling constants at early times, for example, the fine structure constant (e.g., Avelino et al. 2001), or other effects on the relevant cross sections. These variations can be parameterised, their systematic effects on the determination of other cosmological parameters assessed, and we may even be able to quantify them as parameters in their own right.

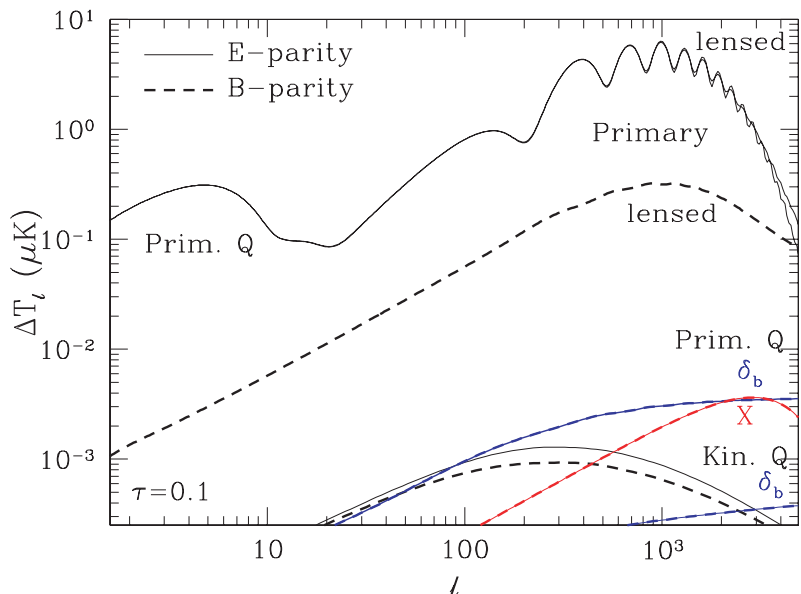

FIG 3.13.—CMB polarisation for the $\Lambda$CDM model with an optical depth from reionisation of $\tau = 0.1$, separated into "electric" ($E$, solid lines) and "magnetic" ($B$, dashed lines) contributions. Secondary anisotropies from the primordial quadrupole are labeled (Prim. $Q$). Homogeneous scattering was assumed for the upper curves, while the scattering was modulated by the baryonic density fluctuations ($\delta_b$) and the ionisation ($X$). For the kinematic quadrupole, the homogeneous and density modulated signals are shown; the ionization modulated and intrinsic quadrupole signals falls below this range. Note that the $B$-type polarization induced by gravitational lensing is much larger than any of these secondary $B$ signals. (Hu 2000)

The effects of early reionisation by the UV light from massive stars and active galactic nuclei can be subsumed in a single parameter, the mean optical depth to the last scattering surface. The effect of the optical depth on CMB temperature anisotropies is degenerate with that of other parameters such as the amplitude and the spectral index of density perturbations (de Bernardis et al. 1997; Griffiths et al. 1999; Stompor et al. 2001); however, polarisation measurements can give a reasonably accurate estimate of the mean optical depth, since reionisation leaves a substantial imprint on the polarisation on the scale of the Hubble radius at the reionisation epoch (Seljak and Zaldarriaga 1997; Kamionkowski and Kosowsky 1998).

This commonly assumed simplicity belies the fact that reionisation is a highly complicated and poorly understood process. It is likely to proceed in an inhomogeneous way, and it may be caused by a wide range of sources, such as active galactic nuclei, early star formation, quasars, and possibly even by relic particles which decay after recombination (Aghanim et al. 1996; Gruzinov and Hu 1998; Knox et al. 1998; Bruscoli et al. 2000; Benson et al. 2001; Liu et al. 2001). Our ability to probe such effects will rely on a sensible parameterisation of theoretical predictions, and a detection of the secondary anisotropies due to inhomogeneous reionisation, the Ostriker-Vishniac (1996) effect, and the SZ foregrounds.

There are at least three effects on CMB temperature and polarisation anisotropies from early reionisation. The first effect is a damping of the primary anisotropies (Sugiyama et al. 1993; Hu and White 1997). The amplitudes of the acoustic peaks in the angular CMB power spectrum are reduced by reionisation. Thus, reionisation adds degrees of freedom which need to be taken into account in the determination of cosmological parameters from the angular CMB power spectrum. The amount of damping of the power spectrum at small angular scales can be parameterised in terms of the optical depth. Reionisation at redshift $\sim 15$ would damp the acoustic peaks by approximately 10% for flat, low density, cold dark matter models with cosmological constant. Noise below $10\,\mu K$ per pixel is required to measure this damping, much better than achieved by *WMAP*. The same damping appears in the polarisation spectrum.

The second effect is the damping and production of polarisation (see e.g., Hu 2000, for a review). Since polarisation traces the last scattering of CMB photons, it will be unique evidence of early reionisation if we find polarisation on large angular scales. We expect to find a new peak in the polarisation power spectrum whose location and height are sensitive to the epoch of reionisation and the optical depth. Therefore, we can easily estimate the damping effect on CMB anisotropies once polarisation has been accurately measured, thus partially breaking the degeneracy of the optical depth with other parameters. *WMAP* has reported a tentative detection of this peak (Kogut et al. 2003). As shown by Holder et al. 2003, *WMAP* is insensitive to even large differences in ionization profiles, but *Planck* will be able to distinguish between various reionisation histories quite well (see Section 2.3.3 and Figure 2.15 for further details).

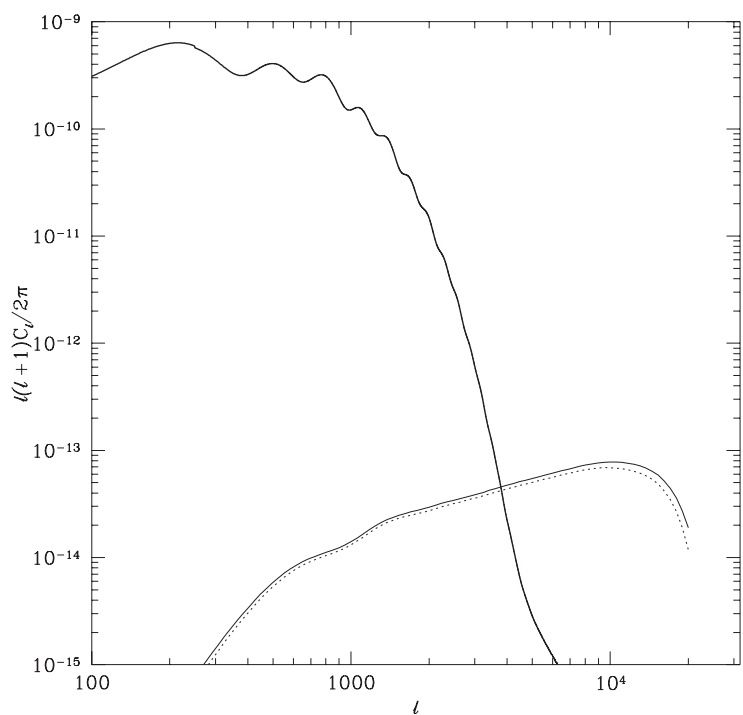

FIG 3.14.—Secondary CMB anisotropies generated by ionised gas escaping from galaxies according to two different models are shown as thin lines. The gas escape fraction was fixed to 10%. The heavy line shows the power spectrum of the primary CMB anisotropies. (Benson et al. 2001).

These two effects are only sensitive to the global signatures of early reionisation. However, the total effect provides information on the spatial distribution of the ionised medium. Each ionised region produces anisotropies in the CMB by its peculiar bulk motion through the kSZ effect (e.g., Valageas et al. 2001). Theoretical studies indicate that it may be possible to obtain information on the reionisation process, such as the escape fraction of UV photons from galaxies into the intergalactic medium and the epoch of reionisation, by using the power spectrum of CMB anisotropies (Benson et al. 2001; cf. Figure 3.14). Since we expect the characteristic size of each ionised region and of the correlation between regions to be imprinted on the primordial CMB anisotropies, measurements with angular resolution better than $10'$ will be crucial. Although the amplitude of CMB anisotropies induced by this effect strongly depends on the epoch and duration of the reionisation process, $1\mu K$ sensitivity is required according to recent theoretical studies. For separating the kSZ effect from the tSZ and other effects, accurately foreground-subtracted temperature maps in several wavebands are required. Moreover, such secondary anisotropies are generated essentially due to the non-linear effect. Thus it is very important to investigate non-Gaussian signatures of these anisotropies.

## 3.4 THE LARGE-SCALE MATTER DISTRIBUTION AND THE CMB

### 3.4.1 Lensing effects on the CMB

CMB photons pass almost the entire Universe on their way to us. They encounter matter inhomogeneities along their paths, whose gravitational pull deflects them. This is the generally weak, large-scale analogue of the gravitational lensing effect discussed earlier in the context of galaxy clusters (see Bartelmann and Schneider 2001 for a review).

Light deflection itself is not an observable effect. If the paths of all CMB photons were deflected in the same way, we would observe a shifted microwave background whose structure would remain unchanged. Any lensing effect on the CMB is thus caused by differential deflection; light paths starting in neighbouring directions experience slightly different deflections (e.g., Cole and Efstathiou 1989; Seljak 1996; Bernardeau 1997).

The deflection angle due to gravitational lensing can be expressed as the gradient of the Newtonian potential of the lensing matter perturbations. Differential deflection is therefore determined by second derivatives of the potential, which can be related by Poisson's equations to the density contrast of the matter perturbations, projected along the line of sight. The variance of the deflection angle due to large-scale structures is

$$\left\langle |\vec{\alpha}|^2 \right\rangle = \frac{9H_0^4\Omega_0^2}{c^4} \int_0^w \frac{W^2(w')}{a^2(w')} \int_0^\infty \frac{P(k,w')}{2\pi k} \, \mathrm{d}k \mathrm{d}w' \, ,$$

where $W(w')$ is a weight function expressing the lensing efficiency at comoving distance $w'$ from the observer, $a$ is the cosmological scale factor, $P(k,w')$ is the dark-matter power spectrum, and $w$ is the comoving distance to the CMB. The effect of these deflections is to modify the CMB power spectrum so that the lensed power spectrum $C_l$ is related to the unlensed power spectrum $\tilde{C}_k$ by

$$C_l = \int_0^\infty \frac{\mathrm{d}k}{\sqrt{2\pi}\epsilon k} \tilde{C}_k \exp\left[ -\frac{(l-k)^2}{2\epsilon^2 k^2} \right] \, ,$$

where $\epsilon \approx 0.01$ is approximately constant. This equation shows explicitly the broadening of the acoustic peaks by convolution with a Gaussian (Seljak 1996).

This leads to a very interesting consequence. Lensing effectively re-distributes power in the CMB, and this leads to a creation of fluctuations on angular scales where the amplitude of the primordial CMB is well suppressed by Silk damping. The lensed CMB should therefore exhibit structure on scales where the primordial CMB would already be featureless (Cole and Efstathiou 1989; Metcalf and Silk 1997; see Figure 3.15).

Differential light deflection also causes structures in the CMB to shift relative to each other. The auto-correlation of hot spots in the CMB, for instance, will be masked in part by the auto-correlation properties of the deflection-angle field, and thus change in a way which is characterised by the matter distribution in front of the CMB (Takada and Futamase 2001).

The polarisation vector of electromagnetic waves is parallel-transported along geodesics, so lensing itself does not change the polarisation properties of the deflected light. However, differential light deflection distorts the polarisation pattern of the background source by solid-angle magnification and de-magnification, thus leading to specific changes in the polarisation of the lensed compared to the unlensed CMB. Specifically, a CMB polarisation pattern which has purely $E$-type (i.e., non-rotational) polarisation modes will, after lensing, also exhibit $B$-type (i.e., rotational) modes. This type of polarisation-mode conversion is a characteristic feature of lensing (Guzik et al. 2000; Benabed et al. 2001).

Since the differential light deflection is related to the deflecting density contrast through the integrated potential and Poisson's equation, it is possible to use the lensed pattern for reconstructing the intervening matter distribution. Suitable linear combinations of temperature gradients are directly related to the projected surface mass density of the lensing matter (Zaldarriaga and Seljak 1999; Bernardeau 1998; Hu 2001).

In the most common scenarios for its origin, the CMB itself is expected to be a Gaussian random field (see § 2.5). Gravitational lensing, however, is caused by large-scale structure at

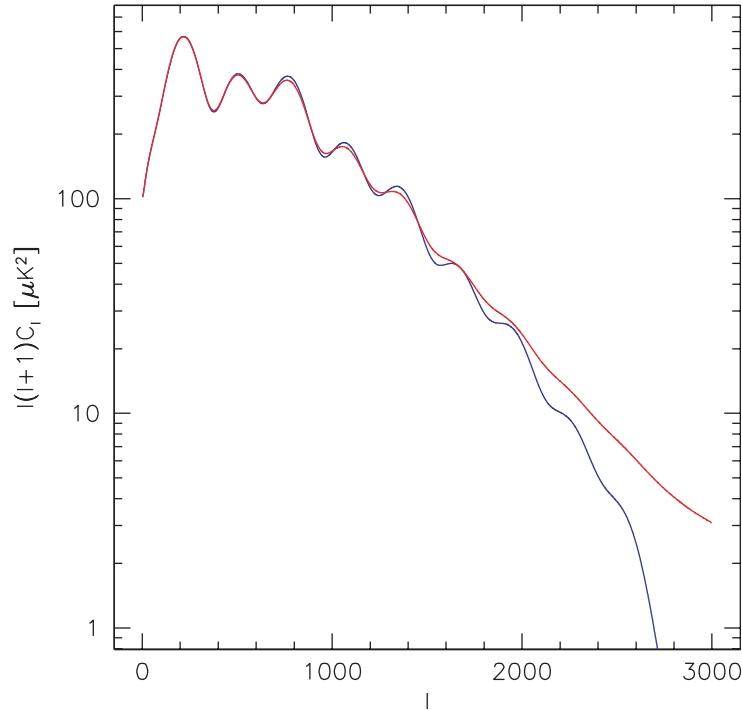

Fɪɢ 3.15.—Unlensed (blue) and lensed (red) CMB power spectra are shown for the ΛCDM model universe. While the lowest-order peaks are unchanged, lensing broadens higher-order peaks and creates power at the smallest scales. (Computed using the CMBfast code, Seljak and Zaldarriaga 1996) much lower redshifts, which has developed a non-linear density contrast. The lensed CMB will therefore exhibit deviations from Gaussian statistics which are characteristic for the late stages of structure growth. Moreover, the statistical properties for morphological measures like the Minkowski functionals are well known for Gaussian random fields, thus their deviation from the Gaussian expectation additionally signals and characterises the non-Gaussianity imprinted by gravitational lensing (Takada and Futamase 2001).

### 3.4.2 Exploitation of lensing effects

These features can be used in various ways for extracting information on the large-scale matter distribution, and possibly also for breaking degeneracies between cosmological parameters. Since lensing mainly affects high-order acoustic peaks in the CMB power spectrum, but leaves the low-order peaks virtually unchanged, it is in principle possible to constrain the lensing signal from the power spectrum alone. However, this is not possible in full generality because the imprint of lensing effects from one cosmological model can to a large degree be mimicked by changing the cosmological parameters, except for the creation of structure on very small angular scales which are well beyond the Silk damping scale. Lensing of the CMB, however, is unavoidable, thus we know that high-order acoustic peaks will be broadened by lensing, and the interpretation of the CMB power spectrum in terms of cosmological parameters depends on our ability to quantify the lensing effect.

The possibility of reconstructing the projected matter distribution from derivatives of the temperature map offers a fascinating additional route towards a quantification of lensing effects on the CMB (see Figure 3.16 for a simulated example). Distortion patterns on individual CMB maps taken at *Planck* resolution and sensitivity will generally be too weak to be directly detected with any significance. However, many maps can be combined either statistically or in other suitable ways that remain to be worked out detail. It should thus be possible to quantify the power spectrum of weak lensing by large-scale structure, and this can then be used for removing effects of lensing from the CMB power spectrum. Furthermore, the lensing power spectrum contains cosmological information of its own, in particular on the shape and amplitude of the dark-matter power spectrum, but also on the cosmological density parameter and, to a lesser degree, on the cosmological constant. Lensing can therefore help break parameter degeneracies in the interpretation of the CMB power spectrum. The lensing distortions in the CMB temperature map, and its creation of $B$ modes from $E$ modes in polarisation maps, are of course tightly related, so their combined analysis should be straightforward. Then, it may be possible to identify polarisation $B$-modes which originated from other processes than lensing, again helping to lift cosmological parameter degeneracies.

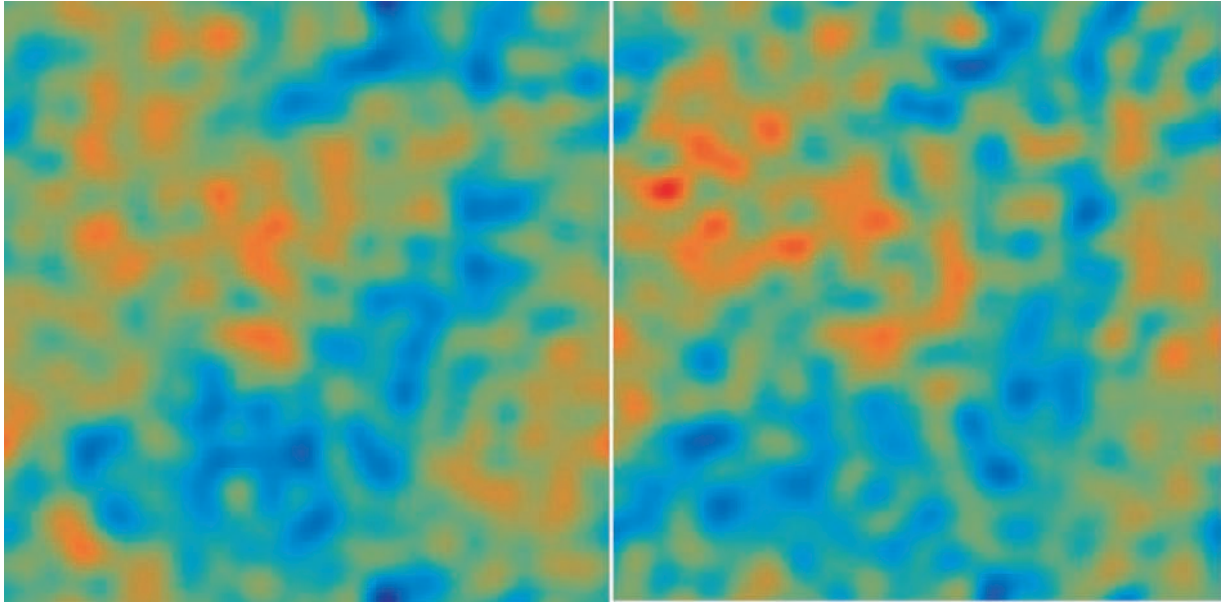

Fig 3.16.— $32° \times 32°$ maps of the lensing deflection field in a $\Lambda$CDM universe (left panel) and its reconstruction from CMB temperature observations (right panel) with realistic assumptions on beam size and noise (from Hu 2001).

An additional and potentially very important way of exploiting lensing effects on the CMB is due to the peculiar geometry of the lensing effect. For maximal effect, lenses have to be placed approximately half-way between the sources and the observer. The redshift where the geometrical lensing efficiency peaks is near unity for the CMB, and near 0.4 for faint background galaxies. Superposed on this geometrical effect is the monotonic growth of cosmic structure. Lenses at lower redshift may be less efficient geometrically, but are more massive and overdense, thus structure growth shifts the peak of lensing efficency for the CMB to redshift 0.6, and for galaxies to redshift 0.35. This means that the CMB and distant galaxies have about two thirds of the effective lensing path length in common, despite their vastly different redshifts (van Waerbeke et al. 2000).

Consequently, there should be a detectable correlation between the matter inhomogeneities responsible for lensing the CMB and the population of faint background galaxies. Planned wide-area surveys in the optical and near infrared like VISTA and PRIME should be sufficiently deep for measuring weak lensing on background galaxies. Combining this signal with lensing patterns in the CMB will considerably increase the significance of both detections.

Large-scale structure lenses are traced by galaxies and galaxy clusters due to biasing. Structures responsible for lensing will thus also be visible in the distribution of galaxies, thus lending additional support to detections of weak lensing on distant galaxies or the CMB.

### 3.4.3 The integrated Sachs-Wolfe effect

At recombination, photons are typically released from regions whose density is over- or underdense compared to the mean. They have to climb out of potential wells, or run down potential hills, and are thus red- or blueshifted compared to the mean CMB temperature. This is the traditional Sachs-Wolfe effect, which provides the only mechanism for creating CMB perturbations on scales larger than the horizon at recombination. The Sachs-Wolfe temperature fluctuations $\delta T$ are related to the potential fluctuations $\delta \Phi$ by

$$\frac{\delta T}{T} = \frac{1}{3} \frac{\delta \Phi}{c^2} \ .$$

As photons propagate to the observer, they keep passing fluctuations in the gravitational potential. If the potential remains static on time scales long compared to the time the photons require to cross a potential fluctuation, the blueshift acquired running into potential well is cancelled by the redshift acquired climbing out of it, and no net effect remains. According to Poisson's equation, the gravitational potential $\Phi$ of the matter fluctuations evolves with time like

$$\Phi = -4\pi G \rho_0 \frac{\delta}{a} \,,$$

where $\rho_0$ is the matter density today and $a$ is the cosmic scale factor. In the early phases of structure formation, the density contrast $\delta$ grows in proportion to the scale factor, so the potential does not change. The evolution at later stages depends on the cosmological model. In low-density universes, structures cease growing after some redshift which depends on the cosmic density parameter, and then the gravitational potential starts decaying, while it remains static in Einstein-de Sitter universes.

Travelling to the observer through an evolving potential, CMB photons can experience a cumulative change in energy: the blueshift of a photon falling into a decaying potential well is not entirely cancelled by the redshift as it climbs out, leading to a net change in photon energy, or temperature, which accumulates along the photon path,

$$\frac{\delta T}{T} = 2 \int \frac{\dot{\Phi}}{c^2} \frac{\mathrm{d}l}{c} \,.$$

This is called the late-time, or integrated Sachs-Wolfe (ISW) effect. It can contribute significantly to the CMB fluctuations on large angular scales. Current measurements of high-redshift supernovae suggest that the Universe is undergoing a period of accelerating expansion, supporting the notion of a non-zero cosmological constant. In such models, the ISW effect is confined to angular scales of order ten degrees and larger.

The ISW effect is also sensitive to possible time-variations of the "quintessence" or dark energy component. If, for simplicity, one models this component by a fluid with constant equation of state, the ISW power is found to be comparable to that induced by the primary anisotropies on those physical scales that re-entered the horizon recently corresponding to angular scales larger than ten degrees. This is because an evolving dark energy component begins to dominate the expansion earlier than in the pure cosmological constant case. The ISW effect can therefore potentially provide a useful discriminant for such "quintessence" models (Baccigalupi et al. 2001).

Since part of the observed CMB anisotropy is associated with the gravitational potential at low redshifts, it must at least in part be correlated with the matter distribution in our vicinity. One may therefore attempt to disentangle the ISW effect from other sources of CMB anisotropy by cross-correlating the *Planck* CMB emission maps with probes which trace the large-scale matter distribution and are thus sensitive to the evolution of the gravitational potential at late times (Bough et al. 1998). Such cross-correlation techniques have previously been applied to COBE-DMR and various extragalactic data-sets (Banday et al. 1996).

A good tracer of the low-redshift matter distribution should satisfy several criteria. First, the tracer objects must be numerous enough for the Poisson error term in any cross-correlation to be small. Second, the tracer must probe as large a volume of redshift space as possible, to maximise the cross-correlation signal. Finally, the tracer survey area should cover a large angular fraction of the sky to minimise the sample variance effect due to incomplete sky coverage in the CMB map.

The Sloan Digital Sky Survey will cover about a quarter of the sky and contain many classes of objects that trace large scale structure. Down to a magnitude limit of $\sim 22.5$ in the $r'$ band, SDSS is expected to find of order 200 million objects, the majority of which will be galaxies. Of those, approximately one million will have spectroscopic redshifts, and photometric redshifts will be available for a substantial fraction of the rest. In addition, the SDSS data will contain of order a million quasars, and spectra will be taken for approximately 100,000 of them. Using spectroscopic and photometric redshifts, one can select a tracer sample to match the redshift window at which the contrast of the ISW signal is most pronounced. Since this redshift is model-dependent, it will be necessary to cross-correlate *Planck* maps with samples covering a range of redshifts. In particular, these objects, after suitable filtering and reconstruction techniques have been applied to recover the underlying projected two-dimensional density distribution, can then be cross-correlated with the *Planck* data sets.

An observed correlation would be important in providing direct support to the standard cosmological paradigm, in which the large-scale distribution of galaxies today form from primordial inhomogeneities amplified by gravitational instability, reflected directly by the effect of the evolving gravitational potentials on the CMB anisotropy. In addition, it may provide additional information to remove degeneracies in the fitting of various cosmological parameters from the CMB anisotropy information.

Since the expected cross-correlation is potentially quite weak (Peiris and Spergel 2000), all parasitic signals must be removed from the *Planck* sky maps. To maximise the desired cross-correlation, the *Planck* maps will necessarily need to be cleaned of nearby sources, to minimise confusion signals. Similarly, since the sought-after correlation arises primarily at angular scales of a few degrees or larger, where the Galaxy can be a serious contaminant, this must be adequately removed prior to performing the cross-correlation analysis. A detailed understanding of foreground issues may also be important from the point-of-view of the impact on, for example, the SDSS sampling and completeness due to dust extinction.

Although the SDSS sets the prototypical example for a suitable wide-area survey, there will also be opportunities from other current and next-generation spectroscopic and photometric redshift surveys, including 2dF, VIRMOS, DEEP, VISTA and WFCAM. Tracing the matter distribution at higher redshift may also be possible using data from the hard X-ray background, or for redshifts around three via the Lyman-break galaxy population.

### 3.4.4 Large-scale structure of warm and hot gas

The thermal and kinetic SZ effects, determined by the thermal energy density of baryonic gas, will be strongest for those lines-of-sight passing through hot, dense electron clouds, e.g., the centres of galaxy clusters. However, since the observed SZ effect integrates along a path of cosmological length, a significant SZ effect may also be detected along lines-of-sight which pass through a medium which is cooler or less dense than a cluster core, but which has a larger volume filling factor in space. Such conditions might prevail in the warm and hot gas seen in some hydrodynamical simulations (e.g., Cen and Ostriker 1999). The SZ effect provides one of the few ways of detecting this gas, which is shock-heated to $10^{5-7}$ K as structures form under gravitational instability, and which, at the present epoch, fills $\sim 10\%$ of space, in mainly unvirialised regions, and contains $\sim$40–50% of all baryons. It is clear that, at the crudest qualitative level, the SZ effect belongs at low redshift, since it will only be relatively late in the evolutionary history of the Universe that gravitational collapse will have proceeded sufficiently far to produce electron clouds sufficiently hot and sufficiently dense to produce an appreciable SZ signal.

The observational parameters of the *Planck* mission constrain the range of SZ science for which it has an advantage over other experiments. The niche within which *Planck* wins will be defined in more detail below, but it is clear that there are two broad goals of the study of large-scale structure (LSS) in *Planck* SZ emission. The first goal is to measure the angular power spectrum produced by galaxy clusters and its evolution with redshift, to produce constraints on cosmology and structure formation theories complementary to those from the primary CMB anisotropy. The second goal is to detect the large-scale filamentary structure of the warm/hot gas of the intergalactic medium, and trace its evolutionary history, thereby constraining models of galaxy formation.

The astrophysical complexity and dynamic range of the set of physical processes affecting SZ emission mean that detailed predictions of what *Planck* will see are challenging goals in themselves. An ideal modelling tool would undertake a simulation of a cosmologically significant volume of space, and include a full, hydrodynamic treatment of the evolution of the baryonic component down to the level of the detailed feedback processes resulting from the formation of massive stars and active galactic nuclei. Such a tool does not exist, due to a combination of restricted computational resources and limited understanding of the relevant astrophysics, and is not likely to become available in the next five years. Simplified methods must be employed.

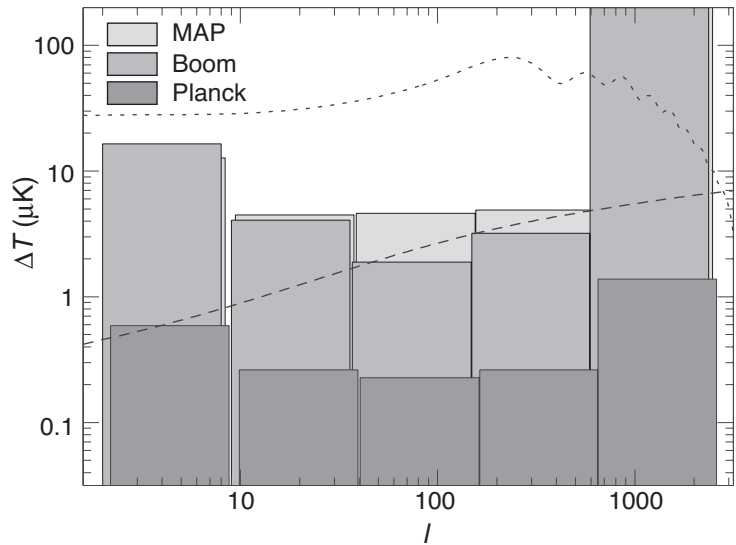

Fig 3.17.—Detection thresholds for the SZ effect due to large-scale structure. Error boxes represent the 1-σ rms residual noise in multipole bands and can be interpreted as the detection threshold. Also shown (dotted) is the level of the primary anisotropies that have been subtracted with the technique and the signal (dashed) expected in a simplified model. (Cooray, Hu and Tegmark 2000).

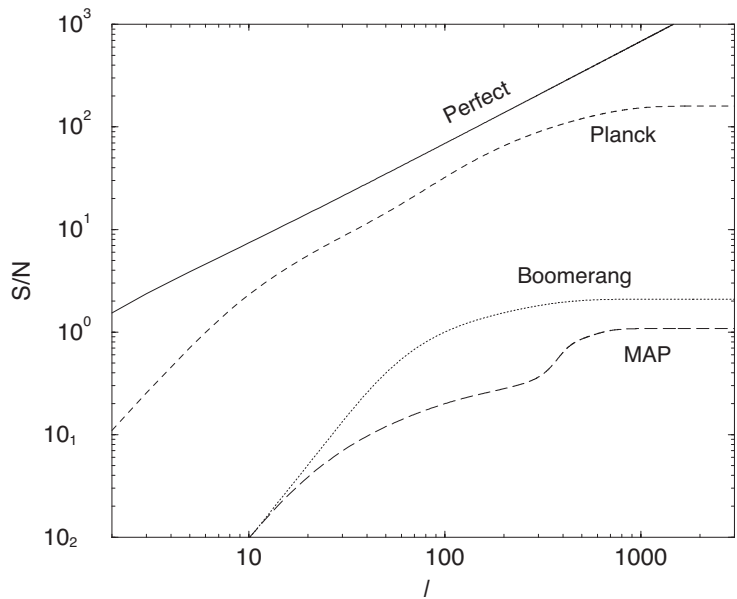

Fig 3.18.— Cumulative signal-to-noise ratio in the measurement of the power spectrum created by the SZ effect due to large-scale structures. Expectations are shown for Boomerang, *WMAP* and *Planck* as a function of multipole order $l$. The solid line is the maximum signal-to-noise ratio achievable in a perfect experiment. The signal-to-noise ratio is adequate in the *Planck* data for a clear detection of the SZ effect, while Boomerang and *WMAP* will only allow reasonable limits to be placed on any SZ contribution. (Cooray, Hu and Tegmark 2000; cf. Figures 3.17 and 3.18).

To date, authors have adopted a mixture of approaches, from purely analytic prescriptions (e.g., Komatsu and Kitayama 1999; Cooray et al. 2000) to full hydrodynamic simulations (e.g., Cen and Ostriker 1999; da Silva et al. 2000; Refregier et al. 2000; Seljak et al. 2000), to mixed schemes which evolve the mass distribution numerically (e.g., Kay et al. 2001) and then use a simple analytic cluster model to convert from halo mass to SZ flux. All three approaches trade off size, resolution, and physical accuracy as required for tractability in finite time, and it is no surprise that they do not yield identical predictions for quantities such as the angular power spectrum of SZ emission. They do, however, agree at the order of magnitude level, which is sufficient to make a robust initial assessment of what *Planck* can achieve.

The power spectrum of the tSZ emission component map contains information about cosmology and the growth of structure that may be extracted via comparison with model predictions. While neither Boomerang nor *WMAP* is likely to detect the SZ power spectrum, *Planck* can detect it at sufficient significance to enable extraction of its shape. This is principally due to *Planck*'s better frequency coverage. The SZ power spectrum can be viewed as the sum of two terms; a Poissonian term, due to bright, nearby clusters, and a clustered term, due to clusters which are fainter or at higher redshift. The first term may be removed by masking out bright nearby clusters, possibly enabling detection of the correlation signal on angular scales of order one degree and measurement of the clustering of clusters at higher redshift or lower mass than probed by the power spectrum of the full map. The amplitudes of both Poissonian and clustered terms produce an independent estimate of the normalisation of the dark-matter fluctuations if the baryon fraction in clusters can be reliably constrained, e.g., by pre-*Planck* ground-based SZ follow-up of clusters studied in detail in the X-ray. The bright sources excised from the SZ map can also yield useful information, and this may be most readily extracted from their redshift space power spectrum or correlation function, once redshifts have been determined from them all.

The SZ signal will be more Gaussian than the density field because it comes from many independent fluctuations along the line-of-sight, but it will still contain a non-Gaussian component induced by the gravitational collapse of the intergalactic gas distribution at low redshift. The simplest indication of non-Gaussianity would be a non-vanishing skewness from the SZ map. This is simple to compute, even at full *Planck* resolution. The skewness should be detectable for *Planck*, but not for *WMAP*, although this prediction is somewhat uncertain. If skewness indicates non-Gaussianity, more detailed information can be derived via the SZ bispectrum, which should also be detectable for *Planck*, although it will be computationally challenging.

Information on the large-scale structure and evolution of the warm/hot gas can be obtained by cross-correlating *Planck* SZ emission and other tracers of the large-scale structures at redshifts below unity, especially where this can be performed in redshift slices, to trace the evolution of the correlated signal. In the simplest terms, in which all LSS tracers are assumed to be linearly biased with respect to the density field, this can be thought of as tracing the evolution of the relative bias between the pressure of the warm/hot gas and the other tracer, although, in practice, more sophisticated biasing models will have to be used.

The first analyses of this kind were undertaken cross-correlating X-ray and COBE-DMR maps (Banday et al. 1996; Kneissl et al. 1997), detecting only a strong signal caused by our Galaxy. With *Planck*, cross-correlating SZ and X-ray maps will aid the identification of clusters present in the *Planck* SZ map at too low significance for robust detection, as well as helping detect the large-scale structure of the warm/hot gas, which will be emitting X-ray flux as well as producing an SZ effect.

Most of the correlated SZ signal on *Planck* scales comes from redshifts below $\sim 0.5$, so it should be possible to compute these cross-correlations in photometric redshift bins, using the Sloan Digital Sky Survey, which will cover $10^4$ square degrees of the high-latitude sky, almost half to be covered in the $J$, $H$, and $K$ bands by the UKIDSS survey. This will facilitate a tracing of the evolution of the warm/hot gas. No detailed analysis has been undertaken for *Planck* yet, but work has been done for the case of *WMAP*. For instance, the cross-correlation between *WMAP* and the photometric redshift catalogue deduced from the sample of SDSS galaxies brighter than $r' = 21$ should be detectable at a signal-to-noise ratio exceeding ten. *Planck*, with its component separation, should do at least an order of magnitude better than this.

The amplitude of any such cross-correlation depends to a factor of a few on the details of the thermal state of the warm/hot gas, which is both an attraction and a concern. In principle, it means that this cross-correlation can distinguish between different models for the heating of the intergalactic medium, but, since this is currently very uncertain, it is difficult to quantify what is expected to be seen with *Planck*. This is definitely one area where a lot more work will be needed between now and *Planck*'s launch.

In addition to computing $N$-point spatial statistics for the SZ emission map, complementary information can be derived by studying the quantitative morphology of the SZ signal via the Minkowski functionals (Sahni et al. 1998; Novikov et al. 1999). These quantities, of which there are $D + 1$ for a $D$-dimensional space, provide a mathematically complete description of the morphology of excursion sets of points above thresholds, and are computationally cheap to compute, scaling as $\mathcal{O}(N)$, where $N$ is the number of pixels above the threshold under consideration.

The variation of the Minkowski functionals as a function of threshold height can be used as a test of Gaussianity in CMB maps, or as a way of quantifying the shapes of contour regions. This could be used in the *Planck* SZ map to search for evidence of the filamentary structure of the warm/hot gas, although that may be washed out in projection, so future work will be needed to assess the usefulness of Minkowski-functional decomposition of *Planck* SZ data.

# CHAPTER 4
# EXTRAGALACTIC SOURCES

## 4.1 OVERVIEW

*Planck* will observe a wide frequency range covering spectral regions difficult or impossible to explore from the ground and only lightly surveyed by present and planned space missions. In the course of this survey, *Planck* will measure many hundreds of radio sources and many thousands of dusty galaxies, providing us with our first complete catalog of submillimetre sources.

*Planck* has a small telescope compared to millimetre and submillimetre antennas on the ground; its angular resolution and sensitivity to point sources are correspondingly lower. Nevertheless, *Planck* observes simultaneously at nine frequencies, covers the entire sky, and will be exquisitely well-calibrated across all nine frequency bands. These advantages will make *Planck* the premiere instrument for finding and studying complete samples of bright, rare objects from 30–850 GHz, and a good complement to present and planned radio and submillimetre instruments, including the Atacama Large Millimetre Array (ALMA), which will produce deeper surveys over limited areas and frequency bands.

In addition, the unresolved cosmic far infrared background (CFIRB) will be observed with high signal to noise ratio at the highest frequencies, allowing accurate separation of this background from infrared cirrus. The CFIRB contains unique information on the correlation properties of weak sources which cannot be studied in any other way before ALMA is in full operations.

This chapter explores *Planck*'s contribution to our understanding of extragalactic sources.

## 4.2 BACKGROUND AND INTRODUCTION

The spectral energy distributions of most extragalactic sources have a minimum at roughly 1–2 mm wavelength, fortuitously close to the peak of the CMB spectrum. This minimum arises from a superposition of two emission mechanisms. The radio emission of most radio sources is predominantly synchrotron with power law spectrum $\nu^\alpha$, where $-1 \lesssim \alpha \lesssim -0.5$. At wavelengths shorter than 1 mm, the emission typically is dominated by thermal dust emission, which increases steeply with frequency ($3 \lesssim \alpha \lesssim 4$). Figure 4.1 shows typical source spectra. A minimum at $\lambda \sim 4$ mm is seen in the spectral energy distribution (SED) of the Milky Way. *Planck*'s frequency bands were selected to cover the millimetre and submillimetre range both to minimize and to characterise Galactic foreground emission.

As described in this Chapter, *Planck* will measure thousands of discrete extragalactic sources. We begin with two general remarks. First, the frequency region spanned by the *Planck* detectors is poorly studied, hence the properties of sources at these frequencies are poorly understood. The fact that *Planck* covers a large frequency range allows it to make a particularly valuable contribution. Second, there are classes of sources known to have spectra rather different from those shown in Figure 4.1. These include sources with "inverted" spectra ($\alpha > 0$) for frequencies up to 1–10 GHz, as well as other sources thought to be among the most recently formed and most luminous extragalactic radio sources. Many such sources are expected to be strongly variable. *Planck* will permit study of the brightest of these objects with unprecedented frequency coverage and over a wide range of time domains. *Planck* will also detect and characterise the sub-millimetre emission of a large number of dusty galaxies, again with unprecedented frequency coverage.

Star forming galaxies radiate in the optical and ultraviolet part of the spectrum, where stars emit most of their energy, but also in the mid and far infrared where the radiation absorbed by dust is reradiated. For the Milky Way (and most local spiral galaxies), about one third of the total output power is emitted in the mid and far infrared. For starburst galaxies the fraction can be as large as 99%. Star forming galaxies thus present a double peaked spectral energy distribution (SED) with a highly variable ratio between the two components. At the source, the infrared part of the SED peaks around 100 µm. For starburst galaxies at redshifts 2 or larger, this peak of the SED is shifted beyond 300 µm, into the *Planck* range.

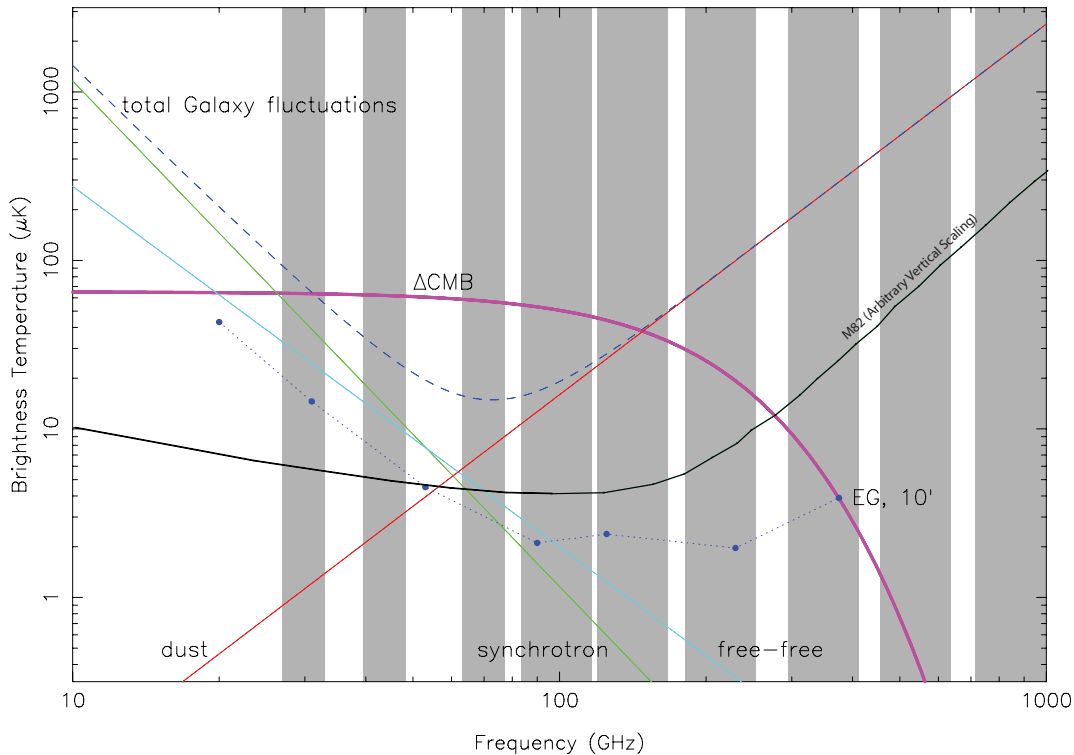

FIG 4.1.— Spectra of sources in "brightness temperature" (in which a Rayleigh-Jeans $\nu^2$ spectrum is flat), superimposed on the *Planck* frequency bands. Spectra of the Galaxy (as measured by *WMAP*, see Fig. 1.3) and M82, a star-forming galaxy, are shown. For the Galaxy, the components contributing to the over-all spectrum are identified. Also shown are the expected level of CMB fluctuations on a 1° scale, and, as a light dashed line (EG), the expected level of fluctuations introduced by all foreground radio sources on a 10′ scale.

Infrared starburst galaxies are often associated with mergers and interacting galaxies. *Planck* will be able to detect the rare high redshift, ultra-luminous, infrared galaxies in the tail of the luminosity function. Furthermore, the cosmic far infrared background (CFIRB) in the submillimetre/millimetre range, made up of the unresolved weaker sources, potentially contains original information on the spatial distribution of mergers, and thus on the galaxy formation process. Only recently have the galactic and the extragalactic components of the far-infrared background been separated in the data of the COBE FIRAS and DIRBE instruments. This background contains power comparable to its optical/UV counterpart. This surprising result (locally, as mentioned above, the integrated infrared emission of galaxies is only one third of the optical) has resulted in a strong interest in the population of sources responsible for this background, but progress has been slow due to the difficulty of observations in this wavelength range. *Planck* will be an important tool for studying the CFIRB.

In §§ 4.3–4.4, we describe some of the science goals to be met using *Planck* observations of extragalactic sources, as well as complementary ground-based observations that will be made. We expect substantial gains in our understanding of extreme radio sources and of star formation processes that drive the thermal re-emission by dusty galaxies.

*Planck* will provide important and novel data on extragalactic sources; however, from the standpoint of the CMB, discrete sources (and the CFIRB) are a foreground contaminant. Hence in § 4.5 we briefly describe how extragalactic sources can be removed from *Planck* images to limit the foreground noise they contribute to CMB images. Proper control of this potential source of error in *Planck*'s cosmological results will require careful pre-launch modeling and observations as well as component separation from the CMB images *Planck* produces.

## 4.3 EXTRAGALACTIC RADIO SOURCES

At present, there are no all-sky surveys of extragalactic discrete sources at wavelengths between 6 cm and 100 µm ($5\,\text{GHz} < \nu < 3\,\text{THz}$). This represents the largest remaining gap in our knowledge of bright extragalactic sources across the electromagnetic spectrum. Throughout this eight octave frequency range, we lack detailed spectral information about known sources, and, more interestingly, we may discover new populations not revealed by lower frequency radio surveys or by the Infrared Astronomical Satellite (IRAS) surveys of two decades ago.

While some progress towards filling this gap in our knowledge is expected from NASA's WMAP mission and from ground-based surveys both at wavelengths down to $\lambda \sim 1\,\text{cm}$ and at submillimetre wavelengths, it is *Planck* that will fill this gap decisively, providing the equivalent of the NGC list for the millimetre and submillimetre sky. At *Planck* frequencies above 217 GHz ($\lambda < 1.4\,\text{mm}$), we expect *Planck* to detect primarily dusty starforming galaxies (see § 4.4 following). Here we concentrate on sources detected at frequencies below 217 GHz (and out of the Galactic plane to minimize contamination by Galactic sources). Modeling of known source counts, extrapolated to higher frequencies (Toffolatti et al. 1998; Cayon et al. 2000; Sokasian et al. 2001), suggests that we will detect from several hundred to a few thousand extragalactic discrete sources in the *Planck* frequency bands from 30 to 217 GHz. (Counts in the highest frequency band may be dominated by starforming galaxies, but we still expect a statistically useful sample of radio sources.) We emphasize that these estimates do not include the possibility of entirely new classes of sources not well represented in lower frequency all-sky surveys at $\nu < 5\,\text{GHz}$. Figure 4.2 shows some examples.

### 4.3.1 The Phenomenology of Radio Sources

The primary emission mechanism in virtually all radio sources is synchrotron emission from relativistic electrons spiraling around magnetic fields in the source. If the electrons have a power law spectrum, with $N(E) \propto E^{-p}$, the synchrotron spectrum is also a power law with $S \propto \nu^{\alpha}$ and $\alpha = (1-p)/2$. For many, but not all, radio sources, $-1.0 \lesssim \alpha \lesssim -0.5$; sources with $\alpha > -0.5$ are referred to as "flat spectrum" sources, and those rare sources with $\alpha > 0$ are referred to as "inverted spectrum" sources.

The observed spectra of radio sources frequently depart from simple power laws with fixed $\alpha$. One cause of a break of $\Delta\alpha$ in the spectral index is the steepening of the electron power law index $p$ by 1.0 or more due to "aging"—the higher the energy of an electron, the less time it takes to radiate a given fraction of that energy. Since high frequency synchrotron emission is dominated by the high energy electrons, the effect of aging is to steepen the radio spectrum at high frequencies. Conversely, very young radio sources may have flatter than average radio spectra.

Another potential cause of curvature in the SEDs of radio sources is opacity: as the optical depth rises, the source spectra invert, with $2 \leq \alpha \leq 2.5$ for fully opaque sources. The result is a peak in the SED at roughly the frequency where the optical depth reaches unity. The frequency at which this spectral peak occurs depends on both the physical parameters of the emitting/absorbing region and its age; younger, more compact sources display this spectral turnover at higher frequencies.

The so-called gigahertz peaked spectrum (GPS) sources (Figure 4.2) are one class that exemplifies this behavior (see, e.g., Stanghellini et al. 1996). We know of sources that have SED peaks at frequencies as high as 100–200 GHz (Grainge and Edge 1998). These are evidently very young or very extreme cases of radio galaxies, and consequently particularly interesting. *Planck* is ideally suited to detect and characterise the brightest of these sources, given its wide frequency response.

Very compact sources can also be variable, with variability time scale roughly limited to $\Delta t > d/c$, where $d$ is the source size. *Planck* will allow us to make essentially simultaneous measurements of a given source in all frequency channels, thus providing single-epoch spectra (a true measure of the SED, unaffected by variability). It will also allow us to monitor variability frequency by frequency, on time scales ranging from hours to 6 months, and hence provide

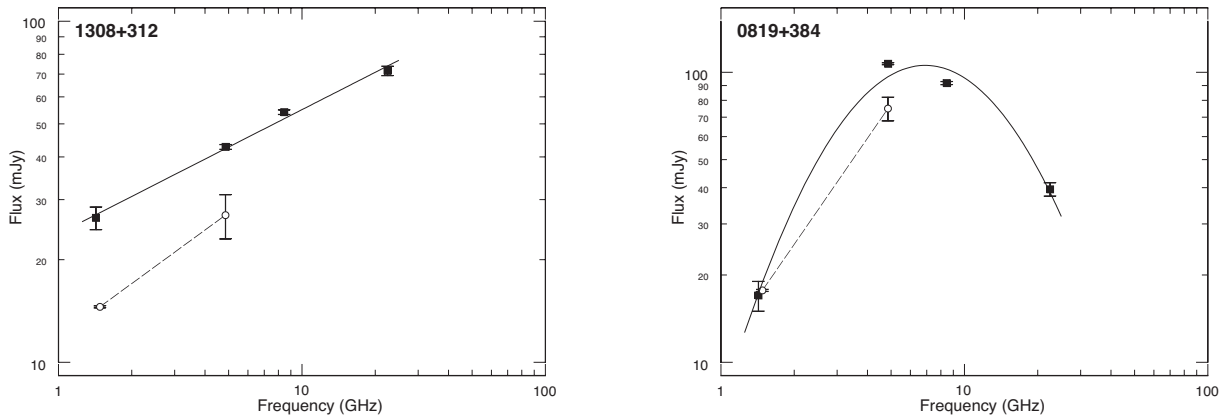

Fɪɢ 4.2.— Radio frequency spectra of two atypical radio sources. One is a GPS source.

information on the size and evolution of extreme radio sources (see §4.3.2 below).

Synchrotron radiation is intrinsically highly polarized. Yet most radio sources observed at MHz or GHz frequencies are not. The lack of polarization may be due to random alignment of the magnetic fields in the source, or to Faraday depolarization of the emergent radiation. The latter effect is frequency dependent (the rotation measure $RM$ varies as $\nu^{-2}$). At *Planck* frequencies, Faraday depolarization will generally be small, so we will be able to measure the Faraday depth as well as to determine the intrinsic polarization of sources.

In some radio sources, free-free emission from ionized hydrogen may play a role, with $S \propto \nu^{-0.1}$ in optically thin sources. This is the case for Galactic radio emission along some lines of sight (see Chapter 5). For most luminous extragalactic sources of the kind *Planck* will detect, on the other hand, we expect the free-free emission to be swamped by synchrotron emission.

Finally, to provide a framework for the following sections, we briefly discuss the current model for the morphology of radio sources. The so-called *unification model* of radio sources permits us to explain a wide variety of extragalactic discrete sources with a common scheme: an active galactic nucleus (AGN) with a black hole, an accretion disk around it, and axial jets (see Figure 4.3, and Urry and Padovani 1995 for a review). Many apparent differences between quasars, double-lobed radio sources, and other classes of radio source are then determined by the orientation of the line of sight to the jet axis. It is clear that orientation to the line of sight cannot explain all apparent differences between classes of radio sources, nor should it be expected to. Intrinsic physical differences, including a variation of isotropic radio luminosity over at least five or six orders of magnitude, are known to exist as well. However, to disentangle the effects of orientation from those of physical differences, extreme cases provide an important diagnostic, and *Planck* is well suited to characterise extreme cases (see §4.3.2). For example, Doppler boosting of photon energies along the jet axis (Blandford et al. 1977) is invoked in unification models to explain the greater apparent brightness of sources viewed along the jet axis; other models invoke absorption in the accretion disk or surrounding dusty torus instead. As we show in §4.3.3, *Planck* observations can help determine the relative importance of these two mechanisms in various classes of radio source.

### 4.3.2 Extreme GPS and Other Strongly Inverted Spectrum Radio Sources

The GPS sources are powerful ($\log P_{1.4\,\mathrm{GHz}} \geq 25\,\mathrm{W\,Hz}^{-1}$), compact ($\lesssim 1\,\mathrm{kpc}$) radio sources with a convex spectrum peaking at GHz frequencies (O'Dea 1998). An example is shown in Figure 4.2.

GPS sources are identified with both galaxies and quasars; however, unification models in which the two populations differ only by orientation do not seem to apply to these extreme sources. Rather, GPS galaxies and GPS quasars appear to be unrelated populations. They have different redshift, rest-frame frequency, linear size, and radio morphology distributions (Stanghellini et al. 1996, 1998; Snellen et al. 1999). Sorting out these differences would be aided by the larger sample of GPS sources, especially those with high peak frequencies, that *Planck* will supply.

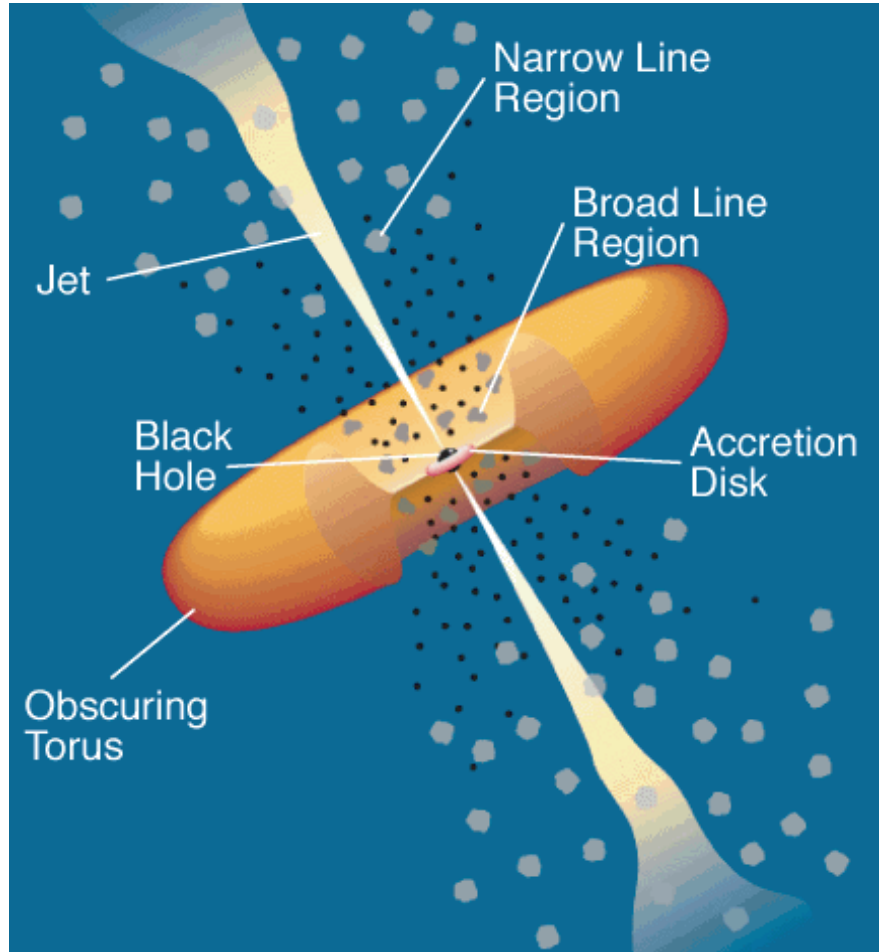

FIG 4.3.— The unified model of radio sources. From Urry and Padovani, 1995.

Any population of sources peaking above 10 GHz (in the observed frame) may be strongly underrepresented in the present samples, because they are relatively weak at the frequencies ($\leq 5$ GHz) where large area surveys are available. Still, scores of such sources are known (Edge et al. 1996, 1998; Grainge & Edge 1998; Tornikoski et al. 2000, 2001; Guerra *et al.*, 2002). Grainge & Edge (1998) report the detection of ~50 GPS sources with spectra still rising above 10 GHz (although the source list is not provided); one of these has the emission peak above 190 GHz in the rest frame. Guerra and his colleagues (private communication) have found another ~100 GPS sources. In addition we know that some sources have been erroneously classified as GPS on the basis of multi-frequency observations made at different epochs. If a source happens to be flaring when a high frequency measurement is made, it can appear to have a gigahertz peak. *Planck*'s *simultaneous* multifrequency measurements allow a straightforward, unambiguous characterisation of GPS sources.

It is now widely agreed that GPS sources correspond to the early stages of the evolution of powerful radio sources, when the radio emitting region grows and expands within the interstellar medium of the host galaxy before plunging into the intergalactic medium and evolving into an extended radio source (Fanti et al. 1995; Readhead et al. 1996; Begelman 1996; Snellen et al. 2000). Conclusive evidence that these sources are young comes from measurements of the propagation velocity of hotspots in a handful of prototype GPS sources. Velocities of up to ~0.4c were measured, implying dynamical ages ~ $10^3$ years (e.g., Owsianik et al. 1998; Tschager et al. 2000). Estimates of the radiative ages of small radio sources are also consistent with their being young (Murgia et al. 1999). The identification and investigation of GPS sources is therefore a key element in the study of the early evolution of radio-loud AGNs.

There is a clear anticorrelation between the peak (turnover) frequency and the projected linear size of GPS sources. This suggests that the physical mechanism (probably synchrotron self-absorption) responsible for the change in spectral slope depends simply on the source size. Although this anticorrelation does not necessarily define an evolutionary track, a decrease of the peak frequency as the emitting region expands is indicated. Thus *Planck*, with its wide frequency coverage extending to high frequencies, may be able to detect these sources very close to the moment when they turn on, perhaps within the first few years.

On the other hand, it is not clear at this stage whether there is a continuity between the low-frequency peaked and the very high-frequency peaked, extreme, GPS sources. The new samples provided by *Planck* and follow-up VLBI observations of extreme GPS sources will allow us to decide. *Planck* might also test the frequency of occurrence of transitions from GPS to blazar spectra, as observed for the quasar PKS0528+134, and allow us to investigate the nature of processes involved.

The self-similar evolution models by Fanti et al. (1995) and Begelman (1996) imply that the radio power drops as the source expands, so that GPS sources evolve into lower luminosity radio sources. With a suitable choice of the parameters, this kind of model may account for the observed counts, and redshift and peak frequency distributions of the currently available samples (De Zotti et al. 2000). On the other hand, Snellen et al. (2000) support a scenario whereby the luminosity of GPS sources *increases* with time until a linear size $\sim 1\,$kpc is reached, and decreases subsequently. The former scenario implies that GPS sources may comprise a quite significant fraction of bright ($S > 1\,$Jy) radio sources at $\nu > 30\,$GHz. Correspondingly, large area, shallow surveys, such as *Planck*'s, are optimally suited to test these models and to select complete samples of these sources. The predicted counts of GPS sources in *Planck* channels depend on the value of the maximum intrinsic peak frequency, $\nu_{p,0}$. Figures 4.4 and 4.5 show predictions for $\nu_{p,0} = 200$ and $1000\,$GHz, respectively.

With multifrequency measurements at two epochs six months apart, *Planck* will allow us to investigate variability on timescales less than $1\,$yr, quite relevant for these very compact sources (the expected linear size of sources peaking at $\sim 100\,$GHz is $\sim 1\,$pc). It is often stated in the literature that GPS sources hardly show any variability (O'Dea 1998; Marecki et al. 1999); however, systematic studies are still lacking. Preliminary results of the study of a complete sample, reported by Stanghellini (1999), indicate low variability for GPS galaxies, while GPS quasars are found to vary as strongly as compact flat-spectrum quasars. The GPS quasars monitored by Tornikoski et al. (2001) also show strong variability (see Figure 4.6). *Planck* data will also yield information on the frequency dependence of variability, thus providing further insight into the physics of these sources. For example, the flux is expected to vary coherently at all frequencies for GPS sources (Snellen et al. 1995; 1998). In contrast, the variability amplitude of flat-spectrum sources is generally found to increase with frequency, consistent with the fact that their SEDs are combinations of different components peaking at different frequencies, with the highest frequency components located closest to the nucleus and most compact.

Another interesting issue is the polarization of GPS sources. Despite the fact that synchrotron emission is intrinsically polarized up to $\sim 75\%$, very low polarization is seen at cm wavelengths ($\sim 0.2\%$ at $6\,$cm; e.g., Stanghellini et al. 1998). At least in some cases (particularly for GPS quasars), these low polarizations may be attributed largely to Faraday depolarization (O'Dea 1998 and references therein). In order to depolarize the synchrotron emission to the observed level, $RM$s must reach really extreme values ($\gtrsim 5 \times 10^5\,$rad m$^{-2}$) or, alternatively, the magnetic field in the circumnuclear region has to be tangled on scales smaller than $1\,$mas. Values of $RM \gtrsim 1000\,$rad m$^{-2}$ are found for only one other class of extragalactic radio source, namely radio galaxies at the centers of cluster cooling flows. This raises the question of whether GPS sources with high $RM$ are also in cooling flow clusters or whether the depolarization is produced in some other way. We note that there are also GPS sources with small measured values of $RM$.

Given these observations, it follows that *Planck* polarization measurements can be very important from two points of view. First, at *Planck* frequencies Faraday depolarization is most likely negligible, so *Planck* measurements will allow us to determine whether the low polarization observed at cm wavelengths is due to very tangled magnetic fields or to large Faraday depths in the sources. From the frequency dependence of polarization, the $RM$s can be determined.

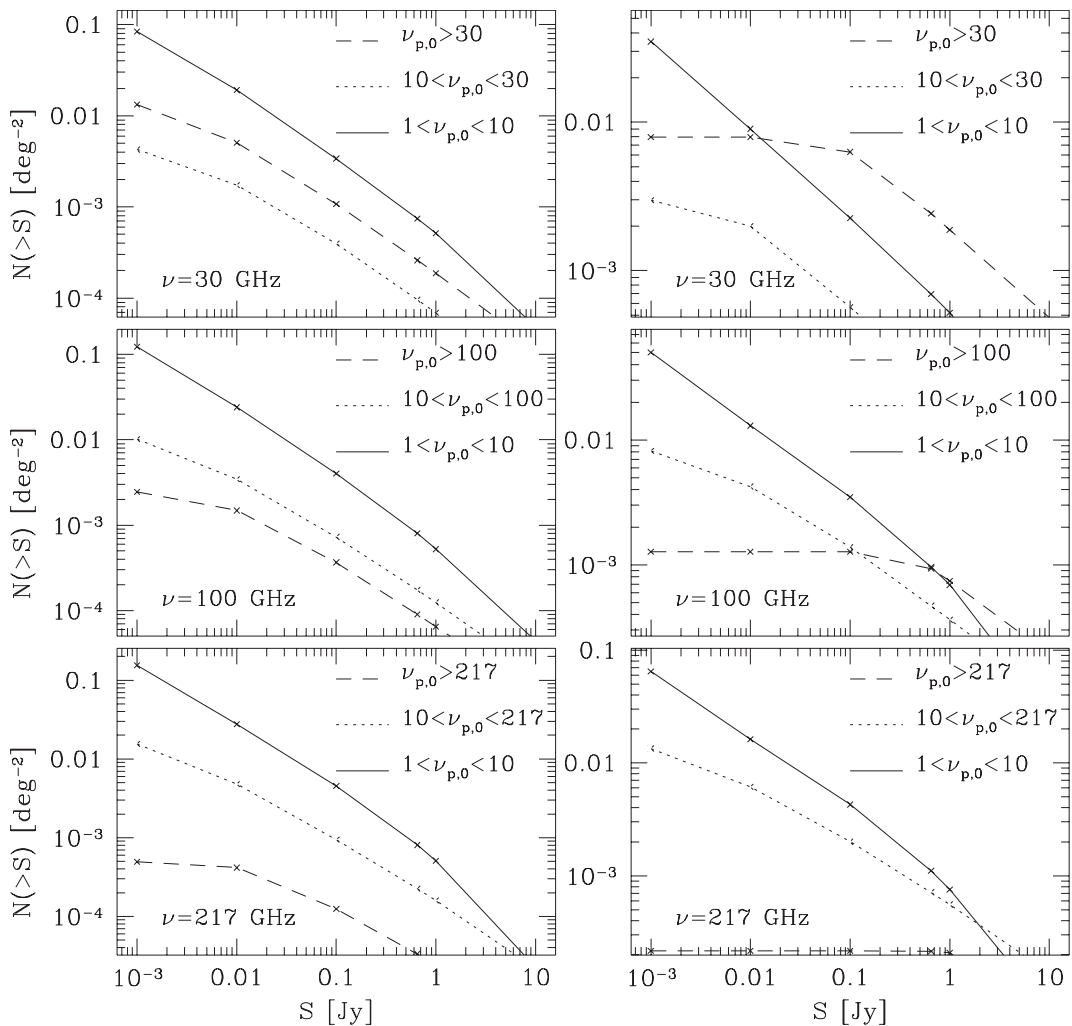

FIG 4.4.— Predicted counts of GPS galaxies (left) and QSOs (right) at three *Planck* observing frequencies. Counts are shown separately for sources whose *observed* peak frequencies $\nu_p$ lie in different ranges. The intrinsic peak frequency $\nu_{p,0}$ (see text) is assumed to be 1000 GHz.

Second, *Planck*'s measurements of the Sunyaev-Zel'dovich effect may also help in investigating the ambient medium around sources and, in particular, in assessing whether or not high-$RM$ GPS sources are associated with cooling flow clusters.

### 4.3.2.1 Other classes of peaked or inverted spectrum sources

*Planck* surveys may reveal two other interesting and "atypical" classes of radio sources.

The recent evidence that essentially all local early-type galaxies host supermassive black holes (e.g., Magorrian et al. 1998), many of which are remarkably underluminous, suggests that they accrete via low radiative efficiency (advection dominated) accretion flows (Rees et al. 1982; Narayan & Yi 1995; Di Matteo et al. 2000). Advection dominated sources are characterised by synchrotron emission with strongly inverted spectrum peaking at millimeter/submillimeter wavelengths. Although such sources would be intrinsically faint, they may be very numerous. *Planck* surveys may detect some advection dominated sources in the high flux density tail (Perna & Di Matteo 2000).

Radio afterglows of gamma-ray bursts are also characterised by an inverted radio spectrum ($S_\nu \propto \nu^{1/3}$, with a turnover at several GHz due to synchrotron self-absorption; Waxman 1997). Although radio fluxes of a few millijanskys or less, far too small to be detectable by *Planck*, have been measured so far, the possibility of exceptionally radio-luminous or nearby gamma-ray bursts cannot be ruled out. Radio frequency measurements provide interesting constraints on the physical properties of these events.

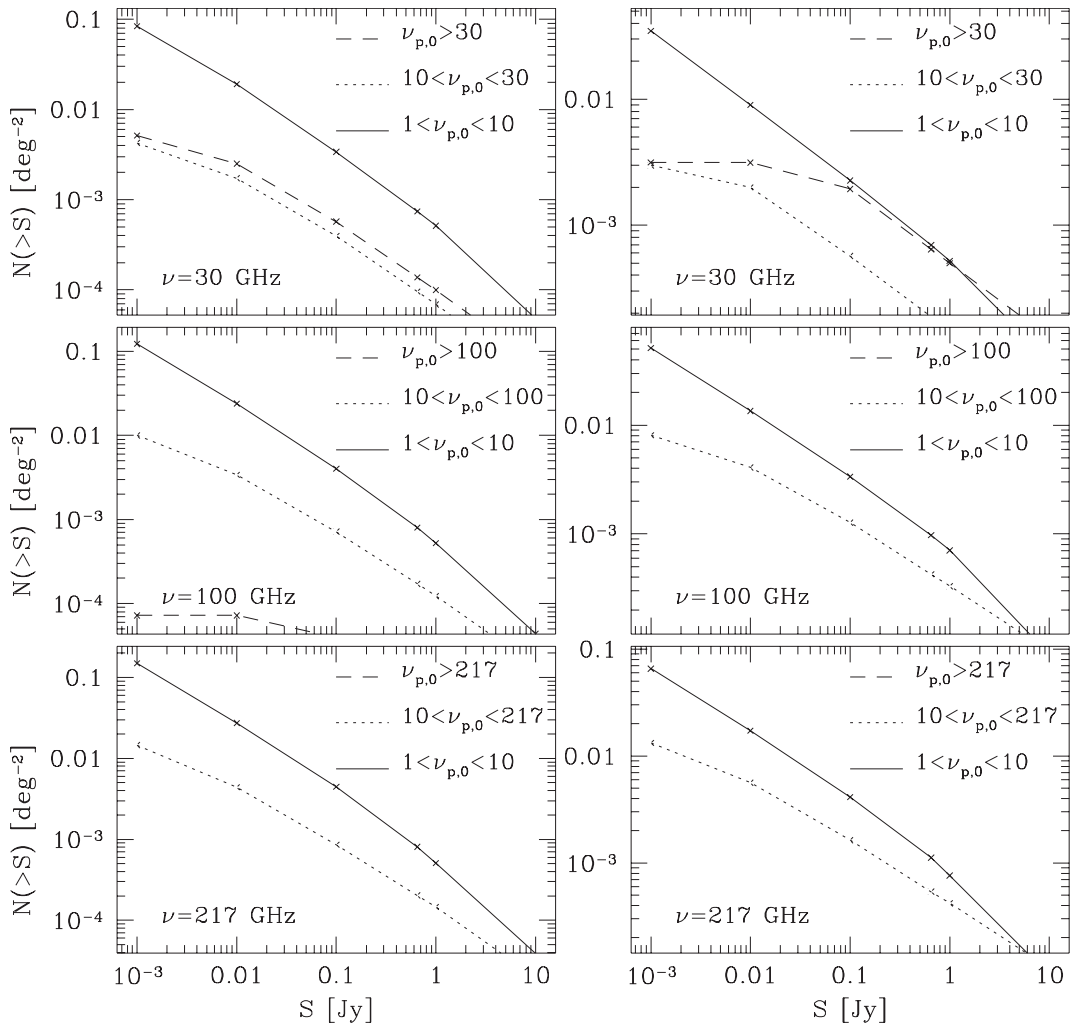

FIG 4.5.— Same as Fig. 4.4, except the assumed intrinsic peak frequency is 200 GHz.

### 4.3.2.2 Complementary ground-based observations

The utility of *Planck* observations will be increased by complementary, lower frequency observations made from the ground. We expect any sources detected by *Planck* to be readily observable at frequencies $\nu < 30$ GHz by existing facilities. For reasons given above, it will be important to make these complementary observations simultaneously with the *Planck* measurements. We have made or will make arrangements with several observatories to have the necessary access during the *Planck* mission.

### 4.3.3 The Astrophysics of Quasars and Blazars

The unique all sky maps of *Planck* will provide data on the radio properties of complete, unbiased samples of various types of AGNs over a wide frequency range. This will enable us to study the basic physical processes and their variations from class to class. Compared to NASA's WMAP satellite, *Planck*'s greater sensitivity will allow it to detect a large number of AGN sources. *Planck* LFI channels are five times more sensitive than comparable frequency channels in WMAP; we thus expect at least 10–12 times more detections.

#### 4.3.3.1 Physical models

Quasars and blazars have strong relativistic radio jets emanating from a compact core, and the jet orientation to the line of sight defines what kind of a source we see (Urry and Padovani 1995). There is, first, Doppler boosting of photon energies along the jet axis. A bright blazar with strong variability is seen if the jet is viewed head-on, and a milder AGN with less dramatic features if the jet is viewed at a larger angle. Second, the appearance of the source may depend dramatically on the viewing angle at larger angles if the core is obscured by a dusty torus (Antonucci and Miller 1985). The radio jets emit synchrotron radiation, and shocks moving in the jet create the typical flat spectrum of AGNs. At higher energies, the inverse Compton mechanism is at work. External photons (from, e.g., the accretion disk or clouds in the broad line region) or synchrotron photons from the jet itself are scattered to higher energies by relativistic electrons in the jets, producing gamma-ray emission. The spectra of such sources have two similar peaks, the synchrotron peak at millimeter-to-optical frequencies and the inverse Compton peak at higher energies.

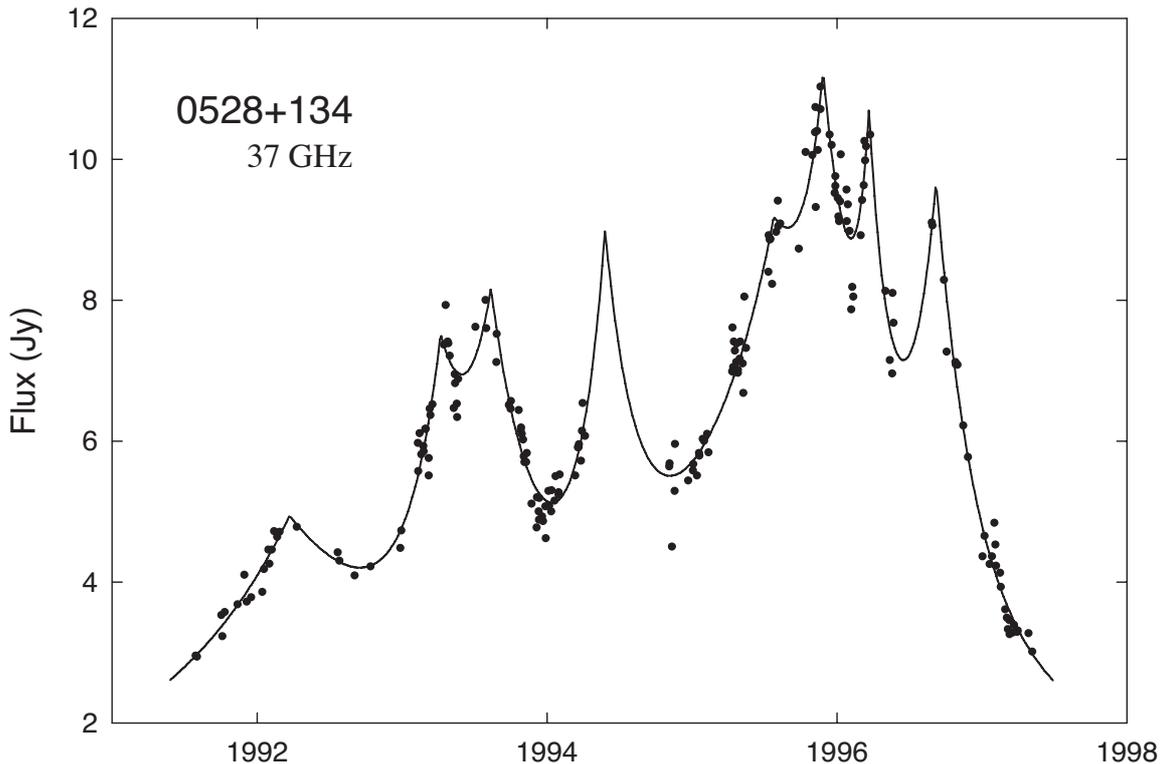

FIG 4.6.— Observations of a highly variable radio source, compared with a fitted model based on exponential flares. From Valtaoja et al. 1999.

*Planck* data will help to determine the relative importance of the Doppler and obscuration mechanisms in AGN, testing the latter with high frequency data sensitive to reprocessed emission in galaxies and quasars. *Planck*'s complete sample of nearly simultaneous radio data at several frequencies can be used for statistical studies of source classification based on observed properties, for the first time including large sets of unbiased, high frequency, radio data.

In the current paradigm, the phenomenology of bright blazars can be accounted for by a sequence in the source power and the intensity of the diffuse radiation field surrounding the relativistic jet. These in turn determine the distribution of the synchrotron and inverse Compton peaks. *Planck* surveys will allow us to determine directly the position and shape of the synchrotron peak in the SED for the most luminous blazars (which appear to peak at the lowest frequencies), a crucial piece of information to determine model parameters such as the distribution and energy of injected particles and the magnetic field strength.

The *Planck* data will also enable us to study the structure and the physics of radio jets in new ways. Using simultaneous spectral, variability, and polarization information, the various jet components (the radio core, standing and evolving shocks, extended remnant emission) can be separated and their basic parameters estimated. This information can be used to construct models of shocked relativistic jets, and also to test unifying schemes, for example by comparing the core/jet strengths in different types of sources.

For a few nearby radio sources (e.g., Centaurus A), *Planck* will resolve the non-thermal synchrotron emission from the radio lobes, permitting a detailed study of their physics.

### 4.3.3.2 BL Lac objects

BL Lac objects (BLOs) are a subset of blazars with moderate to low isotropic radio luminosities. It is not known whether the radio selected BL Lacs (RBLs) and X-ray selected BL Lacs (XBLs) are the two extremes of the general class of BLOs, the observed properties of which are defined by the jet orientation, or whether they have intrinsically different properties. The XBLs have not been actively studied at high radio frequencies; even though they seem to be significantly weaker at radio wavelengths than RBLs, there is no evidence that any are truly radio *silent*, nor could such objects fit in any unified scheme. The newly discovered class of intermediate BLOs (IBLs) seems to consist of sources intermediate to the RBLs and XBLs (Laurent-Muehleisen et al. 1999). *Planck* will give us an unbiased sample of all BLOs, selected at mm wavelengths, which are the optimum wavelengths for identifying and studying these objects. It will enable us to sort out selection biases and to reach sound conclusions for the first time about the interrelationships between different BLO populations.

### 4.3.3.3 Fainter flat-spectrum active galactic nuclei (AGNs)

Studies made at radio frequencies higher than 8.4 GHz have usually excluded fainter (0.1–1 Jy) flat-spectrum sources, and practically no high frequency data exist for sources with flux density $S < 0.5$ Jy at 2.7 GHz. This introduces a selection bias. When selecting candidates for high radio frequency observations, it is generally assumed that the radio spectra peak at low frequencies and fall off, making high frequency observations difficult in the mm range. Also, the selection is often based on single-epoch observations in the low frequency range (this is true especially for southern sources). However, many sources excluded for these reasons have in fact turned out to be much brighter than expected at high radio frequencies (Tornikoski et al. 2000, 2001). This in part is due to variability, and in some cases due to exceptional spectral shapes (inverted spectra up to the mm range). *Planck* will provide a useful census (and spectra) for such sources.

### 4.3.3.4 Variability

During its mission to map the CMB, *Planck* will measure the entire sky at least twice. During this time, *Planck* will provide information on the variability of radio sources on several timescales, from 6 months (successive sky surveys) to a few hours (as sources enter the field of individual receivers). One way of using these data is to decompose the source spectra into source components (underlying jet, evolving shocks, and standing shocks) and study the variability of these individual source elements. This gives an opportunity to measure the constant baseline flux of the source, or to find out if the standing shocks and the core are variable as well.

Another possibility is to examine the intraday variability (IDV) of AGNs. Because of the extended feed arrays, sources will be visible for a period of up to a week, depending on frequency, allowing searches for IDV that would be unreliable and difficult to make from the ground. All IDV sources so far have been identified serendipitously and no statistics on them are available. Studying IDV sources at several radio frequencies will reveal how the IDV phenomenon changes with frequency, and will help determine whether IDV is intrinsic or extrinsic (e.g., due to interstellar scintillation). Interstellar scintillation, which is known to play a role in the observed cases of IDV, is not important at *Planck* frequencies. Therefore a secure detection of IDV by *Planck*, especially if correlated with, e.g., optical measurements, would imply rapid luminosity changes in the source itself. Rapid fluctuations imply such a small emitting region ($d \sim c\Delta t$) the emission could not be explained by the usual incoherent synchrotron process.

### 4.3.3.5 Follow-up satellite and ground-based observations

A wealth of existing and forthcoming missions will provide follow-up opportunities of selected interesting *Planck* sources at several frequencies. The satellite missions include, e.g., the Gamma-ray Large Area Space Telescope (GLAST), Integral, Magic and AGILE (gamma-rays); XMM and Chandra (X-rays); and IRIS/ASTRO-F and *Herschel* (submillimetre/far infrared).

Some of these missions, like GLAST and ASTRO-F, will produce wide-angle surveys; others, like *Herschel*, will target individual sources. Observations in the TeV domain will be available with the Very Energetic Radiation Imaging Telescope Array System (VERITAS) and HESS. In addition, targeted radio and optical observations can be arranged from dedicated facilities on the ground (e.g., Metsahovi Radio Observatory and Tuorla Observatory, and NRAO facilities in the US).

Multifrequency studies of AGNs are extremely useful in many ways. Timelags and intensities of different frequency regions can be studied to model the physics of the sources and various components within them. Source spectra (especially in the *Planck* frequency range) will provide information about the emission process or processes involved. When *Planck* observations are complemented by observations at other energies, we will be able to say, for instance, whether the gamma-rays in AGNs are produced by the external Compton or synchrotron self-Compton scattering process.

To maximize the science yield of multiwavelength observations, they must be nearly simultaneous; otherwise, the variability expected in these extreme sources will confuse determinations of the source SED. Ground and satellite-based observations complementing the *Planck* mission can be planned well in advance, since we will know where *Planck* will be pointing each day. We are planning a quick-look alert system to extract the signals of unexpected sources, or flaring sources, from the time-ordered data stream, to permit prompt follow-up using ground-based instruments.

### 4.3.4 Statistical Properties of Radio Sources

In §§ 4.3.2 and 4.3.3, we discussed known classes of radio sources which *Planck* will detect and characterise. There may, however, be unexpected populations of sources revealed by *Planck*. Hence we plan a systematic study of a range of properties of *all* extragalactic radio sources detected during the mission. This will allow us to search for both unexpected classes and unexpected interrelations between known classes of sources. As noted earlier, to avoid a sample entirely dominated by starforming galaxies, we will select sources at frequencies $\nu \leq 217\,\mathrm{GHz}$, and we will exclude the Galactic plane.

Our program includes the following steps:

*Characterisation*—For each source detected by *Planck*, we will define its spectrum, polarization, and variability using all relevant *Planck* data and auxiliary measurements. We will determine accurate positions using higher resolution (and often lower frequency) surveys or optical identifications. We will complete redshift determinations for the $\sim 300$ brightest sources in each *Planck* band and at least photometric redshift determinations for samples of sources down to the detection limit in each band.

*Classification*—Each detected source will be classified as a blazar, GPS source, compact symmetric object, starburst galaxy, etc. We expect this classification can be done objectively and automatically, but we will also be alert for previously unexpected types of sources. We note that this work will assist the DPCs in preparing their catalogs of discrete sources, and may also assist in component separation (see § 4.5).

*Statistical Analysis*—For each class of radio source, we will define a number of important properties. We will, for instance, analyze the SED, polarization, and variability distributions of each class; the frequency dependence of polarization and variability, if any; and the interrelations between these properties and luminosity and redshift. Since *Planck* provides an unbiased, all-sky survey, our measurements will improve knowledge of the luminosity function of sources at the bright end, where rare objects dominate and all-sky surveys are essential. The results of these statistical analyses may enable us to refine our definitions of the various classes of radio sources, and to determine their underlying physical properties more precisely. The counts and luminosity functions of radio sources in various classes, together with redshift measurements, can be exploited to analyze the evolution of each class of source identified. While we expect most of the bright sources seen by *Planck* to be local, gravitational lensing (and amplification) may in some cases be important for some of the most powerful blazars with synchrotron peaks at millimetre wavelengths. *Planck* will allow us to check this possibility.

*Search for Subliminal Sources*—We expect $\sim 5\sigma$ measurements of sources down to 300–500 mJy, depending on frequency. In the case of known sources below the detection threshold, we may be able to draw additional conclusions by combining measurements of multiple objects of a given class. Another approach to subliminal sources is to use the so-called $P(D)$ technique to constrain the shape of the source counts below the formal detection limit. $P(D)$ analysis, of course, gives no information about specific sources, but may allow us to link our source counts to others made at fainter fluxes but over much smaller regions of the sky.

Although *Planck* is expected to detect only a few hundred sources in some of its frequency bands, clustering may still be detectable. This follows from the fact that *Planck* detections will single out rare high luminosity sources, for which the clustering amplitude may be larger than for the radio source population in general. We will also seek for the effects of clustering by examining the autocorrelation function of fluctuations with an appropriate spectral signature.

## 4.4 DUSTY GALAXIES

A major result of *Planck* will be the first all-sky surveys at submillimetre wavelengths. Above all, *Planck* will obtain the first complete census of bright, dusty galaxies at these wavelengths, as well as the background structures due to weak sources. In this section we discuss low and high redshift dusty galaxies, follow-up with *Herschel* and the Atacama Large Millimetre Array (ALMA), and statistical studies of dusty galaxies.

### 4.4.1 The Planck All-Sky Catalog of Bright Submillimetre Sources

The *Planck* all-sky catalog of bright submillimeter sources will enable both detailed studies of the low redshift infrared galaxy population and exploration of the high luminosity, high redshift tail of the luminosity function. For both types of study, an all-sky survey offers immense advantages. The expected confusion limits, sensitivities and predicted numbers of dusty galaxies for the HFI channels are given in Table 4.1. Note that different models can differ by at least a factor of three either way in these predicted numbers.

TABLE 4.1

Planck Galaxy Surveys

| | Frequency [GHz] | | | | |
|---|---|---|---|---|---|
| | 143 | 217 | 353 | 550 | 850 |
| Confusion limit [mJy, $3\sigma$] . . . . . . . . . . . . . . . . . . . . . | 6.3 | 14.1 | 44.7 | 112 | 251 |
| *Planck* All Sky Survey sensitivity [mJy, $3\sigma$] . . . . . . | 26 | 37 | 75 | 180 | 300 |
| *Planck* Deep Survey sensitivity [mJy, $3\sigma$] . . . . . . . . | 10 | 18.4 | 49 | 170 | 280 |
| Number of galaxies [all sky] . . . . . . . . . . . . . . . . . . . | 570 | 860 | 1700 | 4400 | 35000 |

Some studies will be able to use the Early Release Compact Source Catalog (ERCSC), while others will require the full depth achieved in the final catalogue. In addition, small regions of the sky ($\sim 1\%$) will be covered with 10 times more integration (the *Planck* deep survey) and this will be valuable for some studies.

#### 4.4.1.1 Low redshift dusty galaxies

Although many of the sources detected by *Planck* will be low-redshift spiral galaxies or moderate starburst galaxies already detected in the IRAS survey, *Planck*'s long-wavelength coverage will provide information on the extent of cold dust unavailable from IRAS. Moreover, *Planck* will determine accurately the local luminosity function of galaxies at submillimetre wavelengths.

*Cold dust properties in extended local and Local Group Galaxies*—Local Group galaxies and other nearby galaxies provide a convenient laboratory in which to study the properties of the interstellar medium on galactic scales. *Planck*'s submillimetre observations, in conjunction with already-existing observations of atomic, molecular and ionised gas, will provide a complete picture of the distribution of the cold matter in galaxies. We will be able to address some fundamental questions on ISM properties, on the nature and distribution of the cold component of the ISM, and on its relationship with other galactic components.

Extended cold dust has been detected with deep optical (e.g., Howk and Savage 1999) and submillimetre (SCUBA and ISOPHOT, e.g., Alton et al. 1998) observations. The dust is more extended in some cases than the starlight (Figure 4.7). Mapping of nearby galaxies provides an opportunity to explore fundamental questions related to this cold dust. For example, how extended, massive, or common are these cold components, and how do they interact with the other galactic components? What is the nature and origin of this extended cold dust? How is it related to star formation?

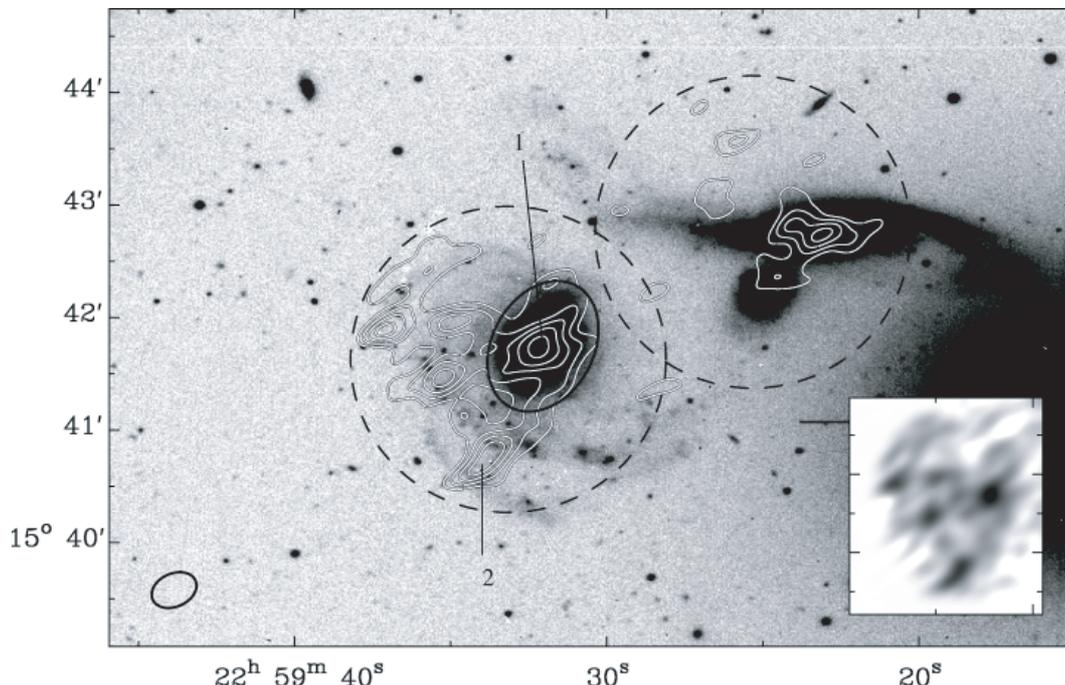

FIG 4.7.— SCUBA 850 $\mu$m maps of NGC7465, overlaid on the greyscale B-band image. Dashed circles indicate the SCUBA field of view, while the solid ellipse indicates the B = 25 mag arcsec$^{-2}$ isophotal size. The dust emission extends considerably beyond this isophote (Thomas et al. 2002). The inset is the SCUBA image in greyscale.

If a massive cold dust component is found to be common in galaxies, it will be necessary to reassess their total mass content. Even with ISOPHOT's limited ability to detect very cold dust, a factor of ten more dust was detected in M31 (Haas et al. 1998) than previously known. A similar finding in the Milky Way came from South Pole submillimeter observations (Pajot et al. 1986). *Planck* could reveal very cold dust components in a variety of other galaxies, including large, well-studied nearby galaxies, such as M51, M83, Centaurus A, NGC1365, and NGC253.

Submillimetre observations with *Planck* are a more effective means of probing the cold component in galaxies than SCUBA observations. Comparison of the cold dust distribution with the atomic and molecular gas will allow us to determine the physical characteristics of the cold material and the relationship to galaxy type, activity and metallicity.

A substantial amount of dark matter is associated with dwarf galaxies. In fact, dark matter seems to be the dominant mass in dwarf galaxies at all radial distances. What is the origin of dark matter in dwarf galaxies? Can cold dust make a significant contribution to the dark matter content? The dwarf galaxies in the Local Group will provide numerous examples to study these questions.

*The submillimetre/millimetre spectral energy distributions of galaxies*—Nearby galaxies are ideal laboratories to study the interplay between the interstellar medium and star formation on global scales. Their emission at *Planck* wavelengths arises from three sources: thermal emission from large grains, synchrotron, and perhaps free-free emission. *Planck* will detect thousands of galaxies sampling all of these physical components. New sources are certain to be detected, and our knowledge of the full spectral energy distribution of starburst galaxies, as well as of AGNs, quasars and radio galaxies, will be expanded.

As previously noted, a very cold ($6 \lesssim T \lesssim 15$ K) dust component is observed in many galaxies. What are the galaxy characteristics which control the mass and properties of the dust? What is the fate of dust as galaxies evolve through different states of merging activity? Based on a limited sample to date, a very cold dust component is found only in the quiescent galaxies, and not in the active galaxies (e.g., Siebenmorgen et al. 1999). Submillimetre SCUBA observations have confirmed the presence of very cold dust in low-metallicity dwarf galaxies (Madden, 2002) and elliptical galaxies (Fich and Hodge 1993), environments where low dust abundance is expected. Combined with IRAS data and ASTRO-F data and dust modelling, *Planck* data will allow characterisation of dust properties for a wide range and number of galaxies. With H<span>I</span> data from various surveys, we can determine dust-to-gas mass ratios as a function of Hubble type, galaxy activity, and metallicity, thus providing valuable statistics to develop models for the formation and evolution of dust in galaxies. To perform any detailed studies at these wavelength ranges, it will be necessary to first separate the individual contributions from thermal dust emission and free-free and synchrotron emission in the *Planck* wavelength bands.

How do the submillimetre/radio properties of the broad-line and narrow-lined objects compare? Can observations be reconciled with AGN unification theories/dusty torus models (see also § 4.4)? One of the biggest challenges in understanding quasars and radio galaxies is characterising the FIR-to-millimetre emission (Figure 4.8) and trying to distinguish contributions from thermal and synchrotron emission. Unified schemes try to link these galaxies together in terms of viewing angle and degrees of obscuration from the surrounding dusty torus. Recent ISOPHOT studies are beginning to characterise the SEDs of quasars out to $200\,\mu$m, finding wide variations in thermal vs. synchrotron emission contributions without firm conclusions yet (e.g., van Bemmel et al. 2000). *Planck* observations combined with IRAS and IRIS data will provide the first large survey of galaxies from which to study possible unification scenarios.

*The local luminosity functions of star-forming galaxies*—The *Planck* observations will determine the luminosity function of star-forming galaxies. This will be the definitive analysis of the local submillimetre luminosity density, obscured volume-averaged star formation rate, and dust mass function, as well as of the newly-discovered luminosity dependence of colour temperature and other multivariate relations. The identifications necessary to complete this task will form the basis of important multi-wavelength follow-up programs.

Most of the galaxies in the final source catalogue will be known from IRAS, largely with redshifts from the Point Source Catalog redshift (PSCz) spectroscopic campaign (Saunders et al. 2000). Previous work on the local submillimetre luminosity function has by necessity been limited to submillimetre follow-ups of known populations, such as the brightest IRAS galaxies (e.g., Dunne et al. 2000; Figure 4.9) needing careful corrections for the incompleteness from the multivariate flux cuts. The depth and all-sky coverage of *Planck* will give the only cleanly flux-limited submillimetre survey of the local Universe, and an unequivocal determination of the local submillimetre luminosity function.

Using multiwavelength LFI/HFI/IRAS data, we will model the spectral energy distributions of these sources with radiative transfer models. One important corollary will be the dust mass function; another will be the exploration of the physical mechanisms underlying the submillimetre luminosity-colour temperature relation recently discovered in local starbursts. With near-UV data from GALEX and optical spectroscopy, we will also be able to make detailed object-by-object comparisons of the UV, H$\alpha$, radio, and far-IR star formation rate indicators in a definitive analysis of the statistical effects of obscuration on star formation in the local Universe.

The multiwavelength coverage will also allow construction of luminosity functions at the same rest-frame wavelengths as high-$z$ submillimetre/mm-wave surveys by SCUBA, MAMBO, and future facilities such as SCUBA-2, BLAST, BOLOCAM, LMT, and APEX. This determination will be central to the study of evolution in these populations.

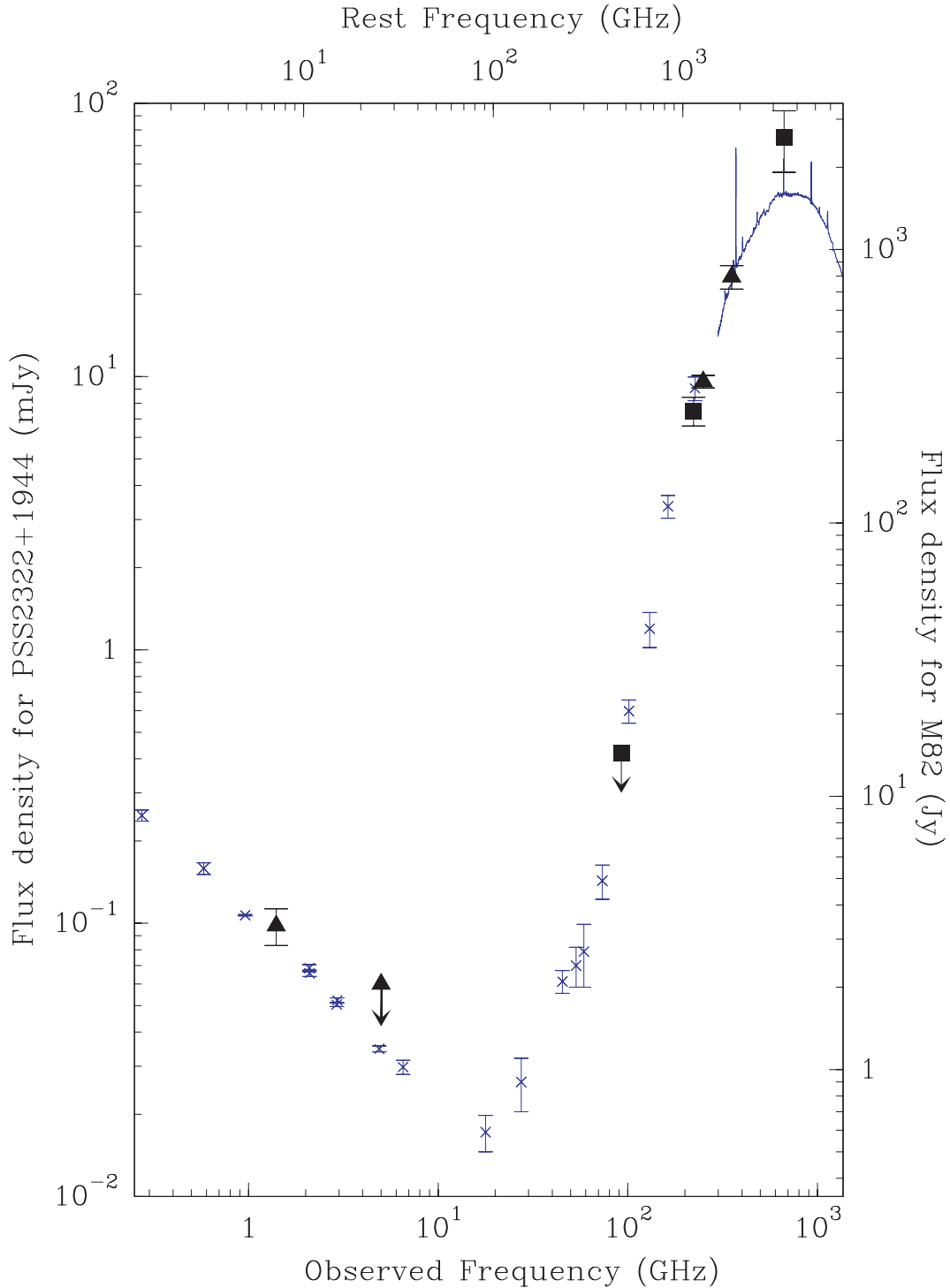

FIG 4.8.— Spectral energy distribution of PSS 2322+1944 (black points, from Cox et al. 2002). For comparison, the radio-to-infrared spectral energy distribution of the starburst galaxy M 82 is shown, redshifted to $z = 4.12$ and normalized to the flux density of PSS 2322+1944 at the observed wavelength of 850 $\mu$m. The crosses show all the currently available photometric data, and the continuous line represents the ISO LWS spectrum. The left- and right-hand flux density scale are adapted for PSS 2322+1944 and M 82, respectively.

To study the gas and dust distributions, morphologies, stellar populations, and AGN content (and in a small number of cases to identify the galaxies), follow-up with VLA/Merlin, optical/near-IR facilities, mm-wave imaging and spectroscopy, and *Herschel* will be important.

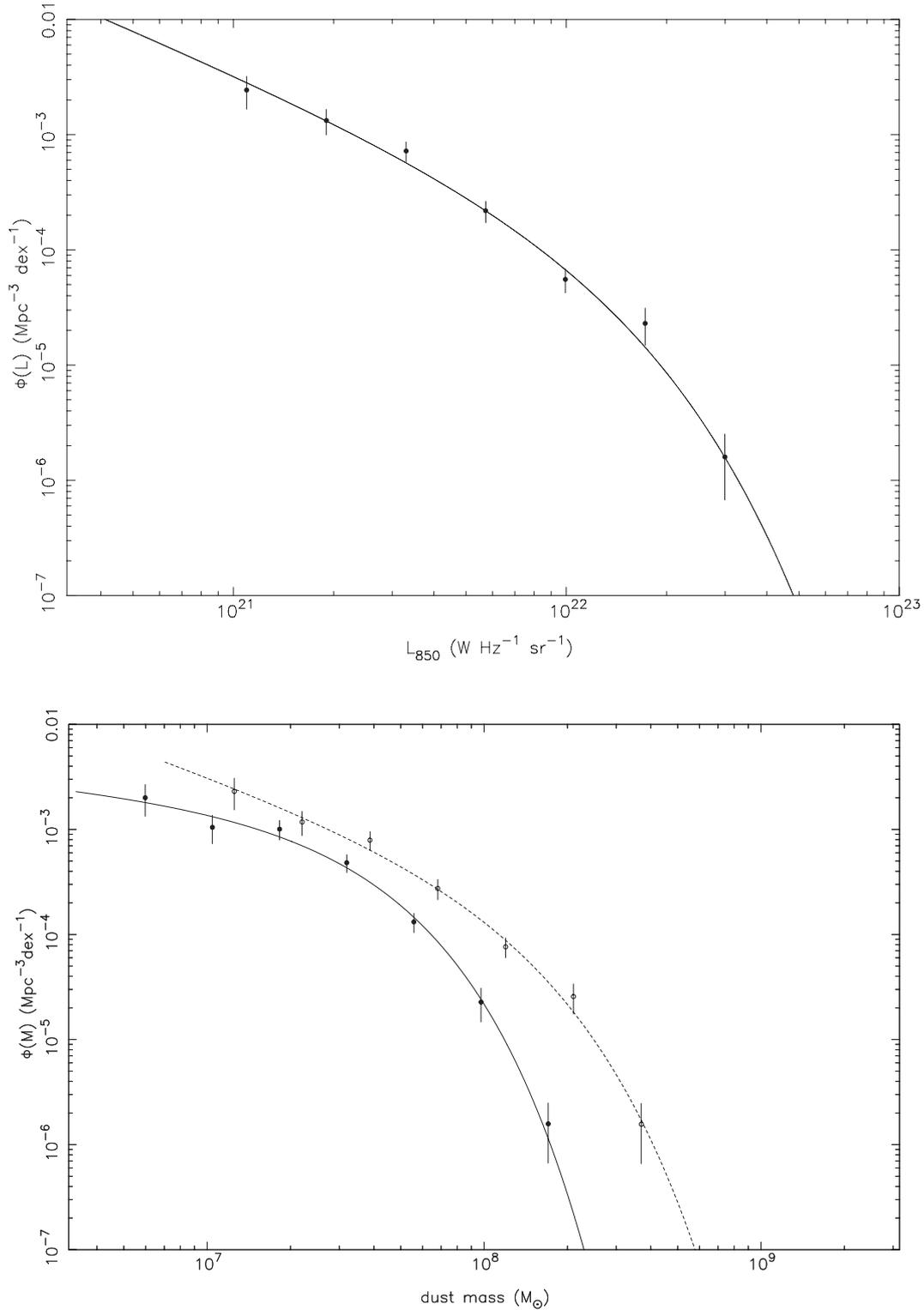

FIG 4.9.— The local luminosity function at $850\,\mu m$ (upper panel) derived by Dunne et al (2000). The lower panel shows their estimates of the dust mass function for local galaxies.

### 4.4.1.2 High redshift dusty galaxies

*Planck* will provide an unbiased all-sky survey of dusty active sources (starbursts and AGNs). The predicted number of infrared/submillimetre galaxies that will be detected (Table 4.1) is strongly sensitive to the models of galaxy formation and to the algorithm used for filtering and source extraction. The current best estimate is several $\times 10^4$ sources at $350\,\mu m$, with a median redshift $z = 0.3$. Detections at 550 and $850\,\mu m$ favour higher redshifts, but the high-$z$ tail depends on the high luminosity cut-off of the luminosity function, which is unknown.

In any case, because of the strong, negative K-corrections, we expect *Planck* to be able to detect the most luminous dusty objects in the observable universe.

To reproduce the galaxy counts obtained by SCUBA (Figure 4.10) and the SED of the CIRB, an extraordinary evolution is required in the star formation rate (SFR; see e.g., Franceschini et al. 1998; Blain et al. 1999; Devriendt and Guiderdoni 2000; Rowan-Robinson 2001; Gispert et al. 2000). It has been proposed recently that the objects detected by SCUBA are elliptical galaxies during their primary and short episode of star formation at $z > 2$, before the onset of QSO activity in their centres takes place (Granato et al. 2000). A distinctive prediction of these models is that the galaxy counts in the far infrared and submillimetre bands should exhibit an extremely steep slope at fluxes $10 \leq S_{850} \leq 100$ mJy at $850\,\mu$m, where the high redshift galaxies start to emerge in the source counts. The question is whether *Planck* can detect them. Since the large majority of these objects are at $z > 2$, their probability to be gravitationally lensed is significant. The effect of gravitational lensing on source counts in the (sub)-mm wavebands is particularly relevant because of the extreme steepness of the counts. From models by Granato et al. (2000), we expect about 0.3–1 sources per square degree brighter than 100 mJy at $850\,\mu$m (taking into account strong lensing).

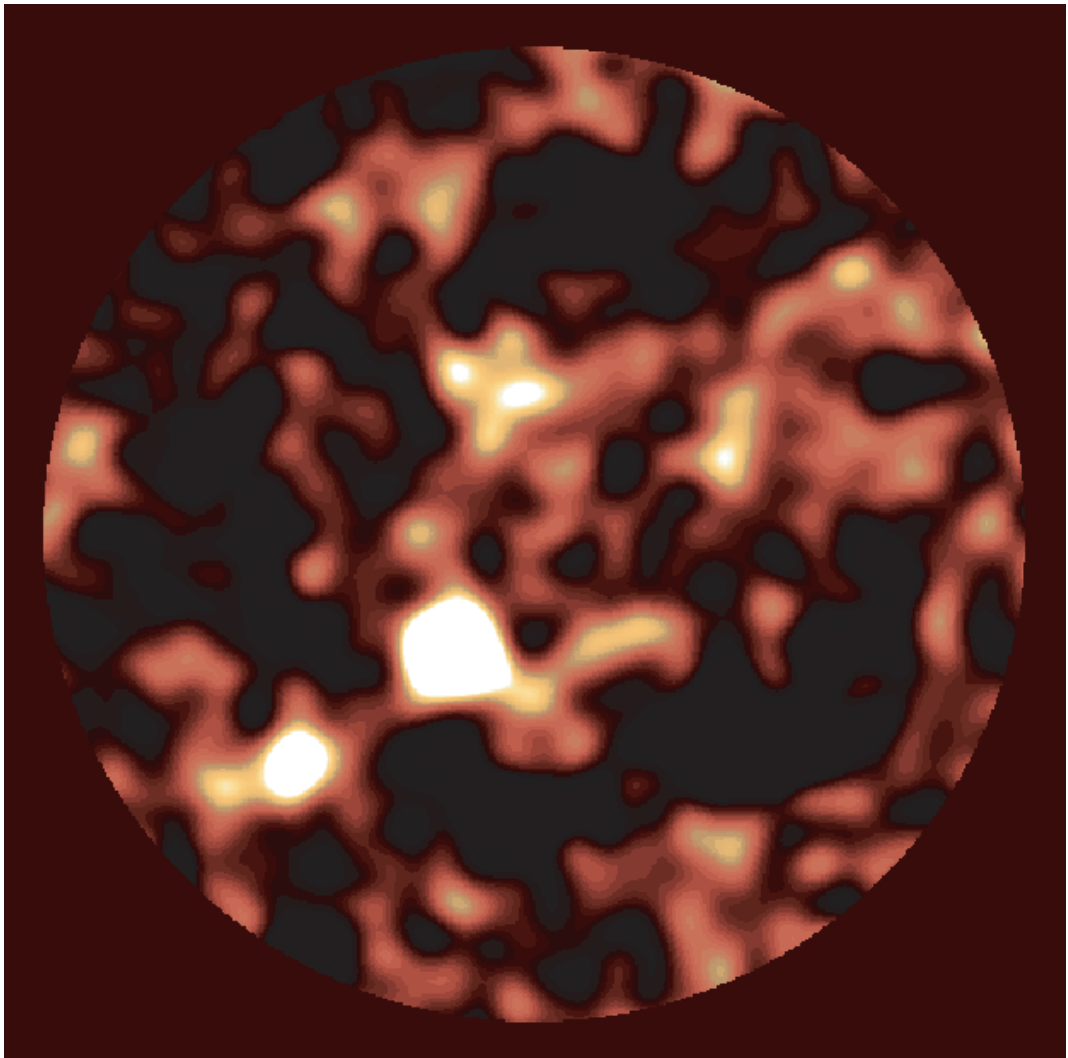

FIG 4.10.— Survey of Hubble Deep Field North with SCUBA at $850\,\mu$m (Hughes et al 1998).

The sources detected by *Planck* at the $3\sigma$ level can be confirmed and more precisely located by cross-correlation with the ASTRO-F source catalogue at $170\,\mu$m. The flux ratio $R = S_{170}/S_{850}$ is an extremely efficient discriminator between local and high $z$ galaxies, since it varies from $10 < R < 100$ for local galaxies to $R \sim 1$ for galaxies at $z = 3$ (see e.g., Carilli and Yun 1998). The areas containing high-$z$ galaxy candidates can then be observed with *Herschel*. The detection of a large number of lensed sources would allow a statistical approach to one of the most interesting cosmological problems, namely that of the matter distribution in the high-$z$ Universe. These sources would also be ideal targets for the James Webb Space Telescope (JWST). The combination of the far infrared observations with optical and near infrared rest-frame data would give an extraordinary contribution to our understanding of the formation of elliptical galaxies.

The sample may also allow a study of clustering of these high redshift massive ellipticals, which will provide useful insights on properties such as the mass of the halo and the duration of the star-formation process (Magliocchetti et al. 2000) and test models for the evolution of the bias factor with $z$ (Moscardini et al. 1998).

Strong clustering also affects the statistics of the intensity fluctuations due to non-resolved sources. The statistics of the fluctuations will be studied with particular attention to the autocorrelation function looking for the source clustering effect. The autocorrelation function due to the sources is expected to change significantly with frequency, and is needed to correct CMB maps for the effects of these foreground sources.

Forthcoming deep surveys at infrared/submillimetre wavelengths with SIRTF and *Herschel* will cover at most $100 \deg^{2}$, and will consequently miss sources rarer than $0.01 \deg^{-2}$, in particular, high-$z$ infrared/submillimetre "monsters." The most luminous high-$z$ galaxies in the Universe can be found among the sources with SEDs peaking at the longest wavelengths between 60 and $850 \,\mu m$. A fraction of these are likely to be lensed objects. For instance, the ERCSC and final CSC, respectively, might include $z = 5$ galaxies as luminous as $L_{IR} = 2 \times 10^{14} L_{\odot}$, or $L_{IR} = 3 \times 10^{13} L_{\odot}$, if they exist. It is expected that the ERCSC sources will mostly be low-$z$ objects, but the selection criteria can already be tested on this sample, with quick check of the procedure and follow-up. From the *Planck* final CSC, a subset '*Planck* Cold Source Sample' of about 500 sources can be generated, optimising the fraction of high-$z$ monsters. To obtain accurate positions, optical identifications, redshift follow-up, and infrared/submillimetre/millimetre fluxes and images of these sources will be needed. *Herschel* SPIRE and PACS can be used to improve the positions of the sources and an extensive follow-up at submillimetre/millimetre wavelengths will be needed with MAMBO, SCUBA-2, or CSO BOLOCAM, plus the LMT and ALMA. Optical identifications will be obtained from these positions, and redshifts will be measured at the VLT, Gemini, and Keck telescopes. The main emission lines will be used for a first classification of the sources (AGN and/or star formation). ALMA sub-arcsec observations of the continuum and CO lines will show evidence of lensing or merging/interaction, and give a direct determination of the redshift and molecular gas content.

### 4.4.1.3 Deriving bright galaxy counts from Planck HFI data

*Planck* will survey the whole sky to a confusion/instrumental noise limit of a few hundred millijanskys in the high-frequency bands of HFI. The submillimetre flux densities of sources at this limit will be about 20–40 times brighter than those found in existing $100 \,\text{arcmin}^{2}$ submillimetre surveys.

The scientific benefits of reaching a source detection threshold of a few hundred millijanskys will be great. The HFI counts are certain to be steep (Figure 4.11), although their exact form at intermediate fluxes is uncertain. This steepness means that a priori a deeper survey is likely to detect many more sources than a shallower one. The shape of the counts will reflect the mode of evolution of dusty galaxies at least at redshifts $z \leq 1$, and depending on the slope of the counts, perhaps to $z \simeq 4$, 5, or even greater. Any underlying subpopulations of either unusually luminous, or untypically distant galaxies, or any significant population of gravitationally lensed galaxies, should be seen as clear humps in the counts. While a confirmation of the nature of any such galaxy population inferred from the counts will require careful follow-up observations, the form of the counts themselves will provide important information about galaxy evolution.

At sub-100 mJy flux densities, too faint for counts to be determined directly, the 545 and 857 GHz maps should be sufficiently deep that the flux distribution of pixels in the sky will be dominated by the approximately log-normal noise distribution expected from point-source confusion rather than by the normal distribution expected from instrumental noise. Details of the noise distribution will reveal information about the form of the sub-100 mJy counts over a large fraction of the sky, and so provide valuable prior information for more accurate extraction of the CMB signal. The angular power spectrum of fluctuations in the HFI images could also be used to investigate the clustering and evolution of faint dusty sources (see Haiman & Knox 2000).

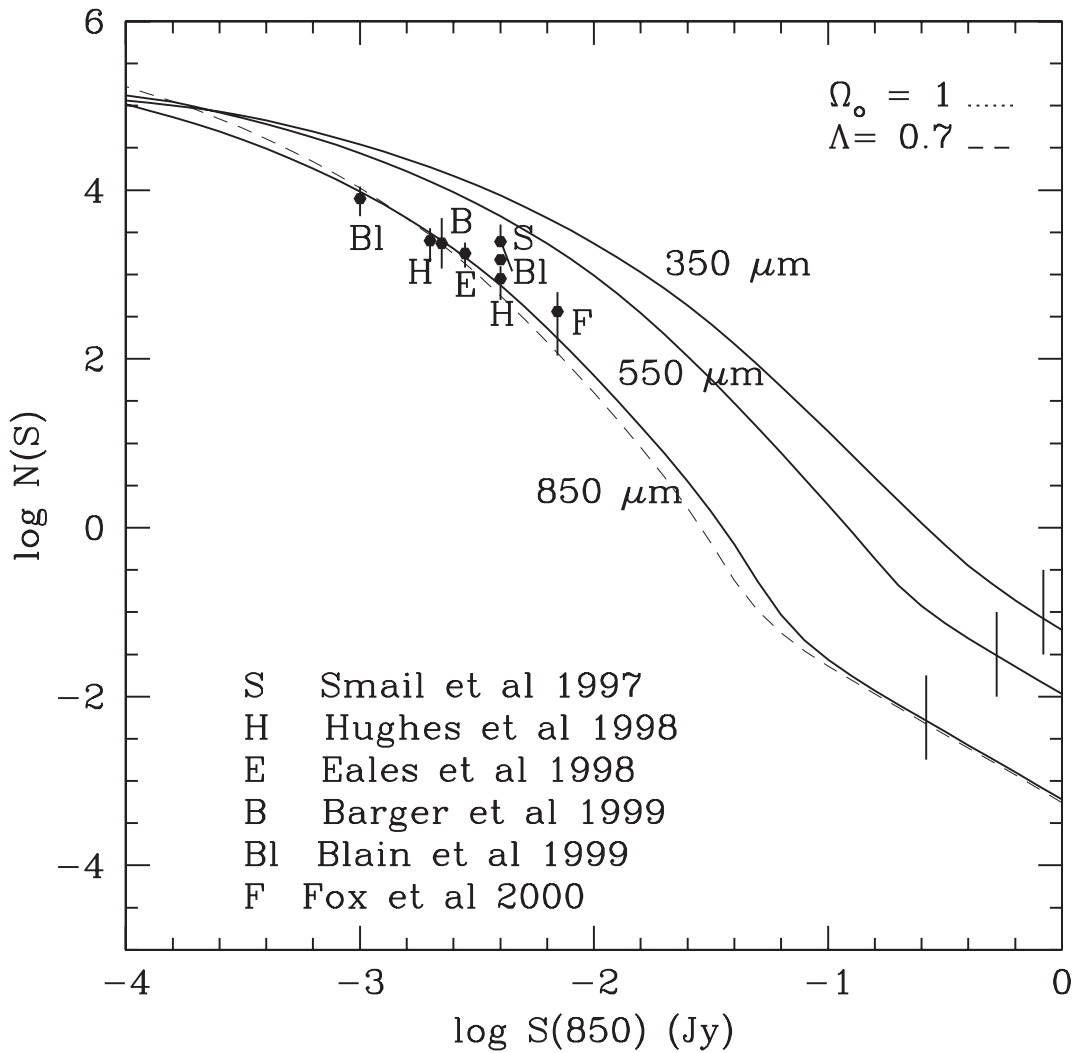

FIG 4.11.— Predicted counts at 350, 550 and 850 μm (Rowan-Robinson 2001). Observed counts at 850 μm are shown, and the
*Planck* sensitivity is indicated by the vertical bars.

### *4.4.1.4 Large-Scale Clustering of Planck Galaxies*

The $10^4$–$10^5$ galaxies discovered by *Planck* will provide a uniform sample exploring a volume of the Universe that is both large (in comparison to IRAS, 2df, and SDSS) and over a significant fraction of the sky. These properties mean that the sample will be of great interest for studies of large scale structure, dynamics, environments and biasing. While SDSS will already have determined the local galaxy power-spectrum over many galaxy types, it will not have been able to confirm the convergence of the dipole, or the local clustering of infrared emitting galaxies, a field that will still be the preserve of far infrared and submillimetre samples. Ultimately these studies will require spectroscopic redshifts, but early studies can be performed using immediately available products from *Planck*.

By launch the existing optical and NIR surveys providing identification and photometric redshifts will be extensive: the SDSS will be complemented by a 3000 deg$^2$ survey in J, H, and K undertaken by UKIRT WFCAM, while in the south VISTA will have undertaken in the *u*, $g$, $r$, $i$, and $z$ bands a $10^4 \deg^2$ survey, with much covered in J, H, and K as well. In addition, follow-up projects to obtain spectroscopic redshifts for the samples will be undertaken (using multi-fibre systems such as 2df and 6df).

### 4.4.1.5 Statistical detections of extragalactic sources

Where fluxes of individual sources are too weak for detection, statistical detections of known classes of extragalactic systems can be achieved by coadding the fluxes at the source positions from the final *Planck* maps. This will be a very powerful tool for the statistical study of distant starbursts and AGN. For example, coadditions of *Planck* 353 GHz maps will produce $> 5\sigma$ statistical detections of $\sim 1000$ examples of any class of object detected to date with SCUBA, or $\sim 100$ of any SCUBA-detected class with a $\nu^3$ spectrum at 343 GHz.

Among the most startling results in submillimetre extragalactic astronomy prior to the advent of SCUBA was the detection of enormous masses of gas and dust in high-redshift AGN. Subsequent submillimetre photometry programs of high-$z$ quasars (e.g., McMahon et al. 1999; Priddey et al. 2003; Omont et al. 2003) and radiogalaxies (e.g., Archibald et al. 2000) have had notable successes. Detections at $\sim 400\,\mu m$ which constrain the dust temperature and mass are hard to obtain, and will remain difficult at least until the advent of SCUBA-2 and *Herschel*. Existing databases (e.g., Veron-Cetty and Veron 2000; Hewett and Burbidge 1993) are more than large enough to seek statistical detections, whether of subsamples (e.g., redshift bins, statistically-complete subsets) or of all. Furthermore, current and future missions (e.g., XMM, GALEX) will detect many thousands of AGN over their mission lifetimes, which we can search for in *Planck* images on a statistical basis.

2MASS and near-UV/optically-selected galaxies will rarely feature in the *Planck* point source catalogs, but may be detectable statistically. For example, coadding at the positions of star-forming galaxies (e.g., from the GALEX mission) will statistically address how the fraction of obscured star formation varies locally as a function of galaxy morphology and spectral type. Little is known about the cosmic star formation history beyond its global average, and the local determinations made here will make illuminating comparisons with ongoing developments at higher redshifts.

## 4.4.2 Far-Infrared Background and Correlations

*Planck*, with its excellent sensitivity to extended sources and its broad frequency coverage, will enable a separation of cirrus and CFIRB anisotropies in the HFI bands. CFIRB anisotropies have now been detected at three wavelengths: $170\,\mu m$ with ISOPHOT (Matsuhara et al.; Lagache & Puget 2000), and $100\,\mu m$ and $60\,\mu m$ with IRAS (Miville-Deschenes et al. 2002). The study of this large scale structure will be the main goal of CFIRB studies with *Planck*. The CFIRB was first detected in 1996 (Puget et al. 1996) and is now well established (Hauser and Dwek 2001). It results from the line of sight accumulation of dust emission associated with more or less distant galaxies. Thus, the CFIRB properties probe galaxy formation and evolution. Observing the CFIRB at better sensitivity and resolution than achieved by the COBE FIRAS and DIRBE instruments should reveal the presence of correlated fluctuations. These fluctuations can be used to determine the physical properties of contributing high-redshift galaxies and their bias relative to the dark matter density field (Knox et al. 2000)

The FIRBACK deep survey at $170\,\mu m$, made with ISOPHOT, detected for the first time fluctuations in the CFIRB (Lagache and Puget 2000). This first detection was limited to Poissonian fluctuations. However, by analysing larger FIRBACK fields, power spectra reveal correlated fluctuations, well above the cirrus contribution (Puget and Lagache 2000). This discovery is only a first step in our understanding of the large scale mass distribution, since the CFIRB at $170\,\mu m$ is dominated by nearby sources. As shown by Gispert et al. (2000), the CFIRB at different wavelengths is made of sources at different redshift, longer wavelengths being dominated by more distant sources. The same applies to the fluctuations: longer wavelengths probe higher redshifts (e.g., Knox et al. 2000). This implies that the CFIRB maps at different wavelengths will not perfectly correlate with each other (Hivon et al. 1997)

We will use this frequency dependence of the CFIRB on the redshift of the underlying sources (Fig. 4.12) to isolate large scale structures by redshift slices. Specifically, removing iteratively the lower frequency maps from the higher ones will give a direct view of the increasing-*z* Universe. In fact, an optimal procedure is more likely to involve different linear combinations of all maps for each redshift range selected. To test the reliability of the redshift information extraction, we plan to use *N*-body simulations, coupled with, for example, the semi-analytical model of Guiderdoni et al. (1998). One of the major challenges for this project will be the removal of the Galactic cirrus contamination, which is the main foreground at high frequency. This contamination will be nearly negligible for $\sim 1\%$ of the sky and very small for $\sim 17\%$ of the sky.)

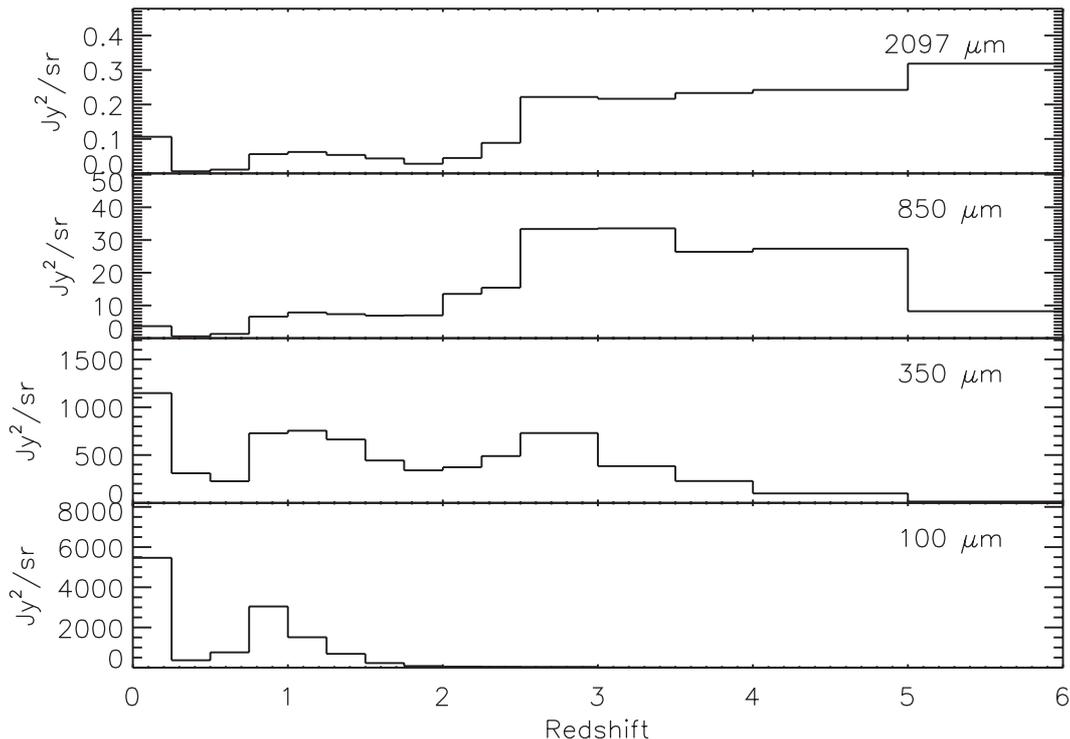

FIG 4.12.—Redshift distribution of the CFIRB fluctuations at 100, 350, 850, and 2097 $\mu$m (from Lagache et al. 2003). It is assumed that sources with flux above 100, 453, 86, and 18 mJy, respectively, can be identified and removed. The flux limits at *Planck* wavelengths come from the combination of the source confusion noise and the instrumental noise in the high-redundancy region.

This analysis of structure as a function of redshift can be done particularly efficiently in areas covered by large cosmological surveys with better angular resolution and sensitivity, as will be made by SIRTF and *Herschel*. In these surveys, a large fraction of the low and intermediate redshift sources dominating the cosmic background at frequencies above 867 GHz will be identified.

### 4.4.3 Follow-up of Unusual Objects with Herschel

We emphasize here the value of rapid follow-up of the early *Planck* survey data with *Herschel*, and the general importance of *Herschel-Planck* synergy.

The *Planck* Early Release Compact Source Catalog (ERCSC) will contain a number of high-*z* galaxies. The strong negative K-correction in the submillimetre means that the most luminous objects anywhere in the visible Universe will be detectable with *Planck* (the ERCSC will see objects with luminosity above $10^{13} L_\odot$ at almost any redshift).

These will be rare objects representing the upper end of the luminosity function at high *z*, and will not be discovered by any other observations. It will be important to follow up the *Planck* detections with *Herschel* before the end of the *Herschel* mission. Most of the (several thousand) sources in the ERCSC will be at low redshift and will have counterparts in existing all-sky surveys (especially IRAS and ASTRO-F). These databases in combination with *Planck* colours can be used quickly to identify a subset of high-$z$ candidates in the ERCSC data. Some will be previously known sources which turn out to have unusual submillimetre colours and some will be previously unknown objects detected only by *Planck*. It will be important to identify high-$z$ candidates early and follow them up with photometry between 170 and $500\,\mu m$ using *Herschel* PACS and SPIRE. This will provide precise positions, wider wavelength coverage, and more accurate flux densities. Further follow up from the ground will allow the full FIR-mm SEDs to be determined. Submillimetre spectroscopy may also allow direct determination of the redshifts and study of the molecular gas conditions.

The Deep Survey area in the ERCSC can also be followed up directly by SPIRE. The total $400\,\mathrm{deg}^2$ area can be observed to the same or lower flux limits with a few weeks of observing time, giving extended spectral coverage and better positions and generating an extensive database with many less extreme objects. This would allow statistical estimation of the redshift distribution, and investigation of the luminosity function at more modest luminosities. It would also provide a large area of sky fully mapped by both *Herschel* and *Planck* for the purposes of *Planck* foreground characterisation.

At most frequencies, *Planck* will provide source positions to an accuracy of a few arcminutes. *Herschel* observations with SPIRE and PACS will both locate the sources much more accurately (to within about $10''$) and define the peak of their spectral energy distributions at wavelengths shortward of those probed by the HFI bands, important for constraining their luminosity once redshifts are known. ALMA will provide detailed sub-arcsecond images of the morphology of the detected galaxies at the HFI frequencies, revealing arc/counterimage structures characteristic of gravitational lensing and morphological evidence of interactions, mergers, and tidal tails. ALMA will also image *Planck* SZ-selected clusters to detect lensed images of background galaxies that could confuse the SZ signal. It will provide many redshifts for detected sources using molecular lines, and will provide line ratios to probe the astrophysics of the sources in detail.

## 4.5 SOURCES AS CONTAMINANTS

The removal of foreground contamination by compact sources is a crucial step towards the main goal of the *Planck* mission, the characterization of primary anisotropies in the CMB. We summarize here some of the steps we will take to do this. These steps can be classified under five headings: pre-launch observations to locate and characterise those extragalactic sources most likely to affect *Planck*; simulation of source counts and properties when direct observations are unavailable; modeling the effect of discrete sources on the temperature and polarization power spectra; the location and classification of extragalactic sources found in *Planck* images; and finally removal of the extragalactic discrete sources (resolved and unresolved) from the images ("component separation"). The first three of these steps are already underway and will be largely complete by launch; these are summarized in §4.5.1. The last two steps will be carried out once *Planck* data arrive (though simulations of source removal are already underway, e.g., Baccigalupi et al. 2000; Bouchet and Gispert 2000; Cayon et al. 2000; Tegmark et al. 2000; and Vielva et al. 2000). These are discussed in §4.5.2.

### 4.5.1 Prelaunch Observations, Modeling, and Simulations

The actual detection threshold for discrete sources in the *Planck* images will depend in large measure on the raw sensitivity at each frequency, but also on source confusion. As a rough guide to the kind of flux densities we need to be concerned with, let us make some simplifying assumptions. First, in the LFI bands and in the low frequency, diffraction limited, HFI bands, $10\,\mu K$ corresponds to $S \approx 50\,\mathrm{mJy}$, independent of frequency. Thus we expect to be able to detect sources with $S > 250\,\mathrm{mJy}$ ($5\sigma$, see Table 1). We would also like to be able to find and remove sources down to some tens of mJy. Note that these estimates do not include the effect of confusion, which may raise the detection thresholds at some frequencies (see Cayon et al. 2000).

Ideally one would know the positions, polarizations, and flux densities at *Planck* frequencies of all sources exceeding some threshold such as 100 mJy *before Planck* flies, but this is not possible. Instead, we will construct pre-launch surveys and catalogs of sources likely to appear in the *Planck* images, monitor known high frequency radio sources, and model source counts, spectra and polarization of extragalactic sources. Limited ground-based surveys at radio wavelengths overlapping some of the *Planck* LFI bands are relatively straightforward compared to similar work in the HFI bands. As a consequence, modeling and simulations will be much more important for the HFI bands.

### 4.5.1.1 Radio surveys

Virtually all of the sky has been surveyed at 5 GHz, and there are wide-angle supporting surveys like the NVSS at 1.4 GHz, as well as lower frequency surveys. In addition there are some surveys at higher frequency of much smaller solid angles which allow us to estimate source counts on a statistical basis. The most comprehensive of these are at 8.4 GHz (Windhorst et al. 1993; and Fomalont et al. 2002), and some work has been undertaken recently at 15 GHz by Taylor et al. (2001).

Several groups are planning to extend large-angle surveys to higher frequencies (e.g., the OCRA group and a group at the Australia Telescope). In addition, a number of groups are surveying hundreds of *known* sources with spectra that appear to rise at centimeter wavelengths.

Available surveys at submillimetre wavelengths have very little impact on our knowledge of the all sky distribution of submillimetre sources because these surveys cover tiny solid angles. In the near future, however, we can expect considerable progress. The SIRTF-Legacy survey will cover several tens of square degrees at 70 and 170 $\mu$m, allowing us to extend the IRAS counts in both wavelength and sensitivity. At longer wavelengths, large area surveys will be carried out by balloon-borne instruments. The ASTRO-F survey will provide all-sky information at 70 and 170 $\mu$m. Finally, WMAP detects the brightest point sources in the sky at frequencies from 23 yo 94 GHz.

As survey catalogs and source lists are identified, they will be folded into a *Planck* pre-launch catalog contained in NASA's Extragalactic Database (NED). This will allow them to be thoroughly cross-identified with other existing NED holdings and citations linked to them. Inclusion in NED also allows us to use the existing infrastructure of searching, data display and permanent storage that NED provides. The survey and catalog results will be included in the NVO as well, when appropriate. We may investigate specialized parameters and functions restricted to the *Planck* data.

### 4.5.1.2 Modeling of discrete source populations

At radio wavelengths, by using source counts available at *Planck* frequencies, by extrapolating faint centimeter source counts to higher frequency, and by modeling radio sources and their spectra, estimates can be made *on a statistical basis* of the foreground contributed by radio sources. One example is the work of Toffolatti et al. (1998). As the various radio surveys referred to above become available, these models can be refined. They need to be extended to the estimation of polarization fluctuations, as well as to the effect of the possible clustering of sources.

For dusty, star-forming galaxies, likely to dominate the counts in the HFI bands, modeling is much more important because ground-based surveys are much more difficult. Fortunately, the modeling of such sources is an active field (e.g., Guiderdoni et al. 1998; Blain et al. 1999; Devriendt, Guiderdoni, & Sadat 1999; Devriendt and Guiderdoni 2000; Takeuchi et al . 2001; Pearson 2001; Franceschini et al. 2000; Rowan-Robinson 2001; and Lagache et al. 2003). These models account for the counts, redshift distribution, and luminosity distribution of the known submillimetre populations. They can be extrapolated to make predictions of the level of fluctuations induced by unresolved sources and confusion. Since the surveys on which they are based cover only small solid angles, however, their predictive ability for the very brightest sources, those most likely to appear in the *Planck* images, is less strong.

*4.5.1.3 Simulations of the effect of sources on CMB maps*

Work is underway to investigate the effect of extragalactic discrete sources on the CMB images, and more particularly on the derivation of power spectra and cosmological parameters from them. We will need accurate simulations of the whole sky, including polarization and clustering effects, and will need to simulate the effect of this background of sources on the time-ordered data stream from the various receivers. We will also need to investigate the effect of unresolved, subliminal sources.

## 4.5.2 Source Removal from Single-Frequency Maps

Methods of separating CMB fluctuations from various foreground components, including extragalactic discrete sources, are under active investigation as well. These include using Wiener filtering and maximum entropy methods as well as wavelet analysis and the use of related filters. Source extraction from simulated multifrequency *Planck* data is already underway by several groups (Tegmark and Efstathiou 1996; Hobson *et al* 1998; Bouchet et al. 1999; Bouchet and Gispert 2000; Cayon et al. 2000; Tegmark et al. 2000; and Vielva et al. 2000). The last group has obtained particularly promising results for the detection of discrete sources by combining maximum entropy methods with a particular form of wavelet analysis. An approach called Independent Component Analysis is being implemented by Baccigalupi et al. (2000). This method, unlike some of those above, does not require a priori assumptions on the spectral properties or the spatial distribution of the various foreground components, but only that they are statistically independent and that all or all but one have a non-Gaussian distribution.

More refined methods must be developped which will take into account the different redshift distributions of the unresolved CFIRB contributing at different frequencies, as discussed in §4.4.2. In addition to work on simulated *Planck* data, some groups are using existing CMB maps to "practice" component separation. While this work is of considerable value, the low sensitivity of these ground-based, balloon-borne, and satellite experiments (WMAP) results in few source detections.

*4.5.2.1 Pointing and calibration issues*

We end with the comment that the characterisation of extragalactic discrete sources and issues of pointing, focal plane geometry, beam determination, and point source photometric calibration of *Planck* are closely intertwined. All these will be carried out through an iterative method which is at the heart of the Level 2 *Planck* data processing. It is essential to build before flight an input catalog of the sources to be used in this work containing all the useful information available.

# CHAPTER 5
# GALACTIC AND SOLAR SYSTEM SCIENCE

## 5.1 OVERVIEW

Once the monopole and dipole terms of the cosmic microwave background (CMB) are removed from the *Planck* maps, emission from the Milky Way will be the dominant feature (Figure 5.1). This emission, which must be further subtracted to obtain the higher order anisotropies of the CMB, will offer a complete portrait of the Galaxy in a virtually unexplored astronomical observing window ranging from the submillimetre at $350\,\mu m$ to the microwave at $1\,cm$. At the longest wavelengths, *Planck* will greatly improve on the only comparable product so far (from WMAP), while at wavelengths shorter than $3\,mm$, *Planck* will provide the first full-sky images. *Planck* therefore enables new and fundamental research on our Galaxy with unprecedented observational advantages: broad spectral coverage; the ability to measure polarised emission; high sensitivity; high photometric accuracy; high spatial resolution (5′) at the shortest wavelengths; and full-sky coverage.

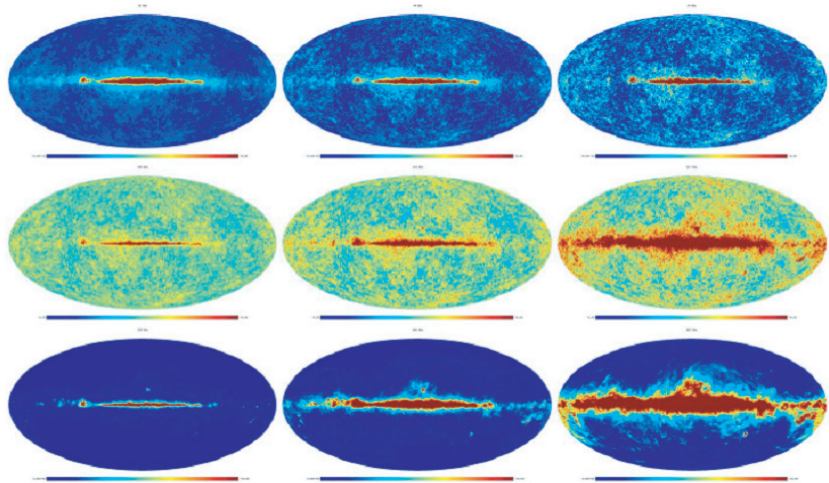

FIG 5.1.— False colour images of the simulated sky in the nine frequency channels of *Planck*, after subtraction of the monopole and dipole CMB components. From top left to bottom right: 30, 44, 70, 100, 143, 217, 353, 545, and 857 GHz channels.

The *Planck* full-sky maps will not only trace the spatial structure of the interstellar medium (ISM) of the Galaxy, but also probe specific conditions such as densities and temperatures. The emission in the *Planck* bands comes mainly from thermal radiation from interstellar dust, synchrotron radiation from high energy particles, and free-free radiation associated with ionized interstellar gas. The ability of *Planck* to measure polarisation will provide unprecedented maps of the magnetic fields in the Galaxy. The broad frequency range gives *Planck* very powerful leverage in this respect, since it will probe simultaneously two different mechanisms coupling the magnetic field to the emission. Since the interstellar gas and dust respond to a variety of physical mechanisms such as gravity, magnetic fields, and hydrodynamical phenomena, we can investigate not only individual objects in the Galaxy, but also the physical processes which act at a global Galactic level. In addition, improved understanding of Galactic foregrounds will lead to significantly better CMB maps.

The spectral coverage of *Planck* will allow separation of the emission components of the ISM based largely on the specific spectral signature of each component. The all-sky coverage of *Planck* is particularly important in the study of the large-scale structure of our Galaxy. Inversion techniques have been developed which will combine *Planck* maps with ancillary data to obtain the 3-dimensional distribution of neutral gas, ionized gas, dust, and magnetic fields throughout the Galaxy. The variations with Galactocentric radius of the properties of these ISM tracers, in particular the contrast between arm and interarm regions, will provide fundamental insights in the evolution of the Galaxy.

The potential of *Planck* is suggested by current multi-frequency studies such as the one illustrated in Figure 5.2, which shows the relation between neutral hydrogen structures, dust complexes, Hɪɪ regions, and supernova remnants in the active star-forming region in Cygnus. Major ground-based surveys are in progress which will complement *Planck* for Galactic studies. These include large-area surveys of Hα, radio recombination lines, Hɪ, CO, and radio emission, including polarisation.

*Planck* will make a major contribution to our understanding of the physics of star formation. It will probe directly the various phases of the ISM, showing the morphology of star forming regions, providing the mass function and the temperatures of the clouds associated with the birth of stars, and also determining the physical conditions of the circumstellar material at the later stages in the stellar life. This is the basic information required to trace the complete evolutionary cycle of stars.

Closer to home, *Planck* will gather the first full-sky picture of the cooler objects in the solar system at submillimetre wavelengths, providing new information on asteroids, planets, comets and the zodiacal emission.

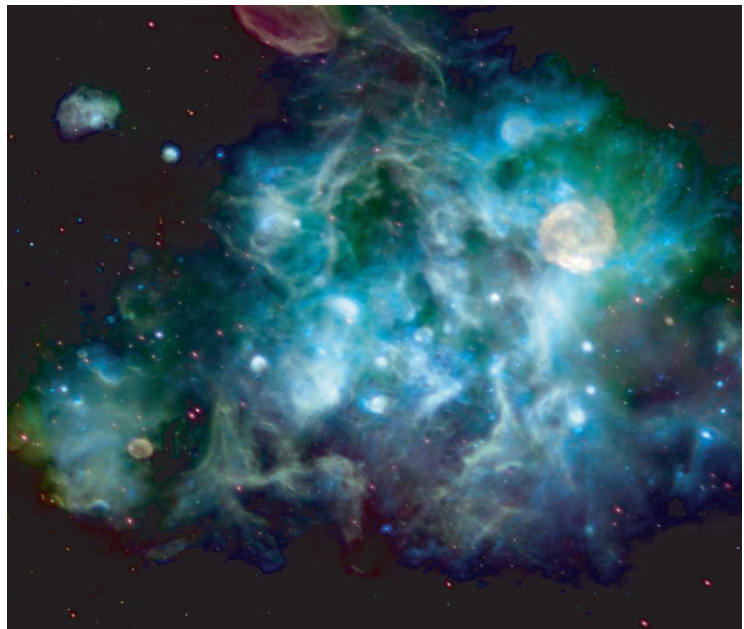

Fɪɢ 5.2.— False colour image of the star-forming complex in Cygnus from the Canadian Galactic Plane Survey (CGPS). The following data have been superposed: 25 and 60 µm image from IRAS (blue and turquoise), H ɪ 21 cm from the DRAO survey (green), and DRAO 74 cm synchrotron (rose). The Hɪɪ regions are seen as white and the supernovae as yellow.

Studies of the morphology of the ISM depend on spatial resolution. For a given spatial resolution, the distance to the object determines the type of investigation. Hence a distinction is made in this Chapter between research on nearby individual objects, which is based on the detailed study of their structure and physical properties, and the study of more distant parts of the Milky Way, which leads to the examination of the large-scale characteristics and distribution of the constituents. The former are described in § 5.2, and the latter in § 5.3. § 5.4 addresses specifically the study of magnetic fields, and § 5.5 is dedicated to Solar System studies.

## 5.2 THE LOCAL INTERSTELLAR MEDIUM

### 5.2.1 Physical Constituents

In the most general picture, the ISM contains cold clouds, either molecular or atomic, an intercloud medium, either neutral or ionized, and hot ionized cavities presumably formed by supernovae. Each of these components forms a phase, i.e., is thermally and dynamically long lived with a similar pressure ($nT \propto 10^3$–$10^4\,\mathrm{cm^{-3}\,K}$) in each phase. Much of the atomic gas (H$\textsc{i}$) is observed to be either the warm neutral medium (WNM; $T \approx 10^4\,\mathrm{K}$) or the cold neutral medium (CNM; $T < 100\,\mathrm{K}$). The warm ionized medium (WIM) is partly ionized gas with $T \approx 10^4\,\mathrm{K}$. A small fraction of the gas is in the hot ionized medium (HIM; $T \approx 10^6\,\mathrm{K}$) with a relatively high filling factor (0.4–0.6). A fraction of the mass of the ISM is molecular (H$_2$) with a filling factor of only a few percent, residing in star-forming Giant Molecular Clouds (GMCs) or in quiescent clouds.

The *Planck* data, with their wide frequency coverage and high sensitivity, will enable the separation of the various phases into distinct maps. The separation will be based not only on the spectral properties of the emission itself, but also on the correlation of the *Planck* maps with spatial templates provided by large-coverage maps of other tracers. The main sources of such templates are listed in Table 5.1, and include the results of ground-based, balloon-based, and space-based surveys. Some specific important examples and their expected use are described in the following subsections.

#### 5.2.1.1 Interstellar Gas

A complete template of the WNM and CNM is provided by the Dwingeloo H$\textsc{i}$ survey (all sky, 36′ resolution) and the Dominion Radio Astrophysical Observatory (DRAO) H$\textsc{i}$ synthesis survey, with its extension to high latitudes. Dedicated DRAO synthesis observations as well as additional single dish observations will enable the faintest regions to be detected.

Moderate density molecular regions will be traced with $^{12}\mathrm{CO}$ (J=1–0) using existing large scale surveys. These surveys will be complemented by new surveys, such as that being undertaken by the Nagoya University group with the Nanten telescope in Chile. High density molecular cores are better traced using rarer isotopes of CO or heavier molecules.

The ionised ISM is of particular importance to *Planck* because of the large contribution of free-free emission in the low frequency channels. Free-free emission can be separated from other diffuse Galactic emission components (synchrotron and anomalous dust emission) by its spectrum. The brightness temperature of free-free emission from H$\textsc{ii}$ regions has a spectral index of approximately $-2.15$. A unique map of the free-free emission will then be available over the whole sky. A new generation of H$\alpha$ surveys (e.g., the WHAM), reaching 0.2–0.5 R on angular scales of $\sim 0°.1$–$0°.2$, should be available by the launch of *Planck* (1 Rayleigh (R) corresponds to a brightness temperature of $5.8\,\mu\mathrm{K}$ at 30 GHz). Together these will provide a powerful resource of data on the ionised gas in the Galaxy.

Comparison of the *Planck* free-free maps, the H$\alpha$ surveys, and the dust distribution will provide a dust absorption correction for the H$\alpha$ data. The corrected H$\alpha$ maps provide estimates of the electron temperature of the ionised gas in various environments. The large-scale distribution of free-free emission is dominated by the Local (Gould Belt) System reaching more than $40°$ from the plane in the center and anti-center regions. The relationship between the ionised gas, the H$\textsc{i}$, the dust, and any synchrotron emission in the environment of this large star-forming region will be investigated. Preliminary correlations of H$\alpha$ surveys with H$\textsc{i}$ and dust show a complex situation that is possibly a consequence of the ionising radiation only penetrating an outer layer of the neutral cloud. Detailed studies of individual regions, as well as statistical studies of classes of regions, will be possible with the *Planck* data supported by H$\alpha$ observations.

*Planck* provides a unique opportunity to map out the three-dimensional distribution of the local interstellar gas covering all phases of the diffuse ISM. The three-dimensional distribution is particularly interesting for the solar neighbourhood within 1 kpc. Some of the specific issues to

TABLE 5.1

Ancillary Surveys (to be used in the analysis of *Planck* maps)

| Survey type | Survey | Resolution ['] | Coverage | Status |
|---|---|---|---|---|
| CO .......... | Composite CFA $^{12}$CO[a] | 8′ | $-10° < b < +10°$ | Completed |
| | FCRAO(CGPS) $^{12}$CO[b] | 0.8 | $+74° < l < +147°$ | In progress |
| | | | $-3°6 < b < +5°6$ | |
| | Nagoya U. $^{12/13}$CO[c] | 2.7 | $-10° < b < +10°$ | In progress |
| HI ............ | Dwingeloo/NFRA[d] | 30′ | Full Sky | Completed |
| | CGPS/DRAO[e] | 1′ | $+74° < l < +147°$ | In progress |
| | | | $-3°6 < b < +5°6$ | |
| | HIPASS/HIJASS | 15′ | Full Sky | In progress |
| IR ........... | 2MASS 1.25/2.2 μm | 0.06′ | Full Sky | Completed |
| | IRAS 12/100 μm[f] | 4′ | Full Sky | Completed |
| | DIRBE 1.25/240 μm | 30′ | Full Sky | Completed |
| | MSX 4/26 μm[g] | 0.3 | $-5° < b < +5°$ | Completed |
| | ISO Serendipitous 170 μm[h] | 2′ | 15% sky | Completed |
| | IRIS/ASTRO-F 50/200 μm[i] | 0.8 | Full Sky | Future |
| | SIRTF 24/160 μm[j] | 0.27 | Maps | In progress |
| | ELISA-balloon 200/600 μm[k] | 3′ | $-20° < b < +20°$ | Future |
| | *Herschel* 100/600 μm | 0.5 | $1000 \deg^2$ | Future |
| H-alpha ....... | WHAM-Fabry-Perot[l] | 60′ | Northern sky | In progress |
| | SHASSA-filter[m] | 5′ | Southern sky | In progress |
| | Manchester WFC-filter[n] | 5′ | | Future |
| Radio ........ | Stockert/Bonn 1.4 GHz[o] | 34′ | Northern sky | Completed |
| | Halslam 408 MHz[p] | 50′ | Full Sky | Completed |
| | FIRAS 100/1000 μm | 420′ | Full Sky | Completed |
| | DMR 90/30 GHz | 420′ | Full Sky | Completed |
| | WMAP 22/90 GHz | 20′ | Full Sky | In progress |
| | Bonn MLS 1.4/2.7 GHz[q] | 10′ | $-10° < b < +10°$ | In progress |
| | HatRAO 2.3 GHz[r] | 20′ | Southern sky | In progress |
| | CGPS/DRAO 408/1420 MHz[s] | 1′ | $+74° < b < +147°$ | In progress |
| | | | $-3.6° < b < +5.6°$ | |
| | Green Bank 8.35/14.35 GHz[t] | 5′ | $-5° < b < +5°$ | In progress |
| X-ray ........ | ROSAT 0.1-4 keV[u] | 12′/2° | Full Sky | Completed |
| γ-ray ......... | CGRO >100 MeV[v] | 120′ | Full Sky | Completed |
| | INTEGRAL <10 MeV | 60′ | $-15° < l < +15°$ | Future |

be addressed are: the relation of Gould's Belt to the nearest starforming molecular clouds (e.g., whether Gould's Belt is the result of a high velocity cloud collision), the nature of the Local Bubble, and the fate of the clouds from which T Tauri stars originate. A tentative detection of a new component of the ISM, the HI Self Absorption features (HISA), has been made based on 21 cm absorption lines (Gibson et al. 2000). Some of these features do not show molecular emission. Their masses are not yet known, and consequently their contribution to the overall mass and temperature distribution of the cool ISM is poorly understood.

### 5.2.1.2 Dust

Our knowledge of interstellar dust is based on extinction observations in the UV to near-infrared, and observations of dust emission in the infrared and beyond. In addition, polarisation measurements have shown that dust is susceptible to the presence of a magnetic field. The dust grains follow a broad size distribution spanning a range from a few nanometres for the smallest, to several hundred nanometres for the biggest grains. Interstellar dust is thought to be composed of a variety of materials which can be classified into three main groups: very small grains with signatures of large polycyclic aromatic hydrocarbon (PAH) molecules, carbonaceous compounds, and silicates. The dust can exist in amorphous or crystalline form, with or without ice mantles, either in (fluffy) aggregates or composites.

*Planck* will provide for the first time a full-sky view of the long wavelength ($\lambda > 350\,\mu m$) emission properties of dust, which are dominated by cool grains at the large end of the size distribution. Due to the scarcity of data at arcminute resolution, the long wavelength emissivity of dust is presently very poorly constrained. *Planck* will give the first complete census of the properties of the big grains in the Galaxy, allowing us to understand how they evolve from the diffuse medium to dense cloud cores, where they can accrete smaller particles and form icy mantles and/or fluffy aggregates. At shorter wavelengths, all-sky data from the Infrared Astronomical Satellite (IRAS) have revealed spatial variations of the infrared colours in the wavelength range $12$–$100\,\mu m$ (see, e.g., Boulanger et al. 1990). This is a strong indication that the composition of the dust in the ISM is not fixed but varies according to changes in the size distribution where transiently heated nanometric particles ("very small grains") dominate the $6$–$60\,\mu m$ range and big grains (larger than about $0.1\,\mu m$ diameter) dominate at $\lambda > 100\,\mu m$ (see Figure 5.2). At submillimetre wavelengths, the pioneering observations performed by PRONAOS have also revealed color variations (more on this below). These can be understood as changes in the temperature and optical properties of the big grains. The observed variations appear to correlate with the abundance of the smallest particles. It is therefore expected that *Planck* will provide new insights in the variation of the emissivity and composition of the big grains.

*Planck* will be much less affected by dust temperature uncertainties than previous far-infrared instruments such as IRAS and the Diffuse Infrared Background Experiment (DIRBE) of COBE because it covers the long wavelength regime where the Rayleigh-Jeans approximation is valid. This is shown in Table 5.2, where the logarithmic slope of the temperature dependence of the dust emission in the IRAS, DIRBE, and HFI bands is indicated. As a consequence of the near-unity slopes, the surface brightness becomes proportional to dust column density. *Planck* will for the first time allow an accurate derivation of the total dust masses in the structures of the ISM, from the high latitude cirrus clouds to nearby star-forming molecular clouds.

TABLE 5.2

Temperature Dependence of Dust Emission at Different Wavelengths[a]

| | $\nu$ [GHz] | | | | |
|---|---|---|---|---|---|
| Temperature | 5000 | 3000 | 857 (HFI) | 545 (HFI) | 353 (HFI) |
| 12 K . . . . . . . . . . . . . . . . . | 12 | 5.0 | 3.4 | 2.2 | 1.4 |
| 17 K . . . . . . . . . . . . . . . . . | 8.5 | 3.6 | 2.6 | 1.9 | 1.6 |
| 30 K . . . . . . . . . . . . . . . . . | 4.8 | 2.3 | 1.8 | 1.5 | 1.3 |

[a] Table entries are values of $\alpha$, where $I_\nu = C T^\alpha$.

*Planck*'s measurements of total dust masses will provide tight constraints on the emissivity index $\beta$. The determination of dust temperatures, however, will require complementary far-infrared/submillimeter data covering the peak of the energy distribution (SED) of the dust at $\lambda \lesssim 300\,\mu m$ (Table 5.2). *Planck* data can be combined with IRAS and Astro-F full-sky survey data to obtain the complete SED of the dust, from which the dust size distribution can be inferred. Data from the ISO Serendipity Survey at $170\,\mu m$, the Spitzer Space Telescope, and *Herschel* can used also over small regions of the sky.

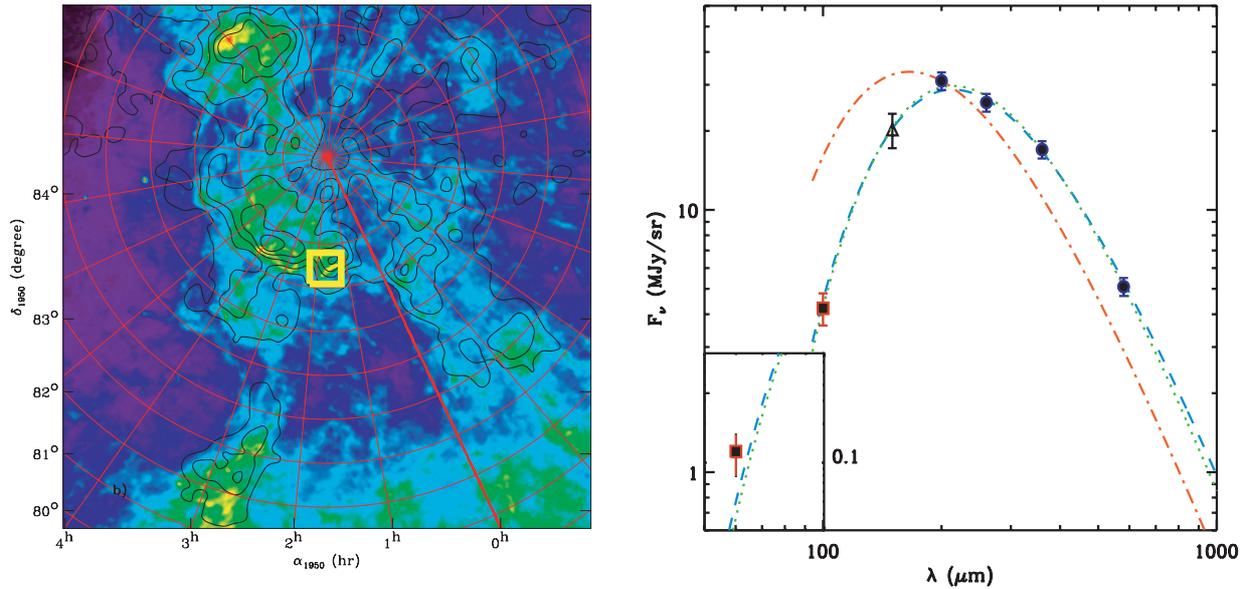

FIG 5.3.— PRONAOS observations of a diffuse cloud in the Polaris Flare. The left panel shows the IRAS $100\,\mu m$ emission of the flare (colors) and the visible extinction measured from star counts (contours starting from $0.2\,\text{mag}$ with a step of $0.2\,\text{mag}$). The region observed with PRONAOS is shown by the square. The right panel shows the spectrum obtained by combining IRAS and PRONAOS data of the diffuse extended emission in this area (blue curve), which reveals the existence of cold dust ($T_d = 13.0\,\text{K}$, $\beta = 2.2$). The spectrum of the average diffuse ISM ($T_d = 17.5\,\text{K}$, $\beta = 2$) is shown for comparison (red curve).

The relationship between dust and gas in the different phases of the ISM is poorly investigated, and only on large scales. The correlation study between far-infrared data from COBE and H I emission has shown that dust associated with the neutral medium (NM) reaches an average temperature of $T_d = 17.5\,\text{K}$ in high Galactic latitude cirrus clouds (Boulanger et al. 1996). This temperature assumes grey body emission, i.e., the dust emission is the product of a blackbody and a power law emissivity, $\nu^\beta$, with $\beta = 2$. The properties of dust associated with the warm ionised medium (WIM) appear similar to those associated with the NM (Lagache et al. 2000). The COBE data suggest the presence of colder regions with $T_d \lesssim 16\,\text{K}$ (Lagache et al. 1998), probably associated with molecular clouds; however, these parameters could only be derived on average over a large fraction of the sky at high latitude. The homogeneity of the dust temperature and emission properties cannot be tested on smaller scales, due to the limited spatial resolution of COBE.

Higher angular resolution observations have been obtained in the far-infrared using the ISOPHOT instrument on the Infrared Space Observatory (ISO) and in the submillimeter with the PRONAOS balloon-borne experiment. These observations indicate unexpected dust properties:

- Low equilibrium temperatures (Figure 5.3) towards diffuse ($A_v <$ few) and moderately opaque ($A_v \gtrsim 2.1$) molecular regions (Bernard et al. 1999, Stepnik et al. 2001), which require a sudden change of the dust properties, possibly due to coagulation of dust into fractal aggregates.

- A significant inverse correlation between the dust emissivity index and temperature, observed toward star forming regions (Dupac et al. 2001), which may reveal new emission properties appearing only at low temperature.

The PRONAOS data, however, are mostly exploratory and possibly biased towards extreme regions. The larger scale maps obtained by the Boomerang balloon experiment lack the higher frequency measurements necessary to constrain sufficiently the dust temperature. It is therefore difficult to assess how ubiquitous the changes are, how they relate to physical conditions such as density, turbulence, etc., and whether they will also be present in more diffuse regions. Understanding how and why the dust properties change is not only fundamental to the description of the ISM, but is also a key question regarding Galactic foreground subtraction in the *Planck* data for cosmology studies.

Such studies are particularly important in the WIM, where modifications of the dust size distribution are expected due to gas-grain and grain-grain collisions in shocks (e.g., Jones, Tielens, and Hollenbach 1996). With *Planck*'s sensitivity, dust studies can be extended to intermediate velocity clouds and even to high velocity clouds whose dust counterparts have yet to be detected (Figure 5.4). Correlations between the dust emission and the velocity and line-width of the atomic and molecular lines will be sought, because the dust, if present in these clouds, could be significantly different from standard Galactic dust.

Departures from grey body laws are also of interest, in particular temperature mixing due to the size distribution of large grains. The possible existence of very cold dust ($4\,\mathrm{K} < T_\mathrm{d} < 7\,\mathrm{K}$) in the diffuse ISM remains to be tested. A search can be made for residual dust emission regions not associated with any of the available templates, since some emission is expected from low density molecular regions where the only molecular species that can form is $H_2$.

The correlation between (sub)millimetre dust emission, extinction, and gas emission will provide a calibration of the various mass tracers and a test of their validity in the various parts of interstellar clouds. Star catalogs (e.g., 2MASS) will be used to trace independently the column density via the extinction from star counts. The comparison between the dust emission traced by *Planck* and molecular line maps (e.g., CO, see Table 5.1) will allow separation of structures at different velocities (and therefore Galactic locations), and further constrain the physical conditions. To derive the density structure and to analyse the evolution of the properties of the dust across selected regions of a molecular cloud, models will be used which take into account dust properties and radiation transfer inside the cloud. Such models have been applied succesfully to IRAS and PRONAOS observations (Bernard et al. 1993; Stepnik et al. 2001).

The study of clouds through the analysis of individual regions will be complemented by statistical investigations of the spatial structure. The power of the brightness fluctuations can be analysed as a function of mean brightness and angular scales (e.g., Abergel et al. 1996). It will be possible to derive a statistically significant angular and mass power spectrum of structure in the ISM, not only in the form of $\log N/\log M$ and $\log N/\log R$ diagrams, but also in the form of structure trees and structure functions. The *Planck* data will enable analysis of spatial structures from the largest scales (Giant Molecular Clouds) to the dense cores out of which stars form, even in opaque regions. Such studies will be particularly fruitful in the case of nearby molecular clouds (e.g., Taurus, Chamaeleon, Ophiuchus, Orion). The goal is to trace the variation of the statistical properties of the emission from diffuse clouds to dense molecular regions, and to understand this variation in terms of the evolution of physical conditions.

### 5.2.1.3 "Anomalous" Dust emission

Cross-correlation studies of CMB and far-IR data have shown a microwave emission component with a spectral index suggestive of free-free emission; however, evidence from COBE, Owens Valley Radio Observatory (OVRO), Saskatoon, and the Tenerife experiments (de Oliveira-Costa et al. 2002; Mukherjee et al. 2001) indicates that this component is correlated with dust emission at a higher level than expected from free-free emission alone. Recent theoretical work suggests that this emission may originate from very small spinning dust grains (Draine & Lazarian 1998; de Oliveira-Costa 2002). However, in diffuse regions it is difficult to reconcile the spectral index of the spinning dust emission with the observed radio spectral index (Kogut 1999). Therefore, the source of the correlated emission is still an open question, and is often referred to as "anomalous" or "spinning" dust emission.

The Wilkinson Microwave Anisotropy Probe (WMAP), which covers the 10–100 GHz range, as well as improved H$\alpha$ surveys, are starting to shed some light on this anomalous emission component (see for example Lagache et al. 2003). The WMAP frequencies are too low to allow a detailed study of the Galactic dust emission temperature. The maximum entropy component separation performed by the WMAP team reveals a dust component with an average spectral index of about $-2.2$ (Bennett et al. 2003). The same analysis reveals a strong synchrotron component that is significantly different in structure from that of the Haslam 408 MHz survey. This component is very well correlated with dust, and it could well be that it includes the so-called anomalous dust emission. Supporting this idea, Lagache et al. (2003) found a strong correlation at high Galactic latitudes between the H I surveys and the dust-subtracted WMAP data. Finkbeiner (2003) has shown that when reasonable foreground template maps are subtracted from the WMAP data a positive residual shows up which is correlated with the square of the dust physical temperature.

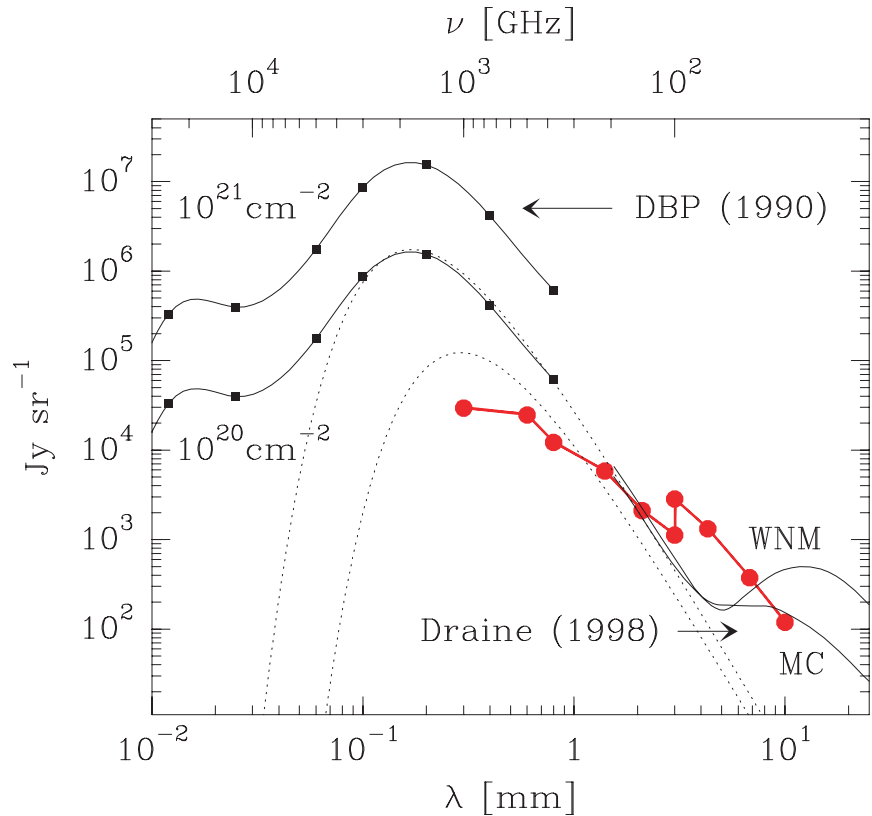

FIG 5.4.— Predicted dust emission from interstellar clouds. In the far-infrared part of the spectrum, solid lines indicate the predicted thermal dust emission for hydrogen column densities $10^{20}$ and $10^{21}\,\mathrm{cm^{-2}}$ (Désert, Boulanger, Puget 1990). At longer wavelengths, predictions for spinning dust emission from the warm neutral medium and molecular clouds are shown for column density $10^{20}\,\mathrm{cm^{-2}}$ (Draine and Lazarian 1998). The dotted lines are $\nu^2 B_\nu(T)$ spectra at temperatures $10\,\mathrm{K}$ (lower curve) and $17\,\mathrm{K}$. The red curve indicates the predicted sensitivity of *Planck* ($1\sigma$ in one year).

*Planck* observations extending from 30 to $900\,\mathrm{GHz}$ will be a key to the confirmation of anomalous dust emission and the determination of its characteristics. Polarisation data may be particularly useful. If the emission does indeed arise in spinning dust grains, centimetre-millimetre wavelengths should correlate with IR dust emission from small grains. High and low frequency *Planck* data can be combined to isolate and remove thermal dust and free-free emission from the channels where the anomalous emission is expected, then the residual can be correlated with IRAS or Astro-F maps.

It has also been hypothesized that the excess millimeter-wave emission is produced by low energy transitions in the solid-state (see Phillips et al. 1972, Agladze et al. 1996). Such an effect has never been fully explored experimentally or theoretically, and work in this area clearly is required in preparation for *Planck* observations. New laboratory measurements will be needed to better characterise the spectroscopic properties of cosmic dust analogues over the *Planck* wavelength range as a function of composition, structure, and temperature. These data will provide the necessary inputs to the modelling of specific dust emission properties as a function of the physical conditions.

### 5.2.2 Compact Stellar Objects

In addition to large structures of gas and dust, the ISM contains a number of classes of compact objects representing different evolutionary stages in star formation, including compact pre-stellar cores, young stellar objects (YSOs) deeply embedded in their progenitor clouds, radio-emitting stars, and supernova remnants. Many local examples of these classes should be bright enough to be detected by *Planck*. Table 5.3 summarizes the properties of the different classes of compact stellar objects observable with *Planck*. The study of these objects using the *Planck* data will be an important contribution to the study of star formation. Details on some of the more interesting classes are provided in the following sections.

TABLE 5.3

PROPERTIES OF COMPACT GALACTIC SOURCES

| | EMISSION MECH. | DETECTABLE MASS $[3\,\sigma/M_\odot]$ | | Planck $\nu$ | DETECTIONS |
|---|---|---|---|---|---|
| TYPE OF OOJECT | $T$ | 2 kpc | 20 kpc | [GHz] | EXPECTED |
| HII regions/star forming complexes . . . . | Dust/Gas(ff) 50/10000 K | 0.11/40 | 11/4000 | 30–857 | >1400 |
| Cold cores/molecular clouds . . . . . . . . . | Dust 15 K | 1.2 | 120 | 353–857 | > $10^4$ confusion |
| Post AGBs/planetary nebulae . . . . . . . . | Dust 100/200 K | 2.3 | 1.8 | 545–857 | about 150 about 40 |
| SNRs . . . . . . . . . . . . . . . . . . . . . . . . . . . | Dust/Synchrotron 100 K/— | | | 30–857 | 230 |

### 5.2.2.1 Cold cores and the early stages of stellar evolution

The earliest phases of low mass star formation are associated with dense and cold cores of molecular clouds supported against gravitational collapse by thermal, magnetic, or turbulent effects, and devoid of embedded objects. Compact molecular clouds and globules are defined here as objects which have dust colour temperature $T_d < 20$ K and angular size less than 20′. Thus as seen by *Planck* they will be unresolved or slightly extended.

Much progress in their study has been made in the last 20 years by means of radio frequency molecular lines; however, it is now clear that dust grains also play a key role, and the knowledge of their physical and chemical properties is of decisive importance. Until recently, optical searches have been the most effective way to detect such compact clouds and globules. They are nonetheless generally difficult to detect, especially at higher Galactic latitudes where the surface density of background stars is low, and at low Galactic latitudes where the more distant ($d > 1$ kpc) small cores are faint. Therefore, it is possible that numerous clouds have remained undetected. Based on optical identifications, the molecular gas properties of a large sample of small molecular clouds and cloud cores have been studied using observations of CO, $NH_3$, and other molecules (e.g., Clemens et al. 1991; Jijina, Myers & Adams 1999). Clemens et al. found typical values for the total cloud mass of 11 $M_\odot$, radius 0.35 pc, density $10^3$ cm$^{-3}$, and far-infrared luminosity 6 $L_\odot$. The total number of globules/small clouds in the Galaxy was estimated to be around $3 \times 10^5$.

Such cold clouds were also observed by ISOGAL, the ISO mid-infrared survey of the Galactic plane (Hennebelle et al. 2001). Resolved observations of their thermal emission in the far infrared and the submillimeter were performed with ISO and PRONAOS. Lehtinen et al. (1998) studied one such globule, heated solely by the interstellar radiation field, with ISO at 100 and 200 $\mu$m. The dust has a temperature of about 13.5 K (assuming $\beta = 2$). The ISOPHOT Serendipity Survey (ISOSS) at 170 $\mu$m covers about 15% of the sky (Tóth et al. 2000; Hotzel et al. 2001), and detected more than a hundred new cold globules and cloud cores. The PRONAOS-SPM observations (1994, 1996, 1999 flights), covering 180–1100 $\mu$m, detected cold dust condensations within star forming regions (Ristorcelli et al. 1998, Dupac et al. 2001) and filamentary and cirrus clouds (Bernard et al. 1999, Stepnik et al. 2001), with typical characteristics of $T_{\mathrm{dust}} = 12$ K, $0.1 < D < 1$ pc, and $5 < M < 30$ $M_\odot$.

The COBE FIRAS and DIRBE surveys, with coarse angular resolution of $7°$ and $0.7°$ respectively, found a cold dust component widely distributed in the Galaxy. This may be, in fact, integrated emission from the numerous cold cores in the Galactic plane. *Planck* should be able to detect a typical cold core of $11\,M_\odot$ up $6\,kpc$ away. Taking confusion into account, *Planck* may be able to identify more than $10^4$ of the estimated $3\times10^5$ such cores in the Galaxy, making them one of the very new class of objects which *Planck* will reveal to astronomers.

There is increasing interest in the idea that Galactic dark matter is composed of very cold ($5 < T < 10\,K$) molecular material in Jupiter-mass clouds, either distributed through the halo or in the outer disk (see, e.g., Pfenniger et al. 1994; Sciama 2000; Lawrence 2001). If such objects contain normal dust and populate the Galactic Plane, they could explain up to a third of the sources seen in deep SCUBA surveys (Lawrence 2001).

The statistical properties of cores are closely related to low-mass star formation and the stellar initial mass function (IMF) because stars are born via the collapse of dense and cold cloud cores or globules. So far, the IRAS Point Source Catalog (PSC) remains the reference database for statistical study of star formation in our Galaxy. The spectral coverage of IRAS has allowed the identification of the evolutionary sequence from deeply embedded YSOs to progressively hotter and older stars (class I, II, III). However, the coldest, youngest accreting protostars (class 0), the protostars in the isothermal contraction phase (class I), and the prestellar sources in the phase preceding gravitational collapse can only be detected at submillimetre wavelengths (André et al. 2000). Similarly, the early stages of massive star formation have not been readily identified by IRAS. Indeed, good candidates are expected to be rare, due to the very short time scales involved. To understand how the IMF comes about, it is thus crucial to investigate the processes by which these pre-stellar condensations form. Except perhaps in the very nearest clouds, *Planck* will not have sufficient angular resolution to resolve individual pre-stellar condensations, but *Planck* will provide the necessary first step, i.e., an unbiased inventory of the sites of on-going and future star formation in the Galaxy, which are precisely the fields to be studied at higher resolution with *Herschel* and the Atacama Large Millimeter Array (ALMA).

Cold cores will be identified using an all-sky catalogue of compact submillimetre and millimetre sources produced by the *Planck* collaboration. The catalog will probably contain a large number of previously undetected objects (especially those with very low temperatures) and possibly also include objects of previously unknown type, which will become visible for the first time as a consequence of the wide wavelength coverage and the high sensitivity of *Planck*.

The list of *Planck*-detected cold cores will form the basis for a wide variety of specific studies. Here are two examples:

- The spatial distribution of the detected cold cores will be correlated with star forming regions and other larger-scale structures in the Galaxy. The *Planck* data will allow determination of the core masses and mass spectrum, and study of its relationship with the local stellar IMF.

- Statistical studies of the structure and morphology of the cores will be performed by combining *Planck* measurements with molecular tracers and dust emission models (see Stepnik et al. 2001); these studies will address issues such as the surface and volume filling factor of the cores, the level of fragmentation, the structure of the envelopes, etc.

### 5.2.2.2 Late type stars

When a low or intermediate mass star is approaching the end of its life, it goes through a period of heavy mass-loss known as the Asymptotic Giant Branch (AGB) phase. Dust formation is very efficient in this period, and the ejected envelopes partially condense into dust grains which completely obscure the central stars. Immediately after the AGB phase, the mass-loss stops and the stars may become optically visible as the dusty shells disperse. Eventually, at a temperature of 20,000–30,000 K, the central star begins to ionize the remaining AGB shell and a Planetary Nebula (PN) forms. The circumstellar envelopes, containing atomic and molecular gas, are characteristic of the post-AGB phases and, as the dusty envelope re-radiates the absorbed stellar light, it will show a clear signature in the far-infrared spectrum of the stars. Moreover, in addition to the IR characteristics, PNs show a strong radio continuum, due to free-free

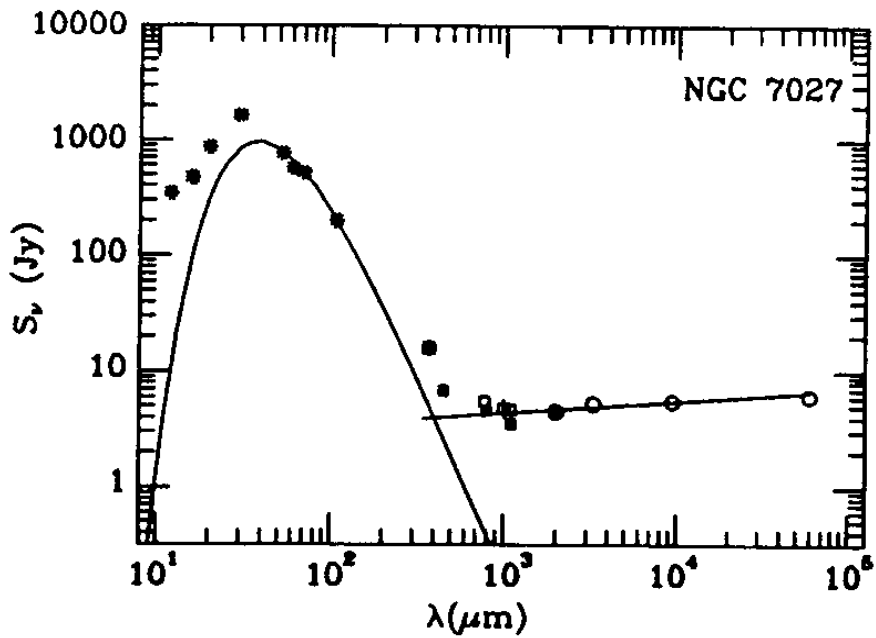

FIG 5.5.— The SED of the young planetary nebula NGC 7027 (reproduced from Hoare, Roche and Clegg, 1992). Thermal emission from dust dominates the IR-mm part of the spectrum, while the contribution of free-free emission from the ionized envelope is evident in the radio.

emission from the part of the circumstellar envelope ionized by the central star (Figure 5.5).

Detailed studies of the spectral energy distributions of a small sample of optically bright post-AGB candidate stars have shown that these objects have a characteristic double-peak distribution, with a broad infrared excess extending from the near-infrared to the millimeter range. For most of the known post-AGB objects, near-infrared as well as far-infrared (IRAS) photometric measurements are available, but more multi-frequency measurements at the *Planck* bands are essential for a better determination of the SED and to put more stringent constraints on the model of their circumstellar envelopes. In particular, the detection of a dust contribution in the sub-millimeter and millimeter would point to the existence of multiple, cold dust shells. By modelling the circumstellar envelope in terms of one or more dusty shells, it should be possible to determine the physical properties of the shells and their mass, which can be related to different mass-loss events suffered by the star during the AGB and post-AGB evolution.

Since PNs and their progenitors are believed to be among the major sources of recycled interstellar matter, determining the properties of dust injected by post-AGB stars in the ISM is very important to any study concerning the origin of diffuse matter in the Galaxy. The *Planck* measurements will be used to determine these physical properties of the post-AGB dusty envelopes.

### 5.2.2.3 Variable Galactic sources

The study of variable Galactic radio sources is particularly important in astronomy since it is generally linked to sudden energy releases. Many variable radio sources, belonging to different classes, will reach, during outbursts, flux levels high enough to be detectable by *Planck* with high signal-to-noise ratio. Particularly interesting are sources whose radio emission is normally very weak or absent and show periods of intense radio emission.

Among these, the so-called Luminous Blue Variable stars, which are among the most luminous stars in the Galaxy, are characterized by sudden outbursts, during which a large amount of mass is ejected from the star. The most active Luminous Blue Variable, $\eta$Carinae, is an impor-tant target for *Planck*. During flares, the flux density can reach several Janskys at centimeter wavelengths, and tens of Janskys at millimeter wavelengths. Over this wavelength range the emission is dominated by two processes: free-free emission from an optically thin nebula and wind, and dust emission at shorter wavelengths.

Massive X-ray binary systems are among the most interesting variable radio sources. One of the most active and variable among these is Cyg X-3, which exhibits quiescent periods and strong outbursts that have been interpreted as synchrotron emission associated with the ejection of relativistic electrons in a jet-like structure. Cyg X-3 shows giant flares reaching up to 20 Jy at 8.4 GHz. The observed spectral index ($S_\nu \propto \nu^\alpha$) at centimetre wavelengths ($\alpha = -0.55$) gives an extrapolated flux density of several Janskys in the region of the *Planck* LFI, and up to 1 Jy at the highest HFI frequencies. However, the measured emission at millimetre wavelengths during quiescent phases is greater than expected, probably due to the presence of a hot and dense wind embedding the system. More than 200 Galactic X-ray binaries are known, and 10% of these are radio loud. They can show superluminal motions like GRS1015+105, V4641 Sgr, and GRO J1655-40, and they can experience strong flares observable in the LFI spectral region.

The radio behaviour of active binary stars such as RS CVn and Algol is characterized by alternating periods of quiescence, with flux densities of few tens of mJy, and active periods with strong flares reaching several Jy at cm wavelengths. The radio spectra show the maximum of emission at frequencies higher than 10 GHz and, during active phases, up to 86 GHz.

Although these sources can be observed from the ground at low frequencies and some of them have been extensively studied also with multi-wavelength campaigns, very little is still known on the mechanisms which drive the observed flares. *Planck* covers an unexplored spectral range which provides unique data for studying these classes of radio sources. The spectral information obtained by *Planck* opens the possibility of modelling the plasma characteristics, the magnetic field, the electron energy distribution, the dissipative processes, and the interaction with the external environment, together with the acceleration mechanisms operating in the different objects during the most energetic events. The fact that *Planck* revisits the sky periodically means that it can detect short-term variability (for the most luminous targets, i.e., Cyg X3) and identify new variables. The latter can be achieved by comparison of the full sky maps taken at different epochs. Special software in the data processing pipeline will be installed for automatic detection of interesting events, which can trigger a response from ground-based observatories within a few days.

### 5.2.2.4 *Galactic Supernova Remnants*

Although Galactic Supernova Remnants (SNRs) have long been studied at radio wavelengths, detailed spectral studies of their radio emission are usually limited to a restricted range of frequencies (typically a few hundred MHz to a few GHz). The *Planck* observations will provide unique data for radio spectral studies of SNRs, extending the available frequency range considerably.

The majority of identified SNRs are 'shell' remnants, showing limb-brightened shells of emission, typically synchrotron radio emission, with a spectral index $\alpha$ of $\sim -0.5$. This spectral index is as expected from simple Fermi shock acceleration theory (e.g., Bell 1978; Drury 1995), although detailed spectral studies of SNRs are required to constrain shock acceleration theories in detail. Two investigations of shell-type remnants can be made with *Planck*, particularly the channel at 100 GHz:

- Searching for spectral flattening at high frequency in bright, young SNRs such as Tycho and Kepler, as predicted by some recent theories.
- Constraining spectral variations across the face of some of the brighter, larger shell SNRs (e.g., IC443 and the $\gamma$Cygni SNR) in order to understand how the differing shock conditions affect the radio spectral index. IC443 has shown interesting spectral variations at lower-frequencies, including a flatter-spectrum region which corresponds to a region of unusually hard X-ray emission where the SNR is interacting with its surroundings.

In addition to 'shell' type SNRs, there is a subset of about a dozen SNRs which are of 'filled-center' type (also called 'plerions'), of which the Crab Nebula is the best known. These remnants show centrally brightened synchrotron radio emission, thought to be due to relativistic particles and magnetic field from a central source (e.g., the well-known pulsar in the case of the Crab Nebula). The radio spectra of these objects, with $-0.3 \lesssim \alpha \lesssim 0$, are significantly flatter than those from shell SNRs. Although the Crab Nebula has a spectral break at $10^4$ GHz, the positions of the spectral breaks in the other filled-center remnants are either ill-defined or at much lower frequencies (tens to hundreds of GHz, see Green & Scheuer 1992). The *Planck* data from LFI and HFI will provide integrated flux densities for several Galactic filled-center remnants in order to define their spectra, and constrain the acceleration and loss mechanisms at work.

An important goal is to discover the plerions in our Galaxy, as well as the brightest in the Magellanic Clouds. Plerionic supernova remnants are nebular sources with an amorphous structure. At radio wavelengths they are characterized by strong non-thermal emission with a flat spectrum. In X-rays, their emission is still non-thermal, with typical spectral indices of about $-1$. From the Galactic distribution of the known plerions one may estimate that about 30 yet unidentified plerions are in our Galaxy, with angular sizes in the range $\sim 0.\!'1\text{--}1'$ and radio fluxes in the range $\sim 0.1\text{--}30$ Jy.

It is possible to take a completely new approach to discriminate between plerions and compact H II regions. In H II regions at $\sim 100$ GHz there is a spectral transition from the (flat) free-free component to the (steep) dust component. In plerions, the radio synchrotron emission is intrinsically much stronger (typically by a factor $\sim 10^3\text{--}10^5$) than the free-free emission from the thermal plasma. Therefore the associated dust component (if any) appears only at much higher frequencies. From the *Planck* survey one can extract spectra up to 850 GHz for all known flat-spectrum radio sources in the Galaxy as well as in the Magellanic Clouds. Those without a spectral steepening near 100 GHz are very likely to be plerions. Increasing by about a factor of three the number of known plerions will help to establish the typical properties of plerions, to test the homogeneity of this class of objects (see Woltjer et al. 1997), as well as to give a better estimate of the iron-core collapse supernova rate.

## 5.3 THE LARGE-SCALE STRUCTURE OF THE MILKY WAY

Although the Galactic plane is crowded by the superposition of interstellar clouds and star formation regions, it is optically thin at all *Planck* wavelengths (Table 5.4). Thus Planck will probe the whole Galactic plane from hundreds of parsecs to a few tens of kiloparsecs. By using velocity information from H I (hyperfine structure), CO (rotational transitions), and H II (recombination lines) emission, the three-dimensional distribution of these components can in principle be recovered. For this purpose it is reasonable to assume that the 3D emissivity of the interstellar medium in each *Planck* band is a linear combination of the three gas components. The linear coefficients can then be retrieved by a global minimisation of this combination on the reprojected Galactic plane maps (Giard et al. 1994). This analysis will yield the local emission spectrum ($W/H_{atom}$) of each ISM phase across the entire Galaxy. Such local spectral emissivities can be further separated in terms of physical components (dust, free-free, and synchrotron). The final products of this complex analysis are the large-scale physical Galactic properties in terms of ISM luminosities, masses, and composition (dust and free-free), star formation activity (dust and free-free), and high energy particle sources and confinement (synchrotron).

TABLE 5.4

Maximum Optical Depths Toward the Galactic Plane

| Component | $\nu$ [GHz] 850 | 100 | 30 |
|---|---|---|---|
| Dust ($A_V = 20$) . . . . . . . . . . . . . | $5 \times 10^{-3}$ | $2 \times 10^{-4}$ | $3 \times 10^{-5}$ |
| Free-free . . . . . . . . . . . . . . . . . . . . | $1 \times 10^{-7}$ | $1 \times 10^{-5}$ | $1 \times 10^{-4}$ |
| Synchrotron . . . . . . . . . . . . . . . . . | | $< 1 \times 10^{-9}$ | |

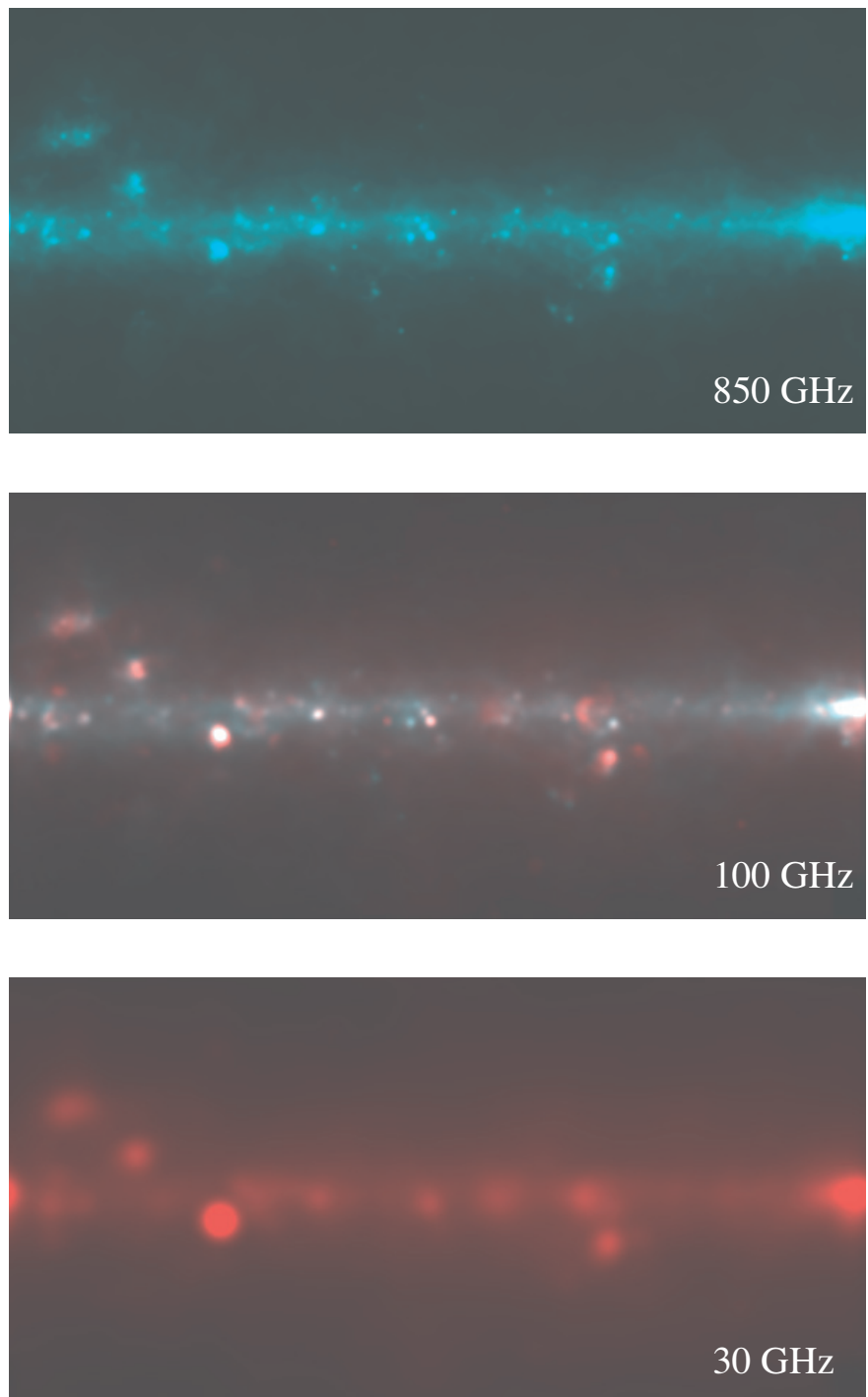

FIG 5.6.— Simulated maps of the dust and free-free emission in *Planck* bands at 850, 100, and 30 GHz. This is a $20° \times 10°$ map centered at $(l, b) = (10°, 0°)$. The free-free contribution (red) is derived from the Bonn 2.7 GHz survey (Reich et al. 1997); the dust contribution (blue) is derived from IRAS $100 \, \mu m$ data. A single dust temperature ($T = 25 \, K$, $n = 1.5$) and a single radio power law ($\alpha = -0.16$) were assumed. The angular resolutions used are respectively $5'$, $9'$, and $30'$.

### 5.3.1  Large-scale Structure of Dust

As in local ISM studies (§ 5.2.1.2), the dust emission in each phase of the interstellar medium can be used to derive dust temperatures and spectral indices throughout the Galactic plane. Meaningful dust parameters can be derived if we complement the *Planck* data with far-infrared surveys (Table 5.1). The correlation between the temperature and the spectral index of the dust can be systematically studied and related to the physical conditions of the mapped regions. The thermal grain emission can also be compared with the emission spectra of PAHs and very small grain infrared emission (from ISO and Spitzer data when available, and from the IRAS 12 and 60 $\mu$m bands) to study the interactions and transitions between the three main dust components in the different phases of grain evolution.

As mentioned previously, dust masses derived from the *Planck* data will be more accurate then previous estimates because *Planck* is less sensitive to the dust temperature than far-infrared experiments (Table 5.1). This should help to resolve the controversy about dust-to-gas gradients across the Galaxy. Currently some groups claim a constant dust-to-gas mass ratio, based on the fact that the observed radial variation in the far-infrared emissivity is consistent with the exponential decrease of the interstellar radiation field which naturally follows the starlight distribution traced by the near infrared surveys (Perault et al. 1989, Giard et al. 1994). Other groups interpret part of this effect as a variation of the dust-to-gas mass ratio from the inner to the outer Galaxy, due to an increase of the star formation activity toward the central regions (Sodroski et al. 1987, 1997). However, the fundamental limitation of all these studies today is that they rely on observations at wavelengths shorter than 240 $\mu$m, where a reliable estimate of both the temperature and mass of the dust is not possible. With the *Planck* data this limitation will be removed.

### 5.3.2 The Ionized ISM

The 3-D distribution of the ionized component of the Galaxy can be derived by combining H$\alpha$ and *Planck* LFI data with kinematic data from radio recombination line surveys. H$\alpha$ observations with high sensitivity are available from WHAM and the Southern Hydrogen Alpha Sky Survey (SHASSA), although they need to be corrected for Galactic dust absorption (Dickinson, Davies, and Davis 2002). Free-free radio emission provides information on the ionized gas along the Galactic plane where dust absorption makes H$\alpha$ surveys blind. The full-scale radio recombination line survey of the Galactic plane in progress using the HIPASS and HIJASS projects at Parkes and Jodrell Bank will give kinematic distances for the free-free emission and will provide valuable input into determining the three-dimensional interstellar radiation field of the Galaxy.

The LFI channels of *Planck* will provide a sensitive and homogeneous survey of the free-free radio emission of the H{\sc ii} regions (associated with massive star formation) in the Galactic disc, including both the compact regions (tracing the younger regions) and the more extended, low density regions (tracing the older sites of star formation). Extensive surveys of compact and extended H{\sc ii} regions are currently limited both by partial spatial coverage and low sensitivity (i.e., Caswell & Haynes 1987; Altenhoff et al. 1970; Felli & Churchwell 1972). The *Planck* data will extend and complement the existing ground-based surveys, leading to a complete picture of both present and past star formation in the Galaxy. The identification of compact and extended H{\sc ii} regions, despite the high confusion level expected in the Galactic plane, will be done by comparing the *Planck* data with ground-based data at millimetre and centimetre wavelengths (following the methods of Paladini et al. 2002) or H$\alpha$ (Reynolds et al. 1999) and radio recombination line surveys (Downes et al. 1980; Caswell & Haynes, 1987; Lockman et al. 1989). The resulting homogeneous database will provide a reliable tool to study the global statistical properties of H{\sc ii} regions, with particular emphasis on their relationship to the dust clouds detected in the *Planck* high frequency channels.

In addition to the ionised gas in H{\sc ii} regions, it should be possible to recover the ionised gas from the low density diffuse medium on large scales within the Galactic plane. The pulsar dispersion measure data in and near the Galactic plane are interpreted as indicating both a narrow and a broad component of the ionised gas distribution. The *Planck*-derived free-free map should provide an independent confirmation or development of the pulsar model, as the dispersion measure depends on the average electron density, while the free-free emission gives the mean-square density. Furthermore, the higher frequency data from *Planck* will allow corrections for the free-free optical depth effects encountered at lower frequencies and radio recombination line data will provide distances.

### 5.3.3 Star Formation Activity

Free-free emission is proportional to the number of ionising photons only, whereas the dust is heated both by ionising and non-ionising photons. The dust to free-free luminosity ratio (infrared excess) of star formation regions can thus be used as an indicator of the number of high mass stars in the complex. *Planck* will enable the study of the dependence of the infrared excess with location in the Galaxy, particularly the arm to inter-arm contrast, and *Planck* can address the question of how the initial mass function can change with the distance to the Galactic centre.

The initial mass function should be closely linked to the distribution of the Giant Molecular Clouds (GMCs), where star formation takes place. This relationship can be investigated, since the statistical properties of GMCs as a function of Galactic position can be determined from the separated three-dimensional components, and in particular with respect to the large-scale spiral structure. The large-scale spatial correlation of dense and diffuse media should allow elucidation of the survival timescales for GMCs in the spiral field and therefore understanding of the star formation processes on a global scale.

Individual star-forming complexes as described in § 5.2.2.1 and 5.2.2.2 are an important focus for the multi-component data from *Planck* supported by $H_I$, CO, and X-ray data. The interaction between $H_{II}$ regions, supernova remnants, and adjacent gas and dust clouds can be mapped, as for example the Cygnus X, $\eta$Carinae, and Vela regions which contain giant $H_{II}$ regions, synchrotron sources, and gas/dust clouds. There will be important input from ground-based observations, particularly high resolution radio continuum and spectral line data. Figure 5.2 shows a false color image of the Cygnus star forming complex obtained from the Canadian Galactic Plane Survey (CGPS), where the $H_I$ (green), IRAS 25 and 60 $\mu$m (blue and turquoise), and synchrotron data (rose) have been superposed. Individual SNRs and $H_{II}$ regions are clearly identified, kinematic data are also available.

The study of the formation of massive stars is difficult due to the fast evolutionary timescales of these objects and the fact that they occur less frequently than lower mass stars (Stahler, Palla, & Ho 2000). One of the most challenging problems is to select complete samples in order to statistically study their Galactic distribution, luminosity function, and physical properties. As in the case of low-mass objects, many investigations of massive star formation have been based on the IRAS PSC (e.g., Wood & Churchwell 1989). IRAS fluxes are sensitive to warm dust emission ($T > 30\,\mathrm{K}$), thus the PSC mostly contains warm and massive starless cores or already formed stars still surrounded by a conspicuous amount of circumstellar matter. In the Galactic plane, especially at the long wavelengths, the PSC is strongly affected by confusion and incompleteness, and the sample of young massive stars is thus far from being complete (Becker et al. 1994; Felli et al. 2000). Global Galactic properties, such as the star formation rate as a function of Galactocentric radius and the luminosity function of compact $H_{II}$ regions, currently derived from the IRAS database, may thus be seriously affected by incompleteness (Ivezic & Elitzur 2000; Comeron & Torra 1996). There are only very few candidates known for accreting massive protostars (e.g., Hunter et al. 2000); they are characterized by strong (sub)millimetre emission peaks with no counterpart in the radio continuum, possibly caused by rapid accretion of infalling matter that quenches the $H_{II}$ region very close to the protostar.

At the high frequencies of HFI, the *Planck* early release compact source catalogue (ERCSC) will have angular resolution similar to the IRAS PSC but will have sensitivity higher by a few orders of magnitude. This will allow us to select complete samples of high mass star forming regions, and to investigate the (sub)millimetre properties of compact $H_{II}$ regions throughout the Galactic disc. It will be possible to detect all high mass star formation regions in most of the *Planck* channels with minimal loss due to confusion. The primary goal will be to produce a reliable catalogue of high mass star forming regions and to examine the Galactic distribution and luminosity function of the sources following a methodology similar to the analysis of Ivezic & Elitzur (2000) and Comeron & Torra (1996). Young massive stars still embedded in parental molecular clouds have a characteristic spectral signature at (sub)millimetre wavelengths where the relative contribution of ionized gas and dust emission is a strong function of wavelength. It is expected that at least a few thousand embedded OB young stars will be detected, of which ∼50% will be new identifications and candidate massive protostars. At *Planck* sensitivities, all embedded O stars in the Galaxy will be detected above the confusion limit.

### 5.3.4 Synchrotron Emission and the Distribution of High Energy Particles

Synchrotron emission arises when electrons spiral along magnetic field lines at relativistic speeds. By nature this emission is directly coupled to the magnetic field, highly polarised, and intimately related to the origin of high energy particles (cosmic rays). Synchrotron surveys at GHz frequencies are strongly affected by Faraday rotation, which scrambles the polarization orientation, eliminating information about the magnetic field orientation in the emission region. Moreover, existing ground-based surveys are either incomplete, confined to low Galactic latitudes, have low angular resolution (e.g., $> 2°$ for the survey of Brouw & Spoelstra, 1979), or are poorly sampled. *Planck* LFI observations therefore provide a unique opportunity to measure the Galactic synchrotron emission over the entire range of latitudes at frequencies high enough that Faraday rotation is negligible (Beck, 2001).

Multi-frequency observations from the ground are still required to provide the total-power synchrotron template and to provide the low-frequency polarization data for Faraday rotation analysis. They should include 5 GHz to minimise atmospheric problems and maximise the observed signal. Existing 25–32 m dishes, under-illuminated to minimise ground pick-up, are suitable for this type of investigation. They need to be on good sites to reduce atmospheric fluctuations, especially at 5 GHz and above. The Bonn Mid-Latitude Survey at 1.4 GHz is planned to cover the northern sky at a resolution of 10′ for $-20° < b < +20°$, $240° > l > 350°$. The Canadian 1.4 GHz Galactic Plane Survey (1′ resolution) will be extended beyond the current region $74° \leq l \leq 147°$, $-3°\!.6 \leq b \leq +5°\!.6$, and a full survey of the Galactic plane with the Very Large Array (VLA) and the Australia Telescope Compact Array (ATCA) will be performed.

Near the Galactic plane, the remnants of old Type I and Type II supernovae produce diffuse synchrotron emission reflecting the relativistic electron and magnetic field distributions. The three-dimensional distribution will be related to the containment time in the Galactic magnetic field and constrained by the field geometry, which should be deducible from the synchrotron polarisation data. The synchrotron disc and halo distributions should be identifiable.

The distinctive synchrotron loops and spurs extending several tens of degrees from the plane will provide important clues about the origin and evolution of the synchrotron disc and halo. Many such features are old supernova remnants whose morphologies merge with those of younger, more compact remnants; however, the origin of the larger loops and spurs is still unclear.

Association with HI, Hα, and X-ray emission will help to clarify the origins of the synchrotron sky. The *Planck* low-frequency channels will provide important spectral information from which ages may be determined. Polarisation data from these features will give an indication of magnetic structures in localized regions within and adjacent to the Galactic plane.

Collisions at relativistic speeds between the high energy particles and other nuclei produce γ-rays. The probability of such collisions is proportional to the density of the ISM. Therefore, a key in the analysis of synchrotron emission will be γ-ray data from INTEGRAL. The diffuse emission of the 511 keV electron-positron annihilation line and the nuclear de-excitation lines of $^{26}$Al are detectable with INTEGRAL. Although the spatial resolution of INTEGRAL is lower than that of *Planck* (60′), the γ-ray data can be used in conjunction with the synchrotron data to derive the magnetic field strength. Finally, the diffuse γ-ray continuum from the Compton Gamma Ray Observatory (CGRO) and possibly INTEGRAL can be used in a multi-component three-dimensional decomposition in order to calibrate the ISM masses.

## 5.4 MAGNETIC FIELDS

### 5.4.1 Polarisation

*Planck* all-sky, multifrequency, polarisation observations will be of great utility for the determination of magnetic field levels and directions, both within nearby objects and at Galactic scales. *Planck*'s unique strength is its ability to probe the same underlying magnetic field simultaneously via its effect on synchrotron and thermal dust emission. Synchrotron emission (see also § 5.3.4) results from the interaction of cosmic ray electrons with a magnetic field and is strongly polarised and sensitive to both the magnitude and direction of the magnetic field. Thermal dust emission can be polarised depending on the properties of the grains and their level of alignment to the magnetic field.

*WMAP* has mapped synchrotron emission over the whole sky up to 90 GHz. In contrast, polarised dust emission has been measured only in limited areas of the Galactic plane (Boomerang, Archeops).

### 5.4.1.1 Polarised synchrotron emission

The intensity of synchrotron radiation is proportional to the density of relativistic electrons in the relevant energy range and (approximately) to $B_\perp^2$, where $B_\perp$ is the plane-of-the sky component of the magnetic field. Assuming energy equipartition, the synchrotron intensity provides a measure of $B_\perp$ and the cosmic ray energy density. Synchrotron radiation is intrinsically highly polarized, up to 70-75% in a completely regular field (e.g., Rybicki & Lightman 1979). The observed synchrotron polarisation depends on the uniformity of the field orientation within the resolution element, and is (under reasonable assumptions) proportional to $B_{reg}^2/(B_{reg}^2 + B_{turb}^2)$, where $B_{turb}$ represents the random (turbulent) component of the field and $B_{reg}$ the regular component in the volume observed. The angle of synchrotron polarisation is perpendicular to $B_\perp$. The minimum synchrotron sky temperature at 30 GHz is $\sim 50\,\mu K$, with typical temperatures of $\sim 100\,\mu K$ at intermediate latitudes. The maximum percentage polarisation at 30 GHz can be close to the theoretical maximum, about 70%. The highest polarized signals can therefore be expected to be of order $35\,\mu K$, depending on the feature observed. Higher values may well found in the loops and spurs (i.e., old supernova remnants close to the plane). At high latitudes away from such local features, the expected degree of polarisation should be only a few percent, i.e., resulting in polarized signals of a few microkelvin. The polarisation in the plane will lie somewhere between these two extremes. The sensitivity of the 30 and 44 Ghz channels of *Planck* is of approximately $10\mu K$ per resolution element, and therefore it can be expected that *Planck* will yield maps of the polarised synchrotron emission with high signal-to-noise ratio over most of the Galactic ISM regions.

### 5.4.1.2 Polarised dust emission

Polarisation of dust emission results from the presence of grains having anisotropic optical properties (e.g., elongated or graphitic) and a preferred orientation with respect to the ambient magnetic field. Several alignment mechanisms have been proposed, but their efficiencies are difficult to quantify and are expected to depend on the environment (e.g., Lazarian et al. 1997). For instance, elongated grains are expected to spin with their long axes perpendicular to the magnetic field. In this case, the direction of polarisation in emission is perpendicular to the sky component $B_\perp$. The degree of polarisation depends on the detailed dust properties, the efficiency of the alignment mechanism, and the uniformity of the magnetic field within the beam. Polarisation of a few percent, up to a maximum of $\sim 9\%$, has been measured at $100\,\mu m$ (Hildebrand 1996).

Stellar extinction data show that diffuse regions in the Galaxy are polarised also. Recently, Archeops (Benoit et al. 2004) detected polarised emission at 353 GHz associated with diffuse regions in the Galaxy (Benoit et al. 2004). The diffuse emission of the Galactic plane in the observed longitude range is polarised at a level of 4-5% with an orientation perpendicular to the plane. Several degree-sized clouds have been observed with much higher levels of polarisation (up to 20%), suggesting a very efficient dust orientation mechanism. Within the Draine and Lee (1984) dust model, where polarisation is associated with silicate grains, the mean polarisation of the far-IR/sub-mm dust emission is estimated to be 5% (Hildebrand and Dragovan 1995).

## 5.4.2 The Galactic Magnetic Field

Galactic magnetic fields are strong enough to influence the dynamics of gas in present day galaxies, and may have played an important role in their formation and early evolution. The large-scale structure of the magnetic field in spiral galaxies is generally thought to be generated by dynamo action on a pre-existing field, but the details of the mechanism and the nature of the seed field are poorly understood. Present knowledge of the structure of the magnetic field on Galactic scales comes mainly from observations of the synchrotron emission from nearby galaxies at frequencies between 1.4 and 5 GHz. In face-on spiral galaxies, the magnetic field is observed to follow the spiral structure with a small pitch angle. The potential information on the vertical structure of the magnetic field is to a large extent erased in edge-on galaxies by Faraday rotation and the line-of-sight averaging of the magnetic field orientation. These effects must also be considered in our own Galaxy. Observations of the Milky Way and external galaxies are therefore complementary: the former provide information on much smaller spatial scales, but are complicated by our position in the plane; the latter give clearer indications of global structure, but are limited in resolution.

A starting point for the construction of a 3-D model of the Galactic magnetic field is provided by Beuermann et al. (1985), who used the 408 MHz Haslam survey to derive a map of the three dimensional structure of synchrotron emissivity. The top view of their model is shown in Figure 5.7. The model accounts for the existence of radio spiral structure, both in the thick non-thermal radio disk and in the thin disk, assuming that the uniform component of the magnetic field ($B_{reg}$) is oriented along the spiral arms, and that the turbulent component ($B_{turb}$) has the same intensity as the uniform component.

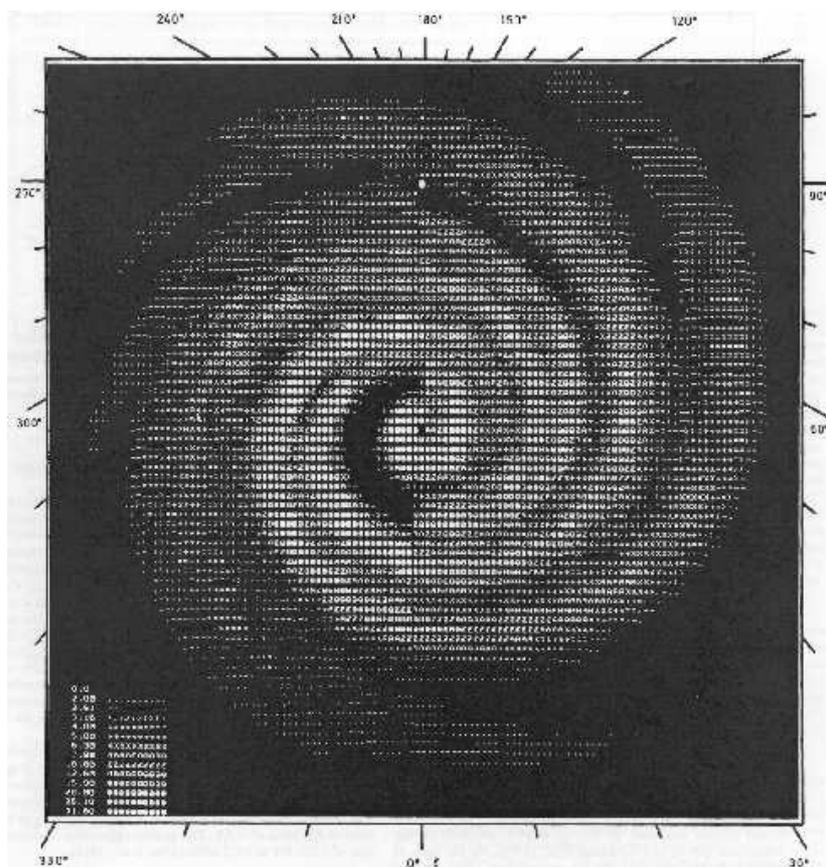

FIG 5.7.— Top view of the 3-D model of synchrotron emissivity derived from the Haslam map by Beuermann et al. (1985). The model contains assumptions on the Galactic magnetic field orientation and degree of alignment along the spiral arm, and therefore can be used to synthesize a map of synchrotron polarisation around the Sun (filled circle).

Rotation measures of discrete extra-galactic sources provide additional information about the Galactic magnetic field, which needs to be collated and incorporated in the magnetic field models. Pulsars, which lie mainly in the Galactic plane, are strongly linearly polarized and can be used to determine the RM and dispersion measure, hence giving information on the line-of-sight component of the magnetic field $B$ and $n_e$ between the pulsar and observer. Extragalactic radio sources provide information at all latitudes, though they have their own intrinsic rotation measures added to those of the Galaxy. In the Local Arm, the magnetic field is directed towards the Sun, while between $l = 0°$ and $l = 45°$ in the Sagittarius arm the field is directed away from the Sun. The field in the outer Perseus arm appears to be in the same direction as the Local Arm (e.g., Simard-Normandin and Kronberg 1980).

Based on these observations, the emerging picture of the Galactic field is that of a bisymmetric spiral with a dynamo mode of odd symmetry contributing to the magnetic field inside the solar circle (e.g., Han et al. 1997).

Further detailed information on the configuration of the large-scale Galactic field comes from studies of the polarisation of starlight in the solar neighborhood ($< 4\,\mathrm{kpc}$). Dust grains are aligned in the Galactic magnetic field, such that starlight is preferentially absorbed along one axis, and is thus left with a small residual linear polarisation parallel to the magnetic field. Mathewson & Ford (1971) presented the first extensive survey of starlight polarisation obtained from 1800 stars. Figure 5.8 shows the polarisation maps derived by Fosalba et al. (2001) from a compilation of 5500 stars. Since the stellar distances can be well-determined, a 3-D picture of the orientation of the magnetic field in the solar neighborhood can be built up. The polarisation at intermediate Galactic latitudes is observed to vary smoothly, with a well-aligned field direction in the Galactic plane.

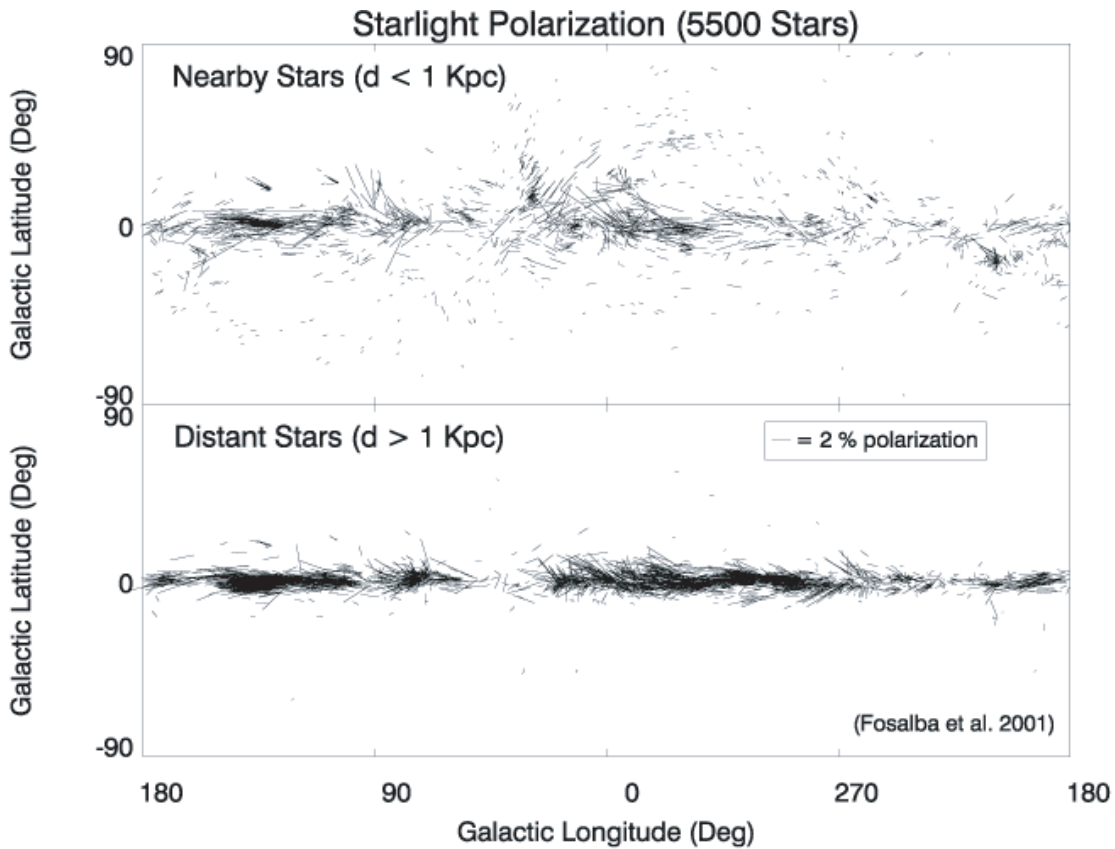

FIG 5.8.— Polarisations derived from 5500 stars, $< 1\,\mathrm{kpc}$ upper, and $> 1\,\mathrm{kpc}$ lower (from Fosalba et al. 2001).

This picture of the large-scale structure of the magnetic field of the Milky Way is clearly incomplete. *WMAP* should soon improve the picture considerably, based on its synchrotron polarisation maps. However, the next quantum step will certainly come from the *Planck* data, with its high sensitivity to polarisation and broad frequency coverage. An important goal for *Planck* is to achieve a robust determination of the large-scale structure of the Galactic magnetic field, using both synchrotron and dust polarised emission, as well as other inputs such as the extensive current data base of pulsar dispersion and rotation measures, extragalactic radio source RMs and the polarisation of individual stars produced by interstellar dust. At the very least, the *Planck* observations of the microwave sky will allow an accurate determination of the orientation of the large scale magnetic field, $B_{\text{reg}}$, and of the relative contribution of $B_{\text{reg}}$ and $B_{\text{turb}}$.

### 5.4.3 Magnetic fields in local objects

The large-scale Galactic magnetic field becomes frozen in the ISM wherever the ionization fraction is significant (i.e., a few $\times 10^{-4}$ for the diffuse ISM). As a consequence, turbulent or gravity-driven motions of the local gas distort the magnetic field lines and generate Alfvén waves, which in turn act on the gas. The importance of this coupling for the dynamical evolution of the ISM depends on the ratio between turbulent and magnetic pressure, which is likely to vary from place to place. In shielded regions, including proto-stellar condensations, the ionization fraction is believed to drop significantly so that the magnetic field decouples from matter; this process may play a key role in regulating the star formation efficiency in the Galaxy.

Several kinds of observations suggest that interstellar clouds may be threaded by local helical magnetic fields (Carlqvist et al. 1998). Helical fields may either support or help confine the gas depending on the relative strength of the poloidal and toroidal components. It would be a major step in the understanding of the evolution of clouds to determine which component dominates where. Carlqvist and Kristen (1997) have computed the kind of polarisation expected from such filamentary clouds threaded by helical fields. One key advantage of the *Planck* all-sky survey will be to allow the separation of the small scale structure and background/foreground contribution to the observed polarisation. Over small regions of the sky it will be possible to complement *Planck* data with high-resolution data from ground-based mm/sub-mm single dish and interferometers (e.g., IRAM and ALMA).

In addition, the *Planck* data will allow statistical studies of the local magnetic field direction and intensity. Chandrasekhar and Fermi (1953) proposed a means to estimate the plane of the sky component of the magnetic field, $B_\perp$, by combining the fluctuations of the gas velocity and that of the magnetic field direction. This method assumes that the Alfven waves propagating along the field are in the linear regime and that the gas motions are isotropic. Under these assumptions, a simple relationship exists between $B_\perp$, the rms line-of-sight velocity, and the dispersion of the polarisation angle. Linking the fluctuations of the polarisation angle with those of the line-of-sight velocity of the gas, on a statistical basis, should provide good estimates of $B_\perp$ in HI and molecular clouds and would complement the determinations of the line-of-sight component of the magnetic field, $B_\parallel$, provided by Zeeman measurements.

## 5.5 SOLAR SYSTEM STUDIES

Except for the Zodiacal Light, comet trails, and passing comets, all known Solar System objects which can be detected by *Planck* are expected to be unresolved. *Planck*'s wavelength range will emphasize the detection of cool objects in the Solar System such as asteroids. The large outer planets will be used as unresolved calibrators for astrometry and photometry at the same time. The wavelength coverage of *Planck* will provide valuable new data on the nature of the emission of these targets. The presence of Solar System objects in the data will be investigated in the routine data reduction pipeline and any object detected will be compared against a known catalogue of objects. Previously unknown objects will be flagged automatically, thus enabling rapid follow-up observations. The science goals and uses of the main classes of Solar System objects are summarized below.

### 5.5.1 Planets

*Planck* will measure the temperature of the planets with high photometric accuracy. This will be a unique opportunity to study the physical processes causing their emission in the *Planck* bands. The accurate determination of their submillimetre and microwave light curves will give valuable extra information regarding the physical processes.

### 5.5.2 Asteroids

*Planck* will measure the diffuse, long wavelength emission from the asteroid belts as well as individual asteroids, which will be correlated with optical and far-infrared catalogues and used to investigate light curves, diameters, albedos, and effective emissivities. The two populations of asteroids that *Planck* has the potential to detect are those located in the Main Belt and, perhaps, Near Earth Asteroids.

#### 5.5.2.1 Main belt asteroids

Most detections of minor bodies will be of main belt asteroids. Several hundred asteroids should be detected by *Planck* (Table 5.5).

TABLE 5.5

Number Of Detectable Main Belt Asteroids[a]

| $R/d$ | Differential Number | Cumulative Number[b] |
|---|---|---|
| $1$–$2 \times 10^{-7}$ . . . . . . . . . . . . | 299 | 397 |
| $2$–$3 \times 10^{-7}$ . . . . . . . . . . . . | 76 | 98 |
| $3$–$4 \times 10^{-7}$ . . . . . . . . . . . . | 15 | 22 |
| $4$–$5 \times 10^{-7}$ . . . . . . . . . . . . | 4 | 7 |
| $R/d > 5 \times 10^{-7}$ . . . . . . . . | 0 | 3 |

[a] The lower threshold considered here for asteroid detection is set to a radius to distance ratio $R/d \sim 10^{-7}$, derived by assuming a typical asteroid temperature of $\sim 150\,\mathrm{K}$ and by taking into account the *Planck* sensitivity at different channels (see Cremonese et al. 2003 for details).

[b] The cumulative number is computed at the lower limit of the range listed for the differential number.

#### 5.5.2.2 Near-Earth asteroids

The ability of *Planck* to detect near-Earth asteroids will depend on their relative motion with respect to the line-of-sight of the detectors. They are moving sufficiently quickly that they must be detected in the time ordered data as part of the *Planck* data reduction pipeline. Preliminary work indicates that it is unlikely that any currently known near-Earth asteroids will be detectable by *Planck* over the course of its mission; however, this does not preclude the discovery of new objects.

### 5.5.3 Comets

Molecular and dust emission from comets has been observed at *Planck* frequencies from both the Earth and space. For *Planck*, the main source of observable radiation from a comet is the continuum due to warm dust. This has been observed from the ground for some periodic and non-periodic comets; however, relatively few objects have been observed at *Planck* wavelengths. Estimates have been made of the expected brightness temperatures of the dust in the comas of a representative sample of comets, indicating that most of the short period comets will not be prominent objects in the *Planck* maps. However, since the physics involved are more complicated than that of asteroids, and given that they show a remarkable and largely unpredictable time variability, it is not straightforward to compile statistics on the numbers of observable objects.

The recent detection by the SWAS satellite of the water emission line at 557 GHz in comet C/1999 H1 (LEE) (Neufeld et al. 2000) suggests the possibility of similar detections by *Planck*. Despite the fact that water is the main volatile in most comets, this was the first detection of this line. The observed comet was not large and its activity, in terms of water production, was quite low. Although the conditions required for water excitation in comet comae are not yet completely understood, it is reasonable to expect that larger comets will generate a proportionally larger signal, and could therefore be detectable by *Planck*.

In spite of the low numbers of comets expected to be detected by *Planck*, a catalogue of potentially observable objects is being prepared. Software in the data processing pipeline will automatically check for such (moving) objects in the acquired data stream.

### 5.5.4 Cometary trails

Cometary Trails are due to the release of large (mm–cm) refractory particles from short period comets and were observed by both IRAS and COBE. Due to the small velocity required for the orbital injection from the cometary nucleus and the other sources of orbital perturbation, these particles have a limited transversal spread across the orbit but a significant longitudinal spread along the orbit of the parent comet. Trails therefore appear as long, narrow features in the sky, which may be up to a few arcminutes in width and tens of degrees in length. The successful observation of cometary trails by *Planck* will complement infrared observations and help to refine the models of grain composition and size distribution.

### 5.5.5 Zodiacal light

The zodiacal dust cloud consists of a population of micron to millimeter size particles sparsely distributed between the Sun and the orbit of the asteroid belt at 2 AU. Depending on the solar distance, these particles attain temperatures between 240 and 280 K, which implies a significant emission component at infrared wavelengths. Infrared and far-infrared space-borne missions like IRAS, COBE, and ISO have made an enormous contribution to our understanding of the complex geometry and dust properties of the zodiacal cloud. COBE observations with FIRAS indicate that the emissivity of the particles starts to fall off as $\lambda^{-2}$ (from $\lambda^0$) at wavelengths in excess of about 150 $\mu$m, suggesting a break in the size distribution: particles larger than about 30 $\mu$m radius are less abundant relative to the smaller ones (Fixsen & Dwek, 2001).

*Planck* should be able to detect zodiacal emission at 857 GHz and possibly 545 GHz. At those wavelengths the zodiacal emission is poorly constrained by observations and its level is uncertain by more than a factor of two. Model calculations indicate a peak surface brightness of about 0.5 MJy sr$^{-1}$ at 857 GHz and about a factor of 2 lower at 545 GHz (Maris et al. 2003). Accurate mapping of the zodiacal dust cloud by *Planck* will not only yield new information of the size distribution and composition, but also of the geometrical distribution of the largest grains in the cloud. The *Planck* satellite will view the zodiacal dust cloud through different lines of sight due to changes in the solar aspect angle and seasonal effects. This will give valuable additional information of the three-dimensional geometrical distribution of the largest grains in the cloud.

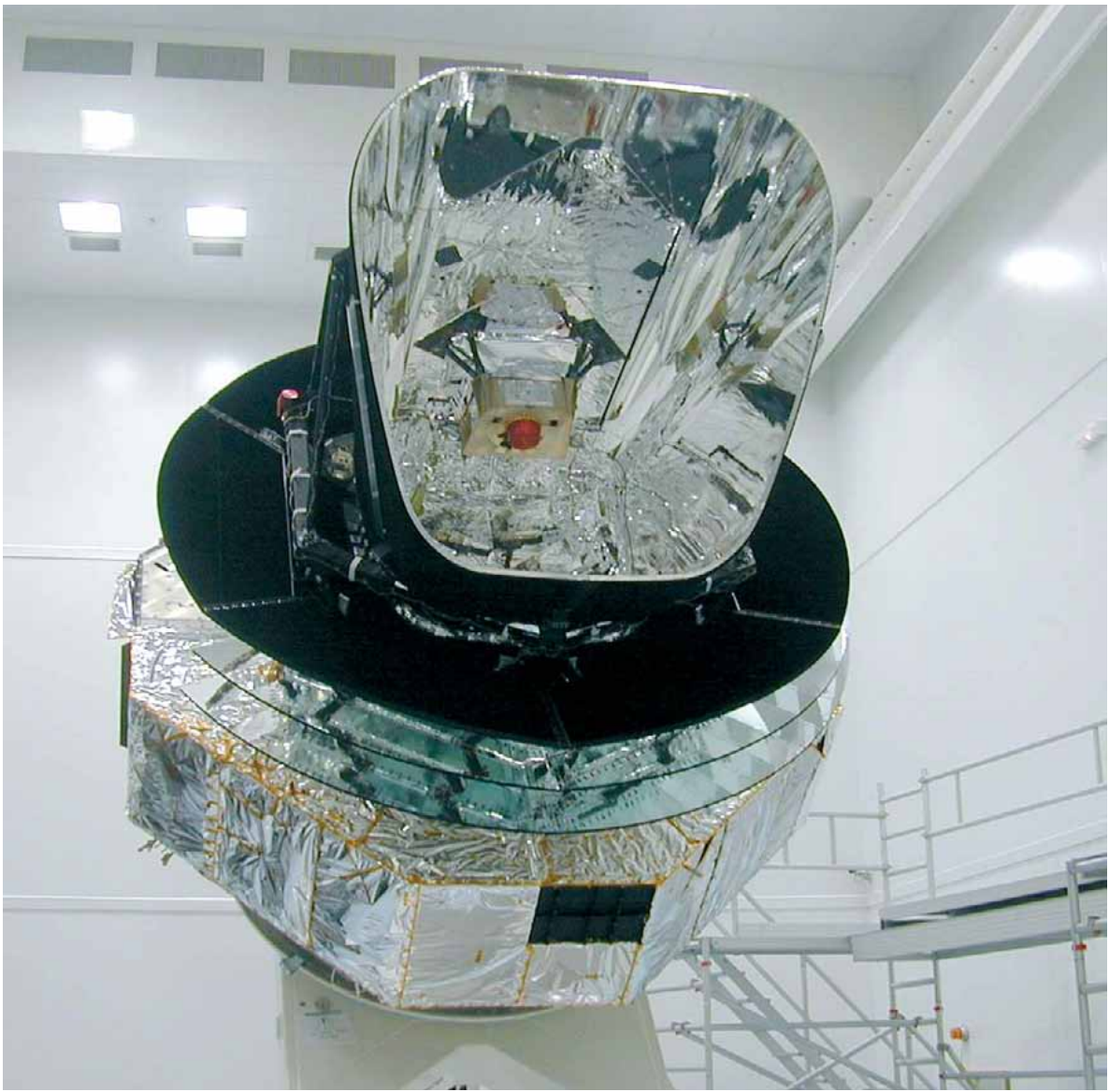

**European Space Agency**
*Agence spatiale européenne*